NATIONAL TECHNICAL UNIVERSITY OF ATHENS
SCHOOL OF NAVAL ARCHITECTURE AND MARINE ENGINEERING
SECTION OF NAVAL AND MARINE HYDRODYNAMICS

# Probabilistic responses of dynamical systems subjected to Gaussian coloured noise excitation. Foundations of a non-Markovian theory

*A thesis submitted for the degree of Philosophiae Doctor by*

**Konstantinos I. Mamis**

Supervisor professor: G. A. Athanassoulis

**Advisory committee**

| | |
|---|---|
| G. Athanassoulis | Professor, National Technical University of Athens (supervisor) |
| C. Papaodysseus | Professor, National Technical University of Athens |
| K. Spyrou | Professor, National Technical University of Athens |

**Examination committee**

| | |
|---|---|
| G. Athanassoulis | Professor, National Technical University of Athens |
| I. Kougioumtzoglou | Associate Professor, Columbia University |
| D. Koutsoyiannis | Professor, National Technical University of Athens |
| M. Loulakis | Associate Professor, National Technical University of Athens |
| C. Papaodysseus | Professor, National Technical University of Athens |
| Th. Sapsis | Associate Professor, Massachusetts Institute of Technology |
| K. Spyrou | Professor, National Technical University of Athens |

*From Saint Thomas Aquinas Commentary on the Metaphysics of Aristotle, book 1, lesson 3*

Constat autem, quod dubitatio et admiratio ex ignorantia provenit. Cum enim aliquos manifestos effectus videamus, quorum causa nos latet, eorum tunc causam admiramur. Et ex quo admiratio fuit causa inducens ad philosophiam, patet quod philosophus est aliqualiter *philomythes*, idest amator fabulae, quod proprium est poetarum. Unde primi, qui per modum quemdam fabularem de principiis rerum tractaverunt, dicti sunt poetae theologizantes, sicut fuit Perseus, et quidam alii, qui fuerunt septem sapientes. Causa autem, quare philosophus comparatur poetae, est ista, quia uterque circa miranda versatur. Nam fabulae, circa quas versantur poetae, ex quibusdam mirabilibus constituuntur. Ipsi etiam philosophi ex admiratione moti sunt ad philosophandum. Et quia admiratio ex ignorantia provenit, patet quod ad hoc moti sunt ad philosophandum ut ignorantiam effugarent. Et sic deinde patet, quod scientiam, *persecuti sunt*, idest studiose quaesierunt, solum ad cognoscendum, et non causa alicuius *usus* idest utilitatis.

Further, he [Aristotle] points out that perplexity and wonder arise from ignorance. For when we see certain obvious effects whose cause we do not know, we wonder about their cause. And since wonder was the motive which led men to philosophy, it is evident that the philosopher is, in a sense, a *philomyth*, i.e. a lover of myth, as is characteristic of the poets. Hence the first men to deal with the principles of things in a mythical way, such as Perseus and certain others, who were the seven sages, were called the theologizing poets. Now, the reason why the philosopher is compared to the poet is that both are concerned with wonders. For the myths with which the poets deal are composed of wonders, and the philosophers themselves were move to philosophize as a result of wonder. And since wonder stems from ignorance, they were obviously moved to philosophize in order to escape from ignorance. It is accordingly evident from this that *they pursued* knowledge, or diligently sought it, only for itself and not for any utility or *usefulness*. [translation by J. P. Rowan]

*Comment*: To our opinion, when the Philosopher and the Angelic Doctor use the term philosophers, they also include what today we call theoretical scientists.

## Acknowledgements

An exciting and quite literal adventure culminates in the present thesis. It is the last stage of a journey in three parts, the other two being my undergraduate studies and master's degree, all three conducted in my alma mater, the National Technical University of Athens. For this reason, and for the moral, academic, as well as financial support, I feel that my first thanks rightly belongs to our university and its people as a whole.

Especially, in the National Technical University of Athens, there is a place very dear to my heart; it is the School of Naval Architecture and Marine Engineering, which I have entered as an undergrad student, and where now I defend my PhD dissertation. Clearly, one could wonder on the relevance of my thesis topic to the title of our School. To this, I feel obliged to answer that it is the long-standing practice of curating theoretical engineering that makes our School to stand out, and to live up to the classical notion of scholarship. Needless to say, dealing with theoretical questions also results in findings that are of great practical significance for demanding real-life engineering applications. It is in such appliactions where innovations in *mathematical technology* are needed on par with breakthroughs in fields that are more commonly regarded as technical, such as computer engineering or material manufacturing.

A main figure in keeping the tradition of theoretical engineering in our School is professor G.A. Athanassoulis, whom I had the privilege to have as supervisor, guiding me along the way. Although thanking him here, in the preface of the work we performed together, seems fitting, it just does not feel enough. From the academic community, I would also like to thank my fellow PhD candidate Z.G. Kapelonis, since his work, on developing the numerical solution scheme for our pdf evolution equations and the Monte Carlo simulations, and my work on deriving these equations, complement each other.

The present, like all PhD theses, stands as a milestone of the time dedicated to this scientific endeavour. However, for the author, it also marks a period of his life of such length and importance that indeed deserves the description formative. For this, I would like to thank my parents, Ioannis and Irene, and all my friends and companions in this journey, for all the things we shared and cherished; for all the time well-spent.

This work has been supported by the ELKE-NTUA Scholarship programme.

Konstantinos I. Mamis

Athens, May 2020

## Synopsis

Determining the probabilistic structure of the response to a dynamical system under random excitations is an important question in various problems of stochastic dynamics, advanced statistical physics, and uncertainty quantification of macroscopic systems. For the special case of systems under white, i.e. delta-correlated, noise excitations, the answer is well-formulated; the evolution of the transition probability density function (pdf) of the response is governed by a partial differential equation (PDE), called the Fokker-Planck-Kolmogorov (FPK) equation. Furthermore, since in this case the response is Markovian, its complete probabilistic structure is defined by means of its transition pdf. Unfortunately, this convenient description, via a single PDE, is not applicable to problems where the random excitations are coloured, i.e. smoothly-correlated, noises, and thus the responses are non-Markovian. What is more, we have to emphasize that, in most real-life applications, random excitations cannot be modelled as delta-correlated in time, e.g. excitations by sea waves, wind or earthquakes, to name a few from engineering. Thus, for their responses, it holds true that "non-Markov is the rule, Markov is the exception" (van Kampen, Braz. J. Phys. 28 (1998) p. 90).

Since, in the case of coloured noise excitation, the response is non-Markovian, its complete probabilistic structure is described by an infinite hierarchy of joint response-excitation pdfs of all orders. The goal of this thesis is to present a methodology for deriving evolution equations for pdfs that are members of the aforementioned hierarchy, namely the one-time response pdf, the joint two-time response pdf, the joint response-initial value pdf, and the joint one-time response-excitation pdf. The dynamical systems considered herein are described by random differential equations (RDEs), which are ordinary differential equations, scalar or multidimensional, under coloured noise excitation. Moreover, their initial values can be random and correlated to excitation. In particular, initial value and excitation are considered as jointly Gaussian. In the introductory Chapter 1, apart from the formulation of the problem, a survey of the relevant literature is performed, and the main contributions of the present thesis are discussed.

In Chapter 2 of the thesis, we rederive the stochastic Liouville equations (SLEs) for the pdfs mentioned above, via the delta projection method, that is, by representing the pdfs as averages of random delta functions. Delta projection method, as presented herein, constitutes a more comprehensive and easily generalizable alternative to the well-known van Kampen's lemma. SLEs are exact, yet non-closed equations, since they contain averaged terms that are expressed via pdfs of higher-order.

In Chapter 3, Novikov-Furutsu (NF) theorem, which is a well-known mathematical tool, used herein for the further evaluation of the averaged terms in SLEs, is proven again in a systematic way, and generalized. NF theorem, in its classical form, evaluates further the mean value of the product of a functional with zero-mean Gaussian random argument, multiplied by the argument itself. First, NF theorem is extended to cases of mappings of two arguments, the one being a random variable and the other a random function, which are jointly Gaussian, and, in general, have non-zero mean values. This extension allows for the derivation of pdf evolution equations corresponding to dynamical systems whose random initial value is correlated to the excitation. Furthermore, NF theorem is also extended for averages in which the aforementioned mapping is multiplied by the temporal derivative of its functional argument. This form of NF theorem is useful in treating the averaged terms of the SLE for the one-time, response-excitation pdf. Last, a substantial generalization of the NF theorem is proven, that evaluates averages containing the functional argument at multiple time instances. This generalized NF theorem is, to the best of our knowledge, new, and contains, as a special case, the well-known Isserlis theorem for the moments of a Gaussian random function. By using the generalized NF theorem, we can evaluate the averaged terms appearing in SLEs that correspond to RDEs excited by a polynomial of Gaussian noise.

In Chapter 4, the final forms of SLEs are obtained, by applying the appropriate form of the NF theorem. We observe that, after the NF theorem, the averages in SLEs are expressed equivalently as nonlocal terms depending on the whole history of both the response and the excitation, a result that is in accordance with the non-Markovian character of the response. Additionally, in Chapter 4, we consider the special cases in which the said nonlocal character is absent from the SLEs, namely the nonlinear RDEs under white noise excitation, and the linear RDEs under coloured noise excitation. For the former case, the classical FPK equations are re-derived, while, for the latter, the derived pdf evolution equations are exactly solved, and their solutions are validated by results obtained from the respective moment (or cumulant) problem. These straightforward applications of the SLEs serve also as a first validation of our methodology so far.

Since SLEs are non-closed due to the presence of nonlocal terms, an approximation closure scheme has to be applied on them, in order to obtain approximate, yet closed pdf evolution equations. This is the topic of Chapter 5. After presenting the most widely-used closure schemes, we introduce a new one, which, in contrast to the rest, retains a tractable amount of the original nonlocality of the SLE, by utilizing the history of appropriate moments of the response (or joint moments of response and excitation in the case of response-excitation SLE). Application of this closure to the SLEs results in a family of novel pdf evolution equations. Furthermore, in Chapter 5, we examine separately the case of multiplicative (or parametric) noise excitation, meaning that, in RDE, excitation is multiplied by function of the response. For this case, the advantages of considering the evolution equation for the joint, one-time response-excitation pdf, instead of the equation for the one-time response pdf, are discussed.

In Chapter 6, the response pdf evolution equations under the novel closure are solved numerically and their results are compared to Monte Carlo (MC) simulations, for the benchmark case of a scalar bistable RDE under Ornstein-Uhlenbeck excitation. As shown in the figures of the Chapter, the novel evolution equations are in good agreement with the MC simulations, even for high noise intensities and large correlation times of the excitation, where the existing pdf evolution equations found in literature fail. What is more, the computational effort for solving the new pdf evolution equations is comparable to the effort required for solving the respective classical FPK equation with one state variable.

Last, in Chapter 7, a recapitulation is performed, and possible directions for future works are discussed.

## Σύνοψη (Synopsis in Greek)

Ο προσδιορισμός της πιθανοθεωρητικής δομής της απόκρισης δυναμικού συστήματος υπό τυχαία διέγερση είναι ένα σημαντικό ερώτημα για πολλά προβλήματα της στοχαστικής δυναμικής, της στατιστικής φυσικής, καθώς και της ποσοτικοποίησης της αβεβαιότητας σε μακροσκοπικά συστήματα. Στην ειδική περίπτωση συστημάτων υπό λευκό, δηλαδή δέλτα-συσχετισμένο θόρυβο, η απάντηση είναι συγκεκριμένη: η εξέλιξη της συνάρτησης πυκνότητας πιθανότητας (σππ) της μετάβασης της απόκρισης υπακούει σε μια μερική διαφορική εξίσωση (ΜΔΕ) που καλείται εξίσωση Fokker-Planck-Kolmogorov (FPK). Επιπλέον, καθώς η απόκριση σε αυτήν την περίπτωση είναι Μαρκοβιανή, όλη της η πιθανοθεωρητική δομή ορίζεται από την εν λογω σππ μετάβασης. Δυστυχώς, αυτή η βολική περιγραφή, μέσω μίας μόνο ΜΔΕ, δεν είναι δυνατή σε προβλήματα υπό διεγέρσεις ομαλώς συσχετισμένων, ή -όπως αλλιώς λέγονται- χρωματισμένων θορύβων. Επιπροσθέτως, θα πρέπει να υπογραμμίσουμε ότι, στις περισσότερες πραγματικές εφαρμογές, οι τυχαίες διεγέρσεις δεν μπορούν να μοντελοποιηθούν ως δέλτα-συσχετισμένες στο χρόνο, βλ. για παράδειγμα διεγέρσεις από θαλάσσιους κυματισμούς, τον άνεμο και τους σεισμούς. Έτσι, για τις αποκρίσεις τους ισχύει ότι «οι μη-Μαρκοβιανές είναι ο κανόνας, οι Μαρκοβιανές είναι η εξαίρεση» (van Kampen, Braz. J. Phys. 28 (1998) σ. 90).

Καθώς λοιπόν στην περίπτωση διέγερσης χρωματισμένου θορύβου η απόκριση είναι μη-Μαρκοβιανή, η συνολική της πιθανοθεωρητική δομή περιγράφεται από μια άπειρη ιεραρχία από κοινού σππ της απόκρισης και της διέγερσης όλων των τάξεων. Ο στόχος της παρούσας διατριβής είναι η παρουσίαση μιας μεθοδολογίας παραγωγής εξελικτικών εξισώσεων για σππ από την παραπάνω ιεραρχία· συγκεκριμένα για την σππ της απόκρισης σε ένα χρόνο, την από κοινού σππ της απόκρισης σε δύο χρόνους, την από κοινού σππ της απόκρισης με την αρχική της τιμή, και την από κοινού σππ απόκρισης και διέγερσης στον ίδιο χρόνο. Τα δυναμικά συστήματα που μελετώνται στο παρόν περιγράφονται από τυχαίες διαφορικές εξισώσεις (ΤΔΕ), οι οποίες και είναι συνήθεις διαφορικές εξισώσεις, βαθμωτές ή πολυδιάστατες, υπό διέγερση χρωματισμένου θορύβου. Επιπλέον, η αρχική τιμή της απόκρισης μπορεί να είναι επίσης τυχαία και συσχετισμένη με την διέγερση. Συγκεκριμένα, η αρχική τιμή και η διέγερση θεωρούνται ως από κοινού Γκαουσιανές. Στο εισαγωγικό Κεφάλαιο 1, εκτός από την διατύπωση του προβλήματος, παρουσιάζεται μια ανασκόπηση της σχετικής βιβλιογραφίας, και συζητούνται οι κύριες συνεισφορές της παρούσας διατριβής.

Στο Κεφάλαιο 2, εξάγουμε εκ νέου τις στοχαστικές εξισώσεις Liouville (ΣEL) για τις σππ που αναφέρουμε ανωτέρω, μέσω της μεθόδου των δέλτα-προβολών, δηλαδή μέσω της αναπαράστασης των σππ ως μέσες τιμές τυχαίων συναρτήσεων δέλτα. Η μέθοδος των δέλτα-προβολών, όπως παρουσιάζεται στο παρόν, αποτελεί μια πιο εμπεριστατωμένη και εύκολα γενικεύσιμη εναλλακτική στο περιώνυμο λήμμα του van Kampen. Οι ΣEL είναι ακριβείς, πλην όμως μη-κλειστές εξισώσεις, καθώς περιέχουν όρους μέσων τιμών που εκφράζονται συναρτήσει σππ ανωτέρων τάξεων.

Στο Κεφάλαιο 3, το θεώρημα Novikov-Furutsu (NF), το οποίο είναι ένα γνωστό μαθηματικό εργαλείο που εδώ χρησιμοποιείται για την περαιτέρω ανάλυση των προαναφερθεισών μέσων τιμών στις ΣEL, αποδεικνύεται εκ νέου με έναν συστηματικό τρόπο, και γενικεύεται. Το θεώρημα NF, στην κλασσική του μορφή, εφαρμόζεται σε μέσες τιμές γινομένων ενός συναρτησιακού ενός Γκαουσιανού τυχαίου ορίσματος με μηδενική μέση τιμή, το οποίο πολλαπλασιάζεται με το εν λόγω όρισμά του. Πρώτα, το θεώρημα NF επεκτείνεται σε περιπτώσεις απεικονίσεων δύο ορισμάτων, με το ένα να είναι τυχαία μεταβλητή και το άλλο τυχαία συνάρτηση, τα οποία είναι από κοινού Γκαουσιανά και έχουν, εν γένει, μη μηδενικές μέσες τιμές. Αυτή η επέκταση επιτρέπει την παραγωγή εξελικτικών εξισώσεων για σππ που αντιστοιχούν σε δυναμικά συστήματα στα οποία η αρχική τιμή της απόκρισης είναι συσχετισμένη με την διέγερση. Επιπρόσθετα, το θεώρημα NF επεκτείνεται για μέσες τιμές εντός των οποίων η προαναφερθείσα απεικόνιση πολλαπλασιάζεται με την χρονική παράγωγο του συναρτησιακού ορίσματός της. Αυτή η μορφή του θεωρήματος NF είναι χρήσιμη στον χειρισμό των μέσων τιμών που εμφανίζονται στην ΣΕL για την από κοινού σπп απόκρισης-διέγερσης. Τέλος, αποδεικνύεται μια ουσιώδης γενίκευση του θεωρήματος NF για μέσες τιμές που περιέχουν το συναρτησιακό όρισμα σε πολλαπλές χρονικές στιγμές. Αυτό το γενικευμένο θεώρημα NF είναι, καθόσον μπορούμε να γνωρίζουμε, καινούργιο, και περιέχει σαν ειδική περίπτωση το γνωστό θεώρημα Isserlis για τις ροπές Γκαουσιανής τυχαίας συνάρτησης. Το γενικευμένο θεώρημα NF επιτρέπει τον χειρισμό των μέσων τιμών που εμφανίζονται στις ΣΕL που αντιστοιχούν σε ΤΔΕ που διεγείρονται από ένα πολυώνυμο Γκαουσιανού θορύβου.

Στο Κεφάλαιο 4, εξάγονται οι τελικές μορφές των ΣΕL, εφαρμόζοντας την κατάλληλη κάθε φορά μορφή του θεωρήματος NF. Παρατηρούμε ότι, μετά το θεώρημα NF, οι μέσες τιμές στις ΣΕL εκφράζονται ισοδύναμα ως μη-τοπικοί όροι που εξαρτώνται από την όλη ιστορία της απόκρισης αλλά και της διέγερσης, ένα αποτέλεσμα που είναι σε συμφωνία με τον μη-Μαρκοβιανό χαρακτήρα της απόκρισης. Επιπροσθέτως, στο Κεφάλαιο 4, μελετούμε τις ειδικές περιπτώσεις κατά τις οποίες ο προαναφερθείς μη-τοπικός χαρακτήρας είναι απών από τις ΣΕL, και συγκεκριμένα για μη-γραμμικές ΤΔΕ υπό διέγερση λευκού θορύβου, καθώς και για γραμμικές ΤΔΕ υπό διέγερση χρωματισμένου θορύβου. Στην πρώτη περίπτωση ανακτάται εκ νέου η κλασσική εξίσωση FPK, ενώ στην δεύτερη οι παραγόμενες εξελικτικές εξίσωσεις για σπп επιλύονται αναλυτικά, και οι λύσεις τους επιβεβαιώνονται από τα αποτελέσματα του αντίστοιχου προβλήματος ροπών (ή σωρευτικών ροπών, ήτοι cumulants[1]). Αυτές οι άμεσες εφαρμογές των ΣΕL αποτελούν και την πρώτη επαλήθευση της μέχρις εδώ μεθοδολογίας μας.

Καθώς οι ΣΕL είναι μη-κλειστές λόγω της παρουσίας των μη-τοπικών όρων, θα πρέπει να εφαρμόσουμε σε αυτούς τους όρους ένα προσεγγιστικό σχήμα ώστε να παραγάγουμε προσεγγιστικές, πλην κλειστές εξελικτικές εξισώσεις για σπп. Αυτό το θέμα πραγματεύεται το Κεφάλαιο 5. Μετά τη παρουσίαση των πιο ευρέως χρησιμοποιούμενων σχημάτων, εισάγουμε ένα νέο σχήμα, το οποίο, σε αντίθεση με τα υπάρχοντα, διατηρεί ένα εύκολα διαχειρίσιμο ποσοστό της αρχικής μη-τοπικότητας των ΣΕL, χρησιμοποιώντας την ιστορία κατάλληλων ροπών της απόκρισης (ή από κοινού ροπών της απόκρισης και της διέγερσης στην περίπτωση της ΣΕL της απόκρισης-διέγερσης). Η εφαρμογή αυτού του σχήματος στις ΣΕL οδηγεί σε μια οικογένεια από νέες εξελικτικές εξισώσεις για σπп. Επιπλέον, στο Κεφάλαιο 5, μελετάται ξεχωριστά η περίπτωση πολλαπλασιαστικού (ή παραμετρικού) θορύβου, το οποίο σημαίνει ότι στην ΤΔΕ, η διέγερση πολλαπλασιάζεται από μια συνάρτηση της απόκρισης. Σε αυτή την περίπτωση, συζητούμε τα πλεονεκτήματα της θεώρησης της εξελικτικής εξίσωσης για την από κοινού σπп απόκρισης-διέγερσης στον ίδιο χρόνο, αντί της εξίσωσης για την σπп της απόκρισης σε ένα χρόνο.

[1] Καθώς ο όρος cumulant συνήθως παραμένει αμετάφραστος στην ελληνική βιβλιογραφία, προτείνουμε την απόδοσή του ως σωρευτική ροπή.

Στο Κεφάλαιο 6, οι νέες εξελικτικές εξισώσεις για την σππ της απόκρισης επιλύονται αριθμητικά και τα αποτελέσματά τους συγκρίνονται με προσομοιώσεις Monte Carlo (MC), για το παράδειγμα αναφοράς μιας δισταθούς ΤΔΕ υπό διέγερση Ornstein-Uhlenbeck. Όπως φαίνεται στα διαγράμματα του κεφαλαίου, οι νέες εξελικτικές εξισώσεις είναι σε καλή συμφωνία με τις προσομοιώσεις MC, ακόμα και για περιπτώσεις διέγερσης από θόρυβο ισχυρής έντασης και μεγάλου χρόνου συσχέτισης, στις οποίες οι εξελικτικές εξισώσεις της υπάρχουσας βιβλιογραφίας αποτυγχάνουν. Συν τοις άλλοις, το υπολογιστικό κόστος για την επίλυση των νέων εξελικτικών εξισώσεων είναι συγκρίσιμο με αυτό της επίλυσης της αντίστοιχης κλασσικής εξίσωσης FPK με μία μεταβλητή κατάστασης.

Τέλος, στο Κεφάλαιο 7, επιχειρείται μια ανακεφαλαίωση της παρούσας διατριβής, και συζητώνται πιθανές κατευθύνσεις για μελλοντικές εργασίες.

# Table of Contents

# Chapter 1: Introduction

## 1.1 Dynamical systems under random excitation

In the present thesis, we consider, as a prototype, the following one-dimensional, random initial-value problem (RIVP):

$$\dot{X}(t;\theta) = h(X(t;\theta)) + q(X(t;\theta))\,\Xi(t;\theta), \tag{1.1}$$

$$X(t_0;\theta) = X_0(\theta). \tag{1.2}$$

In random[2] differential equation (RDE) (1.1), the overdot denotes the differentiation with respect to time argument $t$, $\theta$ denotes the stochastic argument, and $\Xi(t;\theta)$, $X(t;\theta)$ are random functions, called the excitation and the response of the RDE respectively. Deterministic functions $h(x)$, $q(x)$ are considered differentiable[3] with respect to their argument, ensuring thus the pathwise, i.e. for every $\theta$ separately, solution of RIVP (1.1)-(1.2). For the special case where $q(x)$ is constant, $q(x)=\kappa$, the RDE (1.1) is termed as *additively or externally excited*, where in the general case where $q(x)$ is *x*-dependent, RDE (1.1) is under *multiplicative or parametric excitation*.

**Coloured noise excitation.** The distinguishing feature of our study is that excitation $\Xi(t;\theta)$ in RDE (1.1) is considered to be a *smoothly-correlated Gaussian noise*, i.e. its autocovariance function $C_{\Xi\Xi}(t,s)$ is jointly continuous with respect to its two temporal arguments. Such a random process is also term "coloured", in contrast to the delta-correlated white noise, see e.g. (Pugachev & Sinitsyn, 2001, para. 3.3.3). We have to note that, in most of the relevant works, see e.g. (Fox, 1986a; Hänggi & Jung, 1995; Ridolfi, D'Odorico, & Lalo, 2011, sec. 2.5) the term coloured noise is just a metonym for zero-mean valued *Ornstein-Uhlenbeck* (OU) noise, which is a Gaussian process with autocovariance

$$C_{\Xi\Xi}(t,s) = \frac{D}{\tau}\exp\left(-\frac{|t-s|}{\tau}\right), \tag{1.3}$$

where positive coefficients $D$, $\tau$ are the noise intensity and correlation time respectively. In the present thesis however, *the coloured noise is not specified a priori*. For useful generalizations of OU process, see e.g. (Lim & Muniandy, 2003). The importance of coloured noise ex-citation in many advanced and real-life applications, and the theoretical complicacies it induces, are discussed in many works, e.g. (Francescutto & Naito, 2004; Traverso, Vernazza, & Trucco, 2012) for *sea environment*, (Maranò, Edwards, Ferrari, & Fäh, 2017) for *earthquake loads*, (Suvire & Mercado, 2008) for *wind speed*, (Spanio, Hidalgo, & Muñoz, 2017) in *population dynamics*, (Wen et al., 2013) in *nuclear fusion*, (Costa-Filho et al., 2017; De Vega & Alonso, 2017) in *quantum dynamics*, to name a few.

[2] The term "random" is used herein, instead of "stochastic", to distinguish differential equations under coloured noise excitation from Itō stochastic differential equations, where the noise is assumed (usually tacitly) to be white. This terminological distinction has been also used in (Arnold, 1998).

[3] Differentiability of $h(x)$, $q(x)$ is also explicitly dictated by the need of calculating the derivatives of the response with respect to excitation and initial value, see Section 4.1.

**Correlated initial value.** Another feature taken into account in the present study is that the initial value of the response $X_0(\theta)$, in Eq. (1.2), can also be random, correlated to excitation $\Xi(t;\theta)$. In particular, we consider the initial value $X_0(\theta)$ and excitation $\Xi(t;\theta)$ to be *jointly Gaussian* random elements, with (in general, non-zero) mean values $m_{X_0}$, $m_\Xi(t)$, autocovariances $C_{X_0X_0}$, $C_{\Xi\Xi}(t,s)$, and cross-covariance $C_{X_0\Xi}(t)$. In most of the existing works, the initial value is assumed independent from the excitation, or just deterministic. However, for some applications, the importance of taking into account the probabilistic structure of the initial value is significant, and has indeed been considered before (Roerdink, 1981, 1982), since "the statistical properties of $\Xi(t;\theta)$ generally cannot be chosen to be independent of the form of initial probability $f_{X_0}(x)$" (Hänggi, 1989).

**The data and solution of RIVP (1.1)-(1.2).** Having stated the problem, Eqs. (1.1)-(1.2), we have to discuss *what it means to solve an RIVP*. Similarly to the deterministic initial-value problems, initial value $X_0(\theta)$ and excitation $\Xi(t;\theta)$ are considered as data of RIVP (1.1)-(1.2). Since $X_0(\theta)$, $\Xi(t;\theta)$ are random elements correlated with each other, they are fully described by their infinite-dimensional joint probability measure $\mathbf{P}_{X_0\,\Xi(\bullet)}$, which is assumed to possess a well-defined *characteristic functional*, see e.g. (Sobczyk, 1991) definition 1.6′:

$$\varphi_{X_0\Xi}[\upsilon\,;u(\bullet|_{t_0}^{t})] = \mathbb{E}^\theta\left[\exp\left(i\,X_0(\theta)\,\upsilon + i\int_{t_0}^{t}\Xi(s;\theta)\,u(s)\,ds\right)\right], \tag{1.4}$$

where $\mathbb{E}^\theta[\bullet]$ denotes the ensemble average operator with respect to stochastic argument $\theta$. The data measure $\mathbf{P}_{X_0\,\Xi(\bullet)}$ is considered defined over the Borel $\sigma-$algebra of the product space $\mathbb{R}\times\mathscr{Z}$, where $\mathscr{Z}$ is with the space of continuous functions $C\big([t_0,t]\to\mathbb{R}\big)$. The marginal data measure $\mathbf{P}_{\Xi(\bullet)}$ is assumed to have continuous mean value and autocovariance operators (and thus continuous two-point correlation functions), reflecting the coloured-noise character of excitation $\Xi(t;\theta)$.

On the other hand, the complete probabilistic solution of the RIVP (1.1)-(1.2) consists of the infinite-dimensional, joint response-excitation probability measure $\mathbf{P}_{X(\bullet)\,\Xi(\bullet)}$ which is assumed to exist, and also possess a joint characteristic functional. It is assumed that the solution measure $\mathbf{P}_{X(\bullet)\,\Xi(\bullet)}$ is defined over the Borel $\sigma-$algebra of the product space $\mathscr{X}\times\mathscr{Z}$, where $\mathscr{Z}$ is the same as above, and $\mathscr{X}$ coincides with the space $C^1\big([t_0,t]\to\mathbb{R}\big)$, in which the response path functions belong. In addition, the solution measure $\mathbf{P}_{X(\bullet)\,\Xi(\bullet)}$ must obey the compatibility condition that its marginal measure $\mathbf{P}_{X(t_0)\,\Xi(\bullet)}$ is equal to $\mathbf{P}_{X_0\,\Xi(\bullet)}$, i.e. the given data measure. By assuming also that the joint response-excitation probability measure is continuously dis-tributed, all joint response-excitation probability density functions (pdfs) of finite dimension exist, forming thus the infinite hierarchy (Skorokhod, 2005, sec. 2.4)

$$\left( f_{X(\tau_1)\cdots X(\tau_m)\Xi(s_1)\cdots\Xi(s_n)}(x_1,\dots,x_m,u_1,\dots,u_n)\right)_{n,m\in\mathbb{N}}, \quad s_1,\dots,s_n,\tau_1,\dots,\tau_m\in[t_0,t]. \quad (1.5)$$

Thus, having assumed that the underlying infinite-dimensional measure problem is well-defined, we shall focus, in the present thesis, on ***formulating equations governing the evolution of finite-dimensional pdfs*** from the above hierarchy.

**Other approaches of determining the probabilistic structure of the response.** Clearly, the formulation and solution of pdf evolution equations is not the sole answer to the question of quantifying the uncertainty of the response to an RIVP. A “coarser”, yet quite useful method is the formulation and solution of *evolution equations for moments* of the response. The main complicacy of this method is that, for nonlinear RDEs, the system of equations for response moments up to a certain order is non-closed, i.e. it also contains response moments of higher order. This calls for a closure scheme[4] that expresses the said higher-order moments in terms of moments of lower order, see e.g. (Roberts & Spanos, 2003, para. 9.3.2; Tsantili, 2013, Chapter 5). Another way of attacking the problem is by introducing simplifications for the RDE. In this category lies the extensively used and easily applicable method of *equivalent statistical linearization* of the nonlinear RDE (Crandall, 2006; Roberts & Spanos, 2003; West, Rovner, & Lindenberg, 1983). However, the linearization downplays the importance, as well as the mechanisms, of the effects of nonlinearities on the probability structure of the response. Last, another worth-mentioning, popular method is the use of particular representations for the random excitation and/or the response. These are the *Karhunen-Loève* (Ghanem & Spanos, 2003, Chapter 2; Pugachev & Sinitsyn, 2001, sec. 3.9) and the *polynomial chaos expansions* (Creamer, 2008; Ghanem & Spanos, 2003, loc. sit.). The above approaches, despite their significance, are not in the scope of the present thesis.

**Multidimensional random dynamical systems.** The methodology for determining response-excitation pdfs is also applied to the multidimensional counterpart of RIVP (1.1)-(1.2):

$$\dot{X}_n(t;\theta) = h_n\left(\boldsymbol{X}(t;\theta)\right) + \sum_{k=1}^{K} q_{nk}\left(\boldsymbol{X}(t;\theta)\right)\Xi_k(t;\theta), \quad (1.6)$$

$$X_n(t_0;\theta) = X_n^0(\theta), \qquad n=1,\dots,N. \quad (1.7)$$

In multidimensional RDE[5] (1.6), $\boldsymbol{h}(\boldsymbol{x})$ is a deterministic *N*-dimensional vector function, and $\boldsymbol{q}(\boldsymbol{x})$ is a deterministic $N\times K$ matrix function. All first-order partial derivatives of $\boldsymbol{h}(\boldsymbol{x})$, $\boldsymbol{q}(\boldsymbol{x})$ exist and are assumed continuous. Note that the dimension *N* of the response random vector $\boldsymbol{X}(t;\theta)$ does not have to agree with dimension $K$ of the excitation random vector $\boldsymbol{\Xi}(t;\theta)$. The initial value random vector $\boldsymbol{X}^0(\theta)$ and the excitation random vector function $\boldsymbol{\Xi}(t;\theta)$ are considered correlated and, more specifically, jointly Gaussian with (non-zero) mean vectors $\boldsymbol{m}_{\boldsymbol{X}^0}$ and $\boldsymbol{m}_{\boldsymbol{\Xi}}(t)$, autocovariances the $N\times N$ matrix $\boldsymbol{C}_{\boldsymbol{X}^0\boldsymbol{X}^0}$ and $K\times K$ matrix

[4] As discussed subsequently in par. 1.2.3, and at length in sec. 2.3, the pdf evolution equations are also non-closed and in need of a closure scheme, albeit of different nature.

[5] Note that multidimensional RDEs are considered in state-space, being thus always systems of first order RDEs.

$\boldsymbol{C}_{\Xi\Xi}(t, s)$, and cross-covariance the $N \times K$ matrix $\boldsymbol{C}_{X^0\Xi}(t)$. As for scalar RIVP (1.1)-(1.2), RIVP (1.6)-(1.7) is under *additive or external excitation* if $\boldsymbol{q}(\boldsymbol{x})$ equals to a $N \times K$ constant matrix $\boldsymbol{\kappa}$, while for the general case where $\boldsymbol{q}(\boldsymbol{x})$ is $\boldsymbol{x}$ – dependent, excitation termed *multiplicative or parametric*.

**Towards non-Gaussian excitations.** By observing the scalar RDE (1.1), a plausible, from the applications point of view, generalization is to consider the excitation to be non-Gaussian. While this can be performed in a variety of ways, a first step in this direction, is to model the non-Gaussian excitation by a polynomial of a Gaussian process $\Xi(t;\theta)$:

$$\dot{X}(t;\theta) = h(X(t;\theta)) + \sum_{n=1}^{N} q_n(X(t;\theta))\, \Xi^n(t;\theta) \qquad X(t_0;\theta) = X_0(\theta). \tag{1.8a.b}$$

Under the concept of *polynomially Gaussian excitation*, most of the mathematical tools developed for the case of Gaussian excitation can be employed anew. In Section 4.6 of the present thesis, we present first results and validations in formulating response pdf evolution equations for scalar RDEs under quadratic Gaussian excitation.

## 1.2 Survey of the literature on the formulation of pdf evolution equations under coloured excitation

### 1.2.1 Motivation: The classical Fokker-Planck-Kolmogorov equation

At first, let us consider excitation $\Xi(t;\theta)$ in RDE (1.1) to be a *Gaussian white noise*; $C_{\Xi\Xi}(t,s) = 2D(t)\,\delta(t-s)$, where $D(t) > 0$ is its intensity and $\delta(t-s)$ Dirac's delta function. In this case, the response $X(t;\theta)$ to RDE (1.1) is *Markovian* (Øksendal, 2003, para. 7.1), meaning that its complete probabilistic structure is defined in terms of the conditional pdf $f_{X(t)|X_0}(x\,|\,x_0)$. What is more, it is derived, from Chapman-Kolmogorov equation, a partial differential equation (PDE) for the $f_{X(t)|X_0}(x\,|\,x_0)$, called the Fokker-Planck-Kolmogorov (FPK) equation (Cáceres, 2017, secs. 3.18-9; Gardiner, 2004, Chapter 5; Pugachev & Sinitsyn, 2001, sec. 5.6.6; Risken, 1996; C. Soize, 1994; Stratonovich, 1989; Sun, 2006, sec. 6.3):

$$\frac{\partial f_{X(t)|X_0}(x\,|\,x_0)}{\partial t} + \frac{\partial}{\partial x}\Big(A(x,t)\, f_{X(t)|X_0}(x\,|\,x_0)\Big) = \frac{\partial^2}{\partial x^2}\Big(B(x,t)\, f_{X(t)|X_0}(x\,|\,x_0)\Big), \tag{1.9}$$

Analogously, for the multidimensional RDE (1.6) under white noise, i.e. with $\boldsymbol{C}_{\Xi\Xi}(t,s) = 2\boldsymbol{D}(t)\,\delta(t-s)$, where intensity $\boldsymbol{D}(t)$ is a symmetric, positive-definite matrix, we obtain the multidimensional FPK equation

$$\begin{aligned}\frac{\partial f_{\boldsymbol{X}(t)|\boldsymbol{X}_0}(\boldsymbol{x}\,|\,\boldsymbol{x}_0)}{\partial t} &+ \sum_{n=1}^{N} \frac{\partial}{\partial x_n}\Big(A_n(\boldsymbol{x},t)\, f_{\boldsymbol{X}(t)|\boldsymbol{X}_0}(\boldsymbol{x}\,|\,\boldsymbol{x}_0)\Big) = \\ &= \sum_{n_1=1}^{N}\sum_{n_2=1}^{N} \frac{\partial^2}{\partial x_{n_1}\partial x_{n_2}}\Big(B_{n_1 n_2}(\boldsymbol{x},t)\, f_{\boldsymbol{X}(t)|\boldsymbol{X}_0}(\boldsymbol{x}\,|\,\boldsymbol{x}_0)\Big).\end{aligned} \tag{1.10}$$

FPK Eqs. (1.9), (1.10) are ***PDEs of drift-diffusion type***, with their drift, $A(x,t)$ or $\boldsymbol{A}(\boldsymbol{x},t)$, and diffusion, $B(x,t)$ or $\boldsymbol{B}(\boldsymbol{x},t)$, being defined in terms of the RDE characteristics, $h(\bullet)$, $q(\bullet)$ or $\boldsymbol{h}(\bullet)$, $\boldsymbol{q}(\bullet)$, as well as the intensity of the white noise[6].

**Remark 1.1:** It can be easily shown that FPK equations, apart from conditional pdfs $f_{X(t)|X_0}(x\,|\,x_0)$ or $f_{\boldsymbol{X}(t)|\boldsymbol{X}_0}(\boldsymbol{x}\,|\,\boldsymbol{x}_0)$, are also satisfied by the one-time response pdfs, $f_{X(t)}(x)$ or $f_{\boldsymbol{X}(t)}(\boldsymbol{x})$, see e.g. (Gardiner, 2004, secs. 5.2-3). What it changes is the initial condition, from $f_{X(t_0)|X_0}(x\,|\,x_0)=\delta(x-x_0)$ or $f_{\boldsymbol{X}(t_0)|\boldsymbol{X}_0}(\boldsymbol{x}\,|\,\boldsymbol{x}_0)=\delta(\boldsymbol{x}-\boldsymbol{x}_0)$ for the conditional pdfs, to the less singular $f_{X(t_0)}(x)=f_{X_0}(x)$ or $f_{\boldsymbol{X}(t_0)}(\boldsymbol{x})=f_{\boldsymbol{X}_0}(\boldsymbol{x})$ for the response pdfs. In paragraphs 4.1.1, 4.2.2 of the present thesis, where FPK equations are rederived as special cases of pdf evolution equations, response pdfs are considered as their unknowns.

Thus, for the special case of RDEs under white noise excitation, the probabilistic structure of the response is governed by a deterministic PDE. In Gardiner's words: "it is true that there is nothing in a stochastic differential equation that is not in a Fokker-Planck equation"[7]. Unfortunately, this convenient description, via a single PDE, is not applicable to problems where the random excitations are coloured, and thus their responses are *non-Markovian*.[8]

### 1.2.2 The filtering approach to coloured noise

The most straightforward approach, for formulating response pdf evolution equations for systems under coloured noise excitation, is filtering. In this case, the original system of RDEs is augmented by another RDE (or system of RDEs), called the filter, which is excited by white noise and its output is an approximation of the given coloured noise $\Xi(t;\theta)$. The new, augmented system of RDE admits now an exact FPK description, since it is ultimately excited by white noise. This filtering approach is quite popular in engineering applications (Chai, Naess, & Leira, 2015; Francescutto & Naito, 2004; Pugachev & Sinitsyn, 2001, sec. 5.10), and it is also referred to as Markovianization by extension (Krée, 1985), or embedding in a Markovian process of higher dimensions (Hänggi & Jung, 1995, sec. VI.A). Filtering is a generic approach that works well for Gaussian, stationary random excitations, see e.g. (Spanos, 1986). However, for non-Gaussian excitations, non-linear filters should be employed, making the method less straightforward in its use.
As a first example, in the case of zero-mean valued OU excitation, the original RDE system has to be augmented by the linear filter, see e.g. (Hänggi & Jung, 1995, sec. V.C)

---

[6] Since our focus is on coloured noise excitations, delving further into classical FPK equations, with the derivation of the exact forms for their drift and diffusion, is out of the scope of the present paragraph. Such a discussion is lengthy since it is proven that, in the multiplicative case (function *q* being *x*-dependent), the drift of the FPK equation corresponding to RDEs under Itō interpretation, differs from the FPK drift under Stratonovich interpretation, by a term called Wong-Zakai or Sussmann-Doss correction (Doss, 1977; Sun, 2006; Sussmann, 1978; Wong & Zakai, 1965). In paragraphs 4.1.1, 4.2.2, where we rederive the FPK equations, their Stratonovich variant is retrieved. This result and its compatibility with our approach to RDEs is discussed in the said paragraphs.

[7] (Gardiner, 2004), preface to the first edition.

[8] We have to mention that, for the special case of small, but not negligible, correlation time of excitation, we can derive an RDE excited by white noise for the slowly varying quantites of the original dynamical system. For this RDE, we can also formulate the corresponding FPK equation, which has fewer state variables than the orginal dynamical system. This dimension reduction approach is called *stochastic averaging*, see e.g. (Roberts & Spanos, 1986).

$$\dot{\Xi}(t;\theta) = -\frac{1}{\tau}\,\Xi(t;\theta) + \frac{\sqrt{D}}{\tau}\,W(t;\theta), \tag{1.11}$$

where $W(t;\theta)$ is a one-dimensional white noise with unitary intensity, and $D$, $\tau$ are the intensity and correlation time of OU noise, see Eq. (1.3). However, in real-life applications, the filter can be as complicated as the correct representation of the excitation spectrum dictates; e.g. in (Francescutto & Naito, 2004), the filter to a random oscillator, modelling the ship roll motion, is of fourth order, resulting thus in an augmented RDE with six degrees of freedom, which, in turn, corresponds to a six-dimensional FPK equation. This ***inflation of degrees of freedom*** (DOFs) in the FPK equation is the inherent drawback of the filtering approach, posing significant computational challenges.

Usual approaches to the ***numerical solution of the multidimensional FPK equation***[9] are the *finite elements* and *finite differences methods*, see e.g. (Floris, 2013; P. Kumar & Narayanan, 2006; Pichler, Masud, & Bergman, 2013) in which surveys of older results are also present. However, the aforementioned methods are plagued by the curse of dimensionality, making the solution of even the four-dimensional FPK equation a challenge (Masud & Bergman, 2005). In order to overcome this problem, *smooth particle hydrodynamics methods* have been recently applied, see e.g. (Canor & Denoël, 2013). We also have to mention that, in (Psaros, Brudastova, Malara, & Kougioumtzoglou, 2018), the problem of solving multidimensional FPK equations corresponding to augmented RDE systems was tacked within the framework of *path-integral solutions* (Sun, 2006, sec. 6.5).

**On the unified coloured noise approximation.** An approach to RDEs under OU excitation, which is based on the Markovianization by extension, is the unified coloured noise approximation (UCNA), proposed in (Hänggi & Jung, 1987, 1995, sec. V.C). UCNA starts by considering the augmented RDE system (1.1) & (1.11). Then, by using RDE (1.9), OU noise $\Xi(t;\theta)$ is eliminated from RDE (1.1), which, for the case of additive excitation, is expressed equivalently as a nonlinear oscillator excited by white noise

$$\ddot{X}(t;\theta) + \left[\tau^{-1/2} - \tau^{1/2}\,h'(X(t;\theta))\right]\dot{X}(t;\theta) = h(X(t;\theta)) + \frac{D^{1/2}}{\tau^{1/4}}\,W(t;\theta). \tag{1.12}$$

Now, for the case $h'(X(t;\theta)) < 0$, in which the damping coefficient in oscillator (1.12) is positive, an adiabatic elimination, $\ddot{X}(t;\theta) \cong 0$, is performed (Łuczka, 2005, sec. VI.A), that results in the following approximate, yet one-dimensional RDE under white noise excitation:

$$\dot{X}(t;\theta) = \frac{h(X(t;\theta))}{\tau^{-1/2} - \tau^{1/2}\,h'(X(t;\theta))} + \frac{D^{1/2}}{\tau^{1/4}\left[\tau^{-1/2} - \tau^{1/2}\,h'(X(t;\theta))\right]}\,W(t;\theta), \tag{1.13}$$

that admits an FPK description. Thus, the merit of UCNA is that it avoids the DOFs inflation present in the standard filtering approach. Note that UCNA is still used in physical applications under OU excitation, see e.g. the recent work (Wittmann et al., 2017). As stated in (Wio, 2013, sec. 7.1) “UCNA is a reliable Markovian approximation by means of path-integral techniques”. The aforementioned *path-integral techniques*, developed in (Colet, Wio, & Sancho,

[9] Instead of considering the FPK equation, one could alternatively solve the Chapman-Kolmogorov equation corresponding to the augmented dynamical system, by employing numerical path integration, see e.g. (Naess, 2001).

1989; Wio, Colet, & San Miguel, 1989), have also the augmented RDE system as their starting point, and have also resulted in the extension of UCNA for certain classes of non-Gaussian coloured excitation, see (Fuentes, Wio, & Toral, 2002; Wio, 2013, Chapter 8).

### 1.2.3 FPK-like equations for coloured noise excitation

Having noted the computational difficulties of filtering, we move on to a different approach; to try to derive pdf evolution equations analogous to the FPK equation, keeping only the natural degrees of freedom of the RDE, while taking into account the given coloured noise excitation. The main difficulty here, coming from the fact that "non-Markovian processes […] cannot be considered merely as corrections to the class of Markov processes" (van Kampen, 1998), lies in the emergence of terms dependent on the whole time history of the response and excitation, even for the one-time pdf evolution equation, as explained at length subsequently in the present thesis. Therefore, closure techniques for these ***nonlocal terms*** are needed, in order to obtain approximate, yet closed and solvable, forms of the evolution equations. In the literature, the evolution equations for the one-time response pdf are usually called *generalized FPK*, *FPK-like* or of *FPK type/form*, since their white-noise excitation counterpart is the classical FPK equation.[10]

In the rest of the present paragraph, we attempt a chronological survey of the formulation of such equations. When appropriate, we shall refer to particular sections of the main part of the thesis, where more lengthy and technical derivations and proofs are presented. We also have to note that most of the surveyed works were considered with the formulation of FPK-like equations for the case of stationary response pdf, and also for OU excitation. Thus, the results presented below are for this particular case.

**The origins: Kubo and Stratonovich.** The first derivations of pdf equations can be traced back to the pioneering works in the 60's; (Kubo, 1963; Stratonovich, 1963). In the former, the linear, multiplicatively-excited RDE $\dot{X}(t;\theta) = i\,[\omega_0 + \Xi(t;\theta)]\,X(t;\theta)$, called Kubo's oscillator, was studied, while, in Sec. 4.8 of the latter, the nonlinear RDE $\dot{X}(t;\theta) = \varepsilon F(X(t;\theta), \Xi(t;\theta))$ was considered, with $F$ being a known function of response and excitation, and $\varepsilon$ a small parameter. In both cases, the initial value was considered deterministic. The way of proceeding in both works was by considering the characteristic function of the response. First, the solution to RDE was expressed in closed form as $X(t;\theta) = \mathcal{F}[\Xi(\bullet|_{t_0}^{t};\theta)]$, where $\mathcal{F}$ is the appropriate functional on excitation from initial time $t_0$ up to current time $t$.[11] Thus, the characteristic function of the response at current time $t$ is expressed as $\varphi_{X(t)}(u) = \mathbb{E}^{\theta}\Big[\exp\Big((iu)\,\mathcal{F}[\Xi(\bullet|_{t_0}^{t};\theta)]\Big)\Big]$. By employing now the ***cumulant expansion method*** (CEM) (Kubo, 1962), the characteristic function is written as

[10] In (J. Li & Chen, 2009, Chapter 6; Jie Li, 2016), another approach for deriving evolution equations for the one-time, joint response-excitation pdf was proposed, that works well for linear dynamical systems. In contrast to the FPK-like pdf evolution equations, the aforementioned approach results in equations that employ only derivatives of first order. Detailed comparison between the two approaches will be the topic of a future work.

[11] More precisely, $\mathcal{F}$ is the resolvent causal operator of the RDE, which, by fixing current time *t*, can be seen equivalently as a functional on excitation from $t_0$ to *t*.

$\varphi_{X(t)}(u) = \exp\left(\sum_{m=1}^{\infty}(iu)^m \mathcal{G}_m(t)\right)$, where $\mathcal{G}_m$ are *functionals on the cumulants of the excitation*. While CEM will not be used in the present thesis, opting for treating the Gaussian excitations by using the Novikov-Furutsu theorem (see below), we have to note that CEM has been extensively used in the literature, see e.g. (Fox, 1983; Garrido & Sancho, 1982; Roerdink, 1981, 1982; Sirin, 2013; Terwiel, 1974). Furthermore, the proper way of applying CEM, especially for the case of multidimensional RDEs, has been the topic of many works up to now (Bianucci & Bologna, 2019; Fox, 1976; van Kampen, 2007, Chapter XVI).

**Formulation of pdf equations takes off: van Kampen's lemma.** As one can observe in (Stratonovich, 1963, sec. 4.8), obtaining equations for pdfs via the characteristic functions can be rather cumbersome. Thus, what is needed for the easier derivation of response pdf evolution equations is a handy representation of the response pdf *per se*. This was answered by van Kampen's lemma (van Kampen, 1975, p. 269, 1976, p. 209, 2007, p. 411), under which the response pdf is represented as $f_{X(t)}(x) = \mathbb{E}^{\theta}\left[\delta(x - X(t;\theta))\right]$, i.e. as the average of a random delta function. Delta representation of the response pdf can be differentiated with respect to time, resulting into the equation

$$\frac{\partial f_{X(t)}(x)}{\partial t} + \frac{\partial}{\partial x}\left(h(x)\, f_{X(t)}(x)\right) = -\frac{\partial}{\partial x}\left(q(x)\, \mathbb{E}^{\theta}\left[\delta(x - X(t;\theta))\, \Xi(t;\theta)\right]\right). \tag{1.14}$$

Following (Kubo, 1963) and due to its analogy to the classical Liouville equation, Eq. (1.14) is termed the ***stochastic Liouville equation*** (SLE). Note that, all relevant articles succeeding the aforementioned works of van Kampen use his eponymous lemma and consider SLE (1.14), or its multidimensional variant, as their starting point. Van Kampen's lemma and the SLE derivation are revisited in Chapter 2 of the present thesis.

**The advantages of Gaussian excitation: Novikov-Furutsu theorem.** In (Dekker, 1982; Hänggi, 1978; San Miguel & Sancho, 1980; Sancho & San Miguel, 1980)[12] it was observed that the averaged term $\mathbb{E}^{\theta}\left[\delta(x - X(t;\theta))\, \Xi(t;\theta)\right]$ in SLE (1.14) could be further evaluated using the Novikov-Furutsu (NF) theorem; (Furutsu, 1963; Novikov, 1965), see also Section 3.2 of the present thesis. NF theorem, for a functional $\mathcal{F}[\Xi(\bullet|_{t_0}^{t};\theta)]$ whose argument is a *zero-mean valued Gaussian random function*, reads

$$\mathbb{E}^{\theta}\left[\mathcal{F}[\Xi(\bullet|_{t_0}^{t};\theta)]\, \Xi(t;\theta)\right] = \int_{t_0}^{t} C_{\Xi\Xi}(t,\tau)\, \mathbb{E}^{\theta}\left[\frac{\delta \mathcal{F}[\Xi(\bullet|_{t_0}^{t};\theta)]}{\delta \Xi(\tau;\theta)}\right] d\tau. \tag{1.15}$$

Thus, by using the fact that the response is a functional on excitation, SLE (1.14), after the application of NF theorem (1.15), is expressed as

$$\begin{aligned} &\frac{\partial f_{X(t)}(x)}{\partial t} + \frac{\partial}{\partial x}\left(h(x)\, f_{X(t)}(x)\right) = \\ &\qquad = \frac{\partial}{\partial x}\left(q(x)\, \frac{\partial}{\partial x}\int_{t_0}^{t} C_{\Xi\Xi}(t,s)\, \mathbb{E}^{\theta}\left[\delta(x - X(t;\theta))\, \frac{\delta X(t;\theta)}{\delta \Xi(s;\theta)}\right] ds\right). \end{aligned} \tag{1.16}$$

[12] In (Hänggi, 1978) NF theorem is rederived and employed, but not mentioned by name.

Thus, we observe that, in the case of Gaussian excitation, the infinite CEM series is simplified into a single term. SLE (1.16) is equivalent to SLE (1.14); however, in SLE (1.16), the explicit dependence of the averaged quantity on excitation is eliminated, and the ***functional derivative of the response with respect to excitation***, $\delta X(t;\theta)/\delta\Xi(s;\theta)$, appears. Note also that, this step involving NF theorem was performed by two contributors that continued, though the 80's and up to the 90's, to produce central works in the field; these are *P. Hänggi and his group*, and *Barcelona group* of the Department of Theoretical Physics of the University of Barcelona, in which J. M. Sancho and M. San Miguel are main figures.

Barcelona group has also worked towards generalizations of the methodology, producing equations for the joint, two-time pdf of the response (Hernandez-Machado & Sagués, 1983; Hernandez-Machado, Sancho, San Miguel, & Pesquera, 1983), as well as for RDEs under quadratic Gaussian excitation (Sagués, San Miguel, & Sancho, 1984; San Miguel & Sancho, 1981). These generalizations are also considered in Sections 4.5 and 4.6 of the present thesis.

**Near white-noise limit: The small correlation time approximation.** As it is proved and discussed in (Sancho, San Miguel, Katz, & Gunton, 1982), as well as in Chapter 4 of the present thesis, the functional derivative of the response inside SLE (1.13) depends on the whole history of the response and excitation, from initial up to current time. Thus, a ***current-time approximation scheme*** has to be applied on it. A first, straightforward approximation presented in (Hänggi, 1989; Sancho & San Miguel, 1980, 1989; Sancho et al., 1982) is the linear expansion of the derivative $\delta X(t;\theta)/\delta\Xi(s;\theta)$ around current time, using Taylor series. This approximation is called small correlation time (SCT) because, for the stationary case under additive OU excitation "[it] contains all the terms which are of first order in [correlation time] $\tau$" (Sancho & San Miguel, 1989, p. 78):

$$\frac{\partial}{\partial x}\Big(h(x)\, f_X(x)\Big) = D\kappa^2 \frac{\partial^2}{\partial x^2}\Big[\big(1+\tau h'(x)\big)\, f_X(x)\Big], \tag{1.17}$$

where $f_X(x)$ denotes the steady-state limit of $f_{X(t)}(x)$.[13] Also, it has to be noted, that the stationary pdf equation obtained under the SCT approximation, is the same with the one derived in (Stratonovich, 1963). However, the main drawback of SCT Eq. (1.17) is that its diffusion coefficient can take negative values. Thus, the solution $f_X(x)$ of SCT Eq. (1.17) has the unphysical boundary in state-space; $1+\tau h'(x)=0$ (Sancho et al., 1982).

**The small noise intensity approximation.** As a first attempt to break away from SCT regime, small noise intensity approximation was proposed in (Sancho et al., 1982) and reviewed again in (Hänggi, Marchesoni, & Grigolini, 1984; Sancho & San Miguel, 1989). In this approximation, the whole Taylor series of $\delta X(t;\theta)/\delta\Xi(s;\theta)$ around current time is formally considered, with the coefficients in Taylor series calculated under the *small excitation assumption* $\Xi(t;\theta)\cong 0$. This results into the following stationary pdf equation for the case of additive OU excitation, termed the *best Fokker-Planck equation* (BFPE):

$$\frac{\partial}{\partial x}\Big(h(x)\, f_X(x)\Big) = D\kappa \frac{\partial^2}{\partial x^2}\Big(H(x)\, f_X(x)\Big), \tag{1.18}$$

[13] Strangely enough, the stationary FPK-like equations presented in the literature, contain also the time-dependent term $\partial f_{X(t)}(x)/\partial t$. This practice is not followed in the present thesis.

where function $H(x)$ is the solution of ODE; $H(x)+\tau\,[H'(x)h(x)-H(x)h'(x)]=\kappa$. Thus, $H(x)$ is not given in closed form, and, in practice, a $\tau$-expansion up to certain order is employed, instead of its exact calculation, see e.g. (Hänggi et al., 1984). Furthermore, as in SCT Eq. (1.17), the positivity of the diffusion coefficient in Eq. (1.15) is not guaranteed.

**A bona fide FPK equation for small correlation time: Fox's approximation.** Due to the fail of SCT Eq. (1.17) to calculate correctly the mean first-passage time of a bistable RDE, the *effective Fokker-Planck equation* (EFPE) was proposed in (Fox, 1986a, 1987). For the stationary case under additive OU excitation, and under *small correlation time condition* $\tau\,h'(x)<1$, EFPE reads

$$\frac{\partial}{\partial x}\big(h(x)\; f_X(x)\big) = D\kappa^2\,\frac{\partial^2}{\partial x^2}\left[\frac{1}{1-\tau\,h'(x)}\, f_X(x)\right]. \tag{1.19}$$

Thus, in the SCT regime, EFPE (1.19) is a legitimate FPK-like equation, since its diffusion coefficient is positive. Note also that, by expanding the diffusion coefficient of Eq. (1.19) with respect to $\tau$ and keeping only the linear term, stationary SCT Eq. (1.17) is obtained (Hänggi & Jung, 1995) *loc. sit.* We should also mention that Fox, in the aforementioned works, derived EFPE from the original form of SLE, Eq. (1.14), without employing the NF theorem; instead, he performed path integrations with respect to the infinite-dimensional Gaussian probability density functional. In paragraph 5.1.1 of the present thesis, EFPE as well as its time-domain extension, are rederived from Eq. (1.16), that is, the SLE after the application of NF theorem.

**An alternative approach: The projection operator method.** In parallel with the works employing the functional approach of calculating and approximating $\delta X(t;\theta)/\delta\Xi(s;\theta)$ in SLE (1.16), a different direction, called the projection operator method, was followed by *P. Grigolini and his group* (Faetti, Fronzoni, Grigolini, & Mannella, 1988; Faetti & Grigolini, 1987; Grigolini, 1986, 1989; Tsironis & Grigolini, 1988b, 1988a) and the *group of B. J. West* (Lindenberg & West, 1983; J. Masoliver, West, & Lindenberg, 1987; Peacock-López, West, & Lindenberg, 1988). This approach utilized the splitting of the operator of the FPK equation corresponding to the RDE into an unperturbed part, which is identical to the operator of the classical FPK equation, and a perturbation over it, which encapsulates the deviation from the white noise excitation. Finally, the pdf equation derived by Grigolini's group coincides with EFPE, while the pdf equation derived by West's group coincides with BFPE. While projection operator method is not followed in the present thesis, opting for the functional approach of Hänggi and the Barcelona group, it is indeed a very interesting alternative; quite recently, in (Bianucci, 2020), BFPE and EFPE under projection operator method, were revisited.

**Away from the small correlation time regime: Hänggi's ansatz.** In (Hänggi, Mroczkowski, Moss, & McClintock, 1985), a different stationary response pdf equation was proposed:

$$\frac{\partial}{\partial x}\big(h(x)\; f_X(x)\big) = \frac{D\kappa^2}{1-\tau\,\mathbb{E}^\theta\big[h'(X(\infty;\theta))\big]}\,\frac{\partial^2 f_X(x)}{\partial x^2}, \tag{1.20}$$

where $\mathbb{E}^\theta\big[h'(X(\infty;\theta))\big]$ is the steady-state value of response moment $\mathbb{E}^\theta\big[h'(X(t;\theta))\big]$. By comparing Hänggi's Eq. (1.20) to Fox's EFPE (1.19), we see that, in the diffusion coefficient, the $x$-dependent function $h'(x)$ has been substituted by the $x$-independent, yet un-known, response moment $\mathbb{E}^{\theta}\left[h'(X(\infty;\theta))\right]$.[14] As it is discussed in (Hänggi & Jung, 1995), where the name decoupling approximation was also suggested, pdf Eq. (1.17), unlike EFPE (1.19), is not restricted to the SCT regime, since, for globally stable physical systems, $\mathbb{E}^{\theta}\left[h'(X(\infty;\theta))\right] < 0$ and thus the diffusion coefficient in Eq. (1.20) is always positive. However, the justification behind the derivation of Hänggi's stationary Eq. (1.20) is somewhat obscure, even in its rederivation in (Hänggi & Jung, 1995), resulting thus in the characterization of this approximation as "ansatz". We have to also note that, there have been attempts of deriving all the above pdf equations, including Hänggi's one, via a unified way, see e.g. (Faetti & Grigolini, 1987) that employed the operator projection method.

Despite the possible ambiguities in the derivation of Eq. (1.20), Hänggi's ansatz initiates a new direction in the pursuit of formulating FPK-like equations for coloured noise excitation. By recognizing the fact that the response to coloured noise is non-Markovian, *the equation for its pdf cannot be linear and localized, as in the case of the classical FPK equation*. Thus, it is more realistic that the coefficients in pdf evolution equations for coloured noise excitation to depend on the unknown pdf. Such equations are termed ***nonlinear FPK equations*** in the literature (Frank, 2005). This is also the direction followed in Chapter 5 of the present thesis.

**Use of FPK-like equations in applications.** To the best of our knowledge, the quest for new, more accurate equations for the response pdf seems to have come halt by the mid 90's. At this point, we have to make reference to the survey works (Hänggi & Jung, 1995; Łuczka, 2005; Ridolfi et al., 2011, sec. 2.5) that recapitulated the results we have already mentioned. However, the pdf evolution equations derived in the 80's, and especially Fox's and Hänggi's, are used in demanding real-life applications up to now; see e.g. the *works during the last decade:* (Knyaz', 2011) in *complex systems*, (Harne & Wang, 2014) in *energy harvesting*, (Daqaq, 2011) in *sensors design*, (Zhu & Zhu, 2010) in *laser technology*, (Zhang, Yang, Xu, & Xu, 2014) in *stochastic resonance*, (Ridolfi et al., 2011; Spanio et al., 2017; Zeng et al., 2017) in *ecosystems*, (Bose & Trimper, 2011; Idris & Abu Bakar, 2016; Venturi, Sapsis, Cho, & Karniadakis, 2012; Yang et al., 2014; Zeng & Wang, 2010) in *oncology*, (S. H. Li & Zhu, 2018; X. L. Li & Ning, 2016) in *neural systems*, and (Bose & Trimper, 2012; Chattopadhyay & Aifantis, 2016) in *material science*.

## 1.3 Main contributions of the present thesis

In this thesis, the methodology for deriving FPK-like equations corresponding to RDEs under coloured noise excitation is revisited and extended. More specifically, by employing a novel, approximate current-time closure at the SLE, see paragraph 5.1.2, we obtain a family of response pdf evolution equations, corresponding to a scalar, nonlinear, additively excited RDE, whose members have the general form

$$\frac{\partial f_{X(t)}(x)}{\partial t} + \frac{\partial}{\partial x}\left[\left(h(x) + \kappa m_{\Xi}(t)\right) f_{X(t)}(x)\right] = \frac{\partial^2}{\partial x^2}\left[\mathcal{B}[\, f_X \,;\, x, t\,]\, f_{X(t)}(x)\right]. \tag{1.21}$$

[14] The observant reader has already noticed that (Hänggi et al., 1985) precedes (Fox, 1986a). Indeed, one goal of (Fox, 1986a) was to propose a more rational alternative to Hänggi's Eq. (1.17). However, as it is shown in the subsequent discussion, Hänggi's ansatz is indeed significant *per se*, in the direction of obtaining pdf equations that are not restricted to small correlation time. This is why our presentation concludes with Hänggi's equation.

Eq. (1.21) is a *nonlinear and nonlocal evolution equation*, since its diffusion $\mathcal{B}[f_X ; x,t]$, apart from being a function of state and time variables, is also a functional on the unknown response pdf, reflecting the non-Markovian character of the response. Thus, unlike the other existing approaches, our novel technique retains an amount of nonlocality and nonlinearity of the original SLE, albeit of tractable character.

The new pdf evolution Eq. (1.21) exhibits the following plausible features:

- It is valid in both transient and long-time, steady-state regimes,
- It is valid for large correlation times and large noise intensities of the excitation,
- It yields the exact Gaussian solution pdf in the case of linear, additively excited RDE,
- It contains Hänggi's equation as a special case,
- It applies to general Gaussian excitation, characterized by any correlation function,
- It applies to non-zero mean excitation, also correlated with the initial value.

Numerical results, for the benchmark case of a bistable RDE under OU excitation, presented in Sec. 6.2, confirm the validity and the accuracy of the family of the novel response pdf evolution equations, by comparisons with results obtained from Monte Carlo simulations. In Sec. 5.2, the same closure scheme is applied for the derivation of evolution equations for response pdfs of higher order. In Sec. 5.3, the novel closure is also generalized for the SLE corresponding to multidimensional RDEs.

Furthermore, other worth-mentioning points of novelty of the present thesis are:

- In Chapter 2, the derivation of SLEs is revisited, and a more comprehensive and easily generalizable alternative method to van Kampen's lemma is proposed.
- In Chapter 3, the classical NF theorem, Eq. (1.15), is rederived in a more systematic way, and generalized. First, we extend NF theorem for $\mathbb{E}^{\theta}\left[\Xi(s;\theta)\,\mathcal{F}[X_0(\theta);\Xi(\bullet|_{t_0}^{t};\theta)]\right]$, where $\mathcal{F}$ is a function on $X_0(\theta)$ and a functional on $\Xi(\bullet;\theta)$, with its arguments being jointly Gaussian, with, in general, non-zero mean values. This extended NF theorem is employed for the treatment of random initial value, correlated to excitation. Then, NF is extended for $\mathbb{E}^{\theta}\left[\dot{\Xi}(s;\theta)\,\mathcal{F}[X_0(\theta);\Xi(\bullet|_{t_0}^{t};\theta)]\right]$, which is needed for the SLE of $f_{X(t)\Xi(t)}(x,u)$, see Sec. 4.3. Last, NF theorem is generalized for averages containing multiple time instances of excitation $\mathbb{E}^{\theta}\left[\Xi(s_1;\theta)\cdots\Xi(s_n;\theta)\mathcal{F}[X_0(\theta);\Xi(\bullet|_{t_0}^{t};\theta)]\right]$ which contains Isserlis theorem as a special case. This generalization of NF theorem is needed when dealing with polynomially Gaussian excitation, as in RDE (1.8a).
- In Section 4.6, a response pdf evolution equation resembling the Kramers-Moyal expansion, i.e. containing (in principle) all orders of *x*-derivatives, is derived for the case of an RDE under additive quadratic Gaussian excitation. It is also validated that, for the linear case, the aforementioned pdf evolution equation calculates correctly all cumulants of the response.
- In Section 5.4 the case of RDEs under multiplicative excitation is studied. For this case, the advantages of formulating the solving the pdf evolution equation for the one-time joint response-excitation pdf $f_{X(t)\Xi(t)}(x,u)$, instead of the equation for $f_{X(t)}(x)$, are shown.

## 1.4 List of publications

The research presented in the present thesis has also resulted in the following works:

**Publications in scientific journals**

1. K.I. Mamis, G.A. Athanassoulis, Z.G. Kapelonis (2019): "A systematic path to non-Markovian dynamics: New response probability density function evolution equations under Gaussian coloured noise excitation", *Proceedings of the Royal Society of London A*, 471 (20180837).
2. G.A. Athanassoulis, K.I. Mamis (2019): "Extensions of the Novikov-Furutsu theorem, obtained by using Volterra functional calculus", *Physica Scripta*, 94(11) 115217.

**Publications in conference proceedings after full paper review**

1. G.A. Athanassoulis, Z.G. Kapelonis & K.I. Mamis (2018): "Numerical solution of generalized FPK equations corresponding to random differential equations under colored noise excitation. The transient case", in *8th Conference on Computational Stochastic Mechanics*. Paros, Greece.
2. K.I. Mamis, G.A. Athanassoulis & K.E. Papadopoulos (2018): "Generalized FPK equations corresponding to systems of nonlinear random differential equations excited by colored noise. Revisitation and new directions", in *7th International Young Scientists Conference on Computational Science*. Heraklion, Greece. Available at: *Procedia Computer Science*, 136(C), pp. 164–173.

**Presentations in conferences after abstract review**

1. G.A. Athanassoulis and K.I. Mamis (2019): "Uncertainty quantification of responses to nonlinear dynamical systems under coloured noise excitation via pdf evolution equations" in *17$^{th}$ International Probabilistic Workshop*. Edinburgh, Scotland.
2. K.I. Mamis and G.A. Athanassoulis (2019): "Formulation and solution of response pdf evolution equations corresponding to systems under Gaussian coloured noise excitation" in *3$^{rd}$ International Conference on Uncertainty Quantification in Computational Sciences and Engineering.* Heraklion, Greece.
3. K.I. Mamis, Z.G. Kapelonis & G.A. Athanassoulis (2019): "Determining the probabilistic structure of the response to a nonlinear dynamical system under coloured noise excitation" (poster presentation) in *44th Conference of the Middle European Cooperation in Statistical Physics.* Munich, Germany.
4. K.I. Mamis and G.A. Athanassoulis (2018): "Generalized (non-Markovian) FPK equations corresponding to nonlinear random differential equations excited by colored noise. Hänggi's ansatz revisited" (poster presentation) in *43rd Conference of the Middle European Cooperation in Statistical Physics.* Kraków, Poland.

Furthermore, we have also worked on obtaining exact stationary solutions for classical FPK equations. While not directly related to the topic of present thesis, this was indeed a suitable preparatory work on what it means to solve pdf equations corresponding to random dynamical systems. The said study resulted in the following published works:

1. K.I. Mamis and G.A. Athanassoulis (2016): "Exact stationary solutions to Fokker-Planck-Kolmogorov equation for oscillators using a new splitting technique and a new class of stochastically equivalent systems", *Probabilistic Engineering Mechanics*, 45, pp. 22-30. (journal article)
2. K.I. Mamis and G.A. Athanassoulis (2016): "Quantifying the influence of Wong-Zakai correction on a class of exactly solvable generalized Dimentberg oscillators", in *11$^{th}$ HSTAM International Congress on Mechanics: Advances in Theoretical and Applied Mechanics.* Athens, Greece. (publication in conference proceedings)
3. K.I. Mamis and G.A. Athanassoulis (2016): "Emergence of limit cycles in the stationary response probability density functions for a class of exactly solvable nonlinear stochastic oscillators", in *Frontiers of Nonlinear Physics, VI International Conference*. Nizhny Novgorod, Russia. (presentation in conference)
4. K.I. Mamis and G.A. Athanassoulis (2015): "Exact stationary solutions to a class of nonlinear stochastic oscillators. Establishing new benchmark cases for testing numerical solution schemes", in *IV International Young Scientists Conference in computer modeling and simulation*. Athens, Greece. Available at: *Procedia Computer Science*, 66, pp. 33-42. (publication in conference proceedings)

# Chapter 2: Delta projection and the stochastic Liouville equations

**Summary.** In the present chapter[15], the first step in the methodology of deriving response(-excitation) pdf evolution equations is performed. By employing a convenient representation for the pdfs as averages of random delta functions, the following equations, called stochastic Liouville, are derived in a comprehensive way:

$$\frac{\partial f_{X(t)}(x)}{\partial t} + \frac{\partial}{\partial x}\Big(h(x)\, f_{X(t)}(x)\Big) = -\frac{\partial}{\partial x}\Big(q(x)\, \mathbb{E}^{\theta}\big[\delta(x - X(t;\theta))\, \Xi(t;\theta)\big]\Big),$$

$$\frac{\partial f_{X(t)\Xi(t)}(x,u)}{\partial t} + \frac{\partial}{\partial x}\Big[\big(h(x) + q(x)u\big)\, f_{X(t)\Xi(t)}(x,u)\Big] =$$

$$= -\frac{\partial}{\partial u}\, \mathbb{E}^{\theta}\Big[\delta(x - X(t;\theta))\, \delta(u - \Xi(t;\theta))\, \dot{\Xi}(t;\theta)\Big],$$

$$\frac{\partial f_{X(t)X(s)}(x_1,x_2)}{\partial t} + \frac{\partial}{\partial x_1}\Big(h(x_1)\, f_{X(t)X(s)}(x_1,x_2)\Big) =$$

$$= -\frac{\partial}{\partial x_1}\Big(q(x_1)\, \mathbb{E}^{\theta}\Big[\delta(x_1 - X(t;\theta))\, \delta(x_2 - X(s;\theta))\, \Xi(t;\theta)\Big]\Big),\ t \neq s,$$

as well as their counterparts for multidimensional random dynamical systems. Note that, for the derivation of the above equations, ***excitation does need to be Gaussian***. Furthermore, as it will be shown subsequently, the ***stochastic Liouville equations are non-closed***, due to the presence of the averaged terms.

## 2.1 Averages with respect to response-excitation probability measure

Let us consider again the multiplicatively excited, scalar RIVP (1.1)-(1.2), presented in Section 1.1:

$$\dot{X}(t;\theta) = h(X(t;\theta)) + q(X(t;\theta))\, \Xi(t;\theta), \qquad X(t_0;\theta) = X_0(\theta). \qquad \text{(2.1a,b)}$$

In Section 1.1, we also discussed about the existence of the joint response-excitation probability measure $\mathbf{P}_{X(\cdot)\Xi(\cdot)}$, over the Borel $\sigma-$algebra $\mathscr{B}(\mathscr{X} \times \mathscr{Z})$ of the product space $\mathscr{X} \times \mathscr{Z} = C^1\big([t_0,t] \to \mathbb{R}\big) \times C\big([t_0,t] \to \mathbb{R}\big)$, which encapsulates the complete probabil-istic solution of RIVP (2.1a,b). We also assumed that measure $\mathbf{P}_{X(\bullet)\,\Xi(\bullet)}$ is continuously distributed, meaning that the infinite hierarchy of finite dimensional response-excitation pdfs of all orders is well-defined.

[15] A shorter version of the first four Sections was published in the electronic supplementary material of (Mamis, Athanassoulis, & Kapelonis, 2019) as Appendix A.

By virtue of probability measure $\mathbf{P}_{X(\bullet)\,\Xi(\bullet)}$, we define the mean value of a $\mathscr{B}(\mathscr{X}\times\mathscr{Z})-$ measurable mapping (functional) of response and excitation, $\mathcal{G}[X(\bullet|_{t_0}^{t};\theta);\Xi(\bullet|_{t_0}^{t};\theta)]$, as

$$\mathbb{E}^{\theta}_{\mathbf{P}_{X(\bullet)\,\Xi(\bullet)}}\left[\mathcal{G}[X(\bullet|_{t_0}^{t};\theta);\Xi(\bullet|_{t_0}^{t};\theta)]\right] = \\ = \int\limits_{\mathscr{X}\times\mathscr{Z}} \mathcal{G}[\chi(\bullet);\xi(\bullet)]\,\mathbf{P}_{X(\bullet)\,\Xi(\bullet)}\big(d\chi(\bullet)\times d\xi(\bullet)\big). \tag{2.2}$$

In the present work, all mean values appearing are considered as the application of ensemble average operator $\mathbb{E}^{\theta}_{\mathbf{P}_{X(\bullet)\,\Xi(\bullet)}}[\bullet]=\mathbb{E}^{\theta}[\bullet]$, defined by Eq. (2.2).

Let us now consider the $\mathscr{B}(\mathscr{X}\times\mathscr{Z})-$measurable function $G(\boldsymbol{X}(\theta);\boldsymbol{\Xi}(\theta))$, where $\boldsymbol{X}(\theta)$ and $\boldsymbol{\Xi}(\theta)$ are $m-$ and $n-$dimensional random vectors respectively, defined as the response and excitation in multiple (fixed) time instances $s_1,\dots,s_n,\tau_1,\dots,\tau_m\in[t_0,t]$; $\boldsymbol{X}(\theta)=\left[X(\tau_1;\theta)\ \cdots\ X(\tau_m;\theta)\right]^{\mathrm{T}}$ and $\boldsymbol{\Xi}(\theta)=\left[\Xi(s_1;\theta)\ \cdots\ \Xi(s_n;\theta)\right]^{\mathrm{T}}$. Under Volterra's concept of passing from the discrete to continuous (see Appendix A), functional $\mathcal{G}[X(\bullet|_{t_0}^{t};\theta);\Xi(\bullet|_{t_0}^{t};\theta)]$ is the infinite dimensional analogue, for $n,m\to\infty$, of the function $G(\boldsymbol{X}(\theta);\boldsymbol{\Xi}(\theta))$, see also a similar discussion in (Venturi et al., 2012).

By use of Eq. (2.2), the average of random function $G(\boldsymbol{X}(\theta);\boldsymbol{\Xi}(\theta))$ is expressed as

$$\mathbb{E}^{\theta}\left[G(\boldsymbol{X}(\theta);\boldsymbol{\Xi}(\theta))\right] = \\ = \mathbb{E}^{\theta}\left[G\big(X(\tau_1;\theta),\dots,X(\tau_m;\theta);\Xi(s_1;\theta),\dots,\Xi(s_n;\theta)\big)\right] = \\ = \int\limits_{\mathscr{X}\times\mathscr{Z}} G\big(\chi(\tau_1),\dots,\chi(\tau_m;\theta);\xi(s_1),\dots,\xi(s_n)\big)\,\mathbf{P}_{X(\bullet)\,\Xi(\bullet)}\big(d\chi(\bullet)\times d\xi(\bullet)\big). \tag{2.3}$$

Since now the integrand in the right-hand side of Eq. (2.3) depends only on the specific values $\chi(\tau)$ and $\xi(s)$ of the path functions $\chi(\bullet)$ and $\xi(\bullet)$, the infinite-dimensional integral is reduced to a $(n+m)-$dimensional one, with respect to marginal, $(n+m)-$point measure $\mathbf{P}_{X(\tau_1)\cdots X(\tau_m)\,\Xi(s_1)\cdots\Xi(s_n)}$:

$$\mathbb{E}^{\theta}\left[G(\boldsymbol{X}(\theta);\boldsymbol{\Xi}(\theta))\right] = \int\limits_{\mathbb{R}^{n+m}} G(\boldsymbol{w};\boldsymbol{z})\,\mathbf{P}_{X(\tau_1)\cdots X(\tau_m)\,\Xi(s_1)\cdots\Xi(s_n)}\big(d\boldsymbol{w}\times d\boldsymbol{z}\big),$$

which is also written, using the joint pdf $f_{X(\tau_1)\cdots X(\tau_m)\,\Xi(s_1)\cdots\Xi(s_n)}(\boldsymbol{w},\boldsymbol{z})$, whose existence we have previously assumed, as

$$\mathbb{E}^{\theta}\left[G(\boldsymbol{X}(\theta);\boldsymbol{\Xi}(\theta))\right] = \int\limits_{\mathbb{R}^{n+m}} G(\boldsymbol{w};\boldsymbol{z})\,f_{X(\tau_1)\cdots X(\tau_m)\,\Xi(s_1)\cdots\Xi(s_n)}(\boldsymbol{w},\boldsymbol{z})\,d\boldsymbol{w}\,d\boldsymbol{z}. \tag{2.4}$$

By considering now Volterra's passing, in the opposite direction, Eq. (2.4) for $n, m \to \infty$ gives rise to

$$\mathbb{E}^{\theta}_{\mathbf{P}_{X(\bullet)\,\Xi(\bullet)}}\left[\mathcal{G}[X(\bullet|_{t_0}^{t};\theta)\,;\,\Xi(\bullet|_{t_0}^{t};\theta)]\right] = \\ = \int_{\mathcal{X}\times\mathcal{Z}} \mathcal{G}[\chi(\bullet)\,;\,\xi(\bullet)]\, f_{X(\bullet)\,\Xi(\bullet)}[\chi(\bullet)\,;\,\xi(\bullet)]\, d\chi(\bullet)\, d\xi(\bullet), \tag{2.5}$$

which is an equivalent expression of Eq. (2.2), under the assumption that the infinite-dimensional joint response-excitation probability density *functional* $f_{X(\bullet)\,\Xi(\bullet)}[\chi(\bullet)\,;\,\xi(\bullet)]$ exists, see e.g. (Fox, 1986a).

## 2.2 Delta projection method

Representing the one-time pdf of a random function as the average of a random delta function,

$$f_{X(t)}(x) = \left\langle \delta(x - X(t;\theta)) \right\rangle, \tag{2.6}$$

where $\langle \bullet \rangle$ is an appropriate ensemble average operator, is a widely-used practice in statistical mechanics (van Kampen, 2007, Chapter XVI, section 5), stochastic dynamics (Khuri, 2004; Venturi et al., 2012), and the theory of turbulence (Lundgren, 1967), where it is called the ***pdf method***. Herein, the more suggestive term ***delta projection method*** is employed. In this section, we (re)derive formula (2.6) in a more generic way, obtaining also some generalizations of it which are useful in deriving an evolution equation for the response pdf.

Our main point, in this Section, is the identification of operator $\langle \bullet \rangle$ as the ensemble average operator $\mathbb{E}^{\theta}_{\mathbf{P}_{X(\bullet)\,\Xi(\bullet)}}[\bullet] = \mathbb{E}^{\theta}[\bullet]$, defined by Eq. (2.2). Thus, we shall apply Eq. (2.4) to cases where $G(\bullet;\bullet)$ is, or contains, a generalized function, namely the delta function or some derivative of it. Justification of this extension can be made by invoking the theory of generalized stochastic processes; see e.g. (Gelfand & Vilenkin, 1964, Chapter III). Following the tradition in statistical physics, we shall proceed formally, without performing rigorous proofs in the context of the theory of generalized stochastic processes.

Expressing the average of $\delta(x - X(t;\theta))$ via Eq. (2.4), results in

$$\mathbb{E}^{\theta}\left[\delta(x - X(t;\theta))\right] = \int_{\mathbb{R}} \delta(x - w)\, f_{X(t)}(w)\, dw = f_{X(t)}(x), \tag{2.7}$$

which yields the same result as Eq. (2.6), constituting thus its formal derivation. Now, by assuming that functions $g_1(x)$ and $f_{X(t)}(x)$ are continuously differentiable, Eq. (2.4) also provides us with the formula

$$\mathbb{E}^{\theta}\left[\frac{\partial \delta(x - X(t;\theta))}{\partial X(t;\theta)}\, g_1\big(X(t;\theta)\big)\right] = \int_{\mathbb{R}} \frac{\partial \delta(x - w)}{\partial w}\, g_1(w)\, f_{X(t)}(w)\, dw = \\ = -\frac{\partial}{\partial x}\big(g_1(x)\, f_{X(t)}(x)\big). \tag{2.8}$$

By replacing, in the above equation, the first derivative of the delta function by its $k^{\text{th}}$-derivative, we obtain the useful generalization:

$$\mathbb{E}^{\theta}\left[\frac{\partial^k \delta(x-X(t;\theta))}{\partial X^k(t;\theta)}\, g_1\big(X(t;\theta)\big)\right] = (-1)^k \frac{\partial^k}{\partial x^k}\Big(g_1(x)\, f_{X(t)}(x)\Big). \tag{2.9}$$

For Eq. (2.9) to be valid, the functions $g_1(x)$ and $f_{X(t)}(x)$ should possess $k^{\text{th}}$-order continuous derivatives. Let us now examine, via Eq. (2.4), the average of a product of functions that, apart from the $k^{\text{th}}$ derivative of random delta function of $X(t;\theta)$ and the function $g_1\big(X(t;\theta)\big)$, contains also a function of $\Xi(t;\theta)$:

$$\begin{aligned}
&\mathbb{E}^{\theta}\left[\frac{\partial^k \delta(x-X(t;\theta))}{\partial X^k(t;\theta)}\, g_1\big(X(t;\theta)\big)\, g_2\big(\Xi(t;\theta)\big)\right] = \\
&\qquad = \int_{\mathbb{R}^2} \frac{\partial^k \delta(x-w)}{\partial w^k}\, g_1(w)\, g_2(z)\, f_{X(t)\Xi(t)}(w,z)\, dw\, dz = \\
&\qquad = (-1)^k \frac{\partial^k}{\partial x^k}\left(g_1(x) \int_{\mathbb{R}} g_2(z)\, f_{X(t)\Xi(t)}(x,z)\, dz\right) = \\
&\qquad = (-1)^n \frac{\partial^k}{\partial x^k}\Big(g_1(x)\, \mathbb{E}^{\theta}\big[\delta(x-X(t;\theta))\, g_2\big(\Xi(t;\theta)\big)\big]\Big).
\end{aligned} \tag{2.10}$$

The last equality holds true since

$$\int_{\mathbb{R}} g_2(z)\, f_{X(t)\Xi(t)}(x,z)\, dz = \int_{\mathbb{R}^2} \delta(x-w)\, g_2(z)\, f_{X(t)\Xi(t)}(w,z)\, dw\, dz.$$

Note that, for Eq. (2.10) to be valid, functions $g_1(x)$ and $f_{X(t)\Xi(t)}(x,u)$ should possess $k^{\text{th}}$-order continuous derivatives with respect to $x$, while function $g_2(z)$ should only be continuous. Last, we can also consider the random delta function or its derivatives with respect to $\Xi(t;\theta)$, having thus:

$$\begin{aligned}
&\mathbb{E}^{\theta}\left[\frac{\partial^k \delta(x-X(t;\theta))}{\partial X^k(t;\theta)}\, g_1\big(X(t;\theta)\big)\, \frac{\partial^{\ell} \delta(u-\Xi(t;\theta))}{\partial \Xi^{\ell}(t;\theta)}\, g_2\big(\Xi(t;\theta)\big)\right] = \\
&\qquad = \int_{\mathbb{R}^2} \frac{\partial^k \delta(x-w)}{\partial w^k}\, g_1(w)\, \frac{\partial^{\ell} \delta(u-z)}{\partial z^{\ell}}\, g_2(z)\, f_{X(t)\Xi(t)}(w,z)\, dw\, dz = \\
&\qquad = (-1)^{k+\ell} \frac{\partial^{k+\ell}}{\partial x^k \partial u^{\ell}}\Big(g_1(x)\, g_2(u)\, f_{X(t)\Xi(t)}(x,u)\Big).
\end{aligned} \tag{2.11}$$

In Eq. (2.11), $g_1(x)$ should possess $k^{\text{th}}$-order continuous derivatives, $g_2(u)$ should possess $\ell^{\text{th}}$-order continuous derivatives, and the joint pdf $f_{X(t)\Xi(t)}(x,u)$ should posses $k^{\text{th}}$-order continuous derivatives with respect to $x$, and $\ell^{\text{th}}$-order continuous derivatives with respect to $u$.

Let us now consider the case where, in Eq. (2.4), function $G(\boldsymbol{X}(\theta)\,;\boldsymbol{\Xi}(\theta))$ is substituted by

$$\frac{\partial \delta(x - X(\tau_i;\theta))}{\partial X(\tau_i;\theta)} \, G\big(X(\tau_1;\theta), \dots, X(\tau_m;\theta) \,;\, \Xi(s_1;\theta), \dots, \Xi(s_n;\theta)\big),$$

with $\tau_i$ being one of $\tau_1, \tau_2, \dots, \tau_m$. For this case, Eq. (2.4) is specified into

$$\mathbb{E}^{\theta}\left[\frac{\partial \delta(x - X(\tau_i;\theta))}{\partial X(\tau_i;\theta)} \, G(\boldsymbol{X}(\theta) \,;\, \boldsymbol{\Xi}(\theta))\right] =$$

$$= \int_{\mathbb{R}^{n+m}} \frac{\partial \delta(x - w_i)}{\partial w_i} G(\boldsymbol{w};\boldsymbol{z}) \, f_{X(\tau_1)\cdots X(\tau_m)\Xi(s_1)\cdots\Xi(s_n)}(\boldsymbol{w}, \boldsymbol{z}) \, d\boldsymbol{w} \, d\boldsymbol{z}$$

and, by employing the identity for the delta function derivative

$$= -\frac{\partial}{\partial x} \int_{\mathbb{R}^{n+m-1}} G(w_1, \dots, w_{i-1}, x, w_{i+1}, \dots, w_m \,;\, \boldsymbol{z}) \times$$

$$\times \, f_{X(\tau_1)\cdots X(\tau_m)\Xi(s_1)\cdots\Xi(s_n)}(w_1, \dots, w_{i-1}, x, w_{i+1}, \dots, w_m \,;\, \boldsymbol{z}) \, dw_1 \cdots dw_{i-1} dw_{i+1} \cdots dw_m \, d\boldsymbol{z}.$$

The above $(n+m-1)-$integral is easily expressed as an $(n+m)-$integral again

$$= -\frac{\partial}{\partial x} \int_{\mathbb{R}^{n+m}} \delta(x - w_i) \, G(\boldsymbol{w};\boldsymbol{z}) \, f_{X(\tau_1)\cdots X(\tau_m)\Xi(s_1)\cdots\Xi(s_n)}(\boldsymbol{w}, \boldsymbol{z}) \, d\boldsymbol{w} \, d\boldsymbol{z},$$

leading thus to the result

$$\mathbb{E}^{\theta}\left[\frac{\partial \delta(x - X(\tau_i;\theta))}{\partial X(\tau_i;\theta)} \, G(\boldsymbol{X}(\theta) \,;\, \boldsymbol{\Xi}(\theta))\right] = -\frac{\partial}{\partial x} \, \mathbb{E}^{\theta}\Big[\delta(x - X(\tau_i;\theta)) \, G(\boldsymbol{X}(\theta) \,;\, \boldsymbol{\Xi}(\theta))\Big]. \tag{2.12}$$

By using Eq. (2.5), or equivalently considering the continuous analogue, $n, m \to \infty$, of Eq. (2.12), we obtain the following generalization for functionals $\mathcal{G}[X(\bullet|_{t_0}^{t};\theta) \,;\, \Xi(\bullet|_{t_0}^{t};\theta)]$:

$$\mathbb{E}^{\theta}\left[\frac{\partial \delta(x - X(\tau;\theta))}{\partial X(\tau;\theta)} \, \mathcal{G}[X(\bullet|_{t_0}^{t};\theta) \,;\, \Xi(\bullet|_{t_0}^{t};\theta)]\right] = -\frac{\partial}{\partial x} \, \mathbb{E}^{\theta}\Big[\delta(x - X(\tau;\theta)) \, \mathcal{G}[X(\bullet|_{t_0}^{t};\theta) \,;\, \Xi(\bullet|_{t_0}^{t};\theta)]\Big], \tag{2.13}$$

where $\tau \in [t_0, t]$. Following a similar procedure, we also obtain the more general result, for $\tau_1, \tau_2 \in [t_0, t]$, $\tau_1 \neq \tau_2$:

$$\mathbb{E}^{\theta}\left[\frac{\partial \delta(x_1 - X(\tau_1;\theta))}{\partial X(\tau_1;\theta)} \, \delta(x_2 - X(\tau_2;\theta)) \, \mathcal{G}[X(\bullet|_{t_0}^{t};\theta) \,;\, \Xi(\bullet|_{t_0}^{t};\theta)]\right] = -\frac{\partial}{\partial x_1} \, \mathbb{E}^{\theta}\Big[\delta(x_1 - X(\tau_1;\theta)) \, \delta(x_2 - X(\tau_2;\theta)) \, \mathcal{G}[X(\bullet|_{t_0}^{t};\theta) \,;\, \Xi(\bullet|_{t_0}^{t};\theta)]\Big], \tag{2.14}$$

as well as the analogous result for the average containing the random delta function of excitation

$$\mathbb{E}^{\theta}\left[\frac{\partial\delta(u-\Xi(\tau;\theta))}{\partial\Xi(\tau;\theta)}\,\mathcal{G}[X(\bullet|_{t_0}^{t};\theta);\Xi(\bullet|_{t_0}^{t};\theta)]\right]= \\ =-\frac{\partial}{\partial u}\,\mathbb{E}^{\theta}\Big[\delta(u-\Xi(\tau;\theta))\,\mathcal{G}[X(\bullet|_{t_0}^{t};\theta);\Xi(\bullet|_{t_0}^{t};\theta)]\Big], \tag{2.15}$$

**Recapitulation.** In this Section, we derived Eqs. (2.8)-(2.15), which are the identities of averages containing random delta functions. The main convenience in our derivation of the said equations lies in the use of Eq. (2.4), which allowed us to employ the identities of the usual, deterministic delta function and its derivatives. Eqs. (2.8)-(2.15) constitute also a rigorous justification of the ***rules of thumb regarding random delta functions*** employed by e.g. (Cetto, de la Peña, & Velasco, 1984; Klyatskin, 2005; Venturi et al., 2012), which can be summed up as follows:

**i)** The derivative of a random delta function with respect to its random argument is substituted by the derivative with respect to its deterministic argument, and a change in sign.

**ii)** If the argument of a function is the same with the random argument of a random delta function, it is substituted by the deterministic argument of the said random function.

Note that, in (Cetto et al., 1984; Klyatskin, 2005; Sancho et al., 1982), the above rules were applied at non-averaged formulae containing random delta functions, casting thus doubt on their validity.

## 2.3 Derivation of the one-time response stochastic Liouville equation

We shall now apply the delta projection method derived in Section 2.2 to formulate an equation governing the evolution of the pdf $f_{X(t)}(x)$, when $X(t;\theta)$ satisfies RIVP (2.1a,b). By differentiating the first and the last members of Eq. (2.7) with the respect to time $t$, we find

$$\frac{\partial f_{X(t)}(x)}{\partial t}=\frac{\partial}{\partial t}\,\mathbb{E}^{\theta}\big[\delta(x-X(t;\theta))\big]=\mathbb{E}^{\theta}\left[\frac{\partial\delta(x-X(t;\theta))}{\partial X(t;\theta)}\,\dot{X}(t;\theta)\right]. \tag{2.16}$$

The rightmost side of Eq. (2.16) is derived by interchanging differentiation and expectation operators and using chain rule in differentiation. Now, in Eq. (2.16), $\dot{X}(t;\theta)$ is substituted via RDE (2.1a), leading to

$$\frac{\partial f_{X(t)}(x)}{\partial t}=\mathbb{E}^{\theta}\left[\frac{\partial\delta(x-X(t;\theta))}{\partial X(t;\theta)}\,h(X(t;\theta))\right]+ \\ +\mathbb{E}^{\theta}\left[\frac{\partial\delta(x-X(t;\theta))}{\partial X(t;\theta)}\,q(X(t;\theta))\,\Xi(t;\theta)\right]. \tag{2.17}$$

Reformulation of the averages in the right-hand side of Eq. (2.17), using Eqs. (2.9) with $g_1(x)=h(x)$ for the first term, and Eq. (2.11) with $k=1$ and $g_2(u)=h(u)$ for the second term, results in

$$\frac{\partial f_{X(t)}(x)}{\partial t} + \frac{\partial}{\partial x}\left(h(x)\, f_{X(t)}(x)\right) = -\frac{\partial}{\partial x}\left(q(x)\, \mathbb{E}^{\theta}\left[\delta(x - X(t;\theta))\, \Xi(t;\theta)\right]\right). \quad (2.18)$$

Eq. (2.18) is called the ***one-time response stochastic Liouville equation*** (SLE), corresponding to the scalar RIVP (2.1a,b). This equation has been derived by many authors, using various approaches; see e.g. (Cetto et al., 1984; Fox, 1986a; Hänggi, 1978; Sancho & San Miguel, 1980; Sancho et al., 1982; Venturi et al., 2012). Note also that, it is the need for employing Eqs. (2.9), (2.11), for deriving SLE (2.18), that dictates functions $h(x)$, $q(x)$ of RDE (2.1a) to have the additional smoothness of continuous first derivatives.

The initial condition for SLE (2.18) is easily determined, via the initial condition (2.1b), to

$$f_{X(t_0)}(x) = f_{X_0}(x). \quad (2.19)$$

Note that, in most of the existing literature, it is assumed that the initial condition of the response is deterministic; $X_0(\theta) = \alpha = \text{constant}$. This assumption specifies initial condition (2.19) into $f_{X(t_0)}(x) = \delta(x - \alpha)$.

In contrast to the classical Liouville equation (Schwabl, 2006, sec. 1.3.2), (van Kampen, 2007, sec. XVI.5), the SLE (2.18) is not closed, due to the averaged term

$$\mathcal{N}_{\Xi X} = \mathbb{E}^{\theta}\left[\Xi(t;\theta)\, \delta(x - X(t;\theta))\right],$$

appearing in its right-hand side. This fact is better illustrated if the right-hand side of Eq. (2.18) is rewritten as (see the derivation of Eq. (2.11)):

$$\frac{\partial f_{X(t)}(x)}{\partial t} + \frac{\partial}{\partial x}\left(h(x)\, f_{X(t)}(x)\right) = -\frac{\partial}{\partial x}\left(q(x) \int_{\mathbb{R}} z\, f_{X(t)\Xi(t)}(x, z)\, dz\right). \quad (2.20)$$

Eq. (2.20) is clearly not closed, since, apart from $f_{X(t)}(x)$, it also contains the (higher-order) joint response-excitation pdf $f_{X(t)\Xi(t)}(x, z)$. Some ideas for obtaining approximate closures of Eq. (2.20) have been presented in (Athanassoulis, Tsantili, & Kapelonis, 2015; Venturi et al., 2012). Nevertheless, in the present thesis, our goal is to perform an efficient closure for pdf evolution Eq. (2.18) per se, without involving pdfs of higher order, see Chapters 3 and 4.

**Extension for excitation by a polynomial of $\Xi(t;\theta)$.** Despite being unnecessary for the derivations of the present Chapter, excitation $\Xi(t;\theta)$ will be considered subsequently a Gaussian random function. Thus, as we have discussed in Section 1.1, the first step towards non-Gaussian excitations is to be able to consider $\sum_{n=1}^{N} q_n(X(t;\theta))\, \Xi^n(t;\theta)$, instead of $q(X(t;\theta))\, \Xi(t;\theta)$, as excitation to RDE (2.1a). In this case, and by following the same procedure as for deriving SLE (2.18), we easily obtain

$$\frac{\partial f_{X(t)}(x)}{\partial t} + \frac{\partial}{\partial x}\left(h(x)\, f_{X(t)}(x)\right) = -\frac{\partial}{\partial x} \sum_{n=1}^{N} q_n(x)\, \mathbb{E}^{\theta}\left[\delta(x - X(t;\theta))\, \Xi^n(t;\theta)\right]. \quad (2.21)$$

The special case of SLE (2.21) for the case of quadratic noise excitation has been derived in (San Miguel & Sancho, 1981).

## 2.4 Comparison of the stochastic Liouville equation derivation to van Kampen's lemma

A technique popular amongst the physicists, see e.g. (Cetto et al., 1984; San Miguel & Sancho, 1981), for deriving SLE (2.18), is by means of the so-called ***van Kampen's lemma***; see (van Kampen 1975, p.269; van Kampen 1976, p.209; van Kampen 2007, p.411). The rationale behind van Kampen's derivation of the SLE, based on the classical Liouville equation, is distinctively different from the one presented in the previous section, although it leads to the same result. Van Kampen's derivation starts with a pathwise consideration of the given RIVP. That is, Eqs. (21a,b) are considered for every value of the stochastic argument $\theta$ separately:

$$\dot{X}_\theta(t) = h(X_\theta(t)) + q(X_\theta(t))\,\Xi_\theta(t), \quad X_\theta(t_0) = X^0_\theta. \tag{2.22a,b}$$

In this setting, Eqs. (2.22a,b) constitute a ***deterministic initial value problem***, and thus, following (Klyatskin, 2005), Sec. 2.1, we define the indicator function[16]

$$p_\theta(x,t) = \delta\big(x - X_\theta(t)\big),$$

that satisfies the ***classical Liouville equation***

$$\frac{\partial p_\theta(x,t)}{\partial t} = -\frac{\partial}{\partial x}\Big[\big(h(x) + q(x)\,\Xi_\theta(t)\big)\,p_\theta(x,t)\Big], \tag{2.23a}$$

$$p_\theta(x,t_0) = \delta\big(x - X^0_\theta\big); \tag{2.23b}$$

Van Kampen does not follow a formal derivation as Klyatskin's, and does not identify the indicator function as $\delta\big(x - X_\theta(t)\big)$ from the start. Instead, he employs physical arguments after interpreting $p_\theta(x,t)$ as a "flow" and Eq. (2.23a) as a conservation law. Although the argument of van Kampen is somewhat obscure, it can be justified by considering $p_\theta(x,t)$ as a counter of $\theta-$values (outcomes) satisfying the condition $X_\theta(t) = X(t;\theta) = x$ ("flow of outcomes"). By applying, on both Eqs. (2.23a,b) the ensemble average operator $\langle \bullet \rangle$ over all $\theta$, we obtain:

$$\frac{\partial\big\langle p_\theta(x,t)\big\rangle}{\partial t} + \frac{\partial}{\partial x}\Big(h(x)\,\big\langle p_\theta(x,t)\big\rangle\Big) = -\frac{\partial}{\partial x}\Big(q(x)\,\big\langle \Xi(t;\theta)\,p_\theta(x,t)\big\rangle\Big), \tag{2.24a}$$

$$\big\langle p_\theta(x,t_0)\big\rangle = \Big\langle \delta\big(x - X_0(\theta)\big)\Big\rangle. \tag{2.24b}$$

The step that completes this SLE derivation is the identification

$$f_{X(t)}(x) = \big\langle p_\theta(x,t)\big\rangle. \tag{2.25}$$

Eq. (2.25) constitutes the van Kampen's lemma per se. In his works, (van Kampen, *loc. cit.*), van Kampen proves Eq. (2.25) by employing a conservation argument, which is needed since, as we mentioned above, he has to first identify flow $p_\theta(x,t)$ as $\delta\big(x - X_\theta(t)\big)$. Note, how-ever, the Eq. (2.25) has been straightforwardly proven in our approach, see Eq. (2.8), where $\delta\big(x - X(t;\theta)\big)$ has been accepted as a generalized random function from the start.

[16] Note that, in literature, $p_\theta(x,t)$ is also termed as "fine-grained pdf", (Friedrich et al., 2012) Eq. (28), or "raw pdf" (Tatarkovsly & Gremaud, 2017) Eq. (22.8). However, calling a random quantity pdf is somewhat confusing.

From the above discussion, a remarkable conceptual difference between van Kampen's Lemma and our Eq. (2.8) is revealed: in the former approach, the delta function is always treated as a deterministic function, with the probabilistic arguments being provided by the conservation law, while, in the latter, the delta function is interpreted, from the very beginning, as a stochastic function, on which the standard mean-value operator is applied.
In conclusion, although both approaches are interesting, revealing different conceptual settings, we come to believe that delta projection method, presented in Section 2.2, is more comprehensive than van Kampen's lemma, providing a rigorous and systematic way to treat various types of averages occurring in the derivation of the SLE. What is more, delta projection method is easily applicable to deriving other SLEs, as we will discuss subsequently.

## 2.5 Stochastic Liouville equations for higher-order pdfs

Having discussed the way of representing one-time response pdf $f_{X(t)}(x)$ as an average random delta function, Eq. (2.7), as well as formulating the corresponding SLE (2.18), we can now extend the delta projection method for higher-order pdfs, namely the one-time response-excitation pdf $f_{X(t)\Xi(t)}(x,u)$, the response-initial value pdf $f_{X_0 X(t)}(x_0, x_1)$, and the two-times response pdf $f_{X(t)X(s)}(x_1, x_2)$. While, herein, we confine ourselves to the aforementioned higher-order pdfs, the same procedure can be applied for deriving SLEs for any member of the infinite hierarchy of response-excitation pdfs.

### 2.5.1 Stochastic Liouville equation for the one-time response-excitation pdf

By employing Eq. (2.11) for $k = \ell = 1$ and $g_1(\bullet) = g_2(\bullet) = 1$, we can represent the one-time, joint response-excitation pdf as the average of the product of two random delta functions:

$$f_{X(t)\Xi(t)}(x,u) = \mathbb{E}^{\theta}\big[\delta(x - X(t;\theta))\, \delta(u - \Xi(t;\theta))\big]. \tag{2.26}$$

As for the SLE (2.18) of the one-time response pdf, we differentiate both sides of Eq. (2.26) with respect to $t$. This implies that, for formulating the SLE for the one-time response-excitation pdf, $\Xi(t;\theta)$ must belong to $C^1\big([t_0, t] \to \mathbb{R}\big)$, i.e. the paths of excitation must have continuous first time derivatives. Under this additional smoothness requirement for excitation, and by interchanging differentiation with average and using the chain rule, we obtain:

$$\begin{aligned} \frac{\partial f_{X(t)\Xi(t)}(x,u)}{\partial t} &= \mathbb{E}^{\theta}\left[\frac{\partial \delta(x - X(t;\theta))}{\partial X(t;\theta)}\, \delta(u - \Xi(t;\theta))\, \dot{X}(t;\theta)\right] + \\ &\quad + \mathbb{E}^{\theta}\left[\delta(x - X(t;\theta))\, \frac{\partial \delta(u - \Xi(t;\theta))}{\partial \Xi(t;\theta)}\, \dot{\Xi}(t;\theta)\right]. \end{aligned} \tag{2.27}$$

By substituting $\dot{X}(t;\theta)$ via RDE (2.1a), the first average in the right-hand side of Eq. (2.27) is expressed as

$$\mathcal{E}^{\theta}\left[\frac{\partial\delta(x-X(t;\theta))}{\partial X(t;\theta)}\,\delta(u-\Xi(t;\theta))\,\dot{X}(t;\theta)\right]=$$
$$=\mathcal{E}^{\theta}\left[\frac{\partial\delta(x-X(t;\theta))}{\partial X(t;\theta)}\,h(X(t;\theta))\,\delta(u-\Xi(t;\theta))\right]+ \tag{2.28}$$
$$+\mathcal{E}^{\theta}\left[\frac{\partial\delta(x-X(t;\theta))}{\partial X(t;\theta)}\,q(X(t;\theta))\,\delta(u-\Xi(t;\theta))\,\Xi(t;\theta)\right].$$

Now, both averages in the right-hand side of Eq. (2.28) can be further evaluated by using Eq. (2.11), resulting thus into

$$\mathcal{E}^{\theta}\left[\frac{\partial\delta(x-X(t;\theta))}{\partial X(t;\theta)}\,\delta(u-\Xi(t;\theta))\,\dot{X}(t;\theta)\right]=$$
$$=-\frac{\partial}{\partial x}\Big[\big(h(x)+q(x)u\big)\,f_{X(t)\Xi(t)}(x,u)\Big]. \tag{2.29}$$

Let us now move on to the second average in the right-hand side of Eq. (2.27), which contains the first temporal derivative, $\dot{\Xi}(t;\theta)$, of the excitation. First, we have to affirm that the infinite-dimensional response-excitation probability measure $\mathbf{P}_{X(\cdot)\Xi(\cdot)}$, whose existence our approach is based on, contains also the probabilistic structure of $\dot{\Xi}(t;\theta)$. This can be approached in a variety of ways, the most rigorous of which is to invoke the theorem that expresses the $n^{\text{th}}$ order characteristic function of the derivative of a random function via the $(2n)^{\text{th}}$ order characteristic function of the random function per se (Athanassoulis, 2011):

$$\varphi^{(n)}_{\dot{\Xi}(s_1)\cdots\dot{\Xi}(s_n)}(u_1,\dots,u_n)=\lim_{h_1,\dots,h_n\to 0}\varphi^{(2n)}_{\Xi(s_1+h_1)\Xi(s_1)\cdots\Xi(s_n+h_n)\Xi(s_n)}\left(\frac{u_1}{h_1},-\frac{u_1}{h_1},\dots,\frac{u_n}{h_n},-\frac{u_n}{h_n}\right),$$

which holds true for $\Xi(t;\theta)$ being square-differentiable in an open set $T\ni s_1,\dots,s_n$. In our setting, the most suitable, yet formal, representation of $\dot{\Xi}(t;\theta)$ is by using the identity of a temporal delta function:

$$\dot{\Xi}(t;\theta)=\int_{t_0}^{t}\delta'(t-s)\,\Xi(s;\theta)\,ds, \tag{2.30}$$

where $\delta'(t-s)\equiv-\partial\delta(t-s)/\partial s$. Under Eq. (2.30), $\dot{\Xi}(t;\theta)$ is treated formally as a functional of integral type, albeit of singular kernel, depending on the whole history of the excitation. As it will be amply demonstrated in the next Chapter, integral representations like Eq. (2.30) are very convenient when using the Volterra calculus.

Since now $\dot{\Xi}(t;\theta)$ is a functional on $\Xi(t;\theta)$, the second average in the right-hand side of Eq. (2.27) can be further evaluated by employing Eq. (2.15):

$$\mathbb{E}^{\theta}\left[\delta(x-X(t;\theta))\,\frac{\partial\delta(u-\Xi(t;\theta))}{\partial\Xi(t;\theta)}\,\dot{\Xi}(t;\theta)\right]= \\ = -\frac{\partial}{\partial u}\,\mathbb{E}^{\theta}\left[\delta(x-X(t;\theta))\,\delta(u-\Xi(t;\theta))\,\dot{\Xi}(t;\theta)\right]. \tag{2.31}$$

Substitution of Eqs. (2.29) and (2.31) into Eq. (2.27) results into

$$\frac{\partial f_{X(t)\Xi(t)}(x,u)}{\partial t}+\frac{\partial}{\partial x}\left[\left(h(x)+q(x)u\right)\,f_{X(t)\Xi(t)}(x,u)\right]= \\ = -\frac{\partial}{\partial u}\,\mathbb{E}^{\theta}\left[\delta(x-X(t;\theta))\,\delta(u-\Xi(t;\theta))\,\dot{\Xi}(t;\theta)\right]. \tag{2.32}$$

Eq. (2.32) is the ***one-time response-excitation stochastic Liouville equation***. Eq. (2.32) has been also derived, albeit using a different, more convoluted approach, in (Venturi et al., 2012), see equation (2.16). As for SLE (2.18), the initial condition for SLE (2.32) can be easily determined to

$$f_{X(t_0)\Xi(t_0)}(x,u) = f_{X_0\Xi(t_0)}(x,u). \tag{2.33}$$

Having considered the data measure $\mathbf{P}_{X_0\Xi(\cdot)}$ as given, pdf $f_{X_0\Xi(t_0)}(x,u)$ is known. In the special case where the initial values of the response and excitation are independent from one another, initial condition (2.33) is simplified into $f_{X(t_0)\Xi(t_0)}(x,u)=f_{X_0}(x)\,f_{\Xi(t_0)}(u)$.

### 2.5.2 Stochastic Liouville equation for the two-time response pdf

Let us now consider the average of two random delta functions, both of which have as their random argument the response of RIVP (2.1a,b); $\mathbb{E}^{\theta}\left[\delta(x_1-X(t;\theta))\,\delta(x_2-X(s;\theta))\right]$. For the case in which the two time instances are different, $t\neq s$, the above average can be expressed in the usual way, using Eq. (2.4), as

$$\mathbb{E}^{\theta}\left[\delta(x_1-X(t;\theta))\,\delta(x_2-X(s;\theta))\right]= \\ = \int_{\mathbb{R}^2}\delta(x_1-w_1)\,\delta(x_2-w_2)\,f_{X(t)X(s)}(w_1,w_2)\,dw_1dw_2 = f_{X(t)X(s)}(x_1,x_2). \tag{2.34}$$

Thus, Eq. (2.34) constitutes the delta projection representation of the two-time response pdf $f_{X(t)X(s)}(x_1,x_2)$. Note however, that representation (2.34) fails for $t=s$, since

$$\mathbb{E}^{\theta}\left[\delta(x_1-X(t;\theta))\,\delta(x_2-X(t;\theta))\right]=\int_{\mathbb{R}}\delta(x_1-w_1)\,\delta(x_2-w_1)\,f_{X(t)}(w_1)\,dw_1, \tag{2.35}$$

in which the single integral containing the two delta functions is not defined. On the other hand, the fact that delta projection (2.34) fails for $t=s$ does not diminish the importance of formulating evolution equations for $f_{X(t)X(s)}(x_1,x_2)$, since it is the $t\neq s$ case that interests us; for $t=s$, $f_{X(t)X(t)}(x_1,x_2)$ is just a duplication of the one-time response pdf $f_{X(t)}(x)$, whose evolution equation was formulated using the one-time SLE (2.18).

Let us now differentiate the first and the last sides of Eq. (2.34) with respect to time $t$, while time $s$ is treated as parameter:

$$\frac{\partial f_{X(t)X(s)}(x_1, x_2)}{\partial t} = \mathbb{E}^{\theta}\left[\frac{\partial \delta(x_1 - X(t;\theta))}{\partial t}\,\delta(x_2 - X(s;\theta))\right].$$

Then, by using the chain rule

$$\frac{\partial f_{X(t)X(s)}(x_1, x_2)}{\partial t} = \mathbb{E}^{\theta}\left[\frac{\partial \delta(x_1 - X(t;\theta))}{\partial X(t;\theta)}\,\dot{X}(t;\theta)\,\delta(x_2 - X(s;\theta))\right],$$

and expressing $\dot{X}(t;\theta)$ via RDE (2.1a), we obtain

$$\begin{aligned}\frac{\partial f_{X(t)X(s)}(x_1, x_2)}{\partial t} &= \mathbb{E}^{\theta}\left[\frac{\partial \delta(x_1 - X(t;\theta))}{\partial X(t;\theta)}\,h(X(t;\theta))\delta(x_2 - X(s;\theta))\right] + \\ &+ \mathbb{E}^{\theta}\left[\frac{\partial \delta(x_1 - X(t;\theta))}{\partial X(t;\theta)}\,q(X(t;\theta))\,\delta(x_2 - X(s;\theta))\,\Xi(t;\theta)\right].\end{aligned} \tag{2.36}$$

In the right-hand side of Eq. (2.36), both averages can be expressed using Eq. (2.4) and evaluated further by performing the manipulations presented in Section 2.2:

$$\begin{aligned}&\mathbb{E}^{\theta}\left[\frac{\partial \delta(x_1 - X(t;\theta))}{\partial X(t;\theta)}\,h(X(t;\theta))\,\delta(x_2 - X(s;\theta))\right] = \\ &= \int_{\mathbb{R}^2} \frac{\partial \delta(x_1 - w_1)}{\partial w_1}\,h(w_1)\,\delta(x_2 - w_2)\,f_{X(t)X(s)}(w_1, w_2)\,dw_1 dw_2 = \\ &= -\frac{\partial}{\partial x_1}\Big(h(x_1)\,f_{X(t)X(s)}(x_1, x_2)\Big).\end{aligned} \tag{2.37}$$

and

$$\begin{aligned}&\mathbb{E}^{\theta}\left[\frac{\partial \delta(x_1 - X(t;\theta))}{\partial X(t;\theta)}\,q(X(t;\theta))\,\delta(x_2 - X(s;\theta))\,\Xi(t;\theta)\right] = \\ &= \int_{\mathbb{R}^3} \frac{\partial \delta(x_1 - w_1)}{\partial w_1}\,q(w_1)\,\delta(x_2 - w_2)\,z\,f_{X(t)X(s)\Xi(t)}(w_1, w_2, z)\,dw_1 dw_2 dz = \\ &= -\frac{\partial}{\partial x_1}\left(q(x_1)\int_{\mathbb{R}^2}\delta(x_2 - w_2)\,z\,f_{X(t)X(s)\Xi(t)}(x_1, w_2, z)\,dw_2 dz\right) = \\ &= -\frac{\partial}{\partial x_1}\left(q(x_1)\int_{\mathbb{R}^3}\delta(x_1 - w_1)\,\delta(x_2 - w_2)\,z\,f_{X(t)X(s)\Xi(t)}(w_1, w_2, z)\,dw_1 dw_2 dz\right) = \\ &= -\frac{\partial}{\partial x_1}\Big(q(x_1)\,\mathbb{E}^{\theta}\big[\delta(x_1 - X(t;\theta))\,\delta(x_2 - X(s;\theta))\,\Xi(t;\theta)\big]\Big).\end{aligned} \tag{2.38}$$

Substitution of Eqs. (2.37), (2.38) into Eq. (2.36) results into the ***two-time response stochastic Liouville equation***:

$$\frac{\partial f_{X(t)X(s)}(x_1, x_2)}{\partial t} + \frac{\partial}{\partial x_1}\left(h(x_1)\, f_{X(t)X(s)}(x_1, x_2)\right) = \\ = -\frac{\partial}{\partial x_1}\left(q(x_1)\, \mathbb{E}^{\theta}\left[\delta(x_1 - X(t;\theta))\,\delta(x_2 - X(s;\theta))\,\Xi(t;\theta)\right]\right). \tag{2.39}$$

Consistent with the delta representation (2.34), SLE (2.39) is a differential equation with respect to time $t$, while time $s$ is a parameter, for which the condition $t \neq s$ applies. As a consequence, the initial condition of SLE (2.39) should also be parametric with respect to $s$:

$$f_{X(t_0)X(s)}(x_1, x_2) = f_{X_0 X(s)}(x_1, x_2). \tag{2.40}$$

In Eq. (2.40), and contrary to the initial conditions (2.19) and (2.33) of the SLEs for one-time response and one-time response-excitation pdfs, we observe that the pdf in the right-hand side, namely the joint response-initial value pdf $f_{X_0 X(s)}(x_1, x_2)$, is not part of the data of RIVP (2.1a,b), since it models the statistical dependence of the response at any time instance with its initial value. Thus, for determining pdf $f_{X_0 X(s)}(x_1, x_2)$, we have to solve another pdf evolution equation, starting by formulating the stochastic Liouville equation for the response-initial value pdf in the following subsection.

Eq. (2.39) is the same SLE for two-times also derived in (Hernandez-Machado et al., 1983). However, in the aforementioned work, as well as in others of the same research team (Sagués et al., 1984; Sancho & San Miguel, 1989), parameter time $s$ is always considered before evolution time $t$; $s < t$. While such an assumption may be conceptually more convenient, the delta projection method also works for $s > t$. Note also that, under the assumption $s < t$, the aforementioned works consider as initial condition not Eq. (2.40), but $f_{X(s)X(s)}(x_1, x_2) = f_{X(s)}(x_1)\,\delta(x_1 - x_2)$, i.e. for $t = s$. Of course, $f_{X(s)}(x_1)$ in this initial condition is also not known a priori, and has to be calculated by solving another evolution equation, this time for the one-time response pdf. Thus, for two-time response pdfs, the scheme of calculating the initial condition by solving another evolution equation is present both in the existing literature and in our approach.

### 2.5.3 Stochastic Liouville equation for the response-initial value pdf

As usual, we commence again from the delta representation for response-initial value pdf $f_{X_0 X(t)}(x_0, x_1)$ via Eq. (2.4):

$$\mathbb{E}^{\theta}\left[\delta(x_0 - X_0(\theta))\,\delta(x_1 - X(t;\theta))\right] = \\ = \int_{\mathbb{R}^2} \delta(x_0 - w_0)\,\delta(x_1 - w_1)\, f_{X_0 X(t)}(w_0, w_1)\, dw_0\, dw_1 = f_{X_0 X(t)}(x_0, x_1). \tag{2.41}$$

Differentiation of Eq. (2.41) with respect to $t$ results in

$$\frac{\partial f_{X_0 X(t)}(x_0, x_1)}{\partial t} = \mathbb{E}^{\theta}\left[\delta(x_0 - X_0(\theta))\,\frac{\partial \delta(x_1 - X(t;\theta))}{\partial t}\right],$$

and, as previously, we perform the chain rule for the derivative of the delta function and substitute $\dot{X}(t;\theta)$ from RDE (2.1a):

$$\frac{\partial f_{X_0 X(t)}(x_0, x_1)}{\partial t} = \mathbb{E}^{\theta}\left[\delta(x_0 - X_0(\theta))\,\frac{\partial \delta(x_1 - X(t;\theta))}{\partial X(t;\theta)}\,h(X(t;\theta))\right] + \\ + \mathbb{E}^{\theta}\left[\delta(x_0 - X_0(\theta))\,\frac{\partial \delta(x_1 - X(t;\theta))}{\partial X(t;\theta)}\,q(X(t;\theta))\,\Xi(t;\theta)\right]. \tag{2.42}$$

As in the derivation of the two-time SLE (2.39), the averages in the right-hand side of Eq. (2.42) are further evaluated as

$$\mathbb{E}^{\theta}\left[\delta(x_0 - X_0(\theta))\,\frac{\partial \delta(x_1 - X(t;\theta))}{\partial X(t;\theta)}\,h(X(t;\theta))\right] = \\ = \int_{\mathbb{R}^2} \delta(x_0 - w_0)\,\frac{\partial \delta(x_1 - w_1)}{\partial w_1}\,h(w_1)\,f_{X_0 X(t)}(w_0, w_1)\,dw_0\,dw_1 = \\ = -\frac{\partial}{\partial x_1}\Big(h(x_1)\,f_{X_0 X(t)}(x_0, x_1)\Big). \tag{2.43}$$

and

$$\mathbb{E}^{\theta}\left[\delta(x_0 - X_0(\theta))\,\frac{\partial \delta(x_1 - X(t;\theta))}{\partial X(t;\theta)}\,q(X(t;\theta))\,\Xi(t;\theta)\right] = \\ = \int_{\mathbb{R}^3} \delta(x_0 - w_0)\,\frac{\partial \delta(x_1 - w_1)}{\partial w_1}\,q(w_1)\,z\,f_{X_0 X(t)\Xi(t)}(w_0, w_1, z)\,dw_0\,dw_1\,dz = \\ = -\frac{\partial}{\partial x_1}\left(q(x_1)\int_{\mathbb{R}^2} \delta(x_0 - w_0)\,z\,f_{X_0 X(t)\Xi(t)}(w_0, x, z)\,dw_0\,dz\right) = \\ = -\frac{\partial}{\partial x_1}\left(q(x_1)\int_{\mathbb{R}^3} \delta(x_0 - w_0)\,\delta(x_1 - w_1)\,z\,f_{X_0 X(t)\Xi(t)}(w_0, w_1, z)\,dw_0\,dw_1\,dz\right) = \\ = -\frac{\partial}{\partial x_1}\Big(q(x_1)\,\mathbb{E}^{\theta}\big[\delta(x_0 - X_0(\theta))\,\delta(x_1 - X(t;\theta))\,\Xi(t;\theta)\big]\Big). \tag{2.44}$$

Thus, Eq. (2.42), under Eqs. (2.43), (2.44), is expressed as

$$\frac{\partial f_{X_0 X(t)}(x_0, x_1)}{\partial t} + \frac{\partial}{\partial x_1}\Big(h(x_1)\,f_{X_0 X(t)}(x_0, x_1)\Big) = \\ = -\frac{\partial}{\partial x_1}\Big(q(x_1)\,\mathbb{E}^{\theta}\big[\delta(x_0 - X_0(\theta))\,\delta(x_1 - X(t;\theta))\,\Xi(t;\theta)\big]\Big). \tag{2.45}$$

Eq. (2.45) is the ***response-initial time stochastic Liouville equation***. SLE (2.45), in turn, is supplemented by the initial condition:

$$f_{X_0 X(t_0)}(x_0, x_1) = f_{X_0 X_0}(x_0, x_1). \tag{2.46}$$

In the right-hand side of initial condition (2.46), the pdf $f_{X_0 X_0}(x_0, x_1)$ appears, which is the duplication of initial value pdf $f_{X_0}(x_0)$. By identifying the conditional probability distribution as $f_{X_0|X_0}(x_1|\, x_0) = \delta(x_0 - x_1)$, Eq. (2.46) is elaborated as

$$\begin{aligned} f_{X_0 X(t_0)}(x_0, x_1) &= f_{X_0 X_0}(x_0, x_1) = f_{X_0|X_0}(x_1|\, x_0)\, f_{X_0}(x_0) = \\ &= f_{X_0}(x_0)\, \delta(x_0 - x_1). \end{aligned} \tag{2.47}$$

Thus, the initial distribution for SLE (2.45) is the one-dimensional $f_{X_0}(x_0)$, placed on the diagonal $x_0 = x_1$ of the two-dimensional plane $(x_0, x_1)$. Note also that, for SLE (2.45), there is also an obvious long-time behaviour of solution: $f_{X_0 X(t)}(x_0, x_1) = f_{X_0}(x_0)\, f_{X(t)}(x_1)$. This "final" condition states the fact that, in the steady-state, $X(t;\theta)$ is independent from its initial value.

## 2.6 Stochastic Liouville equations for multidimensional dynamical systems

The formulation of SLEs via the delta projection method is also applicable to the multidimensional random dynamical systems discussed in Section 1.1:

$$\dot{X}_n(t;\theta) = h_n\big(\boldsymbol{X}(t;\theta)\big) + \sum_{k=1}^{K} q_{nk}\big(\boldsymbol{X}(t;\theta)\big)\, \Xi_k(t;\theta), \tag{2.48a}$$

$$X_n(t_0;\theta) = X_n^0(\theta), \qquad n = 1, \dots, N, \tag{2.48b}$$

where response $\boldsymbol{X}(t;\theta)$ and excitation $\boldsymbol{\Xi}(t;\theta)$ are *N*- and *K*-dimensional vector random functions respectively. As for its scalar counterpart, Eqs. (2.1a,b), the solution to RIVP (2.48a,b) is associated with infinite-dimensional, joint response-excitation probability measure $\mathbf{P}_{\boldsymbol{X}(\bullet)\boldsymbol{\Xi}(\bullet)}$. Thus, by plausibly assuming again that all finite dimensional response-excitation pdfs exist, the average of a function $G\big(\boldsymbol{X}(\tau;\theta)\,;\boldsymbol{\Xi}(s;\theta)\big)$, $\tau, s \in [t_0, t]$ is expressed as

$$\mathbb{E}^\theta\Big[G\big(\boldsymbol{X}(\tau;\theta)\,;\boldsymbol{\Xi}(s;\theta)\big)\Big] = \int_{\mathbb{R}^{N+K}} G(\boldsymbol{w};\boldsymbol{z})\, f_{\boldsymbol{X}(\tau)\boldsymbol{\Xi}(s)}(\boldsymbol{w}, \boldsymbol{z})\, d\boldsymbol{w}\, d\boldsymbol{z}. \tag{2.50}$$

By employing now formula (2.50) to define the average for the multidimensional random delta function $\delta(\boldsymbol{x} - \boldsymbol{X}(t;\theta)) = \delta(x_1 - X_1(t;\theta)) \cdots \delta(x_N - X_N(t;\theta))$, we obtain

$$\mathbb{E}^\theta\big[\delta(\boldsymbol{x} - \boldsymbol{X}(t;\theta))\big] = \int_{\mathbb{R}^N} \delta(\boldsymbol{x} - \boldsymbol{w})\, f_{\boldsymbol{X}(t)}(\boldsymbol{w})\, d\boldsymbol{w} = f_{\boldsymbol{X}(t)}(\boldsymbol{x}). \tag{2.51}$$

Eq. (2.51) constitutes the delta representation of the response pdf for the multidimensional case. Differentiation of its first and last members with respect to time $t$, and a subsequent interchange of the differentiation and average operators, results in

$$\frac{\partial f_{\boldsymbol{X}(t)}(\boldsymbol{x})}{\partial t} = \mathbb{E}^\theta\left[\frac{\partial \delta(\boldsymbol{x} - \boldsymbol{X}(t;\theta))}{\partial t}\right] = \mathbb{E}^\theta\left[\frac{\partial}{\partial t}\Big(\delta(x_1 - X_1(t;\theta)) \cdots \delta(x_N - X_N(t;\theta))\Big)\right].$$

Using now the product rule of differentiation and the linearity of the average operator,

$$\frac{\partial f_{\boldsymbol{X}(t)}(\boldsymbol{x})}{\partial t} = \sum_{n=1}^{N} \mathbb{E}^{\theta}\left[\frac{\partial \delta(x_n - X_n(t;\theta))}{\partial t} \prod_{\substack{\ell=1\\ \ell\neq n}}^{N} \delta(x_\ell - X_\ell(t;\theta))\right],$$

as well as the chain rule for $\partial \delta(x_n - X_n(t;\theta)/\partial t$,

$$\frac{\partial f_{\boldsymbol{X}(t)}(\boldsymbol{x})}{\partial t} = \sum_{n=1}^{N} \mathbb{E}^{\theta}\left[\dot{X}_n(t;\theta) \frac{\partial \delta(x_n - X_n(t;\theta))}{\partial X_n(t;\theta)} \prod_{\substack{\ell=1\\ \ell\neq n}}^{N} \delta(x_\ell - X_\ell(t;\theta))\right]$$

is obtained. As in the derivation of the scalar SLE, $\dot{X}_n(t;\theta)$ is substituted by using RDE (2.48):

$$\begin{aligned}\frac{\partial f_{\boldsymbol{X}(t)}(\boldsymbol{x})}{\partial t} &= \sum_{n=1}^{N} \mathbb{E}^{\theta}\left[h_n\big(\boldsymbol{X}(t;\theta)\big) \frac{\partial \delta(x_n - X_n(t;\theta))}{\partial X_n(t;\theta)} \prod_{\substack{\ell=1\\ \ell\neq n}}^{N} \delta(x_\ell - X_\ell(t;\theta))\right] + \\ &+ \sum_{n=1}^{N}\sum_{k=1}^{K} \mathbb{E}^{\theta}\left[q_{nk}\big(\boldsymbol{X}(t;\theta)\big)\, \Xi_k(t;\theta) \frac{\partial \delta(x_n - X_n(t;\theta))}{\partial X_n(t;\theta)} \prod_{\substack{\ell=1\\ \ell\neq n}}^{N} \delta(x_\ell - X_\ell(t;\theta))\right].\end{aligned} \tag{2.52}$$

The averages in the right-hand side of Eq. (2.52) are evaluated further using Eq. (2.50) as

$$\mathbb{E}^{\theta}\left[h_n\big(\boldsymbol{X}(t;\theta)\big) \frac{\partial \delta(x_n - X_n(t;\theta))}{\partial X_n(t;\theta)} \prod_{\substack{\ell=1\\ \ell\neq n}}^{N} \delta(x_\ell - X_\ell(t;\theta))\right] = -\frac{\partial}{\partial x_n}\Big(h_n(\boldsymbol{x})\, f_{\boldsymbol{X}(t)}(\boldsymbol{x})\Big), \tag{2.53}$$

and

$$\begin{aligned}&\mathbb{E}^{\theta}\left[q_{nk}\big(\boldsymbol{X}(t;\theta)\big)\, \Xi_k(t;\theta) \frac{\partial \delta(x_n - X_n(t;\theta))}{\partial X_n(t;\theta)} \prod_{\substack{\ell=1\\ \ell\neq n}}^{N} \delta(x_\ell - X_\ell(t;\theta))\right] = \\ &= \int_{\mathbb{R}^{N+1}} q_{nk}(\boldsymbol{w}) \frac{\partial \delta(x_n - w_n)}{\partial w_n} \prod_{\substack{\ell=1\\ \ell\neq n}}^{N} \delta(x_\ell - w_\ell)\, z\, f_{\boldsymbol{X}(t)\Xi_k(t)}(\boldsymbol{w}, z)\, d\boldsymbol{w}\, dz = \\ &= -\frac{\partial}{\partial x_n}\left(q_{nk}(\boldsymbol{x}) \int_{\mathbb{R}} z\, f_{\boldsymbol{X}(t)\Xi_k(t)}(\boldsymbol{x}, z)\, dz\right) = -\frac{\partial}{\partial x_n}\left(q_{nk}(\boldsymbol{x}) \int_{\mathbb{R}^{N+1}} \delta(\boldsymbol{x}-\boldsymbol{w})\, z\, f_{\boldsymbol{X}(t)\Xi_k(t)}(\boldsymbol{x}, z)\, dz\right) = \\ &= -\frac{\partial}{\partial x_n}\Big(q_{nk}(\boldsymbol{x})\, \mathbb{E}^{\theta}\big[\delta(\boldsymbol{x}-\boldsymbol{X}(t;\theta))\, \Xi_k(t;\theta)\big]\Big).\end{aligned} \tag{2.54}$$

By substituting Eqs. (2.53), (2.54) into Eq. (2.52) we have

$$\frac{\partial f_{\boldsymbol{X}(t)}(\boldsymbol{x})}{\partial t} + \sum_{n=1}^{N} \frac{\partial}{\partial x_n}\Big(h_n(\boldsymbol{x})\, f_{\boldsymbol{X}(t)}(\boldsymbol{x})\Big) = = -\sum_{n=1}^{N}\sum_{k=1}^{K} \frac{\partial}{\partial x_n}\Big(q_{nk}(\boldsymbol{x})\, \mathbb{E}^{\theta}\Big[\delta(\boldsymbol{x}-\boldsymbol{X}(t;\theta))\, \Xi_k(t;\theta)\Big]\Big). \tag{2.55}$$

Eq. (2.55) is the ***multidimensional response stochastic Liouville equation***, also derived in (Cetto et al., 1984; Dekker, 1982; Fox, 1983; Garrido & Sancho, 1982; San Miguel & Sancho, 1980). Its initial condition is obviously $f_{\boldsymbol{X}(t)}(\boldsymbol{x}) = f_{\boldsymbol{X}^0}(\boldsymbol{x})$.

For the case of multidimensional RIVP (2.48a,b), SLEs for pdfs of higher order, such as $f_{\boldsymbol{X}(t)\boldsymbol{\Xi}(t)}(\boldsymbol{x},\boldsymbol{u})$, $f_{\boldsymbol{X}(t)\boldsymbol{X}(s)}(\boldsymbol{x}^{(1)},\boldsymbol{x}^{(2)})$ and $f_{\boldsymbol{X}^0\boldsymbol{X}(t)}(\boldsymbol{x}^{(0)},\boldsymbol{x}^{(1)})$ can be also formulated, using the same methodology as above, with the multidimensional two-time response SLE been presented before in (Hernandez-Machado et al., 1983). Here, we state these SLEs, omitting their derivation:

***Multidimensional one-time response-excitation SLE***

$$\frac{\partial f_{\boldsymbol{X}(t)\boldsymbol{\Xi}(t)}(\boldsymbol{x},\boldsymbol{u})}{\partial t} + \sum_{n=1}^{N} \frac{\partial}{\partial x_n}\left[\left(h_n(\boldsymbol{x}) + \sum_{k=1}^{K} q_{nk}(\boldsymbol{x})\, u_k\right) f_{\boldsymbol{X}(t)\boldsymbol{\Xi}(t)}(\boldsymbol{x},\boldsymbol{u})\right] = = -\sum_{k=1}^{K} \frac{\partial}{\partial u_k}\Big(\mathbb{E}^{\theta}\Big[\delta(\boldsymbol{x}-\boldsymbol{X}(t;\theta))\, \delta(\boldsymbol{u}-\boldsymbol{\Xi}(t;\theta))\, \dot{\Xi}_k(t;\theta)\Big]\Big). \tag{2.56}$$

***Multidimensional two-time response SLE***

$$\frac{\partial f_{\boldsymbol{X}(t)\boldsymbol{X}(s)}(\boldsymbol{x}^{(1)},\boldsymbol{x}^{(2)})}{\partial t} + \sum_{n=1}^{N} \frac{\partial}{\partial x_n^{(1)}}\Big(h_n(\boldsymbol{x}^{(1)})\, f_{\boldsymbol{X}(t)\boldsymbol{X}(s)}(\boldsymbol{x}^{(1)},\boldsymbol{x}^{(2)})\Big) = = -\sum_{n=1}^{N}\sum_{k=1}^{K} \frac{\partial}{\partial x_n^{(1)}}\Big(q_{nk}(\boldsymbol{x}^{(1)})\, \mathbb{E}^{\theta}\Big[\delta(\boldsymbol{x}^{(1)}-\boldsymbol{X}(t;\theta))\, \delta(\boldsymbol{x}^{(2)}-\boldsymbol{X}(s;\theta))\, \Xi_k(t;\theta)\Big]\Big). \tag{2.57}$$

***Multidimensional response-initial time SLE***

$$\frac{\partial f_{\boldsymbol{X}^0\boldsymbol{X}(t)}(\boldsymbol{x}^{(0)},\boldsymbol{x}^{(1)})}{\partial t} + \sum_{n=1}^{N} \frac{\partial}{\partial x_n^{(1)}}\Big(h_n(\boldsymbol{x}^{(1)})\, f_{\boldsymbol{X}^0\boldsymbol{X}(t)}(\boldsymbol{x}^{(0)},\boldsymbol{x}^{(1)})\Big) = = -\sum_{n=1}^{N}\sum_{k=1}^{K} \frac{\partial}{\partial x_n^{(1)}}\Big(q_{nk}(\boldsymbol{x}^{(1)})\, \mathbb{E}^{\theta}\Big[\delta(\boldsymbol{x}^{(0)}-\boldsymbol{X}_0(\theta))\, \delta(\boldsymbol{x}^{(1)}-\boldsymbol{X}(t;\theta))\, \Xi_k(t;\theta)\Big]\Big). \tag{2.58}$$

We easily observe that, by considering the scalar case, $N = K = 1$, Eqs. (2.55)-(2.58) simplify into their scalar counterparts, SLEs (2.18), (2.33), (2.39).

# Chapter 3: Correlation splitting and the Novikov-Furutsu theorem

**Summary.** In this chapter[17], we present and generalize the Novikov-Furutsu (NF) theorem, which is the main mathematical tool for evaluating the averages contained in the SLEs, under the assumption that ***excitation and initial value are jointly Gaussian***. Classical NF theorem evaluates further averages of the form $\mathbb{E}^{\theta}\left[\Xi(t;\theta)\,\mathcal{F}[\Xi(\bullet|_{t_0}^{t};\theta)]\right]$, where $\mathcal{F}$ is a functional with zero-mean Gaussian argument. First, in Sec. 3.4, NF theorem is extended for $\mathbb{E}^{\theta}\left[\Xi(s;\theta)\mathcal{F}[X_0(\theta);\Xi(\bullet|_{t_0}^{t};\theta)]\right]$, where $\mathcal{F}$ is a now function on $X_0(\theta)$ and a functional on $\Xi(\bullet;\theta)$, with its arguments being jointly Gaussian, with non-zero mean values. Furthermore, the generalization for $\mathcal{F}$ with vector arguments, as well as the extension for averages of the form $\mathbb{E}^{\theta}\left[\dot{\Xi}(s;\theta)\,\mathcal{F}[X_0(\theta);\Xi(\bullet|_{t_0}^{t};\theta)]\right]$, are also presented. Last, in Sec. 3.5, NF theorem is generalized for averages containing multiple time instances of excitation; $\mathbb{E}^{\theta}\left[\Xi(s_1;\theta)\cdots\Xi(s_n;\theta)\mathcal{F}[X_0(\theta);\Xi(\bullet|_{t_0}^{t};\theta)]\right]$. To the best of our knowledge, this generalization has not been presented before and contains, as a special case, the well-known Isserlis theorem for the moments of a Gaussian process.

## 3.1 The response as a function of initial value and a functional on excitation

Each SLE, derived in Chapter 2, contains an average which, in the scalar case, is one of

$$\mathbb{E}^{\theta}\left[\delta(x-X(t;\theta))\,\Xi(t;\theta)\right],\qquad \mathbb{E}^{\theta}\left[\delta(x-X(t;\theta))\,\delta(u-\Xi(t;\theta))\,\dot{\Xi}(t;\theta)\right],$$

$$\mathbb{E}^{\theta}\left[\delta(x_1-X(t;\theta))\,\delta(x_2-X(s;\theta))\,\Xi(t;\theta)\right],$$

while for the case of excitation by a polynomial of the noise $\Xi(t;\theta)$, averages of the form $\mathbb{E}^{\theta}\left[\delta(x-X(t;\theta))\,\Xi^{n}(t;\theta)\right]$ appear, see SLE (2.21). The goal of the present Chapter, is to evaluate further the above averages, as well as their counterparts in the multidimensional SLEs, in order to be in a suitable form for an approximate closure scheme to be applied.

Our starting point for such an evaluation is a change in how we perceive response $X(t;\theta)$. By utilizing the fact that $X(t;\theta)$ is the solution to RIVP (1.1)-(1.2):

[17] An earlier version of sections 3.3, 3.4 were published as (Athanassoulis & Mamis, 2019).

$$\dot{X}(t;\theta) = h(X(t;\theta)) + q(X(t;\theta))\,\Xi(t;\theta), \qquad X(t_0;\theta) = X_0(\theta), \qquad \text{(3.1a,b)}$$

we may consider it a function of initial value $X_0(\theta)$, as well as a functional on the excitation $\Xi(\bullet;\theta)$ over the time interval $[t_0, t]$ (from the initial time $t_0$ up to current time $t$). Although an explicit form for $X(t;\theta)$ cannot be determined in general, since RDE (3.1a) is nonlinear, the ***function-functional*** (FFℓ) ***character of the response*** can be denoted as $X[X_0(\theta);\Xi(\bullet|_{t_0}^{t};\theta)]$.

Considering the response as an FFℓ of initial value and excitation affects the way of defining the average of a measurable function $G(X(t;\theta);\Xi(t;\theta))$. Previously, in Chapter 2, this average was defined as an integral with respect to the two point, joint ***response-excitation probability measure*** $\mathbf{P}_{X(t)\,\Xi(t)}$, see Eq. (2.3):

$$\mathbb{E}^{\theta}\big[G(X(t;\theta);\Xi(t;\theta))\big] = \int_{\mathbb{R}^2} G(w;z)\,\mathbf{P}_{X(t)\,\Xi(t)}\big(dw\times dz\big). \qquad (3.2)$$

On the other hand, in this Chapter and under $X(t;\theta) = X[X_0(\theta);\Xi(\bullet|_{t_0}^{t};\theta)]$, function $G(X(t;\theta);\Xi(t;\theta))$ is also considered as an FFℓ with random arguments; $\mathcal{G}[X_0(\theta);\Xi(\bullet|_{t_0}^{t};\theta)]$[18]. Thus, its average is alternatively defined as an integral with respect to the joint ***initial value-excitation probability measure*** $\mathbf{P}_{X_0\,\Xi(t)}$:

$$\mathbb{E}^{\theta}\big[G(X(t;\theta);\Xi(t;\theta))\big] =$$
$$= \mathbb{E}^{\theta}\Big[\mathcal{G}[X_0(\theta);\Xi(\bullet|_{t_0}^{t};\theta)]\Big] = \int_{\mathbb{R}\times\mathscr{X}} \mathcal{G}[\chi_0;\xi(\bullet)]\,\mathbf{P}_{X_0\,\Xi(\bullet)}\big(d\chi_0\times d\xi(\bullet)\big). \qquad (3.3)$$

This change of perspective, from Eq. (3.2) to Eq. (3.3), is also implicitly performed in all of our main references, see e.g. (Cetto et al., 1984; Hänggi & Jung, 1995; Sancho et al., 1982), justifying also the statement in (Hänggi, 1978) that "the average $\mathbb{E}^{\theta}\big[\delta(x - X(t;\theta))\big]$ is over all the realizations of the stochastic driving force $\Xi(t;\theta)$ and over the initial probability $f_{X_0}(x)$ of the distributed starting value $X_0(\theta)$".

In this way, the averages in the SLEs shown above, can be seen special cases of

$$\mathbb{E}^{\theta}\Big[\mathcal{F}[X_0(\theta);\Xi(\bullet|_{t_0}^{t};\theta)]\,\Xi(s;\theta)\Big], \quad \mathbb{E}^{\theta}\Big[\mathcal{F}[X_0(\theta);\Xi(\bullet|_{t_0}^{t};\theta)]\,\dot{\Xi}(s;\theta)\Big].$$

Such averages, as well as more general forms, e.g. $\mathbb{E}^{\theta}\Big[\Xi(s_1;\theta)\cdots\Xi(s_n;\theta)\,\mathcal{F}[\cdots]\Big]$, will be evaluated subsequently in the present Chapter, under the assumption that random arguments $X_0(\theta)$, $\Xi(\bullet;\theta)$ are jointly Gaussian.

---

[18] A more detailed notation would be $G(X(t;\theta);\Xi(t;\theta)) = \mathcal{G}[X_0(\theta), \Xi(t;\theta);\Xi(\bullet|_{t_0}^{t};\theta)]$, denoting also the explicit dependence on $\Xi(t;\theta)$. However, this is unnecessary, since the possibility of an explicit dependence on a value $\Xi(s;\theta)$, $s\in[t_0, t]$ will considered covered by notation $\mathcal{G}[X_0(\theta);\Xi(\bullet|_{t_0}^{t};\theta)]$.

**On the smoothness of FFℓs.** An FFℓ with random arguments, $\mathcal{G}[X_0(\theta)\,;\,\Xi(\bullet|_{t_0}^{t}\,;\theta)]$, will be henceforth called a random function-functional (RFFℓ), while the term FFℓ will be reserved for its deterministic counterpart, $\mathcal{G}[\upsilon\,;\,u(\bullet|_{t_0}^{t})]:\mathbb{R}\times\mathcal{Z}\to\mathbb{R}$, where $\mathcal{Z}=C\big([t_0,t]\to\mathbb{R}\big)$ is the space of continuous functions. FFℓ $\mathcal{G}[\upsilon\,;\,u(\bullet|_{t_0}^{t})]$ is assumed to have derivatives of any order, both with respect to the scalar argument $\upsilon$ and the function argument $u(\bullet)$, and thus it is expandable in Volterra-Taylor series, jointly with respect to $\upsilon$ and $u(\bullet)$, around a fixed pair $\big(\upsilon_0\,;\,u_0(\bullet)\big)$. Here, functional derivatives are considered in the sense of Volterra (Volterra, 1930), which are reconsidered in a rigorous manner by (Donsker & Lions, 1962). For the sake of brevity, the above smoothness conditions of $\mathcal{G}[\upsilon\,;\,u(\bullet|_{t_0}^{t})]$ will be denoted as $C^{\infty}$.

Furthermore, RFFℓ $\mathcal{G}[X_0(\theta)\,;\,\Xi(\bullet|_{t_0}^{t}\,;\theta)]$ is obtained by replacing the argument $\big(\upsilon\,;\,u(\bullet)\big)$ of $\mathcal{G}[\upsilon\,;\,u(\bullet|_{t_0}^{t})]$ by the random element $\big(X_0(\theta)\,;\,\Xi(\bullet\,;\theta)\big)$. The latter is fully described by the infinite-dimensional joint probability measure $\mathbf{P}_{X_0\,\Xi(\bullet)}$, defined over the Borel $\sigma-$algebra $\mathcal{B}(\mathbb{R}\times\mathcal{Z})$. Since $\big(X_0(\theta)\,;\,\Xi(\bullet;\theta)\big):\theta\to\mathbb{R}\times\mathcal{Z}$ is Borel measurable, and $\mathcal{G}[\upsilon\,;\,u(\bullet|_{t_0}^{t})]$ is $C^{\infty}$, the composition $\mathcal{G}[X_0(\theta)\,;\,\Xi(\bullet|_{t_0}^{t}\,;\theta)]$ is also Borel measurable. Thus, under the above description, RFFℓ $\mathcal{G}[X_0(\theta)\,;\,\Xi(\bullet|_{t_0}^{t}\,;\theta)]$, whose deterministic counterpart $\mathcal{G}[\upsilon\,;\,u(\bullet|_{t_0}^{t})]$ is $C^{\infty}$, will be subsequently termed as "sufficiently smooth".

At this point, we also have to mention the discrepancy between the requirement for $C^{\infty}$ smoothness and the fact that, in SLE averages, the RFFℓ is a random delta function. A rigorous treatment of this point requires the formulation of the results in the context of generalized random functions, see (Gelfand & Vilenkin, 1964) Chapter 3, and their proof by reduction to the dual space of $C^{\infty}$ spaces. This analysis involves Gelfand-Shilov spaces and will not be performed herein. Alternatively, and in accordance with our practice in the previous Chapter 2, we shall proceed formally by considering derivatives of the delta function as usual ones.

## 3.2 The classical Novikov-Furutsu theorem. Novikov's proof

The most widely-used formula for evaluating averages of form $\mathbb{E}^{\theta}\Big[\mathcal{F}[\Xi(\bullet|_{t_0}^{t}\,;\theta)]\,\Xi(t\,;\theta)\Big]$ is the ***Novikov-Furutsu*** (NF) ***theorem*** derived independently in (Furutsu, 1963; Novikov, 1965):[19]

$$\mathbb{E}^{\theta}\Big[\mathcal{F}[\Xi(\bullet|_{t_0}^{t}\,;\theta)]\,\Xi(t\,;\theta)\Big]=\int_{t_0}^{t}C_{\Xi\Xi}(t,\tau)\,\mathbb{E}^{\theta}\left[\frac{\delta\mathcal{F}[\Xi(\bullet|_{t_0}^{t}\,;\theta)]}{\delta\Xi(\tau\,;\theta)}\right]d\tau\,, \tag{3.4}$$

[19] It is also interesting to see the classical NF theorem, Eq. (3.4), as the infinite-dimensional analogue of *Stein's lemma*, (Barbour & Chen, 2005) which states that a scalar Gaussian random variable $X(\theta)$ with zero mean value and unit variance satisfies the equation $\mathbb{E}^{\theta}[\,f'(X(\theta))-X(\theta)\,f(X(\theta))]=0$ for every $f\in C^{1}$.

where $\Xi(\bullet;\theta)$ is a ***zero-mean Gaussian random function*** with autocovariance function $C_{\Xi\Xi}(t,s)$, $\mathcal{F}[\Xi(\bullet|_{t_0}^{t};\theta)]$ is a ***functional*** of $\Xi(\bullet;\theta)$ over $[t_0,t]$, and $\delta/\delta\Xi(\tau;\theta)$ is the Volterra functional derivative with respect to $\Xi(\bullet;\theta)$ at $\tau$.

We observe that, by using the NF theorem, the explicit dependence of the averaged quantity on $\Xi(t;\theta)$ has been eliminated, resulting in an integral expression that "measures" (via the functional derivative) the effect of variations in $\Xi(\tau;\theta)$, $\tau\in[t_0,t]$, on $\mathcal{F}[\Xi(\bullet|_{t_0}^{t};\theta)]$, weighted by the autocorrelation $C_{\Xi\Xi}(t,\tau)$ between $\Xi(\tau;\theta)$ at any time instance $\tau\in[t_0,t]$, and the value $\Xi(t;\theta)$ that appeared explicitly in the left-hand side of Eq. (3.4).

In other words, NF theorem constitutes a formula of ***correlation splitting*** (Klyatskin, 2005) ch. 4. By this term, we describe the evaluation of averages containing functionals of random arguments, in terms of the probabilistic structure of their arguments and the analytic (deterministic) properties of the functional per se, see e.g. (Bochkov, Dubkov, & Malakhov, 1977), (M. Scott, 2013) sec. 11.5. The possibility of correlation splitting is a commonly recurring question in many fields of physics and engineering where smoothly-correlated (coloured) noises are encountered, e.g. *turbulent diffusion* (Cook, 1978; Hyland, McKee, & Reeks, 1999; Krommes, 2015; Martins Afonso, Mazzino, & Gama, 2016; Shrimpton, Haeri, & Scott, 2014), *random waves* (Bobryk, 1993; Creamer, 2008; Konotop & Vazquez, 1994; Rino, 1991; Sobczyk, 1985), and *stochastic dynamics* (Kliemann & Sri Namachchivaya, 2018; Klyatskin, 2005, 2015; Protopopescu, 1983; M. Scott, 2013).

**Use of discrete NF theorem in statistical linearization.** The *N*-dimensional analogue of NF theorem (3.4), formulated for the Gaussian random vector $\boldsymbol{X}(\theta)$ as $\mathbb{E}^{\theta}[\boldsymbol{X}(\theta)\,f(\boldsymbol{X}(\theta))]=$ $\boldsymbol{C}_{\boldsymbol{XX}}\,\mathbb{E}^{\theta}[\nabla f(\boldsymbol{X}(\theta))]$, with $\nabla$ being the gradient operator, is used in statistical linearization for determining the characteristics of a linear element that is equivalent to a nonlinear, memoryless element under Gaussian random excitation (Roberts & Spanos, 2003, sec. 5.2.1). There, it is not mentioned as NF theorem, and its derivation is attributed to (Kazakov, 1965).

**Use of NF theorem in deriving pdf evolution equations.** NF theorem, Eq. (3.4), has been employed for evaluating SLEs further in various previous works, e.g. in (Hänggi & Jung, 1995; Peacock-López et al., 1988; Sancho et al., 1982). In these works, excitation $\Xi(\bullet;\theta)$ was assumed as zero-mean valued, while the initial value was assumed deterministic. Thus the response could be treated as a functional on excitation only, and not as an RFFℓ. These assumptions made the NF theorem, in its classical form of Eq. (3.4), to be readily applicable to SLE averages. In the following Sections, NF theorem is extended for RFFℓs and for non-zero mean valued $\Xi(\bullet;\theta)$.

**Novikov's proof.** Now, for comparison purposes with our proof presented subsequently, we present how Novikov proved formula (3.4) in (Novikov, 1965), repeated also in (M. Scott, 2013).[20] The first step is to expand functional $\mathcal{F}[\Xi(\bullet|_{t_0}^{t};\theta)]$ into ***Volterra-Taylor series*** (see Appendix A) around $\Xi_0(\bullet;\theta)=0$:

[20] At this point, and for the sake of completeness, we have to mention that the derivation of NF theorem by Furutsu poses the following peculiarity. NF theorem is equation (5.18) of (Furutsu, 1963), where it is stated that it

$$\mathcal{F}[\Xi(\bullet|_{t_0}^{t};\theta)] = \sum_{m=0}^{\infty}\frac{1}{m!}\int_{t_0}^{t}\overset{(m)}{\cdots}\int_{t_0}^{t}\left.\frac{\delta^m \mathcal{F}[\Xi_0(\bullet|_{t_0}^{t};\theta)]}{\delta\Xi_0(\tau_1;\theta)\cdots\delta\Xi_0(\tau_m;\theta)}\right|_{\Xi_0(\bullet;\theta)=0}\times$$
$$\times\,\Xi(\tau_1;\theta)\cdots\Xi(\tau_m;\theta)\,d\tau_1\cdots d\tau_m. \tag{3.5}$$

Then, by multiplying both sides of Eq. (3.5) with $\Xi(t;\theta)$ and taking the average, we obtain

$$\mathbb{E}^{\theta}\Big[\mathcal{F}[\Xi(\bullet|_{t_0}^{t};\theta)]\,\Xi(t;\theta)\Big] = \sum_{m=0}^{\infty}\frac{1}{m!}\int_{t_0}^{t}\overset{(m)}{\cdots}\int_{t_0}^{t}\left.\frac{\delta^m \mathcal{F}[\Xi_0(\bullet|_{t_0}^{t};\theta)]}{\delta\Xi_0(\tau_1;\theta)\cdots\delta\Xi_0(\tau_m;\theta)}\right|_{\Xi_0(\bullet;\theta)=0}\times$$
$$\times\,\mathbb{E}^{\theta}[\Xi(t;\theta)\,\Xi(\tau_1;\theta)\cdots\Xi(\tau_m;\theta)]\,d\tau_1\cdots d\tau_m. \tag{3.6}$$

Note that, in the right-hand side of Eq. (3.6), the functional derivative of $\mathcal{F}[\Xi_0(\bullet|_{t_0}^{t};\theta)]$ is outside the average operator, since it is evaluated for the deterministic argument $\Xi_0(\bullet;\theta)=0$. Now, moment $\mathbb{E}^{\theta}[\Xi(t;\theta)\,\Xi(\tau_1;\theta)\cdots\Xi(\tau_m;\theta)]$ is evaluated further by employing ***Isserlis theorem*** (Isserlis, 1918) for the moments of a zero-mean valued Gaussian process. For an odd number of terms inside the average (i.e. for even $m$ in our case) the said moment equals to zero, while for even number of terms (odd $m$) the moment is expressed as a sum of products of second-order moments, in which $t,\tau_1,\dots,\tau_m$ are arranged in all possible pairwise combinations. Thus, for both the even and odd cases, and for $m\geq 1$, we may use the recursive relation:

$$\mathbb{E}^{\theta}[\Xi(t;\theta)\,\Xi(\tau_1;\theta)\cdots\Xi(\tau_m;\theta)] =$$
$$= \sum_{\ell=1}^{m} C_{\Xi\Xi}(t,\tau_\ell)\,\mathbb{E}^{\theta}[\Xi(\tau_1;\theta)\cdots\Xi(\tau_{\ell-1};\theta)\,\Xi(\tau_{\ell+1};\theta)\cdots\Xi(\tau_m;\theta)], \tag{3.7}$$

by which the $(m+1)-$order moment of the zero-mean valued Gaussian process is expressed via the $(m-1)-$order moment. By substituting Eq. (3.7), as well as $\mathbb{E}^{\theta}[\Xi(t;\theta)]=0$ (for the term with $m=0$), into Eq. (3.6), and after some algebraic manipulations, we have

$$\mathbb{E}^{\theta}\Big[\mathcal{F}[\Xi(\bullet|_{t_0}^{t};\theta)]\,\Xi(t;\theta)\Big] =$$
$$= \sum_{m=1}^{\infty}\frac{1}{m!}\sum_{\ell=1}^{m}\int_{t_0}^{t}C_{\Xi\Xi}(t,\tau_\ell)\int_{t_0}^{t}\overset{(m-1)}{\cdots}\int_{t_0}^{t}\left.\frac{\delta^m \mathcal{F}[\Xi_0(\bullet|_{t_0}^{t};\theta)]}{\delta\Xi_0(\tau_1;\theta)\cdots\delta\Xi_0(\tau_m;\theta)}\right|_{\Xi_0(\bullet;\theta)=0}\times$$
$$\times\,\mathbb{E}^{\theta}[\Xi(\tau_1;\theta)\cdots\Xi(\tau_{\ell-1};\theta)\,\Xi(\tau_{\ell+1};\theta)\cdots\Xi(\tau_m;\theta)]\,d\tau_1\cdots d\tau_m. \tag{3.8}$$

In each term of the $\ell-$sum, we perform the following changes of integration variables: $\tau_\ell=\tau$, $\tau_k=\tau_{k-1}$ for all $k=\ell+1,\dots m$. Under these changes, it is easy to see that all terms in the $\ell-$sum are equal, resulting thus into

---

will be proven in the next paper of this series. However, the proof of NF theorem is absent from his next paper (Furutsu, 1964).

$$\mathbb{E}^{\theta}\left[\mathcal{F}[\Xi(\bullet|_{t_0}^{t};\theta)]\,\Xi(t;\theta)\right] =$$

$$= \int_{t_0}^{t} C_{\Xi\Xi}(t,\tau) \left[ \sum_{m=1}^{\infty} \frac{1}{(m-1)!} \int_{t_0}^{t} \overset{(m-1)}{\cdots} \int_{t_0}^{t} \left. \frac{\delta^m \mathcal{F}[\Xi_0(\bullet|_{t_0}^{t};\theta)]}{\delta\Xi_0(\tau_1;\theta)\cdots\delta\Xi_0(\tau_{m-1};\theta)\,\delta\Xi_0(\tau;\theta)} \right|_{\Xi_0(\bullet;\theta)=0} \times \right.$$

$$\left. \times\, \mathbb{E}^{\theta}[\Xi(\tau_1;\theta)\cdots\Xi(\tau_{m-1};\theta)]\, d\tau_1 \cdots d\tau_{m-1} \right] d\tau. \tag{3.9}$$

By performing the change in index $k = m-1$ and using the linearity of average operator, Eq. (3.9) is written equivalently as

$$\mathbb{E}^{\theta}\left[\mathcal{F}[\Xi(\bullet|_{t_0}^{t};\theta)]\,\Xi(t;\theta)\right] =$$

$$= \int_{t_0}^{t} C_{\Xi\Xi}(t,\tau)\, \mathbb{E}^{\theta} \left[ \sum_{k=0}^{\infty} \frac{1}{k!} \int_{t_0}^{t} \overset{(k)}{\cdots} \int_{t_0}^{t} \left. \frac{\delta^k}{\delta\Xi_0(\tau_1;\theta)\cdots\delta\Xi_0(\tau_k;\theta)} \frac{\delta\mathcal{F}[\Xi_0(\bullet|_{t_0}^{t};\theta)]}{\delta\Xi_0(\tau;\theta)} \right|_{\Xi_0(\bullet;\theta)=0} \times \right.$$

$$\left. \times\, \Xi(\tau_1;\theta)\cdots\Xi(\tau_k;\theta)\, d\tau_1 \cdots d\tau_k \right] d\tau. \tag{3.10}$$

Finally, the infinite series in the right-hand side of Eq. (3.10) is identified, via Eq. (3.5), as the Volterra-Taylor expansion for $\delta\mathcal{F}[\Xi(\bullet|_{t_0}^{t};\theta)]/\delta\Xi(\tau;\theta)$, resulting thus into NF theorem, Eq. (3.4). This completes Novikov's proof.

### 3.3 The average of a random function-functional as a pseudo-differential operator

Our approach in proving an extension of the NF theorem, Eq. (3.4), for averages of the form $\mathbb{E}^{\theta}\left[\mathcal{F}[X_0(\theta);\Xi(\bullet|_{t_0}^{t};\theta)]\,\Xi(s;\theta)\right]$, and for $\Xi(s;\theta)$ being a Gaussian process with non-zero mean value, is based on a more general result, stated as follows.

**Theorem 3.1: The average of an RFFℓ.** The average of a general, nonlinear RFFℓ, $\mathcal{G}[X_0(\theta);\Xi(\bullet|_{t_0}^{t};\theta)]$, whose random arguments are possibly non-Gaussian, can be expressed as a pseudo-differential operator, defined in terms of the characteristic function-functional of its arguments:

$$\mathbb{E}^{\theta}\left[\mathcal{G}[X_0(\theta);\Xi(\bullet|_{t_0}^{t};\theta)]\right] = \varphi_{\hat{X}_0\hat{\Xi}(\bullet)}\left(\frac{\partial}{i\,\partial\upsilon};\frac{\delta}{i\,\delta u(\bullet)}\right)\mathcal{G}[\upsilon;u(\bullet|_{t_0}^{t})]\Big|_{\substack{\upsilon=m_{X_0}\\ u(\bullet)=m_{\Xi}(\bullet)}} =$$

$$= \mathbb{E}^{\theta}\left[\exp\left(\hat{X}_0(\theta)\frac{\partial}{\partial\upsilon} + \int_{t_0}^{t} ds\,\hat{\Xi}(s;\theta)\frac{\delta}{\delta u(s)}\right)\right]\mathcal{G}[\upsilon;u(\bullet|_{t_0}^{t})]\Big|_{\substack{\upsilon=m_{X_0}\\ u(\bullet)=m_{\Xi}(\bullet)}} \tag{3.11}$$

where $m_{X_0}$ and $m_{\Xi}(s)$ are the mean values of $X_0(\theta)$ and $\Xi(\bullet;\theta)$, respectively, $\partial/\partial\upsilon$ denotes partial differentiation with respect to $\upsilon$, and $\delta/\delta u(s)$ denotes Volterra functional differentiation with respect to the function $u(\bullet)$ at $s$. Further, the quantities

$$\hat{X}_0(\theta) = X_0(\theta) - m_{X_0} \qquad \hat{\Xi}(s;\theta) = \Xi(s;\theta) - m_{\Xi}(s) \qquad \text{(3.12a,b)}$$

are the fluctuations of the random elements $X_0(\theta)$ and $\Xi(s;\theta)$ around their mean values, and $\varphi_{\hat{X}_0 \hat{\Xi}(\bullet)}[\upsilon\,;u(\bullet)]$ is the ***joint characteristic FFℓ*** of the said fluctuations:

$$\varphi_{\hat{X}_0 \hat{\Xi}(\bullet)}[\upsilon\,;u(\bullet|_{t_0}^{t})] = \mathbb{E}^{\theta}\left[\exp\left(i\,\hat{X}_0(\theta)\,\upsilon + i\int_{t_0}^{t}\hat{\Xi}(s;\theta)\,u(s)\,ds\right)\right], \qquad (3.13)$$

where $i$ is the imaginary unit, $i = \sqrt{-1}$. ■

**Remark 3.1:** By comparing formula (3.11) to Eq. (3.3), we observe that, by Theorem 3.1, we compute an integration over an infinite-dimensional space by the action of a pseudo-differential operator. A similar practice of ***integration by differentiation in functional spaces***, albeit with quite different applications, has been recently introduced by (Jia, Tang, & Kempf, 2017; Kempf, Jackson, & Morales, 2014, 2015) in quantum field theory.

**Remark 3.2:** A formula similar to Eq. (3.11) for a random functional of the form $\mathcal{G}[\Xi(\bullet\,;\theta)] = \mathcal{G}_1[\Xi(\bullet\,;\theta)]\,\mathcal{G}_2[\Xi(\bullet\,;\theta)]$ is given by (Klyatskin, 2005), Ch. 4. The proof given by Klyatskin is concise and unmotivated, thus difficult to follow for a reader not well-acquainted with Volterra functional calculus. Eq. (3.11) includes Klyatskin's result as a special case. The proof presented herein is fairly detailed and motivated by easily understood analogues in the discrete setting. Then, the Volterra technique of passing from the discrete to continuous reveals the appropriate forms in the function-functional setting.

The proof of Theorem 3.1 is completed in paragraph 3.3.2, and is based on the expansion of the RFFℓ $\mathcal{G}[X_0(\theta)\,;\Xi(\bullet|_{t_0}^{t}\,;\theta)]$ around the mean values $m_{X_0}$ and $m_{\Xi}(\bullet)$, by employing the function-functional shift operator. The latter is introduced in the following paragraph 3.3.1, via Volterra's concept of passing from the discrete to continuous. This principle motivates a calculus for functionals that is analogous to the calculus for functions of many variables, which can be (and has been) rigorous; see also Appendix A for relevant discussion and references.

### 3.3.1 The function-functional shift operator and its exponential form

As a first step for constructing the shift operator for FFℓs, we introduce its discrete analogue, the multivariate shift operator $\mathcal{T}_{\hat{x}}$ for functions of many variables, $g(\boldsymbol{x}_0 + \hat{\boldsymbol{x}}) = \mathcal{T}_{\hat{x}}\, g(\boldsymbol{x}_0)$, which is in fact a compact form of expressing the Taylor series expansion.

Let $g(\bullet) \in C^{\infty}(\mathbb{R}^{N+1} \to \mathbb{R})$. Then, the Taylor expansion of $g(\boldsymbol{x}) = g(\boldsymbol{x}_0 + \hat{\boldsymbol{x}})$ around the point $\boldsymbol{x}_0 \in \mathbb{R}^{N+1}$, reads as

$$g(\boldsymbol{x}_0 + \hat{\boldsymbol{x}}) = \left(1 + \sum_{m=1}^{\infty}\frac{1}{m!}\sum_{n_1=1}^{N+1}\overset{(m)}{\cdots}\sum_{n_m=1}^{N+1}\hat{x}_{n_1}\cdots\hat{x}_{n_m}\frac{\partial^m}{\partial x_{n_1}\cdots\partial x_{n_m}}\right) g(\boldsymbol{x}_0). \qquad (3.14)$$

To write the above equation in a more concise form, the following identity is used

$$\left(\sum_{n=1}^{N+1} h_n\,u_n\right)^m \equiv \prod_{i=1}^{m}\sum_{n_i=1}^{N+1} h_{n_i}\,u_{n_i} = \sum_{n_1=1}^{N+1}\overset{(m)}{\cdots}\sum_{n_m=1}^{N+1} h_{n_1}\cdots h_{n_m}\cdot u_{n_1}\cdots u_{n_m}\,, \qquad (3.15)$$

which is valid under the assumption that all symbols $h_{n_i}, u_{n_i}$ commute, or their order is definitely prescribed, as dictated by the specific case. Setting $h_n = \hat{x}_n$ and $u_n = \partial \bullet / \partial x_n$, and imposing the order convention that all $\hat{x}_n$ precede of all $\partial \bullet / \partial x_n$, we can apply Eq. (3.15) to express the differential operator appearing in the right-hand side of Eq. (3.14), recasting the latter in the following compact form:

$$g(\boldsymbol{x}_0 + \hat{\boldsymbol{x}}) = \left[ \sum_{m=0}^{\infty} \frac{1}{m!} \left( \sum_{n=1}^{N+1} \hat{x}_n \frac{\partial}{\partial x_n} \right)^m \right] g(\boldsymbol{x}_0). \tag{3.16}$$

In Eq. (3.16) the additional (trivial) convention $\left( \sum_{n=1}^{N+1} \hat{x}_n \, \partial \bullet / \partial x_n \right)^0 = 1$ is used. Then, we observe that the infinite series appearing in the right-hand side of Eq. (3.16), under the said order assumption, is identified as the Taylor series of the symbol $\exp\left( \sum_{n=1}^{N+1} \hat{x}_n \, \partial \bullet / \partial x_n \right)$, leading to the nice, compact formula

$$g(\boldsymbol{x}_0 + \hat{\boldsymbol{x}}) = \exp\left( \sum_{n=1}^{N+1} \hat{x}_n \frac{\partial}{\partial x_n} \right) g(\boldsymbol{x}_0) = \exp\left( \hat{\boldsymbol{x}} \cdot \nabla_{\boldsymbol{x}} \right) g(\boldsymbol{x}_0). \tag{3.17}$$

Comparing Eq. (3.17) with $g(\boldsymbol{x}_0 + \hat{\boldsymbol{x}}) = \mathcal{T}_{\hat{\boldsymbol{x}}} \, g(\boldsymbol{x}_0)$, we see that the symbol $\exp\left( \hat{\boldsymbol{x}} \cdot \nabla_{\boldsymbol{x}} \bullet \right)$ can be identified as the shift operator for the case of functions of many variables, with $\hat{\boldsymbol{x}}$ being its shift argument. This is the multivariate analogue of the shift (translation) operator $\exp(\hat{x} \, d \bullet / dx)$ for functions of one variable, first introduced by Lagrange; see also (Glaeske, Prudnikov and Skòrnik, 2006, Sec. 1.1). Denoting the first component of $\boldsymbol{x}$ by (the scalar) $\upsilon$, and the remaining ones with the $N-$dimensional vector $\boldsymbol{u}$, Eq. (3.17) is recast as

$$G(\upsilon; \boldsymbol{u}) = G(\upsilon_0 + \hat{\upsilon}; \boldsymbol{u}_0 + \hat{\boldsymbol{u}}) = \mathcal{T}_{\hat{\upsilon}\hat{\boldsymbol{u}}} G(\upsilon_0; \boldsymbol{u}_0) =$$
$$= \exp\left( \hat{\upsilon} \frac{\partial}{\partial \upsilon} + \sum_{n=1}^{N} \hat{u}_n \frac{\partial}{\partial u_n} \right) G(\upsilon_0; \boldsymbol{u}_0) = \exp\left( \hat{\upsilon} \frac{\partial}{\partial \upsilon} + \hat{\boldsymbol{u}} \cdot \nabla_{\boldsymbol{u}} \right) G(\upsilon_0; \boldsymbol{u}_0), \tag{3.18}$$

providing us with the shift operator for a function of a scalar and a vector argument, $G(\upsilon; \boldsymbol{u})$. Now, by setting in the above equation $u_n = u(t_n)$, where $u(t)$ is a continuous function, and applying Volterra's approach to pass from the discrete to continuous, we observe that the function $G(\upsilon; \boldsymbol{u})$ tends to the function-functional $\mathcal{G}[\upsilon; u(\bullet|_{t_0}^{t})]$ and the sum $\sum_{n=1}^{N} \hat{u}_n \, \partial / \partial u_n$ tends to the integral $\int_{t_0}^{t} \hat{u}(s) \, \delta / \delta u(s) \, ds$, where $\delta / \delta u(s)$ is the Volterra functional derivative. Thus, Eq. (3.18) suggests the following exponential form of the shift operator $\mathcal{T}_{\hat{\upsilon}\hat{u}(\bullet)} \bullet$ when applied to the FFℓ $\mathcal{G}[\upsilon; u(\bullet|_{t_0}^{t})]$:

$$\mathcal{G}[\upsilon; u(\bullet|_{t_0}^{t})] = \mathcal{G}[\upsilon_0 + \hat{\upsilon}; u_0(\bullet|_{t_0}^{t}) + \hat{u}(\bullet|_{t_0}^{t})] =$$
$$= \mathcal{T}_{\hat{\upsilon}\hat{u}(\bullet)} \mathcal{G}[\upsilon_0; u_0(\bullet|_{t_0}^{t})] = \tag{3.19}$$
$$= \exp\left( \hat{\upsilon} \frac{\partial}{\partial \upsilon} + \int_{t_0}^{t} \hat{u}(s) \frac{\delta}{\delta u(s)} ds \right) \mathcal{G}[\upsilon_0; u_0(\bullet|_{t_0}^{t})].$$

By omitting the scalar argument $\upsilon$, Eq. (3.19) defines the functional shift operator $\mathcal{T}_{\hat{u}(\bullet)}\,\bullet$, derived in (Sobczyk 1985, chap.I, Eq. 1.116'), where Volterra's passing was also employed. A rigorous derivation of Eq. (3.19) can be obtained by direct use of the Volterra-Taylor series expansion of $\mathcal{G}[\upsilon\,;u(\bullet|_{t_0}^{t})]$; see Appendix A.

### 3.3.2 Proof of Theorem 3.1

Equation (3.19) is the essential deterministic prerequisite for the proof of Theorem 3.1. Substituting, in Eq. (3.19), the arguments $\upsilon$, $u(\bullet)$ by the random arguments $X_0(\theta)$, $\Xi(\bullet;\theta)$, we obtain the following representation of the RFFℓ $\mathcal{G}[X_0(\theta)\,;\Xi(\bullet|_{t_0}^{t};\theta)]$:

$$\mathcal{G}[X_0(\theta)\,;\Xi(\bullet|_{t_0}^{t};\theta)] \equiv \mathcal{G}[m_{X_0}+\hat{X}_0(\theta)\,;m_{\Xi}(\bullet|_{t_0}^{t})+\hat{\Xi}(\bullet|_{t_0}^{t};\theta)] \equiv$$

$$\equiv \mathcal{T}_{\hat{X}_0\hat{\Xi}(\bullet)}(\theta)\Big[\,\mathcal{G}[m_{X_0}\,;m_{\Xi}(\bullet|_{t_0}^{t})]\Big] =$$

$$= \exp\left(\hat{X}_0(\theta)\,\frac{\partial}{\partial\upsilon}+\int_{t_0}^{t} ds\;\hat{\Xi}(s;\theta)\,\frac{\delta}{\delta u(s)}\right)\mathcal{G}[m_{X_0}\,;m_{\Xi}(\bullet|_{t_0}^{t})]. \tag{3.20}$$

Recall that $m_{X_0}$, $m_{\Xi}(\bullet)$ are the mean values, and $\hat{X}_0(\theta)$, $\hat{\Xi}(\bullet;\theta)$ are the fluctuations of the random elements $X_0(\theta)$, $\Xi(\bullet;\theta)$ around their mean values; see Eqs. (3.12a,b).

Eq. (3.20) provides us with a decomposition of the RFFℓ $\mathcal{G}[X_0(\theta)\,;\Xi(\bullet;\theta)]$ in the form of a *shift operator with random shift arguments* $\hat{X}_0(\theta)$, $\hat{\Xi}(\bullet;\theta)$, *"times" the deterministic FFℓ* $\mathcal{G}[m_{X_0}\,;m_{\Xi}(\bullet)]$. By averaging both sides of Eq. (3.20), we obtain

$$\mathbb{E}^{\theta}\Big[\mathcal{G}[X_0(\theta)\,;\Xi(\bullet|_{t_0}^{t};\theta)]\Big] =$$
$$= \mathbb{E}^{\theta}\left[\exp\left(\hat{X}_0(\theta)\,\frac{\partial}{\partial\upsilon}+\int_{t_0}^{t} ds\;\hat{\Xi}(s;\theta)\,\frac{\delta}{\delta u(s)}\right)\right]\mathcal{G}[m_{X_0}\,;m_{\Xi}(\bullet|_{t_0}^{t})]. \tag{3.21}$$

Finally, by recalling the form of the joint characteristic FFℓ of fluctuations, Eq. (3.13), we see that Eq. (3.21) can be also written as

$$\mathbb{E}^{\theta}\Big[\mathcal{G}[X_0(\theta)\,;\Xi(\bullet|_{t_0}^{t};\theta)]\Big] = \varphi_{\hat{X}_0\hat{\Xi}(\bullet)}\left[\frac{\partial}{i\,\partial\upsilon}\,;\frac{\delta}{i\,\delta u(s)}\right]\mathcal{G}[m_{X_0}\,;m_{\Xi(\bullet)}(\bullet|_{t_0}^{t})],$$

which is exactly Eq. (3.11). Thus, the proof of Theorem 3.1 is completed.

### 3.3.3 Cumulant expansion of the characteristic function-functional and the averaged shift operators

Formula (3.11) for averages of RFFℓs can be evaluated further, by expressing $\varphi_{\hat{X}_0\hat{\Xi}(\bullet)}[\upsilon\,;u(\bullet|_{t_0}^{t})]$ as the exponential of an infinite series. As in paragraph 3.3.1, we first consider the finite-dimensional analogue of $\varphi_{\hat{X}_0 \hat{\Xi}(\cdot)}[\upsilon\,;u(\bullet|_{t_0}^{t})]$, and then proceed by employing the Volterra's passing again. The characteristic function $\varphi_{\boldsymbol{X}}(\boldsymbol{\upsilon})$ of an $(N+1)-$dimensional random vector $\boldsymbol{X}(\theta)$ is defined as

$$\varphi_{\boldsymbol{X}}(\boldsymbol{\upsilon}) = \mathbb{E}^{\theta}\left[\exp\left(i\sum_{n=1}^{N+1} X_n(\theta)\,\upsilon_n\right)\right]. \tag{3.22}$$

Under appropriate conditions, we may expand the exponential function under the average sign around zero:

$$\varphi_{\boldsymbol{X}}(\boldsymbol{\upsilon}) = \mathbb{E}^{\theta}\left[\exp\left(i\sum_{n=1}^{N+1} X_n(\theta)\,\upsilon_n\right)\right] = \mathbb{E}^{\theta}\left[\sum_{m=0}^{\infty}\frac{i^m}{m!}\left(\sum_{n=1}^{N+1} X_n(\theta)\,\upsilon_n\right)^m\right] \Rightarrow$$

$$\varphi_{\boldsymbol{X}}(\boldsymbol{\upsilon}) = \mathbb{E}^{\theta}\left[1+\sum_{m=1}^{\infty}\frac{i^m}{m!}\sum_{n_1=1}^{N+1}\overset{(m)}{\cdots}\sum_{n_m=1}^{N+1} X_{n_1}(\theta)\cdots X_{n_m}(\theta)\,\upsilon_{n_1}\cdots\upsilon_{n_m}\right],$$

and by employing the linearity of the average operator, we obtain

$$\varphi_{\boldsymbol{X}}(\boldsymbol{\upsilon}) = 1+\sum_{m=1}^{\infty}\frac{i^m}{m!}\sum_{n_1=1}^{N+1}\overset{(m)}{\cdots}\sum_{n_m=1}^{N+1}\mathbb{E}^{\theta}\left[X_{n_1}(\theta)\cdots X_{n_m}(\theta)\right]\upsilon_{n_1}\cdots\upsilon_{n_m}. \tag{3.23}$$

Quantities $\left(R_{\boldsymbol{X}}^{(m)}\right)_{n_1\cdots n_m} = \mathbb{E}^{\theta}\left[X_{n_1}(\theta)\cdots X_{n_m}(\theta)\right]$ are the moments of the random vector $\boldsymbol{X}(\theta)$, and thus Eq. (3.23) constitutes the ***moment expansion*** for the characteristic function of $\boldsymbol{X}(\theta)$. By observing Eq. (3.23), we can identify the moments as the partial derivatives of the characteristic function, multiplied by an appropriate power of the imaginary unit

$$\left(R_{\boldsymbol{X}}^{(m)}\right)_{n_1\cdots n_m} = i^{-m}\left.\frac{\partial^m \varphi_{\boldsymbol{X}}(\boldsymbol{\upsilon})}{\partial\upsilon_{n_1}\cdots\partial\upsilon_{n_m}}\right|_{\boldsymbol{\upsilon}=\boldsymbol{0}}, \qquad \text{for } m\in\mathbb{N}. \tag{3.24}$$

Now, instead of expanding the characteristic function per se, we can alternatively formally expand its natural logarithm, $\ln\left(\varphi_{\boldsymbol{X}}(\boldsymbol{\upsilon})\right)$, see e.g. (Cramér, 1946). In this case, $\varphi_{\boldsymbol{X}}(\boldsymbol{\upsilon})$ is expressed as

$$\varphi_{\boldsymbol{X}}(\boldsymbol{\upsilon}) = \exp\left(\sum_{m=1}^{\infty}\frac{i^m}{m!}\sum_{n_1=1}^{N+1}\overset{(m)}{\cdots}\sum_{n_m=1}^{N+1}\left(\varkappa_{\boldsymbol{X}}^{(m)}\right)_{n_1\cdots n_m}\upsilon_{n_1}\cdots\upsilon_{n_m}\right). \tag{3.25}$$

In Eq. (3.25), quantities $\left(\varkappa_{\boldsymbol{X}}^{(m)}\right)_{n_1\cdots n_m}$ are called the ***cumulants*** of $\boldsymbol{X}(\theta)$, and Eq. (3.25) constitutes the ***cumulant expansion*** of $\varphi_{\boldsymbol{X}}(\boldsymbol{\upsilon})$. Via Eq. (3.25), the cumulants are defined as

$$\left(\varkappa_{\boldsymbol{X}}^{(m)}\right)_{n_1\cdots n_m} = i^{-m}\left.\frac{\partial^m \ln\left(\varphi_{\boldsymbol{X}}(\boldsymbol{\upsilon})\right)}{\partial\upsilon_{n_1}\cdots\partial\upsilon_{n_m}}\right|_{\boldsymbol{\upsilon}=\boldsymbol{0}}, \qquad \text{for } m\subset\mathbb{N}. \tag{3.26}$$

Thus, by comparing Eqs. (3.24), (3.26), we observe that moments and cumulants are related to each other by the Faà di Bruno's formula for higher order derivatives of composite functions, see e.g. (Johnson, 2002).

By denoting now the first component of $\boldsymbol{X}(\theta)$ as $X_0(\theta)$, and its corresponding argument $\upsilon_1$ as just $\upsilon$, while the remaining components of $\boldsymbol{X}(\theta)$, $\boldsymbol{\upsilon}$ are denoted as the $N-$dimensional random vector $\Xi(\theta)$ and the deterministic vector $\boldsymbol{u}$ respectively, Eq. (3.25) is easily recast as the joint characteristic function

$$\varphi_{X_0\Xi}(\upsilon\,;\boldsymbol{u}) = \exp\left(\sum_{m=1}^{\infty}\frac{i^m}{m!}\sum_{k=0}^{m}\binom{m}{k}\sum_{n_1=1}^{N}\overset{(k)}{\cdots}\sum_{n_k=1}^{N}\left(\varkappa_{X_0\Xi}^{(m-k,\,k)}\right)_{n_1\cdots n_k}\upsilon^{m-k}u_{n_1}\cdots u_{n_k}\right), \quad (3.27)$$

where $\binom{m}{k}=\dfrac{m!}{k!\,(m-k)!}$ is the binomial coefficient, and $\left(\varkappa_{X_0\Xi}^{(m-k,\,k)}\right)_{n_1\cdots n_k}$ is the $m-$order joint cumulant of $X_0(\theta)$ and $\Xi(\theta)$, which is of $(m-k)-$order with respect to $X_0(\theta)$ and of $k-$order with respect to $\Xi(\theta)$. By taking now the Volterra limit of Eq. (3.27), we obtain the following formula for the joint characteristic FFℓ $\varphi_{X_0\Xi(\bullet)}[\upsilon\,;u(\bullet|_{t_0}^{t})]$:

$$\varphi_{X_0\Xi(\bullet)}[\upsilon\,;u(\bullet|_{t_0}^{t})] =$$
$$= \exp\left(\sum_{m=1}^{\infty}\frac{i^m}{m!}\sum_{k=0}^{m}\binom{m}{k}\upsilon^{m-k}\int_{t_0}^{t}\overset{(k)}{\cdots}\int_{t_0}^{t}\varkappa_{X_0\Xi}^{(m-k,\,k)}(\tau_1,\dots,\tau_k)\,u(\tau_1)\cdots u(\tau_k)\,d\tau_1\cdots d\tau_k\right)$$

and after simple manipulations

$$\varphi_{X_0\Xi(\bullet)}[\upsilon\,;u(\bullet|_{t_0}^{t})] =$$
$$= \prod_{m=1}^{\infty}\prod_{k=0}^{m}\exp\left(\frac{i^m\upsilon^{m-k}}{k!(m-k)!}\int_{t_0}^{t}\overset{(k)}{\cdots}\int_{t_0}^{t}\varkappa_{X_0\Xi}^{(m-k,\,k)}(\tau_1,\dots,\tau_k)\,u(\tau_1)\cdots u(\tau_k)\,d\tau_1\cdots d\tau_k\right). \quad (3.28)$$

For the zero-mean fluctuations $\hat{X}_0$, $\hat{\Xi}(\bullet)$, the first order cumulants are zero, since they coincide with the mean values. Furthermore, using the ***semi-invariance property*** of the cumulants (McCullagh & Kolassa, 2009), that reads $\varkappa_{\hat{X}_0\hat{\Xi}}^{(m-k,\,k)}=\varkappa_{X_0\Xi}^{(m-k,\,k)}$ for $m\geq 2$, the characteristic FFℓ of the fluctuations is specified, via Eq. (3.28), to

$$\varphi_{\hat{X}_0\hat{\Xi}(\bullet)}[\upsilon\,;u(\bullet|_{t_0}^{t})] =$$
$$= \prod_{m=2}^{\infty}\prod_{k=0}^{m}\exp\left(\frac{i^m\upsilon^{m-k}}{k!(m-k)!}\int_{t_0}^{t}\overset{(k)}{\cdots}\int_{t_0}^{t}\varkappa_{X_0\Xi}^{(m-k,\,k)}(\tau_1,\dots,\tau_k)\,u(\tau_1)\cdots u(\tau_k)\,d\tau_1\cdots d\tau_k\right). \quad (3.29)$$

By substituting the expression of Eq. (3.29) for $\varphi_{\hat{X}_0\hat{\Xi}(\bullet)}[\upsilon\,;u(\bullet|_{t_0}^{t})]$ into Eq. (3.11) of Theorem 3.1, we obtain

$$\mathbb{E}^{\theta}\left[\mathcal{G}[X_0(\theta);\Xi(\bullet|_{t_0}^{t};\theta)]\right] = \prod_{m=2}^{\infty}\prod_{k=0}^{m}\bar{\mathcal{T}}_{X_0\Xi}^{(m-k,k)}\,\mathcal{G}[\upsilon;u(\bullet|_{t_0}^{t})]\Bigg|_{\substack{\upsilon=m_{X_0}\\ u(\bullet)=m_{\Xi}(\bullet)}}, \tag{3.30}$$

where operators $\bar{\mathcal{T}}_{X_0\Xi}^{(m-k,k)}$ are defined as

$$\bar{\mathcal{T}}_{X_0\Xi}^{(m-k,k)}\bullet = \exp\left(\frac{1}{k!(m-k)!}\int_{t_0}^{t}\overset{(k)}{\cdots}\int_{t_0}^{t}\varkappa_{X_0\Xi}^{(m-k,k)}(\tau_1,\ldots,\tau_k)\,\frac{\partial^{m-k}\bullet}{\partial\upsilon^{m-k}}\,\frac{\delta^k\bullet}{\delta u(\tau_1)\cdots\delta u(\tau_k)}\,d\tau_1\cdots d\tau_k\right). \tag{3.31}$$

By Eq. (3.30), the pseudo-differential operator $\varphi_{\hat{X}_0\hat{\Xi}(\bullet)}[\partial/i\partial\upsilon\,;\delta/i\delta u(s)]$ of Eq. (3.11) can be expressed as the infinite product of operators $\bar{\mathcal{T}}_{X_0\Xi}^{(m-k,k)}\bullet$. By comparing Eq. (3.31) to the definition relation (3.19) of the deterministic shift operator $\mathcal{T}_{\hat{\upsilon}\hat{u}(\bullet)}\bullet$, operators $\bar{\mathcal{T}}_{X_0\Xi}^{(m-k,k)}$ can be considered as higher-order versions of shift operators, since they are also in exponential form, containing derivatives of order higher than the first. Thus, they will henceforth be called ***averaged shift operators***.

## 3.4 Extensions of the Novikov-Furutsu theorem for random function-functionals

### 3.4.1 The extended Novikov-Furutsu theorem and its proof

Having the general formula for the average of an RFFℓ in its final form, Eq. (3.30), we prove now the following theorem, by specifying the form of RFFℓ $\mathcal{G}[X_0(\theta);\Xi(\bullet|_{t_0}^{t};\theta)]$, as well as the probabilistic structure of its arguments.

**Theorem 3.2: The extended Novikov-Furutsu theorem.** For a sufficiently smooth RFFℓ $\mathcal{F}[X_0(\theta)\,;\Xi(\bullet|_{t_0}^{t};\theta)]\equiv\mathcal{F}[\cdots]$, whose arguments $X_0(\theta)$, $\Xi(\bullet\,;\theta)$ are ***jointly Gaussian***, the following formula holds true

$$\begin{aligned}\mathbb{E}^{\theta}\left[\Xi(s;\theta)\,\mathcal{F}[\cdots]\right] &= \\ &= m_{\Xi}(s)\,\mathbb{E}^{\theta}\left[\mathcal{F}[\cdots]\right] + C_{X_0\Xi}(s)\,\mathbb{E}^{\theta}\left[\frac{\partial\mathcal{F}[\cdots]}{\partial X_0(\theta)}\right] + \\ &\quad + \int_{t_0}^{t}C_{\Xi\Xi}(s,\tau)\,\mathbb{E}^{\theta}\left[\frac{\delta\mathcal{F}[\cdots]}{\delta\Xi(\tau;\theta)}\right]d\tau,\end{aligned} \tag{3.32}$$

where $m_{\Xi}(s)$, $C_{\Xi\Xi}(s,\tau)$ are the mean value and autocovariance functions of $\Xi(s;\theta)$, and $C_{X_0\Xi}(s)$ denotes its cross-covariance with $X_0(\theta)$. Note that, for $s=t$, the extension of the classical NF theorem, Eq. (3.4), for a non-zero mean $\Xi(t;\theta)$ uncorrelated to $X_0(\theta)$, has also been given in (Hänggi, 1989; Cáceres, 2017, Sec. 1.12.2). The general form of Eq. (3.32), corresponding to correlated $X_0(\theta)$, $\Xi\,(t;\theta)$, is novel, to the best of our knowledge. ■

In the proof that follows, we assume that time instance $s$ belongs to $[t_0, t]$. However, as we will discuss after the proof, the same procedure can be followed for $s > t$, resulting in the same formula (3.33). This feature, in spite of not been usually mentioned in the existing literature, is indeed hinted in the original Novikov's proof (Novikov, 1965), where the range of $s$ is not clearly specified.

**Proof of Theorem 3.2.** For jointly Gaussian arguments, all cumulants $\varkappa_{X_0 \Xi}^{(m-k,k)}(\tau_1, \dots, \tau_k)$ of order higher than the second are zero, and the second order cumulants are equal to the second order central moments: $\varkappa_{X_0 \Xi}^{(2,0)} = C_{X_0 X_0}$, $\varkappa_{X_0 \Xi}^{(0,2)}(\tau_1, \tau_2) = C_{\Xi\Xi}(\tau_1, \tau_2)$, $\varkappa_{X_0 \Xi}^{(1,1)}(\tau) = C_{X_0 \Xi}(\tau)$ (McCullagh & Kolassa, 2009). Thus, the characteristic FFℓ of the fluctuations is specified into

$$\varphi^{\text{Gauss}}_{\hat{X}_0 \hat{\Xi}(\bullet)}[\upsilon ; u(\bullet|_{t_0}^{t})] = \exp\left( -\frac{1}{2} \int_{t_0}^{t}\int_{t_0}^{t} C_{\Xi\Xi}(\tau_1, \tau_2) u(\tau_1) u(\tau_2)\, d\tau_1 d\tau_2 \right) \times$$
$$\times \exp\left( -\frac{1}{2} C_{X_0 X_0} \upsilon^2 \right) \cdot \exp\left( -\upsilon \int_{t_0}^{t} C_{X_0 \Xi}(\tau) u(\tau)\, d\tau \right), \tag{3.33}$$

derived also in Appendix A using Volterra's passing. Thus, in Eq. (3.30), for the case of Gaussian arguments, all averaged shift operators of order higher than the second vanish, resulting into

$$\mathbb{E}^{\theta}\left[ \mathcal{G}[X_0(\theta); \Xi(\bullet|_{t_0}^{t}; \theta)] \right] = \bar{\mathcal{T}}_{X_0\Xi}^{(2,0)}\, \bar{\mathcal{T}}_{X_0\Xi}^{(1,1)}\, \bar{\mathcal{T}}_{X_0\Xi}^{(0,2)}\, \mathcal{G}[\upsilon; u(\bullet|_{t_0}^{t})] \Big|_{\substack{\upsilon = m_{X_0} \\ u(\bullet) = m_{\Xi}(\bullet)}} \tag{3.34}$$

with the ***quadratic averaged shift operators***, denoted collectively as $\bar{\mathcal{T}}^{(2)}$, being defined via Eq. (3.31) as

$$\bar{\mathcal{T}}_{X_0\Xi}^{(2,0)} \bullet = \exp\left( \frac{1}{2} C_{X_0 X_0} \frac{\partial^2 \bullet}{\partial \upsilon^2} \right), \tag{3.35a}$$

$$\bar{\mathcal{T}}_{X_0\Xi}^{(1,1)} \bullet = \exp\left( \int_{t_0}^{t} C_{X_0\Xi}(\tau) \frac{\partial \bullet}{\partial \upsilon} \frac{\delta \bullet}{\delta u(\tau)}\, d\tau \right), \tag{3.35b}$$

$$\bar{\mathcal{T}}_{X_0\Xi}^{(0,2)} \bullet = \exp\left( \frac{1}{2} \int_{t_0}^{t}\int_{t_0}^{t} C_{\Xi\Xi}(\tau_1, \tau_2) \frac{\delta^2 \bullet}{\delta u(\tau_1)\, \delta u(\tau_2)}\, d\tau_1 d\tau_2 \right). \tag{3.35c}$$

**Properties of operators $\bar{\mathcal{T}}^{(2)}$.** On $C^{\infty}$ FFℓs, operators $\bar{\mathcal{T}}^{(2)}$ are well-defined and they have the following properties, which are needed subsequently for the proof of the extended NF theorem. As the proofs in (Jia et al., 2017), where integration by differentiation is also performed, the following Lemmata are proven using operators $\bar{\mathcal{T}}^{(2)}$ in series form, which are obtained by expanding the exponential in the right-hand sides of Eqs. (3.35a,b,c):

$$\bar{\mathcal{T}}^{(2,0)}_{X_0\Xi}\bullet = \sum_{p=0}^{\infty}\frac{1}{p!}\frac{1}{2^p}C^p_{X_0X_0}\frac{\partial^{2p}\bullet}{\partial\upsilon^{2p}}, \tag{3.36a}$$

$$\bar{\mathcal{T}}^{(1,1)}_{X_0\Xi}\bullet = $$
$$= \sum_{p=0}^{\infty}\frac{1}{p!}\int_{t_0}^{t}\overset{(p)}{\cdots}\int_{t_0}^{t}C_{X_0\Xi}(\tau^{(1)})\cdots C_{X_0\Xi}(\tau^{(p)})\frac{\partial^p\,\delta^p\,\bullet}{\partial\upsilon^p\,\delta u(\tau^{(1)})\cdots\delta u(\tau^{(p)})}d\tau^{(1)}\cdots d\tau^{(p)}, \tag{3.36b}$$

$$\bar{\mathcal{T}}^{(0,2)}_{X_0\Xi}\bullet = \sum_{p=0}^{\infty}\frac{1}{p!}\frac{1}{2^p}\int_{t_0}^{t}\overset{(2p)}{\cdots}\int_{t_0}^{t}\Big[C_{\Xi\Xi}(\tau_1^{(1)},\tau_2^{(1)})\cdots C_{\Xi\Xi}(\tau_1^{(p)},\tau_2^{(p)})\times$$
$$\times\frac{\delta^{2p}\,\bullet}{\delta u(\tau_1^{(1)})\delta u(\tau_2^{(1)})\cdots\delta u(\tau_1^{(p)})\delta u(\tau_2^{(p)})}d\tau_1^{(1)}d\tau_2^{(1)}\cdots d\tau_1^{(p)}d\tau_2^{(p)}\Big]. \tag{3.36c}$$

**Lemma 3.1: Operators $\bar{\mathcal{T}}^{(2)}$ are linear.** That is, for any two $C^\infty$ FFℓs, $\mathcal{G}[\upsilon\,;u(\bullet|_{t_0}^{t})]$, $\mathcal{F}[\upsilon\,;u(\bullet|_{t_0}^{t})]$ it holds true that

$$\bar{\mathcal{T}}^{(2)}\Big[\alpha\,\mathcal{G}[\upsilon\,;u(\bullet|_{t_0}^{t})]+\beta\,\mathcal{F}[\upsilon\,;u(\bullet|_{t_0}^{t})]\Big] =$$
$$= \alpha\,\bar{\mathcal{T}}^{(2)}\Big[\mathcal{G}[\upsilon\,;u(\bullet|_{t_0}^{t})]\Big]+\beta\,\bar{\mathcal{T}}^{(2)}\Big[\mathcal{F}[\upsilon\,;u(\bullet|_{t_0}^{t})]\Big], \tag{3.37}$$

where $\bar{\mathcal{T}}^{(2)}\bullet$ stands for any of the three operators $\bar{\mathcal{T}}^{(2,0)}_{X_0\Xi}\bullet$, $\bar{\mathcal{T}}^{(1,1)}_{X_0\Xi}\bullet$, $\bar{\mathcal{T}}^{(0,2)}_{X_0\Xi}\bullet$, and $\alpha$, $\beta$ are constants or scalar functions having argument(s) different than the differentiation argument(s) appearing in the corresponding operator $\bar{\mathcal{T}}^{(2)}\bullet$.

**Proof.** The action of operator $\bar{\mathcal{T}}^{(1,1)}_{X_0\Xi}$ on $\alpha\,\mathcal{G}[\upsilon\,;u(\bullet|_{t_0}^{t})]+\beta\,\mathcal{F}[\upsilon\,;u(\bullet|_{t_0}^{t})]$ is expressed via Eq. (3.36b) as

$$\bar{\mathcal{T}}^{(1,1)}_{X_0\Xi}\Big[\alpha\,\mathcal{G}[\upsilon\,;u(\bullet|_{t_0}^{t})]+\beta\,\mathcal{F}[\upsilon\,;u(\bullet|_{t_0}^{t})]\Big] =$$
$$=\sum_{p=0}^{\infty}\frac{1}{p!}\int_{t_0}^{t}\overset{(p)}{\cdots}\int_{t_0}^{t}C_{X_0\Xi}(\tau^{(1)})\cdots C_{X_0\Xi}(\tau^{(p)})\frac{\partial^p\,\delta^p\Big[\alpha\mathcal{G}[\upsilon;u(\bullet|_{t_0}^{t})]+\beta\mathcal{F}[\upsilon;u(\bullet|_{t_0}^{t})]\Big]}{\partial\upsilon^p\,\delta u(\tau^{(1)})\cdots\delta u(\tau^{(p)})}d\tau^{(1)}\cdots d\tau^{(p)}.$$

Employing the linearity of derivatives and integrals, in conjunction with the assumption that $\alpha$, $\beta$ are independent from the differentiation arguments $\upsilon$ and $u(\bullet)$ of operator $\bar{\mathcal{T}}^{(1,1)}_{X_0\Xi}$, each term of the right-hand side of the above equation is linearly decomposed, resulting in the linearity of the operator $\bar{\mathcal{T}}^{(1,1)}_{X_0\Xi}$. The proofs for $\bar{\mathcal{T}}^{(2,0)}_{X_0\Xi}$, $\bar{\mathcal{T}}^{(0,2)}_{X_0\Xi}$ are similar. ∎

**Lemma 3.2: Operators $\bar{\mathcal{T}}^{(2)}$ commute with $\upsilon-$ and $u(\tau)-$differentiations.** That is, for a $C^\infty$ FFℓ $\mathcal{G}[\upsilon\,;u(\bullet|_{t_0}^{t})]$, and for $\bar{\mathcal{T}}^{(2)} \in \left\{ \bar{\mathcal{T}}^{(2,0)}_{X_0\Xi}, \bar{\mathcal{T}}^{(1,1)}_{X_0\Xi}, \bar{\mathcal{T}}^{(0,2)}_{X_0\Xi} \right\}$,

$$\frac{\partial}{\partial \upsilon}\left[ \bar{\mathcal{T}}^{(2)} \mathcal{G}[\upsilon\,;u(\bullet|_{t_0}^{t})] \right] = \bar{\mathcal{T}}^{(2)} \left[ \frac{\partial \mathcal{G}[\upsilon\,;u(\bullet|_{t_0}^{t})]}{\partial \upsilon} \right], \tag{3.38a}$$

and

$$\frac{\delta}{\delta u(\tau)}\left[ \bar{\mathcal{T}}^{(2)} \mathcal{G}[\upsilon\,;u(\bullet|_{t_0}^{t})] \right] = \bar{\mathcal{T}}^{(2)} \left[ \frac{\delta \mathcal{G}[\upsilon\,;u(\bullet|_{t_0}^{t})]}{\delta u(\tau)} \right]. \tag{3.38b}$$

**Proof.** Using Eq. (3.36b), we may write

$$\frac{\partial}{\partial \upsilon}\left[ \bar{\mathcal{T}}^{(1,1)}_{X_0\Xi} \mathcal{G}[\upsilon\,;u(\bullet|_{t_0}^{t})] \right] =$$

$$= \frac{\partial}{\partial \upsilon} \sum_{p=0}^{\infty} \frac{1}{p!} \int_{t_0}^{t} \overset{(p)}{\cdots} \int_{t_0}^{t} C_{X_0\Xi}(\tau^{(1)}) \cdots C_{X_0\Xi}(\tau^{(p)}) \frac{\partial^p \, \delta^p \, \mathcal{G}[\upsilon\,;u(\bullet|_{t_0}^{t})]}{\partial \upsilon^p \, \delta u(\tau^{(1)}) \cdots \delta u(\tau^{(p)})} \, d\tau^{(1)} \cdots d\tau^{(p)}$$

[using the linearity and continuity of the derivative]

$$= \sum_{p=0}^{\infty} \frac{1}{p!} \int_{t_0}^{t} \overset{(p)}{\cdots} \int_{t_0}^{t} C_{X_0\Xi}(\tau^{(1)}) \cdots C_{X_0\Xi}(\tau^{(p)}) \frac{\partial^{p+1} \, \delta^p \, \mathcal{G}[\upsilon\,;u(\bullet|_{t_0}^{t})]}{\partial \upsilon^{p+1} \, \delta u(\tau^{(1)}) \cdots \delta u(\tau^{(p)})} \, d\tau^{(1)} \cdots d\tau^{(p)}$$

[which is identified, via Eq. (3.36b), as]

$$= \bar{\mathcal{T}}^{(1,1)}_{X_0\Xi} \left[ \frac{\partial \mathcal{G}[\upsilon\,;u(\bullet|_{t_0}^{t})]}{\partial \upsilon} \right].$$

Proof of the lemma for $\frac{\delta}{\delta u(s)}\left[ \bar{\mathcal{T}}^{(1,1)}_{X_0\Xi} \mathcal{G}[\upsilon\,;u(\bullet|_{t_0}^{t})] \right]$ is similar, as it is also for the other two operators $\bar{\mathcal{T}}^{(2,0)}_{X_0\Xi}$, $\bar{\mathcal{T}}^{(0,2)}_{X_0\Xi}$. ■

**Lemma 3.3: Operators $\bar{\mathcal{T}}^{(2)}$ commute with each other.** That is, for any $C^\infty$ FFℓ $\mathcal{G}[\upsilon\,;u(\bullet|_{t_0}^{t})]$

$$\bar{\mathcal{T}}^{(2,0)}_{X_0\Xi} \bar{\mathcal{T}}^{(1,1)}_{X_0\Xi} \bar{\mathcal{T}}^{(0,2)}_{X_0\Xi} \, \mathcal{G}[\upsilon\,;u(\bullet|_{t_0}^{t})] = \bar{\mathcal{T}}^{(1,1)}_{X_0\Xi} \bar{\mathcal{T}}^{(2,0)}_{X_0\Xi} \bar{\mathcal{T}}^{(0,2)}_{X_0\Xi} \, \mathcal{G}[\upsilon\,;u(\bullet|_{t_0}^{t})] = \cdots$$

$$\cdots = \bar{\mathcal{T}}^{(0,2)}_{X_0\Xi} \bar{\mathcal{T}}^{(2,0)}_{X_0\Xi} \bar{\mathcal{T}}^{(1,1)}_{X_0\Xi} \, \mathcal{G}[\upsilon\,;u(\bullet|_{t_0}^{t})]. \tag{3.39}$$

In other words, the product of the three operators $\bar{\mathcal{T}}^{(2)}$, under any permutation of their order, has the same action on $\mathcal{G}[\upsilon\,;u(\bullet|_{t_0}^{t})]$.

**Proof.** For the sake of simplicity, we shall prove the commutativity of operators $\bar{\mathcal{T}}^{(1,1)}_{X_0\Xi}$ and $\bar{\mathcal{T}}^{(2,0)}_{X_0\Xi}$. We begin from

$$\bar{\mathcal{T}}^{(1,1)}_{X_0\Xi}\bar{\mathcal{T}}^{(2,0)}_{X_0\Xi}\mathcal{G}[\upsilon\,;u(\bullet|_{t_0}^{t})] = \qquad \text{[by using Eqs. (3.36a,b)]}$$

$$=\sum_{p=0}^{\infty}\frac{1}{p!}\int_{t_0}^{t}\overset{(p)}{\cdots}\int_{t_0}^{t}C_{X_0\Xi}(\tau^{(1)})\cdots C_{X_0\Xi}(\tau^{(p)})\times$$

$$\times\frac{\partial^p\,\delta^p}{\partial\upsilon^p\,\delta u(\tau^{(1)})\cdots\delta u(\tau^{(p)})}\left[\sum_{m=0}^{\infty}\frac{1}{m!}\frac{1}{2^m}C^m_{X_0X_0}\frac{\partial^{2m}\mathcal{G}[\upsilon\,;u(\bullet|_{t_0}^{t})]}{\partial\upsilon^{2m}}\right]d\tau^{(1)}\cdots d\tau^{(p)}=$$

[rearranging the order of summations and using the linearity of derivatives and integrals]

$$=\sum_{m=0}^{\infty}\frac{1}{m!}\frac{1}{2^m}C^m_{X_0X_0}\times$$

$$\times\frac{\partial^{2m}}{\partial\upsilon^{2m}}\left[\sum_{p=0}^{\infty}\frac{1}{p!}\int_{t_0}^{t}\overset{(p)}{\cdots}\int_{t_0}^{t}C_{X_0\Xi}(\tau^{(1)})\cdots C_{X_0\Xi}(\tau^{(p)})\,\frac{\partial^p\,\delta^p\mathcal{G}[\upsilon\,;u(\bullet|_{t_0}^{t})]}{\partial\upsilon^p\,\delta u(\tau^{(1)})\cdots\delta u(\tau^{(p)})}\,d\tau^{(1)}\cdots d\tau^{(p)}\right]$$

[which, by Eqs. (3.36a,b), is identified as]

$$=\bar{\mathcal{T}}^{(2,0)}_{X_0\Xi}\bar{\mathcal{T}}^{(1,1)}_{X_0\Xi}\mathcal{G}[\upsilon\,;u(\bullet|_{t_0}^{t})]. \qquad \blacksquare$$

After proving Lemmata 3.1-3.3, we return to Eq. (3.34), and consider the case where $\mathcal{G}[\cdots]=\Xi(s;\theta)\,\mathcal{F}[\cdots]$:

$$\mathbb{E}^{\theta}\left[\Xi(s;\theta)\mathcal{F}[X_0(\theta);\Xi(\bullet|_{t_0}^{t};\theta)]\right]=$$

$$=\left\{\bar{\mathcal{T}}^{(2,0)}_{X_0\Xi}\,\bar{\mathcal{T}}^{(1,1)}_{X_0\Xi}\,\bar{\mathcal{T}}^{(0,2)}_{X_0\Xi}\left[u(s)\mathcal{F}[\upsilon;u(\bullet|_{t_0}^{t})]\right]\right\}_{\substack{\upsilon=m_{X_0}\\ u(\bullet)=m_{\Xi}(\bullet)}}. \tag{3.40}$$

The main part of the proof for the extended NF theorem lies in the successive applications of the three operators $\bar{\mathcal{T}}^{(2)}$ on $u(s)\,\mathcal{F}[\upsilon\,;u(\bullet|_{t_0}^{t})]$. Since, by virtue of Lemma 3.3, the order of application of operators $\bar{\mathcal{T}}^{(2)}$ does not alter the final result, we choose the order shown in Eq. (3.40). The action of each operator $\bar{\mathcal{T}}^{(2)}$ is given by the following Lemmata 3.4-3.6.

**Lemma 3.4.** The action of operator $\bar{\mathcal{T}}^{(2,0)}_{X_0\Xi}$ on $u(s)\,\mathcal{F}[\upsilon\,;u(\bullet|_{t_0}^{t})]$ is given by

$$\bar{\mathcal{T}}^{(2,0)}_{X_0\Xi}\left[u(s)\,\mathcal{F}[\upsilon\,;u(\bullet|_{t_0}^{t})]\right]=u(s)\,\bar{\mathcal{T}}^{(2,0)}_{X_0\Xi}\left[\mathcal{F}[\upsilon\,;u(\bullet|_{t_0}^{t})]\right]. \tag{3.41}$$

**Proof.** Eq. (3.41) holds true due to the linearity of $\bar{\mathcal{T}}^{(2,0)}_{X_0\Xi}$, since the factor $u(s)$ is independent from $\upsilon$, which is the differentiation argument of $\bar{\mathcal{T}}^{(2,0)}_{X_0\Xi}$. $\blacksquare$

In the right-hand side of Eq. (3.41), $\bar{\mathcal{T}}^{(2,0)}_{X_0\Xi}\left[\mathcal{F}[\upsilon\,;u(\bullet)]\right]$ is a new FFℓ, which is denoted by $\mathcal{F}_1[\upsilon\,;u(\bullet)]$. By applying operator $\bar{\mathcal{T}}^{(1,1)}_{X_0\Xi}$ on both sides of Eq. (3.41) we obtain

$$\bar{\mathcal{T}}^{(1,1)}_{X_0\Xi}\,\bar{\mathcal{T}}^{(2,0)}_{X_0\Xi}\left[u(s)\,\mathcal{F}[\upsilon\,;u(\bullet|_{t_0}^{t})]\right] = \bar{\mathcal{T}}^{(1,1)}_{X_0\Xi}\left[u(s)\,\mathcal{F}_1[\upsilon\,;u(\bullet|_{t_0}^{t})]\right]. \tag{3.42}$$

The right-hand side of Eq. (3.42) is calculated by using the following result:

**Lemma 3.5.** The action of operator $\bar{\mathcal{T}}^{(1,1)}_{X_0\Xi}$ on $u(t)\,\mathcal{F}_1[\upsilon\,;u(\bullet|_{t_0}^{t})]$ is given by

$$\begin{aligned}\bar{\mathcal{T}}^{(1,1)}_{X_0\Xi}\left[u(s)\,\mathcal{F}_1[\upsilon\,;u(\bullet|_{t_0}^{t})]\right] &= u(s)\,\bar{\mathcal{T}}^{(1,1)}_{X_0\Xi}\left[\mathcal{F}_1[\upsilon\,;u(\bullet|_{t_0}^{t})]\right]+\\ &\quad + C_{X_0\Xi}(s)\,\bar{\mathcal{T}}^{(1,1)}_{X_0\Xi}\left[\frac{\partial\mathcal{F}_1[\upsilon\,;u(\bullet|_{t_0}^{t})]}{\partial\upsilon}\right].\end{aligned} \tag{3.43}$$

**Proof.** First, the following *product rule for higher-order Volterra derivatives* is formulated, for any $k\in\mathbb{N}$:

$$\begin{aligned}\frac{\delta^k\left[u(s)\,\mathcal{F}[\upsilon\,;u(\bullet|_{t_0}^{t})]\right]}{\delta u(\tau^{(1)})\cdots\delta u(\tau^{(k)})} &= u(s)\,\frac{\delta^k\,\mathcal{F}[\upsilon\,;u(\bullet|_{t_0}^{t})]}{\delta u(\tau^{(1)})\cdots\delta u(\tau^{(k)})}+\\ &\quad+\sum_{n=1}^{k}\delta(s-\tau^{(n)})\,\frac{\delta^{k-1}\,\mathcal{F}[\upsilon\,;u(\bullet|_{t_0}^{t})]}{\prod_{\substack{\ell=1\\ \ell\neq n}}^{k}\delta u(\tau^{(\ell)})},\end{aligned} \tag{3.44}$$

with $\prod_{\substack{\ell=1\\ \ell\neq n}}^{k}\delta u(\tau^{(\ell)})=\delta u(\tau^{(1)})\cdots\delta u(\tau^{(n-1)})\,\delta u(\tau^{(n+1)})\cdots\delta u(\tau^{(k)})$. Eq. (3.44) is easily proven via mathematical induction on index $k$, commencing from the relation for $k=1$ (product rule for the first-order derivative):

$$\frac{\delta\left[u(s)\,\mathcal{F}[\upsilon\,;u(\bullet|_{t_0}^{t})]\right]}{\delta u(\tau)} = u(s)\,\frac{\delta\mathcal{F}[\upsilon\,;u(\bullet|_{t_0}^{t})]}{\delta u(\tau)}+\frac{\delta u(s)}{\delta u(\tau)}\,\mathcal{F}[\upsilon\,;u(\bullet|_{t_0}^{t})],$$

in which $\delta u(s)/\delta u(\tau)=\delta(s-\tau)$. Now, by using the series expansion of Eq. (3.36b), and employing the product rule, Eq. (3.44), we obtain

$$\bar{\mathcal{T}}^{(1,1)}_{X_0\Xi}\left[u(s)\,\mathcal{F}_1[\upsilon\,;u(\bullet|_{t_0}^{t})]\right] = u(s)\,\mathrm{A}+C_{X_0\Xi}(s)\,\mathrm{B}, \tag{3.45}$$

where $\mathrm{A}$, $\mathrm{B}$ are

$$\mathrm{A} = \sum_{p=0}^{\infty}\frac{1}{p!}\int_{t_0}^{t}\overset{(p)}{\cdots}\int_{t_0}^{t}C_{X_0\Xi}(\tau^{(1)})\cdots C_{X_0\Xi}(\tau^{(p)})\,\frac{\partial^p\,\delta^p\,\mathcal{F}_1[\upsilon\,;u(\bullet|_{t_0}^{t})]}{\partial\upsilon^p\,\delta u(\tau^{(1)})\cdots\delta u(\tau^{(p)})}\,d\tau^{(1)}\cdots d\tau^{(p)},$$

$$\mathrm{B} = \sum_{p=1}^{\infty}\frac{1}{p!}\sum_{n=1}^{p}\int_{t_0}^{t}\overset{(p-1)}{\cdots}\int_{t_0}^{t}\prod_{\substack{m=1\\ m\neq n}}^{p}C_{X_0\Xi}(\tau^{(m)})\,\frac{\partial^p\,\delta^{p-1}\,\mathcal{F}_1[\upsilon\,;u(\bullet|_{t_0}^{t})]}{\partial\upsilon^p\prod_{\substack{m=1\\ m\neq n}}^{p}\delta u(\tau^{(\ell)})}\prod_{\substack{m=1\\ m\neq n}}^{p}d\tau^{(m)}.$$

By the symmetry of integration arguments, all terms of the $n-$sum are equal, thus,

$$\mathrm{B} = \sum_{p=1}^{\infty} \frac{1}{(p-1)!} \int_{t_0}^{t} \overset{(p-1)}{\cdots} \int_{t_0}^{t} C_{X_0\Xi}(\tau^{(1)}) \cdots C_{X_0\Xi}(\tau^{(p-1)}) \frac{\partial^p \delta^{p-1} \mathcal{F}_1[\upsilon\,; u(\bullet|_{t_0}^{t})]}{\partial \upsilon^p \, \delta u(\tau^{(1)}) \cdots \delta u(\tau^{(p-1)})} \, d\tau^{(1)} \cdots d\tau^{(p-1)}.$$

In view of Eq. (3.36b), series A is identified as

$$\mathrm{A} = \bar{\mathcal{T}}^{(1,1)}_{X_0\Xi} \left[ \mathcal{F}_1[\upsilon\,; u(\bullet|_{t_0}^{t})] \right]. \tag{3.46}$$

For series B, we perform the index change $k = p-1$, resulting in

$$\mathrm{B} = \sum_{k=0}^{\infty} \frac{1}{k!} \int_{t_0}^{t} \overset{(k)}{\cdots} \int_{t_0}^{t} C_{X_0\Xi}(\tau^{(1)}) \cdots C_{X_0\Xi}(\tau^{(k)}) \frac{\partial^k \delta^k}{\partial \upsilon^k \, \delta u(\tau^{(1)}) \cdots \delta u(\tau^{(k)})} \left[ \frac{\partial \mathcal{F}_1[\upsilon\,; u(\bullet|_{t_0}^{t})]}{\partial \upsilon} \right] d\tau^{(1)} \cdots d\tau^{(k)}. \tag{3.47}$$

Via Eq. (3.36b), the right-hand side of Eq. (3.47) is identified as

$$\mathrm{B} = \bar{\mathcal{T}}^{(1,1)}_{X_0\Xi} \left[ \frac{\partial \mathcal{F}_1[\upsilon\,; u(\bullet|_{t_0}^{t})]}{\partial \upsilon} \right]. \tag{3.48}$$

Substitution of Eqs. (3.46), (3.48) into Eq. (3.45) results in Eq. (3.43), completing thus the proof of Lemma 3.5 ■

Now, by denoting the term $\bar{\mathcal{T}}^{(1,1)}_{X_0\Xi} \left[ \mathcal{F}_1[\upsilon\,; u(\bullet)] \right]$ by $\mathcal{F}_2[\upsilon\,; u(\bullet)]$, and combining Eqs. (3.43) and (3.42), we obtain

$$\bar{\mathcal{T}}^{(1,1)}_{X_0\Xi} \bar{\mathcal{T}}^{(2,0)}_{X_0\Xi} \left[ u(s) \mathcal{F}[\upsilon\,; u(\bullet|_{t_0}^{t})] \right] = u(s)\, \mathcal{F}_2[\upsilon\,; u(\bullet|_{t_0}^{t})] + \\ + C_{X_0\Xi}(s)\, \bar{\mathcal{T}}^{(1,1)}_{X_0\Xi} \left[ \frac{\partial \mathcal{F}_1[\upsilon\,; u(\bullet|_{t_0}^{t})]}{\partial \upsilon} \right].$$

Applying the operator $\bar{\mathcal{T}}^{(0,2)}_{X_0\Xi}$ on both sides of the above equation, leads to

$$\bar{\mathcal{T}}^{(0,2)}_{X_0\Xi} \bar{\mathcal{T}}^{(1,1)}_{X_0\Xi} \bar{\mathcal{T}}^{(2,0)}_{X_0\Xi} \left[ u(s) \mathcal{F}[\upsilon\,; u(\bullet|_{t_0}^{t})] \right] = \bar{\mathcal{T}}^{(0,2)}_{X_0\Xi} \left[ u(s)\, \mathcal{F}_2[\upsilon\,; u(\bullet|_{t_0}^{t})] \right] + \\ + \bar{\mathcal{T}}^{(0,2)}_{X_0\Xi} \left\{ C_{X_0\Xi}(s)\, \bar{\mathcal{T}}^{(1,1)}_{X_0\Xi} \left[ \frac{\partial \mathcal{F}_1[\upsilon\,; u(\bullet|_{t_0}^{t})]}{\partial \upsilon} \right] \right\}. \tag{3.49}$$

We shall now elaborate on the two terms appearing in the right-hand side of Eq. (3.49). By using the linearity of $\bar{\mathcal{T}}^{(0,2)}_{X_0\Xi}$ (Lemma 3.1) and the commutation of $\bar{\mathcal{T}}^{(2,0)}_{X_0\Xi}$ with the $\upsilon-$derivative (Lemma 3.2) in conjunction with the definition of $\mathcal{F}_1[\upsilon\,; u(\bullet)]$, the second term in the right-hand side of Eq. (3.49) is expressed in terms of the FFℓ $\mathcal{F}[\upsilon\,; u(\bullet)]$ as follows:

$$\bar{\mathcal{T}}^{(0,2)}_{X_0\Xi}\left\{C_{X_0\Xi}(s)\,\bar{\mathcal{T}}_{\hat{X}_0\hat{\Xi}(\bullet)}\left[\frac{\partial\mathcal{F}_1[\upsilon\,;u(\bullet|^t_{t_0})]}{\partial\upsilon}\right]\right\} = \tag{3.50}$$

$$= C_{X_0\Xi}(s)\,\bar{\mathcal{T}}^{(0,2)}_{X_0\Xi}\,\bar{\mathcal{T}}^{(1,1)}_{X_0\Xi}\,\bar{\mathcal{T}}^{(2,0)}_{X_0\Xi}\left[\frac{\partial\mathcal{F}[\upsilon\,;u(\bullet|^t_{t_0})]}{\partial\upsilon}\right].$$

Concerning the first term in the right-hand side of Eq. (3.49), we need the following Lemma:

**Lemma 3.6.** The action of operator $\bar{\mathcal{T}}^{(0,2)}_{X_0\Xi}$ on $u(s)\,\mathcal{F}_2[\upsilon\,;u(\bullet|^t_{t_0})]$ is given by

$$\bar{\mathcal{T}}^{(0,2)}_{X_0\Xi}\left[u(s)\,\mathcal{F}_2[\upsilon\,;u(\bullet|^t_{t_0})]\right] = u(s)\,\bar{\mathcal{T}}^{(0,2)}_{X_0\Xi}\left[\mathcal{F}_2[\upsilon\,;u(\bullet|^t_{t_0})]\right] + \tag{3.51}$$

$$+\int_{t_0}^{t} C_{\Xi\Xi}(s,\tau)\,\bar{\mathcal{T}}^{(0,2)}_{X_0\Xi}\left[\frac{\delta\mathcal{F}_2[\upsilon\,;u(\bullet|^t_{t_0})]}{\delta u(\tau)}\right]d\tau.$$

**Proof.** By expanding $\bar{\mathcal{T}}^{(0,2)}_{X_0\Xi}\left[u(s)\,\mathcal{F}_2[\upsilon\,;u(\bullet|^t_{t_0})]\right]$ in series using Eq. (3.36c), and employing the product rule, Eq. (3.44), we obtain

$$\bar{\mathcal{T}}^{(0,2)}_{X_0\Xi}\left[u(s)\,\mathcal{F}_2[\upsilon\,;u(\bullet|^t_{t_0})]\right] = u(s)\,\Gamma + \Delta, \tag{3.52}$$

where $\Gamma$ is

$$\Gamma = \sum_{p=0}^{\infty}\frac{1}{p!}\frac{1}{2^p}\int_{t_0}^{t}\overset{(2p)}{\cdots}\int_{t_0}^{t}\Big[C_{\Xi\Xi}(\tau_1^{(1)},\tau_2^{(1)})\cdots C_{\Xi\Xi}(\tau_1^{(p)},\tau_2^{(p)})\times$$

$$\times\frac{\delta^{2p}\mathcal{F}_2[\upsilon\,;u(\bullet|^t_{t_0})]}{\delta u(\tau_1^{(1)})\delta u(\tau_2^{(1)})\cdots\delta u(\tau_1^{(p)})\delta u(\tau_2^{(p)})}\,d\tau_1^{(1)}d\tau_2^{(1)}\cdots d\tau_1^{(p)}d\tau_2^{(p)}\Big],$$

while $\Delta$ can be written as follows, after taking into account the symmetry of autocorrelation function $C_{\Xi\Xi}(\tau_1,\tau_2)$:

$$\Delta = \sum_{p=1}^{\infty}\frac{1}{p!}\frac{1}{2^p}\sum_{n=1}^{2p}\int_{t_0}^{t}\overset{(2p-1)}{\cdots}\int_{t_0}^{t}\Big[C_{\Xi\Xi}(s,\tau_2^{(n)})\prod_{\substack{m=1\\ m\neq n}}^{p}C_{\Xi\Xi}(\tau_1^{(m)},\tau_2^{(m)})\times$$

$$\times\frac{\delta^{2p-1}\mathcal{F}_2[\upsilon\,;u(\bullet|^t_{t_0})]}{\delta u(\tau_2^{(n)})\prod_{\substack{m=1\\ m\neq n}}^{p}\delta u(\tau_1^{(m)})\delta u(\tau_2^{(m)})}\,d\tau_2^{(n)}\prod_{\substack{m=1\\ m\neq n}}^{p}d\tau_1^{(m)}d\tau_2^{(m)}\Big].$$

By virtue of Eq. (3.36c), series $\Gamma$ is identified as

$$\Gamma = \bar{\mathcal{T}}^{(0,2)}_{X_0\Xi}\left[\mathcal{F}_2[\upsilon\,;u(\bullet|^t_{t_0})]\right]. \tag{3.53}$$

We turn now our attention to the series $\Delta$. By performing, in each term of the $n-$sum $\sum_{n=1}^{2p}\cdots$, the change of integration variables: $\tau=\tau_2^{(n)}$ and $\tau_i^{(\ell)}=\tau_i^{(m)}$ for $m<n$, $\tau_i^{(\ell)}=\tau_i^{(m-1)}$ for $m>n$, $i=1,2$, it is easy to see that all $2p$ terms in $n-$sum are equal:

$$\Delta = \sum_{p=1}^{\infty} \frac{1}{(p-1)!} \frac{1}{2^{p-1}} \int_{t_0}^{t} \overset{(2p-1)}{\cdots} \int_{t_0}^{t} \Big[ C_{\Xi\Xi}(s,\tau)\, C_{\Xi\Xi}(\tau_1^{(1)},\tau_2^{(1)}) \cdots C_{\Xi\Xi}(\tau_1^{(p-1)},\tau_2^{(p-1)}) \times$$
$$\times \frac{\delta^{2p-1} \mathcal{F}_2[\upsilon\,;u(\bullet|_{t_0}^{t})]}{\delta u(\tau)\delta u(\tau_1^{(1)})\delta u(\tau_2^{(1)})\cdots \delta u(\tau_1^{(p-1)})\delta u(\tau_2^{(p-1)})} d\tau\, d\tau_1^{(1)} d\tau_2^{(1)} \cdots d\tau_1^{(p-1)} d\tau_2^{(p-1)} \Bigg].$$

Further, we perform the index change $k = p-1$ and interchange $\tau-$integration with summation, resulting in

$$\Delta = \int_{t_0}^{t} C_{\Xi\Xi}(s,\tau) \sum_{k=0}^{\infty} \frac{1}{k!} \frac{1}{2^k} \int_{t_0}^{t} \overset{(2k)}{\cdots} \int_{t_0}^{t} \Big[ C_{\Xi\Xi}(\tau_1^{(1)},\tau_2^{(1)}) \cdots C_{\Xi\Xi}(\tau_1^{(k)},\tau_2^{(k)}) \times$$
$$\times \frac{\delta^{2k}}{\delta u(\tau_1^{(1)})\delta u(\tau_2^{(1)})\cdots \delta u(\tau_1^{(k)})\delta u(\tau_2^{(k)})} \left[ \frac{\delta \mathcal{F}_2[\upsilon\,;u(\bullet|_{t_0}^{t})]}{\delta u(\tau)} \right] d\tau_1^{(1)} d\tau_2^{(1)} \cdots d\tau_1^{(k)} d\tau_2^{(k)} \Bigg] d\tau. \tag{3.54}$$

The sum in the right-hand side of Eq. (3.54) is identified, via Eq. (3.36c), as

$$\Delta = \int_{t_0}^{t} C_{\Xi\Xi}(s,\tau)\, \bar{\mathcal{T}}_{X_0\Xi}^{(0,2)} \left[ \frac{\delta \mathcal{F}_2[\upsilon\,;u(\bullet|_{t_0}^{t})]}{\delta u(\tau)} \right] d\tau. \tag{3.55}$$

By substituting Eqs. (3.53), (3.55) into Eq. (3.52), we obtain Eq. (3.51). Thus, proof of Lemma 3.6 is completed. ■

**Finalization of the proof of Theorem 3.2:** Substituting Eqs. (3.50), (3.51) into Eq. (3.49), recalling that $\mathcal{F}_2[\upsilon\,;u(\bullet)] = \bar{\mathcal{T}}_{X_0\Xi}^{(1,1)}\big[\mathcal{F}_1[\upsilon\,;u(\bullet)]\big]$, $\mathcal{F}_1[\upsilon\,;u(\bullet)] = \bar{\mathcal{T}}_{X_0\Xi}^{(2,0)}\big[\mathcal{F}[\upsilon\,;u(\bullet)]\big]$, and employing the commutation of operators $\bar{\mathcal{T}}_{X_0\Xi}^{(1,1)}$, $\bar{\mathcal{T}}_{X_0\Xi}^{(2,0)}$ with the Volterra derivative $\delta \bullet / \delta u(s)$ (Lemma 3.2), we obtain

$$\begin{aligned}
\bar{\mathcal{T}}_{X_0\Xi}^{(0,2)} \bar{\mathcal{T}}_{X_0\Xi}^{(1,1)} \bar{\mathcal{T}}_{X_0\Xi}^{(2,0)} \Big[ u(s)\, \mathcal{F}[\upsilon\,;u(\bullet|_{t_0}^{t})] \Big] &= \\
= u(s)\, \bar{\mathcal{T}}_{X_0\Xi}^{(0,2)} \bar{\mathcal{T}}_{X_0\Xi}^{(1,1)} \bar{\mathcal{T}}_{X_0\Xi}^{(2,0)} \Big[ \mathcal{F}[\upsilon\,;u(\bullet|_{t_0}^{t})] \Big] &+ \\
+ C_{X_0\Xi}(s)\, \bar{\mathcal{T}}_{X_0\Xi}^{(0,2)} \bar{\mathcal{T}}_{X_0\Xi}^{(1,1)} \bar{\mathcal{T}}_{X_0\Xi}^{(2,0)} \left[ \frac{\partial \mathcal{F}[\upsilon\,;u(\bullet|_{t_0}^{t})]}{\partial \upsilon} \right] &+ \\
+ \int_{t_0}^{t} C_{\Xi\Xi}(s,\tau)\, \bar{\mathcal{T}}_{X_0\Xi}^{(0,2)} \bar{\mathcal{T}}_{X_0\Xi}^{(1,1)} \bar{\mathcal{T}}_{X_0\Xi}^{(2,0)} \left[ \frac{\delta \mathcal{F}[\upsilon\,;u(\bullet|_{t_0}^{t})]}{\delta u(\tau)} \right] d\tau.&
\end{aligned} \tag{3.56}$$

By setting $\upsilon = m_{X_0}$, and $u(\bullet) = m_{\Xi}(\bullet)$ in Eq. (3.56), and applying Eq. (3.34) to each term of the form $\bar{\mathcal{T}}_{X_0\Xi}^{(0,2)} \bar{\mathcal{T}}_{X_0\Xi}^{(1,1)} \bar{\mathcal{T}}_{X_0\Xi}^{(2,0)}[\cdots]$ in both sides of Eq. (3.56), we obtain the extended NF theorem, Eq. (3.32). The proof is now completed.

**Remark 3.3: Comparison to other proofs.** Apart from the extensions, the proof of NF theorem presented herein is significantly different from the ones presented by other authors, being more systematic and easily generalizable. The original proof by Novikov (Novikov, 1965), found also in (Ishimaru, 1978; Konotop & Vazquez, 1994; M. Scott, 2013) lies in expanding the functional $\mathcal{F}[\Xi(\bullet;\theta)]$ into a Volterra-Taylor series with respect to $\Xi(\bullet;\theta)$ around zero, multiplying both sides by $\Xi(t;\theta)$ and then taking the average, see also Section 3.2. For the completion of the proof, Isserlis' theorem (Isserlis, 1918) for computing higher order moments in terms of the autocorrelation for the Gaussian case is invoked, which is not needed in our proof. Our derivation is based on a generic result (Theorem 3.1), which is valid for any type of random arguments, and the specification of the characteristic (function-) functional as Gaussian, allowing for generalizations to other cases, as it will be shown in the following sections. In (Klyatskin, 2005; Sobczyk, 1985) the starting point of the proof is the functional shift operator in exponential form, as in the present work. However, a different operator formalism is used afterwards, making the proof less transparent, to our opinion.

**Remark 3.4: On the range of $s$ in $\mathbb{E}^{\theta}\left[\Xi(s;\theta)\,\mathcal{F}[X_0(\theta);\Xi(\bullet|_{t_0}^{t};\theta)]\right]$.** As we have mentioned after stating the extended NF theorem, Eq. (3.32), argument $s$ does not need to be confined in the range $[t_0, t]$ of the RFFℓ $\mathcal{F}[\cdots]$, provided that $s$ belongs to the domain of definition of $\Xi(\bullet;\theta)$. This can be easily shown by following the above proof for $s > t$, in which case $\Xi(s;\theta)\,\mathcal{F}[X_0(\theta);\Xi(\bullet|_{t_0}^{t};\theta)]$ is an RFFℓ $\mathcal{G}[X_0(\theta);\Xi(\bullet|_{t_0}^{s};\theta)]$, i.e. the range of the functional dependence on $\Xi(\bullet;\theta)$ is now $[t_0, s] \supseteq [t_0, t]$. This change in range of the RFFℓ alters the range of all integrations appearing in the proof, which should also be $[t_0, s]$ in this case. This means that the integral term in Eq. (3.56) would be altered into

$$\int_{t_0}^{s} C_{\Xi\Xi}(s,\tau)\,\bar{\mathcal{T}}^{(0,2)}_{X_0\Xi}\,\bar{\mathcal{T}}^{(1,1)}_{X_0\Xi}\,\bar{\mathcal{T}}^{(2,0)}_{X_0\Xi}\left[\frac{\delta\mathcal{F}[\upsilon;u(\bullet|_{t_0}^{t})]}{\delta u(\tau)}\right]d\tau .$$

However, since the range of FFℓ $\mathcal{F}[\cdots]$ remains $[t_0, t]$, its functional derivative $\delta\mathcal{F}[\upsilon;u(\bullet|_{t_0}^{t})]/\delta u(\tau)$ is zero for $\tau\in[t, s]$, and thus the range of integral is restricted again to $[t_0, t]$. With these observations, we see that Eq. (3.56), and thus extended NF theorem, Eq. (3.32) is also valid for $s > t$.

### 3.4.2 Extension for the multidimensional case

In this paragraph, we concisely present the generalizations of Theorems 3.1 and 3.2 for the case where $\boldsymbol{X}^0(\theta)$ is an $N-$dimensional random vector and $\boldsymbol{\Xi}(\bullet;\theta)$ is a $K-$dimensional random function. Note that, in general, dimensions $N$ and $K$ do not necessarily agree. Since the proofs of the generalized theorems are similar to those of Theorems 3.1, 3.2, their details are omitted.

**Theorem 3.3: The average of an RFFℓ with vector arguments.** Under appropriate smoothness assumptions, the mean value of the RFFℓ $\mathcal{G}[\boldsymbol{X}^0(\theta);\boldsymbol{\Xi}(\bullet;\theta)]$ is expressed as

$$\mathbb{E}^{\theta}\Big[\mathcal{G}[\boldsymbol{X}^0(\theta)\,;\,\Xi(\bullet\,;\theta)]\Big] = \varphi_{\hat{X}^0\hat{\Xi}(\cdot)}\left(\frac{\partial}{i\,\partial \boldsymbol{v}}\,;\,\frac{\delta}{i\,\delta \boldsymbol{u}(\bullet)}\right)\mathcal{G}[\boldsymbol{v}\,;\boldsymbol{u}(\bullet|_{t_0}^{t})]\Big|_{\substack{\boldsymbol{v}=\boldsymbol{m}_{X^0}\\ \boldsymbol{u}(\bullet)=\boldsymbol{m}_{\Xi(\cdot)}(\bullet)}} =$$

$$= \mathbb{E}^{\theta}\left[\exp\left(\sum_{n=1}^{N}\hat{X}_n^0(\theta)\frac{\partial}{\partial \upsilon_n} + \sum_{k=1}^{K}\int_{t_0}^{t} ds\,\hat{\Xi}_k(s;\theta)\frac{\delta}{\delta u_k(s)}\right)\right]\mathcal{G}[\boldsymbol{v}\,;\boldsymbol{u}(\bullet|_{t_0}^{t})]\Big|_{\substack{\boldsymbol{v}=\boldsymbol{m}_{X^0}\\ \boldsymbol{u}(\bullet)=\boldsymbol{m}_{\Xi}(\bullet)}} \tag{3.57}$$

where $\hat{\boldsymbol{X}}^0(\theta) = \boldsymbol{X}^0(\theta) - \boldsymbol{m}_{X^0}$, $\hat{\Xi}(s;\theta) = \Xi(s;\theta) - \boldsymbol{m}_{\Xi}(s)$ are the fluctuations of the random elements $\boldsymbol{X}^0(\theta)$ and $\Xi(s;\theta)$ around their mean values, and $\varphi_{\hat{X}^0\hat{\Xi}(\cdot)}[\boldsymbol{v}\,;\,\boldsymbol{u}(\bullet)]$ is the joint characteristic FFℓ of the said fluctuations. ■

As in Theorem 3.1, Theorem 3.3 is proved by identifying the similarity between the average of the random shift operator $\mathcal{T}_{\hat{X}^0\hat{\Xi}(\cdot)}(\theta)[\bullet]$ for the multidimensional case, defined by

$$\mathcal{G}[\boldsymbol{X}^0(\theta)\,;\,\Xi(\bullet|_{t_0}^{t}\,;\theta)] \equiv \mathcal{G}[\boldsymbol{m}_{X^0} + \hat{\boldsymbol{X}}^0(\theta)\,;\,\boldsymbol{m}_{\Xi}(\bullet|_{t_0}^{t}) + \hat{\Xi}(\bullet|_{t_0}^{t}\,;\theta)] \equiv$$

$$\equiv \mathcal{T}_{\hat{X}^0\hat{\Xi}(\cdot)}(\theta)\Big[\mathcal{G}[\boldsymbol{m}_{X^0}\,;\boldsymbol{m}_{\Xi}(\bullet|_{t_0}^{t})]\Big] =$$

$$= \exp\left(\sum_{n=1}^{N}\hat{X}_n^0(\theta)\frac{\partial}{\partial \upsilon_n} + \sum_{k=1}^{K}\int_{t_0}^{t} ds\,\hat{\Xi}_k(s;\theta)\frac{\delta}{\delta u_k(s)}\right)\mathcal{G}[\boldsymbol{m}_{X^0}\,;\boldsymbol{m}_{\Xi}(\bullet|_{t_0}^{t})], \tag{3.58}$$

and the definition relation of joint characteristic FFℓ

$$\varphi_{\hat{X}^0\hat{\Xi}(\cdot)}[\boldsymbol{v}\,;\,\boldsymbol{u}(\bullet)] = \mathbb{E}^{\theta}\left[\exp\left(i\sum_{n=1}^{N}\hat{X}_n^0(\theta)\,\upsilon_n + i\sum_{k=1}^{K}\int_{t_0}^{t}\hat{\Xi}_k(s;\theta)\,u_k(s)\,ds\right)\right]. \tag{3.59}$$

Eqs. (3.58), (3.59) can be obtained using Volterra’s passing from the discrete to continuous, as has been done before for their scalar counterparts, Eqs. (3.20), (3.13) respectively.

**Theorem 3.4: The extended Novikov-Furutsu theorem for the multidimensional case.** Under appropriate smoothness assumptions, for the RFFℓ $\mathcal{F}[\boldsymbol{X}^0(\theta)\,;\,\Xi(\bullet|_{t_0}^{t}\,;\theta)] \equiv \mathcal{F}[\cdots]$, with jointly Gaussian arguments $\boldsymbol{X}^0(\theta)$, $\Xi(\bullet\,;\theta)$, the following formula holds true, for $k = 1, \ldots, K$:

$$\mathbb{E}^{\theta}\Big[\Xi_k(s;\theta)\,\mathcal{F}[\cdots]\Big] =$$

$$= m_{\Xi_k}(s)\,\mathbb{E}^{\theta}\big[\mathcal{F}[\cdots]\big] + \sum_{n=1}^{N} C_{X_n^0\Xi_k}(s)\,\mathbb{E}^{\theta}\left[\frac{\partial\mathcal{F}[\cdots]}{\partial X_n^0(\theta)}\right] + \tag{3.60}$$

$$+ \sum_{\ell=1}^{K}\int_{t_0}^{t} C_{\Xi_k\Xi_\ell}(s,\tau)\,\mathbb{E}^{\theta}\left[\frac{\delta\mathcal{F}[\cdots]}{\delta\Xi_\ell(\tau;\theta)}\right]d\tau. \quad ■$$

Eq. (3.60) is proved by substituting $\mathcal{G}[\boldsymbol{X}^0(\theta)\,;\,\Xi(\bullet|_{t_0}^{t}\,;\theta)] = \Xi_k(s;\theta)\,\mathcal{F}[\cdots]$ in Eq. (3.58), and employing the joint characteristic FFℓ of jointly Gaussian $\hat{\boldsymbol{X}}^0(\theta)$, $\hat{\Xi}(\bullet\,;\theta)$:

$$\varphi^{\text{Gauss}}_{\hat{X}^0\hat{\Xi}(\cdot)}[\boldsymbol{\upsilon}\,;\boldsymbol{u}(\bullet|_{t_0}^{t})] =$$
$$= \exp\left(-\frac{1}{2}\sum_{k_1=1}^{K}\sum_{k_2=1}^{K}\int_{t_0}^{t}\int_{t_0}^{t} C_{\Xi_{k_1}\Xi_{k_2}}(s_1,s_2)\,u_{k_1}(s_1)\,u_{k_2}(s_2)\,ds_1 ds_2\right)\times$$
$$\times\exp\left(-\frac{1}{2}\sum_{n_1=1}^{N}\sum_{n_2=1}^{N} C_{X^0_{n_1}X^0_{n_2}}\,\upsilon_{n_1}\upsilon_{n_2}\right)\cdot\exp\left(-\sum_{n=1}^{N}\upsilon_n\sum_{k=1}^{K}\int_{t_0}^{t} C_{X^0_n\Xi_k}(s)\,u_k(s)\,ds\right). \quad (3.61)$$

Eq. (3.61) can be easily obtained by using Volterra's approach. For proving Eq. (3.60), a procedure similar to the proof of Theorem 3.2 is followed, by defining and studying the appropriate quadratic averaged shift operators $\bar{\mathcal{T}}^{(2)}$.

### 3.4.3 Extension for averages containing the temporal derivative of the argument

As we have mentioned in the introductory Section 3.1, another class of averages needing further evaluation is $\mathbb{E}^{\theta}\left[\mathcal{F}[X_0(\theta)\,;\Xi(\bullet|_{t_0}^{t};\theta)]\,\dot{\Xi}(s;\theta)\right]$, where overdot denotes the first temporal derivative. Such averages, for jointly Gaussian arguments, are expressed via Eq. (3.40) as

$$\mathbb{E}^{\theta}\left[\dot{\Xi}(s;\theta)\mathcal{F}[X_0(\theta);\Xi(\bullet|_{t_0}^{t};\theta)]\right] =$$
$$= \left\{\bar{\mathcal{T}}^{(2,0)}_{X_0\Xi}\,\bar{\mathcal{T}}^{(1,1)}_{X_0\Xi}\,\bar{\mathcal{T}}^{(0,2)}_{X_0\Xi}\left[\dot{u}(s)\mathcal{F}[\upsilon;u(\bullet|_{t_0}^{t})]\right]\right\}_{\substack{\upsilon=m_{X_0}\\ u(\bullet)=m_{\Xi}(\bullet)}}. \quad (3.62)$$

Thus, in this case, we have to determine the actions of quadratic averaged shift operators $\bar{\mathcal{T}}^{(2)}$ on the deterministic counterpart of $\dot{\Xi}(s;\theta)\mathcal{F}[\cdots]$, that is $\dot{u}(s)\mathcal{F}[\upsilon;u(\bullet|_{t_0}^{t})]$. First, as in Lemma 3.4, action of $\bar{\mathcal{T}}^{(2,0)}_{X_0\Xi}$ is easily determined, by employing its linearity, into

$$\bar{\mathcal{T}}^{(2,0)}_{X_0\Xi}\left[\dot{u}(s)\mathcal{F}[\upsilon;u(\bullet|_{t_0}^{t})]\right] = \dot{u}(s)\,\bar{\mathcal{T}}^{(2,0)}_{X_0\Xi}\left[\mathcal{F}[\upsilon;u(\bullet|_{t_0}^{t})]\right]. \quad (3.63)$$

Now, for the actions of $\bar{\mathcal{T}}^{(1,1)}_{X_0\Xi}\,\bullet$, $\bar{\mathcal{T}}^{(0,2)}_{X_0\Xi}\,\bullet$, a product rule for higher-order Volterra derivatives has to be derived, for $\dot{u}(s)\,\mathcal{F}[\upsilon\,;u(\bullet|_{t_0}^{t})]$. Following the discussion in paragraph 2.5.1, and under the assumption that $u(\bullet)$ is $C^1\left([t_0,t]\to\mathbb{R}\right)$, like the paths of $\Xi(\bullet;\theta)$, $\dot{u}(s)$ is expressed as a linear functional of integral type with a singular kernel:

$$\dot{u}(s) = \int_{t_0}^{s}\delta'(s-\tau)\,u(\tau)\,d\tau, \quad (3.64)$$

where $\delta'(s-\tau)\equiv-\partial\delta(s-\tau)/\partial\tau$. Expression (3.64) is formal, yet it makes the Volterra derivative of $\dot{u}(s)$ easily computable to

$$\frac{\delta \dot{u}(s)}{\delta u(\tau)} = \delta'(s-\tau). \tag{3.65}$$

Under Eq. (3.65), the product rule (3.44) of Volterra derivatives for $\dot{u}(s)\,\mathcal{F}[\upsilon\,;u(\bullet|_{t_0}^{t})]$ is expressed, for $k \in \mathbb{N}$ as

$$\frac{\delta^k\left[\dot{u}(s)\,\mathcal{F}[\upsilon\,;u(\bullet|_{t_0}^{t})]\right]}{\delta u(\tau^{(1)})\cdots\delta u(\tau^{(k)})} = \dot{u}(s)\,\frac{\delta^k\,\mathcal{F}[\upsilon\,;u(\bullet|_{t_0}^{t})]}{\delta u(\tau^{(1)})\cdots\delta u(\tau^{(k)})} + \\ + \sum_{n=1}^{k}\delta'(s-\tau^{(n)})\,\frac{\delta^{k-1}\,\mathcal{F}[\upsilon\,;u(\bullet|_{t_0}^{t})]}{\prod_{\ell=1,\ \ell\neq n}^{k}\delta u(\tau^{(\ell)})}, \tag{3.66}$$

By employing now Eq. (3.66) and following the same procedure as for Lemmata 3.5, 3.6, the actions of $\bar{\mathcal{T}}_{X_0\Xi}^{(1,1)}\,\bullet$, $\bar{\mathcal{T}}_{X_0\Xi}^{(0,2)}\,\bullet$ are determined into

$$\bar{\mathcal{T}}_{X_0\Xi}^{(1,1)}\left[\dot{u}(s)\,\mathcal{F}[\upsilon\,;u(\bullet|_{t_0}^{t})]\right] = \dot{u}(s)\,\bar{\mathcal{T}}_{X_0\Xi}^{(1,1)}\left[\mathcal{F}[\upsilon\,;u(\bullet|_{t_0}^{t})]\right] + \\ + \dot{C}_{X_0\Xi}(s)\,\bar{\mathcal{T}}_{X_0\Xi}^{(1,1)}\left[\frac{\partial\mathcal{F}[\upsilon\,;u(\bullet|_{t_0}^{t})]}{\partial\upsilon}\right], \tag{3.67}$$

and

$$\bar{\mathcal{T}}_{X_0\Xi}^{(0,2)}\left[\dot{u}(s)\,\mathcal{F}[\upsilon\,;u(\bullet|_{t_0}^{t})]\right] = \dot{u}(s)\,\bar{\mathcal{T}}_{X_0\Xi}^{(0,2)}\left[\mathcal{F}[\upsilon\,;u(\bullet|_{t_0}^{t})]\right] + \\ + \int_{t_0}^{t}\partial_s C_{\Xi\Xi}(s,\tau)\,\bar{\mathcal{T}}_{X_0\Xi}^{(0,2)}\left[\frac{\delta\mathcal{F}[\upsilon\,;u(\bullet|_{t_0}^{t})]}{\delta u(\tau)}\right]d\tau, \tag{3.68}$$

where $\partial_s C_{\Xi\Xi}(s,\tau) = \partial C_{\Xi\Xi}(s,\tau)/\partial s$. By employing Eqs. (3.63), (3.67), (3.68) and following the proof in paragraph 3.4.1, we finally obtain the generalization of NF theorem:

$$\mathbb{E}^{\theta}\left[\dot{\Xi}(s;\theta)\,\mathcal{F}[\cdots]\right] = \\ = \dot{m}_{\Xi}(s)\,\mathbb{E}^{\theta}\left[\mathcal{F}[\cdots]\right] + \dot{C}_{X_0\Xi}(s)\,\mathbb{E}^{\theta}\left[\frac{\partial\mathcal{F}[\cdots]}{\partial X_0(\theta)}\right] + \\ + \int_{t_0}^{t}\partial_s C_{\Xi\Xi}(s,\tau)\,\mathbb{E}^{\theta}\left[\frac{\delta\mathcal{F}[\cdots]}{\delta\Xi(\tau;\theta)}\right]d\tau\,. \tag{3.69}$$

In formula (3.69), the moments including temporal derivatives $\dot{m}_{\Xi}(s) = \mathbb{E}^{\theta}[\dot{\Xi}(s;\theta)]$, $\dot{C}_{X_0\Xi}(s) = C_{X_0\dot{\Xi}}(s) = \mathbb{E}^{\theta}[X_0(\theta)\,\dot{\Xi}(s;\theta)] - m_{X_0}\,\dot{m}_{\Xi}(s)$, $\partial_s C_{\Xi\Xi}(s,\tau) = C_{\dot{\Xi}\Xi}(s,\tau) = \mathbb{E}^{\theta}[\dot{\Xi}(s;\theta)\,\Xi(\tau;\theta)] - \dot{m}_{\Xi}(s)\,m_{\Xi}(\tau)$ are considered known, as data of the problem.

### 3.5 Generalization of Novikov-Furutsu theorem for averages with multiple time factors of the argument

A last, but quite significant generalization of the NF theorem, is to consider averages that explicitly contain $\Xi(\bullet;\theta)$ in multiple time instances; $\mathbb{E}^{\theta}\left[\Xi(s_1;\theta)\cdots\Xi(s_n;\theta)\,\mathcal{F}[\cdots]\right]$. As it will be shown in Chapter 4, the above form of averages is the most general we shall encounter. To the best of our knowledge, this generalization has not been presented before.

**Theorem 3.5: The generalized Novikov-Furutsu theorem.** Let us consider again the smooth enough RFFℓ $\mathcal{F}[X_0(\theta);\Xi(\bullet|_{t_0}^{t};\theta)]=\mathcal{F}[\cdots]$ with jointly Gaussian arguments, and $n$ time instances $s_1, s_2, \dots, s_n$ belonging to the domain of definition of $\Xi(\bullet;\theta)$[21]. In this case, it holds true that

$$\mathbb{E}^{\theta}\left[\Xi(s_1;\theta)\cdots\Xi(s_n;\theta)\,\mathcal{F}[\cdots]\right]=$$

$$=\sum_{I_1\cup I_2\cup I_3=I^{(n)}}\prod_{i_1\in I_1}m_{\Xi}(s_{i_1})\prod_{i_2\in I_2}C_{X_0\Xi}(s_{i_2})\sum_{k=0}^{\lfloor |I_3|/2\rfloor}\sum_{\{P,S\}\in\wp_k(I_3)}\prod_{\{j_1,j_2\}\in P}C_{\Xi\Xi}(s_{j_1},s_{j_2})\times$$

$$\times\overbrace{\int_{t_0}^{t}\cdots\int_{t_0}^{t}}^{(|I_3|-2k)}\prod_{i_3\in S}C_{\Xi\Xi}(s_{i_3},\tau_{i_3})\,\mathbb{E}^{\theta}\left[\frac{\partial^{|I_2|}\delta^{|I_3|-2k}\mathcal{F}[\cdots]}{\partial X_0^{|I_2|}(\theta)\prod_{i_3\in S}\delta\Xi(\tau_{i_3};\theta)}\right]\prod_{i_3\in S}d\tau_{i_3}, \tag{3.70}$$

where $I^{(n)}=\{1,2,3,\dots,n\}$, that is, the index set of time instances $s_1, s_2, \dots, s_n$, while the sum $\sum_{I_1\cup I_2\cup I_3=I^{(n)}}\bullet$ extends over all partitions of $I^{(n)}$ into three disjoint subsets $I_1$, $I_2$, $I_3$. Furthermore, $\lfloor\bullet\rfloor$ denotes the floor function, $|\bullet|$ the cardinality of a set, and the sum $\sum_{\{P,S\}\in\wp_k(I_3)}\bullet$ extends over the set $\wp_k(I_3)$ of all unordered partitions of subset $I_3$ into $k$ unordered pairs and $|I_3|-2k$ singletons. For every partition of $I_3$, the set containing the pairs is denoted as $P$, while the set of singletons is denoted as $S$.

**The NF theorem for averages containing powers of the Gaussian argument.** Let us write the special case of Eq. (3.70), in which all time instances are equal: $s_1=s_2=\cdots=s_n=s$:

$$\mathbb{E}^{\theta}\left[\Xi(s;\theta)\cdots\Xi(s;\theta)\,\mathcal{F}[\cdots]\right]\equiv\mathbb{E}^{\theta}\left[\Xi^{n}(s;\theta)\,\mathcal{F}[\cdots]\right]=$$

$$=\sum_{I_1\cup I_2\cup I_3=I^{(n)}}m_{\Xi}^{|I_1|}(s)\,C_{X_0\Xi}^{|I_2|}(s)\sum_{k=0}^{\lfloor |I_3|/2\rfloor}\sum_{\{P,S\}\in\wp_k(I_3)}C_{\Xi\Xi}^{k}(s,s)\times$$

$$\times\overbrace{\int_{t_0}^{t}\cdots\int_{t_0}^{t}}^{(|I_3|-2k)}\prod_{i=1}^{|I_3|-2k}C_{\Xi\Xi}(s,\tau_i)\,\mathbb{E}^{\theta}\left[\frac{\partial^{|I_2|}\delta^{|I_3|-2k}\mathcal{F}[\cdots]}{\partial X_0^{|I_2|}(\theta)\prod_{i=1}^{|I_3|-2k}\delta\Xi(\tau_i;\theta)}\right]\prod_{i=1}^{|I_3|-2k}d\tau_i. \tag{3.71}$$

---

[21] As in the proof of the extended NF theorem, Eq. (3.32), the proof of Eq. (3.70) will be performed for $s_1, s_2, \dots, s_n\in[t_0,t]$. However, by following the arguments of Remark 3.4, the validity of Eq. (3.70) can be extended for time instances after $t$.

In Eq. (3.71), for a given $k$, it is easily observed that all terms of the sum $\sum_{\{P,S\}\in\wp_k(I_3)}\bullet$ are equal. The number of partitions of $I_3$ with $k$ pairs is calculated to $|I_3|!/\left(2^k\,k!\,(|I_3|-2k)!\right)$ (Knuth, 1973) p. 65. This number is identified as $H_{|I_3|,k}$, since by

$$H_{n,k} = \frac{n!}{2^k\,k!\,(n-2k)!}, \tag{3.72}$$

we denote the absolute values of the coefficients appearing in the $n-$***th probabilist's Hermite polynomial*** $He_n(x)$ (Abramowitz & Stegun, 1964) table 22.3:

$$He_n(x) = \sum_{k=0}^{\lfloor n/2\rfloor}(-1)^k\,H_{n,k}\,x^{n-2k}. \tag{3.73}$$

Thus, Eq. (3.71) is simplified to

$$\mathbb{E}^\theta\left[\Xi^n(s;\theta)\,\mathcal{F}[\cdots]\right] =$$

$$= \sum_{I_1\cup I_2\cup I_3=I^{(n)}} m_\Xi^{|I_1|}(s)\,C_{X_0\Xi}^{|I_2|}(s)\sum_{k=0}^{\lfloor |I_3|/2\rfloor}H_{|I_3|,k}\,C_{\Xi\Xi}^k(s,s)\times$$

$$\times\int_{t_0}^{t}\overset{(|I_3|-2k)}{\cdots}\int_{t_0}^{t}\prod_{i=1}^{|I_3|-2k}C_{\Xi\Xi}(s,\tau_i)\,\mathbb{E}^\theta\left[\frac{\partial^{|I_2|}\delta^{|I_3|-2k}\mathcal{F}[\cdots]}{\partial X_0^{|I_2|}(\theta)\prod_{i=1}^{|I_3|-2k}\delta\Xi(\tau_i;\theta)}\right]\prod_{i=1}^{|I_3|-2k}d\tau_i. \tag{3.74}$$

Observe now that, in the sum $\sum_{I_1\cup I_2\cup I_3=I^{(n)}}\bullet$ of the right-hand side of Eq. (3.74), the terms corresponding to partitions of $I^{(n)}$ with the same cardinalities of sets $I_1$, $I_2$, $I_3$, $\left(|I_1|,|I_2|,|I_3|\right)=(m_1,m_2,m_3)$, are equal. Since the number of partitions of $n$ elements into 3 subsets with $m_1$ elements in the first set, $m_2$ elements in the second and $m_3$ in the third, is the multinomial coefficient with three factors $\binom{n}{m_1,m_2,m_3}$, the sum $\sum_{I_1\cup I_2\cup I_3=I^{(n)}}\bullet$ can be substituted by $\sum_{m_1+m_2+m_3=n}\binom{n}{m_1,m_2,m_3}\bullet$, resulting thus in

$$\mathbb{E}^\theta\left[\Xi^n(s;\theta)\,\mathcal{F}[\cdots]\right] = \sum_{m_1+m_2+m_3=n}\binom{n}{m_1,m_2,m_3}m_\Xi^{m_1}(s)\,C_{X_0\Xi}^{m_2}(s)\sum_{k=0}^{\lfloor m_3/2\rfloor}H_{m_3,k}\,C_{\Xi\Xi}^k(s,s)\times$$

$$\times\int_{t_0}^{t}\overset{(m_3-2k)}{\cdots}\int_{t_0}^{t}\prod_{i=1}^{m_3-2k}C_{\Xi\Xi}(s,\tau_i)\,\mathbb{E}^\theta\left[\frac{\partial^{m_2}\delta^{m_3-2k}\mathcal{F}[\cdots]}{\partial X_0^{m_2}(\theta)\prod_{i=1}^{m_3-2k}\delta\Xi(\tau_i;\theta)}\right]\prod_{i=1}^{m_3-2k}d\tau_i. \tag{3.75}$$

Eq. (3.75) is also derived independently in Appendix B, by following a constructive proof based on the application of quadratic averaged shift operators, as shown in the previous paragraph 3.4.1.

**Remark 3.5: Deduction of Isserlis theorem from Eq. (3.70).** First, let us specify Eq. (3.70) for a ***functional*** $\mathcal{F}[\Xi(\bullet|_{t_0}^{t};\theta)]$ ***with zero-mean Gaussian argument***:

$$\mathbb{E}^{\theta}\Big[\Xi(s_1;\theta)\cdots\Xi(s_n;\theta)\,\mathcal{F}[\Xi(\bullet|_{t_0}^{t};\theta)]\Big] = \sum_{k=0}^{\lfloor n/2\rfloor}\sum_{\{P,S\}\in\wp_k(I^{(n)})}\prod_{\{j_1,j_2\}\in P}C_{\Xi\Xi}(s_{j_1},s_{j_2})\times$$

$$\times\int_{t_0}^{t}\overset{(n-2k)}{\cdots}\int_{t_0}^{t}\prod_{i\in S}C_{\Xi\Xi}(s_i,\tau_i)\,\mathbb{E}^{\theta}\left[\frac{\delta^{n-2k}\mathcal{F}[\Xi(\bullet|_{t_0}^{t};\theta)]}{\prod_{i\in S}\delta\Xi(\tau_i;\theta)}\right]\prod_{i\in S}d\tau_i. \tag{3.76}$$

Eq. (3.76) can be seen as the generalization of the classical NF theorem, Eq. (3.4), for averages with multiple times instances of the argument. Consider now that $\mathcal{F}[\Xi(\bullet|_{t_0}^{t};\theta)]=1$. In this case, all the derivatives appearing in the right-hand side of Eq. (3.76) are zero, except for the case $n-2k=0$ where it equals to 1. Since the maximum value of $k$ is $\lfloor n/2\rfloor$, $n-2k=0$ cannot be achieved for odd $n=2\ell-1$, resulting thus in

$$\mathbb{E}^{\theta}\Big[\Xi(s_1;\theta)\cdots\Xi(s_{2\ell-1};\theta)\Big] = 0,\quad \ell\in\mathbb{N}. \tag{3.77}$$

For even $n=2\ell$, $n-2k=0$ is achieved for $k=\lfloor n/2\rfloor=\ell$. In this case, $\wp_\ell(I^{(2\ell)})$ denotes the set of all different partitions of $I^{(2\ell)}$ into $\ell$ pairs, which means that all elements of $I^{(2\ell)}$ are paired; $S=\varnothing$. Thus, the notation can be simplified by omitting the set of singletons, leading to

$$\mathbb{E}^{\theta}\Big[\Xi(s_1;\theta)\cdots\Xi(s_{2\ell};\theta)\Big] = \sum_{P\in\wp_\ell(I^{(2\ell)})}\prod_{\{j_1,j_2\}\in P}C_{\Xi\Xi}(s_{j_1},s_{j_2}).\quad \ell\in\mathbb{N}. \tag{3.78}$$

The results given by Eqs. (3.77) and (3.78) constitute the well-known Isserlis theorem (Isserlis, 1918; Song & Lee, 2015).

### 3.5.1 Proof of the generalized Novikov-Furutsu theorem by mathematical induction

For $n=1$, $I^{(1)}=\{1\}$, and thus its possible partitions $(I_1,I_2,I_3)$, over which the sum in the right-hand side of Eq. (3.70), are $(\{1\},\varnothing,\varnothing)$, $(\varnothing,\{1\},\varnothing)$, $(\varnothing,\varnothing,\{1\})$. In this case, Eq. (3.70) is simplified into

$$\mathbb{E}^{\theta}\big[\Xi(s;\theta)\,\mathcal{F}[\cdots]\big] = m_{\Xi}(s)\,\mathbb{E}^{\theta}\big[\mathcal{F}[\cdots]\big]+$$

$$+\,C_{X_0\Xi}(s)\,\mathbb{E}^{\theta}\left[\frac{\partial\mathcal{F}[\cdots]}{\partial X_0}\right]+\int_{t_0}^{t}C_{\Xi\Xi}(s,\tau)\,\mathbb{E}^{\theta}\left[\frac{\delta\mathcal{F}[\cdots]}{\delta\Xi(\tau;\theta)}\right]d\tau,$$

which is the extended NF theorem, Eq. (3.32), that was proved previously in paragraph 3.4.1. As a next step, we assume that Eq. (3.70) holds true for a fixed *n*. Thus, for $n+1$:

$$\mathbb{E}^{\theta}\Big[\Xi(s_1;\theta)\cdots\Xi(s_{n+1};\theta)\,\mathcal{F}[\cdots]\Big] = \mathbb{E}^{\theta}\Big[\Xi(s_1;\theta)\cdots\Xi(s_n;\theta)\Big[\Xi(s_{n+1};\theta)\,\mathcal{F}[\cdots]\Big]\Big].$$

By considering $\Xi(s_{n+1};\theta)\,\mathcal{F}[\cdots]$ as another RFFℓ, and using the inductive hypothesis, we obtain

$$\mathbb{E}^{\theta}\left[\prod_{i\in I^{(n+1)}}\Xi(s_i;\theta)\,\mathcal{F}[\cdots]\right]=\mathbb{E}^{\theta}\left[\prod_{i\in I^{(n)}}\Xi(s_i;\theta)\left[\Xi(s_{n+1};\theta)\,\mathcal{F}[\cdots]\right]\right]=$$

$$=\sum_{I_1\cup I_2\cup I_3=I^{(n)}}\prod_{i_1\in I_1}m_{\Xi}(s_{i_1})\prod_{i_2\in I_2}C_{X_0\Xi}(s_{i_2})\sum_{k=0}^{\lfloor |I_3|/2\rfloor}\sum_{\{P,S\}\in\wp_k(I_3)}\prod_{\{j_1,j_2\}\in P}C_{\Xi\Xi}(s_{j_1},s_{j_2})\times$$

$$\times\int_{t_0}^{t}\overset{(|I_3|-2k)}{\cdots}\int_{t_0}^{t}\prod_{i_3\in S}C_{\Xi\Xi}(s_{i_3},\tau_{i_3})\,\mathbb{E}^{\theta}\left[\frac{\partial^{|I_2|}\delta^{|I_3|-2k}\left[\Xi(s_{n+1};\theta)\,\mathcal{F}[\cdots]\right]}{\partial X_0^{|I_2|}(\theta)\prod_{i_3\in S}\delta\Xi(\tau_{i_3};\theta)}\right]\prod_{i_3\in S}d\tau_{i_3},\tag{3.79}$$

The derivative appearing in the right-hand side of Eq. (3.79) is elaborated further using the following formula, which is valid for any $q\in\mathbb{N}$:

$$\frac{\delta^{|P|}\left[\Xi(s;\theta)\,\mathcal{F}[\cdots]\right]}{\prod_{i\in P}\delta\Xi(\tau_i;\theta)}=\Xi(s;\theta)\,\frac{\delta^{|P|}\mathcal{F}[\cdots]}{\prod_{i\in P}\delta\Xi(\tau_{i_3};\theta)}+$$
$$+\sum_{\ell\in P}\delta(s-\tau_\ell)\,\frac{\delta^{|P|-1}\mathcal{F}[\cdots]}{\prod_{i\in P\setminus\{\ell\}}\delta\Xi(\tau_i;\theta)}.\tag{3.80}$$

where $P\setminus\{\ell\}$ denotes the set $P$ from which the element $\ell$ is excluded. Eq. (3.80) is easily proven via mathematical induction, commencing from the relation for $|P|=1$ (product rule for the first-order derivative (Parr & Yang, 1989), Eq. (A.4)):

$$\frac{\delta\left[\Xi(s;\theta)\,\mathcal{F}[\cdots]\right]}{\delta\Xi(\tau;\theta)}=\Xi(s;\theta)\,\frac{\delta\mathcal{F}[\cdots]}{\delta\Xi(\tau;\theta)}+\frac{\delta\Xi(s;\theta)}{\delta\Xi(\tau;\theta)}\,\mathcal{F}[\cdots],$$

and noting that $\delta\Xi(s;\theta)/\delta\Xi(\tau;\theta)=\delta(s-\tau)$ (Parr & Yang, 1989), Eq. (A.27). Furthermore, in order for the case of $|P|=0$ to be incorporated in Eq. (3.80), we introduce the assumption that $\sum_{\ell\in\varnothing}\bullet=0$. By using Eq. (3.80), the linearity of the involved operators (derivative, mean value, integrals and summations), and some algebraic manipulations, we obtain the following decomposition:

$$\mathbb{E}^{\theta}\left[\prod_{i\in I^{(n+1)}}\Xi(s_i;\theta)\,\mathcal{F}[\cdots]\right]=E_1+E_2,\tag{3.81}$$

where

$$E_1=\sum_{I_1\cup I_2\cup I_3=I^{(n)}}\prod_{i_1\in I_1}m_{\Xi}(s_{i_1})\prod_{i_2\in I_2}C_{X_0\Xi}(s_{i_2})\sum_{k=0}^{\lfloor |I_3|/2\rfloor}\sum_{\{P,S\}\in\wp_k(I_3)}\prod_{\{j_1,j_2\}\in P}C_{\Xi\Xi}(s_{j_1},s_{j_2})\times$$
$$\times\int_{t_0}^{t}\overset{(|I_3|-2k)}{\cdots}\int_{t_0}^{t}\prod_{i_3\in S}C_{\Xi\Xi}(s_{i_3},\tau_{i_3})\,\mathbb{E}^{\theta}\left[\Xi(s_{n+1};\theta)\,\mathcal{F}_1[\cdots]\right]\prod_{i_3\in S}d\tau_{i_3},\tag{3.82}$$

with $\mathcal{F}_1[\cdots]=\partial^{|I_2|}\delta^{|I_3|-2k}\mathcal{F}[\cdots]\Big/\left(\partial X_0^{|I_2|}(\theta)\prod_{i_3\in S}\delta\Xi(\tau_{i_3};\theta)\right)$ being a new RFFℓ, and

$$E_2=\sum_{I_1\cup I_2\cup I_3=I^{(n)}}\prod_{i_1\in I_1}m_{\Xi}(s_{i_1})\prod_{i_2\in I_2}C_{X_0\Xi}(s_{i_2})\times$$

$$\times \sum_{k=0}^{\lfloor |I_3|/2 \rfloor} \sum_{\{P,S\}\in \wp_k(I_3)} \sum_{\ell \in S} \prod_{\{j_1,j_2\}\in P\cup\{\ell,n+1\}} C_{\Xi\Xi}(s_{j_1}, s_{j_2}) \times \tag{3.83}$$

$$\times \int_{t_0}^{t} \overset{(|I_3|-2k-1)}{\cdots} \int_{t_0}^{t} \prod_{i_3\in S\setminus\{\ell\}} C_{\Xi\Xi}(s_{i_3}, \tau_{i_3})\, \exists^{\theta}\left[ \frac{\partial^{|I_2|}\delta^{|I_3|-2k-1}\mathcal{F}[\cdots]}{\partial X_0^{|I_2|}(\theta) \prod_{i_3\in S\setminus\{\ell\}} \delta\Xi(\tau_{i_3};\theta)} \right] \prod_{i_3\in S\setminus\{\ell\}} d\tau_{i_3}\,.$$

Term $E_1$ is elaborated further by applying the first step of the induction, Eq. (3.32), to the term $\exists^{\theta}\left[\Xi(s_{n+1};\theta)\,\mathcal{F}_1[\cdots]\right]$ that appears in the right-hand side of Eq. (3.82). This, results into

$$E_1 = E_{11} + E_{12} + E_{13} \tag{3.84}$$

with

$$E_{11} = \sum_{I_1\cup I_2\cup I_3 = I^{(n)}} \prod_{i_1\in I_1\cup\{n+1\}} m_{\Xi}(s_{i_1}) \prod_{i_2\in I_2} C_{X_0\Xi}(s_{i_2}) \times$$

$$\times \sum_{k=0}^{\lfloor |I_3|/2 \rfloor} \sum_{\{P,S\}\in \wp_k(I_3)} \prod_{\{j_1,j_2\}\in P} C_{\Xi\Xi}(s_{j_1}, s_{j_2}) \times \tag{3.85}$$

$$\times \int_{t_0}^{t} \overset{(|I_3|-2k)}{\cdots} \int_{t_0}^{t} \prod_{i_3\in S} C_{\Xi\Xi}(s_{i_3}, \tau_{i_3})\, \exists^{\theta}\left[ \frac{\partial^{|I_2|}\delta^{|I_3|-2k}\mathcal{F}[\cdots]}{\partial X_0^{|I_2|}(\theta) \prod_{i_3\in S} \delta\Xi(\tau_{i_3};\theta)} \right] \prod_{i_3\in S} d\tau_{i_3}\,,$$

$$E_{12} = \sum_{I_1\cup I_2\cup I_3 = I^{(n)}} \prod_{i_1\in I_1} m_{\Xi}(s_{i_1}) \prod_{i_2\in I_2\cup\{n+1\}} C_{X_0\Xi}(s_{i_2}) \times$$

$$\times \sum_{k=0}^{\lfloor |I_3|/2 \rfloor} \sum_{\{P,S\}\in \wp_k(I_3)} \prod_{\{j_1,j_2\}\in P} C_{\Xi\Xi}(s_{j_1}, s_{j_2}) \times \tag{3.86}$$

$$\times \int_{t_0}^{t} \overset{(|I_3|-2k)}{\cdots} \int_{t_0}^{t} \prod_{i_3\in S} C_{\Xi\Xi}(s_{i_3}, \tau_{i_3})\, \exists^{\theta}\left[ \frac{\partial^{|I_2|+1}\delta^{|I_3|-2k}\mathcal{F}[\cdots]}{\partial X_0^{|I_2|+1}(\theta) \prod_{i_3\in S} \delta\Xi(\tau_{i_3};\theta)} \right] \prod_{i_3\in S} d\tau_{i_3}\,,$$

$$E_{13} = \sum_{I_1\cup I_2\cup I_3 = I^{(n)}} \prod_{i_1\in I_1} m_{\Xi}(s_{i_1}) \prod_{i_2\in I_2} C_{X_0\Xi}(s_{i_2}) \times$$

$$\times \sum_{k=0}^{\lfloor |I_3|/2 \rfloor} \sum_{\{P,S\}\in \wp_k(I_3)} \prod_{\{j_1,j_2\}\in P} C_{\Xi\Xi}(s_{j_1}, s_{j_2}) \times \tag{3.87}$$

$$\times \int_{t_0}^{t} \overset{(|I_3|+1-2k)}{\cdots} \int_{t_0}^{t} \prod_{i_3\in S\cup\{n+1\}} C_{\Xi\Xi}(s_{i_3}, \tau_{i_3})\, \exists^{\theta}\left[ \frac{\partial^{|I_2|}\delta^{|I_3|+1-2k}\mathcal{F}[\cdots]}{\partial X_0^{|I_2|}(\theta) \prod_{i_3\in S\cup\{n+1\}} \delta\Xi(\tau_{i_3};\theta)} \right] \prod_{i_3\in S\cup\{n+1\}} d\tau_{i_3}\,.$$

By combining Eqs. (3.81), (3.84) we have

$$\exists^{\theta}\left[ \prod_{i\in I^{(n+1)}} \Xi(s_i;\theta)\,\mathcal{F}[\cdots] \right] = E_{11} + E_{12} + E_{13} + E_2\,. \tag{3.88}$$

**Lemma 3.7.** The following formula holds true

$$E_{13}+E_2=\sum_{I_1\cup I_2\cup I_3=I^{(n)}}\prod_{i_1\in I_1}m_{\Xi}(s_{i_1})\prod_{i_2\in I_2}C_{X_0\Xi}(s_{i_2})\times$$
$$\times\sum_{k=0}^{\lfloor(|I_3|+1)/2\rfloor}\sum_{\{P,S\}\in\wp_k(I_3\cup\{n+1\})}\prod_{\{j_1,j_2\}\in P}C_{\Xi\Xi}(s_{j_1},s_{j_2})\times \tag{3.89}$$
$$\times\int_{t_0}^{t}\overset{(|I_3|+1-2k)}{\cdots}\int_{t_0}^{t}\prod_{i_3\in S}C_{\Xi\Xi}(s_{i_3},\tau_{i_3})\,\Xi^{\theta}\left[\frac{\partial^{|I_2|}\delta^{|I_3|+1-2k}\mathcal{F}[\cdots]}{\partial X_0^{|I_2|}(\theta)\prod_{i_3\in S}\delta\Xi(\tau_{i_3};\theta)}\right]\prod_{i_3\in S}d\tau_{i_3}.$$

Due to its length, proof of Lemma 3.7 is given in Appendix C. ■

In the right-hand sides of Eqs. (3.85), (3.86) and (3.89), we observe that $n+1$, which is the additional element of $I^{(n+1)}$ compared to $I^{(n)}$, is adjoined to $I_1$, $I_2$, $I_3$ subsets of $I^{(n)}$ respectively. Thus, by introducing $I'_1$, $I'_2$, $I'_3$ as the subsets partitioning $I^{(n+1)}$, the aforementioned equations can be expressed equivalently in terms of the subsets of $I^{(n+1)}$ as

$$E_{11}=\sum_{\substack{I'_1\cup I'_2\cup I'_3=I^{(n+1)}\\ n+1\in I'_1}}\prod_{i_1\in I'_1}m_{\Xi}(s_{i_1})\prod_{i_2\in I'_2}C_{X_0\Xi}(s_{i_2})\sum_{k=0}^{\lfloor|I'_3|/2\rfloor}\sum_{\{P,S\}\in\wp_k(I'_3)}\prod_{\{j_1,j_2\}\in P}C_{\Xi\Xi}(s_{j_1},s_{j_2})\times$$
$$\times\int_{t_0}^{t}\overset{(|I'_3|-2k)}{\cdots}\int_{t_0}^{t}\prod_{i_3\in S}C_{\Xi\Xi}(s_{i_3},\tau_{i_3})\,\Xi^{\theta}\left[\frac{\partial^{|I'_2|}\delta^{|I'_3|-2k}\mathcal{F}[\cdots]}{\partial X_0^{|I'_2|}(\theta)\prod_{i_3\in S}\delta\Xi(\tau_{i_3};\theta)}\right]\prod_{i_3\in S}d\tau_{i_3}, \tag{3.90}$$

$$E_{12}=\sum_{\substack{I'_1\cup I'_2\cup I'_3=I^{(n+1)}\\ n+1\in I'_2}}\prod_{i_1\in I'_1}m_{\Xi}(s_{i_1})\prod_{i_2\in I'_2}C_{X_0\Xi}(s_{i_2})\sum_{k=0}^{\lfloor|I'_3|/2\rfloor}\sum_{\{P,S\}\in\wp_k(I'_3)}\prod_{\{j_1,j_2\}\in P}C_{\Xi\Xi}(s_{j_1},s_{j_2})\times$$
$$\times\int_{t_0}^{t}\overset{(|I'_3|-2k)}{\cdots}\int_{t_0}^{t}\prod_{i_3\in S}C_{\Xi\Xi}(s_{i_3},\tau_{i_3})\,\Xi^{\theta}\left[\frac{\partial^{|I'_2|}\delta^{|I'_3|-2k}\mathcal{F}[\cdots]}{\partial X_0^{|I'_2|}(\theta)\prod_{i_3\in S}\delta\Xi(\tau_{i_3};\theta)}\right]\prod_{i_3\in S}d\tau_{i_3}, \tag{3.91}$$

$$E_{13}+E_2=\sum_{\substack{I'_1\cup I'_2\cup I'_3=I^{(n+1)}\\ n+1\in I'_3}}\prod_{i_1\in I_1}m_{\Xi}(s_{i_1})\prod_{i_2\in I_2}C_{X_0\Xi}(s_{i_2})\sum_{k=0}^{\lfloor|I'_3|/2\rfloor}\sum_{\{P,S\}\in\wp_k(I'_3)}\prod_{\{j_1,j_2\}\in P}C_{\Xi\Xi}(s_{j_1},s_{j_2})\times$$
$$\times\int_{t_0}^{t}\overset{(|I'_3|-2k)}{\cdots}\int_{t_0}^{t}\prod_{i_3\in S}C_{\Xi\Xi}(s_{i_3},\tau_{i_3})\,\Xi^{\theta}\left[\frac{\partial^{|I_2|}\delta^{|I'_3|-2k}\mathcal{F}[\cdots]}{\partial X_0^{|I_2|}(\theta)\prod_{i_3\in S}\delta\Xi(\tau_{i_3};\theta)}\right]\prod_{i_3\in S}d\tau_{i_3}. \tag{3.92}$$

Since $\sum_{I'_1\cup I'_2\cup I'_3=I^{(n+1)}}\bullet=\sum_{\substack{I'_1\cup I'_2\cup I'_3=I^{(n+1)}\\ n+1\in I'_1}}\bullet+\sum_{\substack{I'_1\cup I'_2\cup I'_3=I^{(n+1)}\\ n+1\in I'_2}}\bullet+\sum_{\substack{I'_1\cup I'_2\cup I'_3=I^{(n+1)}\\ n+1\in I'_3}}\bullet$, by substituting Eqs. (3.90)-(3.92) into Eq. (3.88), Eq. (3.70) for $n+1$ is obtained, completing thus the proof of Theorem 3.5 by mathematical induction.

# Chapter 4: SLEs after the NF theorem. Formulation of exact pdf evolution equations for linear, additively excited systems as first examples

**Summary.** In the present chapter, we employ the generalizations of NF theorem, Chapter 3, in order to evaluate further the SLEs derived in Chapter 2. In this way, we obtain the final forms of SLEs corresponding to dynamical systems whose random initial value and excitation are jointly Gaussian. After the NF theorem, the averaged terms in the original SLEs are expressed equivalently as ***nonlocal terms depending on the whole history of the response and excitation***. Thus, in order for SLEs to be of practical use, an approximate, current-time closure has to be employed on the said terms, which is the topic of the next chapter. However, in this chapter, we also consider, as first validations of our methodology, some ***special cases in which SLEs are in closed form***. The first such case is the nonlinear dynamical systems under white noise excitation, for which we retrieve the classical Fokker-Planck-Kolmogorov equation from the one-time response SLEs. Another case in which SLEs are, in fact, exact pdf equations in closed form are the linear systems under additive coloured noise excitation. In each of the Secs. 4.1-5, it is validated that, for linear systems under additive Gaussian excitation, the expected Gaussian pdf is the unique solution to the respective evolution equation. Last, in Sec. 4.6, the case of quadratic Gaussian excitation is examined. For this case, and by employing the last generalization of NF theorem of Sec. 3.5, SLE takes a form that contains, in principle, all *x*-derivatives, resembling thus the Kramers-Moyal expansion. It is also validated that, for the linear-additive case, the said pdf evolution equation calculates correctly all cumulants of the response.

## 4.1 The final form of one-time response stochastic Liouville equation

In Section 2.3, SLE (2.18) for the one-time response pdf, was formulated

$$\frac{\partial f_{X(t)}(x)}{\partial t} + \frac{\partial}{\partial x}\Big(h(x)\, f_{X(t)}(x)\Big) = -\frac{\partial}{\partial x}\Big(q(x)\, \mathbb{E}^{\theta}\big[\delta(x-X(t;\theta))\,\Xi(t;\theta)\big]\Big), \tag{4.1}$$

corresponding to the scalar RIVP (1.1)-(1.2):

$$\dot{X}(t;\theta) = h(X(t;\theta)) + q(X(t;\theta))\,\Xi(t;\theta), \qquad X(t_0;\theta) = X_0(\theta). \tag{4.2a,b}$$

By considering the response $X(t;\theta)$ as a RFFℓ of the initial value $X_0(\theta)$ and excitation $\Xi(\bullet;\theta)$; $X(t;\theta) = X[X_0(\theta);\Xi(\bullet|_{t_0}^{t};\theta)]$, and under the assumption that $X_0(\theta)$, $\Xi(\bullet;\theta)$ are jointly Gaussian, the average in the right-hand side of SLE (4.1) can be further evaluated by using the extended NF theorem (3.32), specified for this case into

$$\begin{aligned}
&\mathbb{E}^{\theta}\Big[\,\Xi(t;\theta)\,\delta(x-X[X_0(\theta);\Xi(\bullet|_{t_0}^{t};\theta)])\Big] = \\
&\quad = m_{\Xi}(t)\,\mathbb{E}^{\theta}\Big[\delta(x-X[X_0(\theta);\Xi(\bullet|_{t_0}^{t};\theta)])\Big] + \\
&\quad + C_{X_0\Xi}(t)\,\mathbb{E}^{\theta}\left[\frac{\partial\delta(x-X[X_0(\theta);\Xi(\bullet|_{t_0}^{t};\theta)])}{\partial X_0(\theta)}\right] + \\
&\quad + \int_{t_0}^{t} C_{\Xi\Xi}(t,s)\,\mathbb{E}^{\theta}\left[\frac{\delta\delta(x-X[X_0(\theta);\Xi(\bullet|_{t_0}^{t};\theta)])}{\delta\Xi(\tau;\theta)}\right] ds.
\end{aligned} \tag{4.3}$$

For the last two averages appearing in the rightmost side of NF theorem, Eq. (4.3), the chain rule for the derivatives of random delta function is applied, resulting in

$$\begin{aligned}
&\mathbb{E}^{\theta}\Big[\,\Xi(t;\theta)\,\delta(x-X[X_0(\theta);\Xi(\bullet|_{t_0}^{t};\theta)])\Big] = \\
&\quad = m_{\Xi}(t)\,\mathbb{E}^{\theta}\Big[\delta(x-X[X_0(\theta);\Xi(\bullet|_{t_0}^{t};\theta)])\Big] + \\
&\quad + C_{X_0\Xi}(t)\,\mathbb{E}^{\theta}\left[\frac{\partial\delta(x-X[X_0(\theta);\Xi(\bullet|_{t_0}^{t};\theta)])}{\partial X(t;\theta)}\,V_{X_0}(t;\theta)\right] + \\
&\quad + \int_{t_0}^{t} C_{\Xi\Xi}(t,s)\,\mathbb{E}^{\theta}\left[\frac{\partial\delta(x-X[X_0(\theta);\Xi(\bullet|_{t_0}^{t};\theta)])}{\partial X(t;\theta)}\,V_{\Xi(s)}(t;\theta)\right] ds,
\end{aligned} \tag{4.4}$$

where $V_{X_0}(t;\theta)$, $V_{\Xi(s)}(t;\theta)$ are defined as the derivatives of the response with respect to initial value and excitation respectively

$$V_{X_0}(t;\theta) = \frac{\partial X[X_0(\theta);\Xi(\bullet|_{t_0}^{t};\theta)]}{\partial X_0(\theta)}, \quad V_{\Xi(s)}(t;\theta) = \frac{\delta X[X_0(\theta);\Xi(\bullet|_{t_0}^{t};\theta)]}{\delta\Xi(s;\theta)}, \tag{4.5a,b}$$

which are collectively called the ***variational derivatives of the response***. Despite the response being the solution to a nonlinear RDE, its variational derivatives are easily calculated:

**On the calculation of variational derivatives in closed form.** The derivatives $V_{X_0}(t;\theta)$, $V_{\Xi(s)}(t;\theta)$ can be calculated by formulating and solving the corresponding variational equations. The latter are formally derived from RIVP (4.2a,b), by applying the differential operators $\partial\bullet/\partial X_0(\theta)$ and $\delta\bullet/\delta\Xi(s;\theta)$, respectively, see also (Anosov & Arnold, 1987) Sec. 2.7, (Amann, 1990) Ch. II Sec. 9, (Grigorian, 2008) Sec. 2.10:

$$\dot{V}_{X_0}(t;\theta) = \big[h'(X(t;\theta)) + q'(X(t;\theta))\,\Xi(t;\theta)\big]\,V_{X_0}(t;\theta), \tag{4.6a}$$

$$V_{X_0}(t_0;\theta) = 1, \tag{4.6b}$$

$$\dot{V}_{\Xi(s)}(t;\theta) = \left[h'(X(t;\theta)) + q'(X(t;\theta))\,\Xi(t;\theta)\right] V_{\Xi(s)}(t;\theta), \tag{4.7a}$$

$$V_{\Xi(s)}(s;\theta) = q(X(s;\theta)). \tag{4.7b}$$

A more detailed derivation of variational equations is presented in paragraph 4.2.1 of the next section, where we consider the more general case of determining the variational derivatives of the response to a multidimensional random dynamical system. Note that, the initial value problem (4.7a,b) for $V_{\Xi(s)}(t;\theta)$ is defined for $t > s$ since, by causality, any perturbation $\delta\Xi(s;\theta)$, acting at time $s$, cannot result in a perturbation $\delta X(t;\theta)$ for $t < s$; thus, $V_{\Xi(s)}(t;\theta) = 0$ for $t < s$. Since Eqs. (4.6a), (4.7b) are linear ODEs with respect to $t$, their solutions are explicitly determined to

$$V_{X_0}(t;\theta) = \exp\left(\int_{t_0}^{t}\left[h'(X(u;\theta)) + q'(X(u;\theta))\,\Xi(u;\theta)\right] du\right), \tag{4.8}$$

$$V_{\Xi(s)}(t;\theta) = q(X(s;\theta)) \exp\left(\int_{s}^{t}\left[h'(X(u;\theta)) + q'(X(u;\theta))\,\Xi(u;\theta)\right] du\right). \tag{4.9}$$

By observing variational derivatives in their closed forms, Eqs. (4.8), (4.9), they are identified as ***functionals of response and excitation***; $V_{X_0}(t;\theta) = V_{X_0}[X(\bullet|_{t_0}^{t};\theta);\Xi(\bullet|_{t_0}^{t};\theta)]$, $V_{\Xi(s)}(t;\theta) = V_{\Xi(s)}[X(\bullet|_{s}^{t};\theta);\Xi(\bullet|_{s}^{t};\theta)]$.

Let us now return to Eq. (4.4). Having considered the response as an FFℓ with Gaussian arguments; $X[X_0(\theta);\Xi(\bullet|_{t_0}^{t};\theta)]$, in order to be able to employ the NF theorem and the chain rule, we shall now revert to considering the response as a random function per se; $X(t;\theta)$. Thus, as we have discussed in Section 3.1, the average $\mathbb{E}^{\theta}[\bullet]$ is considered once again with respect to joint response-excitation probability measure $\mathbf{P}_{X(\cdot)\Xi(\cdot)}$, and the notation of Eq. (4.4) is simplified into

$$\begin{aligned}\mathbb{E}^{\theta}\left[\Xi(t;\theta)\,\delta(x-X(t;\theta))\right] &= m_{\Xi}(t)\,\mathbb{E}^{\theta}\left[\delta(x-X(t;\theta))\right] + \\ &+ C_{X_0\Xi}(t)\,\mathbb{E}^{\theta}\left[\frac{\partial\delta(x-X(t;\theta))}{\partial X(t;\theta)} V_{X_0}[X(\bullet|_{t_0}^{t};\theta);\Xi(\bullet|_{t_0}^{t};\theta)]\right] + \\ &+ \int_{t_0}^{t} C_{\Xi\Xi}(t,s)\,\mathbb{E}^{\theta}\left[\frac{\partial\delta(x-X(t;\theta))}{\partial X(t;\theta)} V_{\Xi(s)}[X(\bullet|_{\tau}^{t};\theta);\Xi(\bullet|_{s}^{t};\theta)]\right] ds.\end{aligned} \tag{4.10}$$

By this change in perspective with regard to the response, the averages in the right-hand side of Eq. (4.10) can be further evaluated using the formulae (2.7), (2.13) of delta projection:

$$\begin{aligned}\mathbb{E}^{\theta}\left[\Xi(t;\theta)\,\delta(x-X(t;\theta))\right] &= m_{\Xi}(t)\,f_{X(t)}(x) - \\ &- C_{X_0\Xi}(t)\,\frac{\partial}{\partial x}\mathbb{E}^{\theta}\left[\delta(x-X(t;\theta))\,V_{X_0}[X(\bullet|_{t_0}^{t};\theta);\Xi(\bullet|_{t_0}^{t};\theta)]\right] - \\ &- \frac{\partial}{\partial x}\int_{t_0}^{t} C_{\Xi\Xi}(t,s)\,\mathbb{E}^{\theta}\left[\delta(x-X(t;\theta))\,V_{\Xi(s)}[X(\bullet|_{s}^{t};\theta);\Xi(\bullet|_{s}^{t};\theta)]\right] ds.\end{aligned} \tag{4.11}$$

Substitution of Eq. (4.11) into Eq. (4.1) results into the ***final form of SLE*** for the one-time response pdf:

$$\frac{\partial f_{X(t)}(x)}{\partial t} + \frac{\partial}{\partial x}\Big[\big(h(x) + q(x)\, m_{\Xi}(t)\big)\, f_{X(t)}(x)\Big] =$$
$$= C_{X_0\Xi}(t)\,\frac{\partial}{\partial x}\left(q(x)\,\frac{\partial}{\partial x}\,\mathbb{E}^{\theta}\Big[\delta(x-X(t;\theta))\, V_{X_0}[X(\bullet|_{t_0}^{t};\theta)\,;\,\Xi(\bullet|_{t_0}^{t};\theta)]\Big]\right) + \tag{4.12}$$
$$+ \frac{\partial}{\partial x}\left(q(x)\,\frac{\partial}{\partial x}\int_{t_0}^{t} C_{\Xi\Xi}(t,s)\,\mathbb{E}^{\theta}\Big[\delta(x-X(t;\theta))\, V_{\Xi(s)}[X(\bullet|_{s}^{t};\theta)\,;\,\Xi(\bullet|_{s}^{t};\theta)]\Big]\, ds\right).$$

By comparing SLE in its final form, Eq. (4.12), to its original one, Eq. (4.1), we observe that the use of NF theorem, and the subsequent manipulations, resulted in **(i)** an augmented drift term, which can be identified as the right-hand side of RDE (4.2a) with excitation replaced by its mean value, **(ii)** the appearance of second-order $x$ – derivatives in the right-hand side of the equation, casting thus SLE into a form that bears resemblance to a drift-diffusion equation like the classical FPK, and **(iii)** the appearance of averages of the random delta function multiplied by the variational derivatives, which depend on the whole history of response and excitation. The latter can be seen as an ample manifestation of the non-Markovian character of the response, coming from the smoothly-correlated, i.e. non-white, character of the excitation. Thus, SLE (4.12) may be exact, since the tools used for its derivation (delta projection and NF theorem) do not include any approximation, but it is a non-closed evolution equation nevertheless.

**Comparison to SLEs found in the literature.** In the special case of initial value not being correlated with the excitation, $C_{X_0\Xi}(t) = 0$, and for zero-mean excitation, $m_{\Xi}(t) = 0$, Eq. (4.12) reduces to the SLE for scalar RDEs under coloured Gaussian excitation derived in various previous works (Fox, 1987; Hänggi et al., 1985; Peacock-López et al., 1988; Sancho et al., 1982), and (Hänggi & Jung, 1995) equation (3.27), where it is called the coloured noise master equation. While the problem of correlation between initial value and excitation has been studied before, e.g. in (Roerdink, 1981, 1982) (for linear RDEs under general, possibly non Gaussian excitation), our approach explicitly incorporates the effect of initial value correlation into the final SLE (4.12), by using the extended NF theorem, Eq. (4.3). Thus, all pdf evolution equations derived from SLE (4.12) will inherit this effect in a straightforward way.

At this point, we shall derive, from final SLE (4.12), the classical FPK equation, as well as an exact response pdf evolution equation for linear RDEs under additive coloured noise excitation.

### 4.1.1 Rederivation of the classical Fokker-Planck-Kolmogorov equation

Consider SLE (4.12) for the nonlinear RIVP (4.2a,b) under ***zero-mean white noise excitation*** with intensity $D(t) > 0$, and for uncorrelated initial value to excitation:

$$m_{\Xi}(t) = 0, \quad C^{\mathrm{WN}}_{\Xi\Xi}(t,s) = 2D(t)\,\delta(t-s), \quad C_{X_0\Xi}(t) = 0\,.$$

In this case, in the right-hand side of SLE (4.12), the upper time limit $t$ of the integral coincides with the singular point of the delta function $\delta(t-s)$, making the value of this integral ambiguous. To resolve this ambiguity, we approximate the singular autocovariance function of the excitation, $C_{\Xi\Xi}^{\mathrm{WN}}(t, s)$, by a weighted delta family,

$$C_{\Xi\Xi}^{(\varepsilon)}(t, s) = 2D(t)\,\delta_\varepsilon(t-s), \tag{4.13}$$

where $\delta_\varepsilon(t-s) = (1/\varepsilon)\,\gamma\big((t-s)/\varepsilon\big)$, with $\gamma(\bullet)$ being a non-negative smooth kernel function ((Cheney & Light, 2009), Ch. 20). In addition, in the present case, $\gamma(\bullet)$ should be even, in order for $C_{\Xi\Xi}^{(\varepsilon)}(t, s)$ to be a valid autocovariance function. The last requirement implies that $\lim_{\varepsilon\downarrow 0}\int_{t_0}^{t}\delta_\varepsilon(t-s)\,ds = 1/2$, which, after the standard proof procedure for kernel functions ((Bandyopadhyay, 2002) Sec. 12.4, (Cheney & Light, 2009), *loc. cit.*), leads to identity

$$\lim_{\varepsilon\downarrow 0}\int_{t_0}^{t}\delta_\varepsilon(t-s)\,g(s)\,ds = \frac{1}{2}\,g(t), \tag{4.14}$$

for any continuous function $g(\bullet)$. Since $V_{\Xi(s)}$ is a continuous function with respect to its argument $s$, see Eq. (4.9), identity (4.14) can be employed, resulting in the following calculation of the integral in SLE (4.12):

$$\int_{t_0}^{t} C_{\Xi\Xi}^{\mathrm{WN}}(t, \tau)\,V_{\Xi(s)}[X(\bullet|_s^t;\theta)\,;\,\Xi(\bullet|_s^t;\theta)]\,ds \equiv$$

$$\equiv \lim_{\varepsilon\downarrow 0}\int_{t_0}^{t} C_{\Xi\Xi}^{(\varepsilon)}(t, s)\,V_{\Xi(s)}[X(\bullet|_s^t;\theta)\,;\,\Xi(\bullet|_s^t;\theta)]\,ds =$$

$$= 2D(t)\lim_{\varepsilon\downarrow 0}\int_{t_0}^{t}\delta_\varepsilon(t-s)\,V_{\Xi(s)}[X(\bullet|_s^t;\theta)\,;\,\Xi(\bullet|_s^t;\theta)]\,ds =$$

$$= D(t)\,V_{\Xi(t)}[X(\bullet|_t^t;\theta)\,;\,\Xi(\bullet|_t^t;\theta)],$$

and by substituting $V_{\Xi(t)}[X(\bullet|_t^t;\theta)\,;\,\Xi(\bullet|_t^t;\theta)] = V_{\Xi(t)}(t;\theta)$ from Eq. (4.9), we obtain

$$\int_{t_0}^{t} C_{\Xi\Xi}^{\mathrm{WN}}(t, s)\,V_{\Xi(s)}[X(\bullet|_s^t;\theta)\,;\,\Xi(\bullet|_s^t;\theta)]\,ds = D(t)\,q(X(t;\theta)). \tag{4.15}$$

Substitution of Eq. (4.15) into SLE (4.12) for $m_\Xi(t) = 0$ and $C_{X_0\Xi}(t) = 0$, results into

$$\frac{\partial f_{X(t)}(x)}{\partial t} + \frac{\partial}{\partial x}\Big(h(x)\,f_{X(t)}(x)\Big) = D(t)\,\frac{\partial}{\partial x}\left(q(x)\,\frac{\partial}{\partial x}\,\mathbb{E}^\theta\big[\delta(x-X(t;\theta))q(X(t;\theta))\big]\right)$$

and by employing delta projection formula (2.9) for the average, we obtain

$$\frac{\partial f_{X(t)}(x)}{\partial t} + \frac{\partial}{\partial x}\Big(h(x)\,f_{X(t)}(x)\Big) = D(t)\,\frac{\partial}{\partial x}\left[q(x)\,\frac{\partial}{\partial x}\Big(q(x)\,f_{X(t)}(x)\Big)\right]. \tag{4.16}$$

Eq. (4.16) is the scalar FPK equation under the Stratonovich interpretation, corresponding to stochastic differential equations under white noise, see e.g. (Öttinger, 1996) equation (3.107). The fact that, in the white-noise limit, SLE (4.12) gives rise to the Stratonovich, and not the Itō variant of FPK equation, is consistent with our approximation of the delta-correlation by a smooth delta family, since as it is stated in (Öttinger, 1996), *loc. sit.*, “if the delta-correlated noise in a stochastic differential equation is an idealization of a noise with a very short but nonzero correlation time, […] Stratonovich’s calculus should be applied.”

**Remark 4.1: On factor 2 multiplying white noise intensity.** In this paragraph, we considered the white-noise autocovariance function to be $2D(t)\,\delta(t-s)$. However, in many works, see e.g. (Gardiner, 2004, sec. 3.5.2; Soong & Grigoriu, 1993, sec. 4.75; Sun, 2006, sec. 6.2), factor 2 multiplying white noise intensity is absent. While this discrepancy is not essential, the reason for us to consider the factor 2 is that, if we choose a weighted delta family whose members are the usual Ornstein-Uhlenbeck autocorrelation functions

$$C^{(\varepsilon)}_{\Xi\Xi}(t,s) = \frac{D(t)}{\varepsilon}\exp\left(-\frac{|t-s|}{\varepsilon}\right), \tag{4.17}$$

the limit of Eq. (4.17) for $\varepsilon \to 0$ is $2D(t)\,\delta(t-s)$ and not $D(t)\,\delta(t-s)$, see e.g. (Toral & Colet, 2014, sec. 6.6).

### 4.1.2 The case of linear RDE under additive Gaussian noise[22]

By considering $h(x) = \eta x$ with $\eta < 0$ for stability purposes, and $q(x) = \kappa =$constant, RIVP (4.2a,b) becomes linear and additively excited:

$$\dot{X}(t;\theta) = \eta X(t;\theta) + \kappa\Xi(t;\theta), \qquad X(t_0;\theta) = X_0(\theta). \tag{4.18a,b}$$

For linear-additive RIVP (4.18a,b), formulae (4.8), (4.9) for the variational derivatives of the response are simplified into

$$V_{X_0}(t;\theta) = e^{\eta(t-t_0)}, \qquad V_{\Xi(s)}(t;\theta) = \kappa\, e^{\eta(t-s)} \tag{4.19a,b}$$

Thus, in this case, the variational derivatives are independent from the response and the excitation, and so SLE (4.12) is, in fact, the *exact* response pdf evolution equation:

$$\frac{\partial f_{X(t)}(x)}{\partial t} + \frac{\partial}{\partial x}\Big[\big(\eta x + \kappa m_\Xi(t)\big) f_{X(t)}(x)\Big] = D^{\text{eff}}(t)\,\frac{\partial^2 f_{X(t)}(x)}{\partial x^2}, \tag{4.20}$$

where the term $D^{\text{eff}}(t)$, called the ***effective noise intensity***, is given by

$$D^{\text{eff}}(t) = \kappa\, e^{\eta(t-t_0)}\, C_{X_0\Xi}(t) + \kappa^2\int_{t_0}^{t} e^{\eta(t-s)}\, C_{\Xi\Xi}(t,s)\, ds. \tag{4.21}$$

Pdf evolution Eq. (4.21) is also supplemented with the initial Gaussian condition:

$$f_{X(t_0)}(x) = f_{X_0}(x) = \frac{1}{\sqrt{2\pi\sigma^2_{X_0}}}\exp\left[-\frac{1}{2}\frac{(x-m_{X_0})^2}{\sigma^2_{X_0}}\right], \tag{4.22}$$

where $m_{X_0}$, $\sigma^2_{X_0}$ are the initial mean value and variance respectively. It is a well-known result that the response process of any linear system, with Gaussian initial distribution, to an additive Gaussian excitation (either coloured or white) is also a Gaussian process. Furthermore, the mean value $m_X(t)$ and variance $\sigma_X^2(t)$ of the response $X(t;\theta)$ can be determined as the solutions to the respective moment equations, derived directly from RIVP (4.18a,b). This task is performed in Appendix D, in which Eqs. (D.3), (D.27) read

$$m_X(t) = m_{X_0} e^{\eta(t-t_0)} + \kappa \int_{t_0}^{t} m_\Xi(\tau)\, e^{\eta(t-\tau)}\, d\tau, \tag{4.23}$$

and

$$\sigma_X^2(t) = \sigma_{X_0}^2\, e^{2\eta(t-t_0)} + 2\kappa \int_{t_0}^{t} C_{X\Xi}(\tau,\tau)\, e^{2\eta(t-\tau)}\, d\tau. \tag{4.24}$$

[22] A previous version of this paragraph was published as section 7 of (Athanassoulis & Mamis, 2019).

**Remark 4.2: Connection between effective noise intensity and cross-correlation.** In Appendix D, Eq. (D.22) for response-excitation cross-covariance reads

$$C_{X\Xi}(t,s) = C_{X_0\Xi}(s)\, e^{\eta(t-t_0)} + \kappa \int_{t_0}^{t} C_{\Xi\Xi}(\tau,s)\, e^{\eta(t-\tau)}\, d\tau. \tag{4.25}$$

Comparing Eq. (4.25) to Eq. (4.21), effective noise intensity $D^{\text{eff}}(t)$ can be expressed in terms of the one-time response-excitation cross-covariance as

$$D^{\text{eff}}(t) = \kappa C_{X\Xi}(t,t). \tag{4.26}$$

Under Eq. (4.26), Eq. (4.24) for the variance of the response is expressed equivalently as

$$\sigma_X^2(t) = \sigma_{X_0}^2\, e^{2\eta(t-t_0)} + 2 \int_{t_0}^{t} D^{\text{eff}}(\tau)\, e^{2\eta(t-\tau)}\, d\tau. \tag{4.27}$$

**Solution of Eq. (4.20).** Following (Sun, 2006) sec. 6.3, pdf evolution Eq. (4.20) can be solved by employing the ***Fourier transform***; $\varphi_{X(t)}(y) = \int_{\mathbb{R}} e^{iyx} f_{X(t)}(x)\, dx$, which leads to the following equation of first partial derivatives for the characteristic function $\varphi_{X(t)}(y)$:

$$\frac{\partial \varphi_{X(t)}(y)}{\partial t} = \eta y \frac{\partial \varphi_{X(t)}(y)}{\partial y} + \left( i\kappa m_\Xi(t)\, y - D^{\text{eff}}(t)\, y^2 \right) \varphi_{X(t)}(y), \tag{4.28a}$$

supplemented with the transformed initial condition (4.22)

$$\varphi_{X(t_0)}(y) = \exp\left( i m_{X_0}\, y - \frac{1}{2} \sigma_{X_0}^2\, y^2 \right). \tag{4.28b}$$

Initial value problem (4.28a,b) is solved (Polyanin, Zaitsev, & Moussiaux, 2001) sec. 4.1, by first determining the characteristic curve $w(y,t) = ye^{\eta t}$ as the solution of the characteristic equation $dt = -dy/(\eta y)$. Then, we seek a solution of the form $g(w) \exp\left( \int_{t_0}^{t} h(w,t)\, dt \right)$, where $g(w)$ is a function of the characteristic curve, to be defined by the initial condition (4.28b), and $h(w,t) = i\kappa m_\Xi(t)\, we^{-\eta t} - D^{\text{eff}}(t)\, w^2 e^{-2\eta t}$, that is the coefficient multiplying $\varphi_{X(t)}(y)$ in Eq. (4.28a), rewritten in terms of $w$, $t$. Finally, by returning to the original variables $y$, $t$, we obtain the solution

$$\varphi_{X(t)}(y) = \exp\left[ i \left( m_{X_0} e^{\eta(t-t_0)} + \kappa \int_{t_0}^{t} m_{\Xi}(\tau)\, e^{\eta(t-\tau)} d\tau \right) y \right] \times$$
$$\times \exp\left[ -\frac{1}{2} \left( \sigma_{X_0}^2 e^{2\eta(t-t_0)} + 2 \int_{t_0}^{t} D^{\text{eff}}(\tau)\, e^{2\eta(t-\tau)}\, d\tau \right) y^2 \right]. \tag{4.29}$$

Employing the inverse Fourier transform, and utilizing Eqs. (4.23), (4.27) for $m_X(t)$ and $\sigma_X^2(t)$, results in

$$f_{X(t)}(x) = \frac{1}{\sqrt{2\pi\sigma_X^2(t)}} \exp\left[ -\frac{1}{2} \frac{\left(x - m_X(t)\right)^2}{\sigma_X^2(t)} \right], \tag{4.30}$$

which is the expected Gaussian distribution. This result constitutes the verification of the response pdf evolution Eq. (4.20).

**Remark 4.3:** The ***uniqueness of Gaussian solution*** (4.30) is ensured by the injectivity of Fourier transform for absolutely integrable functions, and the uniqueness of solution for transformed problem (4.28a,b), see (Polyanin et al., 2001), Sec. 10.1.2. What is more, the uniqueness of solution for Eq. (4.20) is also proven directly, without resorting to Fourier transform, in Appendix E.

**Remark 4.4:** While solving Eq. (4.20) via Fourier transform is preferable, since it also ensures the uniqueness of solution, Eq. (4.20) can be also solved directly using the ansatz and change of variables of (Shtelen & Stogny, 1989). This solution procedure, presented in Appendix F, reduces the pdf evolution Eq. (4.20) to the ***one-dimensional heat equation***.

## 4.2 The final one-time stochastic Liouville equation for multidimensional systems

The same procedure presented in Section 4.1, can be followed for the multidimensional, one-time, response SLE (2.55):

$$\frac{\partial f_{X(t)}(\boldsymbol{x})}{\partial t} + \sum_{n=1}^{N} \frac{\partial}{\partial x_n} \left( h_n(\boldsymbol{x})\, f_{X(t)}(\boldsymbol{x}) \right) =$$
$$= -\sum_{n=1}^{N} \sum_{k=1}^{K} \frac{\partial}{\partial x_n} \left( q_{nk}(\boldsymbol{x})\, \mathbb{E}^{\theta}\left[ \delta(\boldsymbol{x} - \boldsymbol{X}(t;\theta))\, \Xi_k(t;\theta) \right] \right), \tag{4.31}$$

corresponding to the multidimensional RIVP

$$\dot{X}_n(t;\theta) = h_n\left(\boldsymbol{X}(t;\theta)\right) + \sum_{k=1}^{K} q_{nk}\left(\boldsymbol{X}(t;\theta)\right) \Xi_k(t;\theta), \tag{4.32a}$$

$$X_n(t_0;\theta) = X_n^0(\theta), \qquad n = 1, \dots, N, \tag{4.32b}$$

in which initial value $\boldsymbol{X}^0(\theta)$ and excitation $\boldsymbol{\Xi}(\bullet;\theta)$ are considered jointly Gaussian. In SLE (4.31), the average $\mathbb{E}^{\theta}\left[ \delta(\boldsymbol{x} - \boldsymbol{X}(t;\theta))\, \Xi_k(t;\theta) \right]$ can be further evaluated by using the multidimensional variant of the extended NF theorem, Eq. (3.60), which is specified into

$$\mathbb{E}^{\theta}\left[\delta(\boldsymbol{x}-\boldsymbol{X}(t;\theta))\,\Xi_k(t;\theta)\right] = m_{\Xi_k}(t)\,\mathbb{E}^{\theta}\left[\delta(\boldsymbol{x}-\boldsymbol{X}(t;\theta))\right]+ \\ +\sum_{n=1}^{N} C_{X_n^0\Xi_k}(t)\,\mathbb{E}^{\theta}\left[\frac{\partial\delta(\boldsymbol{x}-\boldsymbol{X}[\boldsymbol{X}^0(\theta);\boldsymbol{\Xi}(\bullet|_{t_0}^{t};\theta)])}{\partial X_n^0(\theta)}\right]+ \\ +\sum_{\ell=1}^{K}\int_{t_0}^{t} C_{\Xi_k\Xi_\ell}(t,s)\,\mathbb{E}^{\theta}\left[\frac{\delta\delta(\boldsymbol{x}-\boldsymbol{X}[\boldsymbol{X}^0(\theta);\boldsymbol{\Xi}(\bullet|_{t_0}^{t};\theta)])}{\delta\Xi_\ell(s;\theta)}\right]ds. \tag{4.33}$$

By substituting Eq. (4.33) into SLE (4.31), and employing the chain rule for random delta function derivatives, as well as the delta projection formulae for the multidimensional case, see Section 2.6, we obtain the ***final form of the multidimensional SLE***:

$$\frac{\partial f_{\boldsymbol{X}(t)}(\boldsymbol{x})}{\partial t}+\sum_{n=1}^{N}\frac{\partial}{\partial x_n}\left[\left(h_n(\boldsymbol{x})+\sum_{k=1}^{K} q_{nk}(\boldsymbol{x})\,m_{\Xi_k}(t)\right)f_{\boldsymbol{X}(t)}(\boldsymbol{x})\right]= \\ =\sum_{n_1=1}^{N}\sum_{k=1}^{K}\frac{\partial}{\partial x_{n_1}}\left(q_{n_1k}(\boldsymbol{x})\sum_{n_2=1}^{N}\frac{\partial}{\partial x_{n_2}}\sum_{n_3=1}^{N} C_{X_{n_3}^0\Xi_k}(t)\,\mathbb{E}^{\theta}\left[\delta(\boldsymbol{x}-\boldsymbol{X}(t;\theta))\,V_{n_2n_3}^{\boldsymbol{X}^0}(t;\theta)\right]\right)+ \\ +\sum_{n_1=1}^{N}\sum_{k_1=1}^{K}\frac{\partial}{\partial x_{n_1}}\left(q_{n_1k_1}(\boldsymbol{x})\sum_{n_2=1}^{N}\frac{\partial}{\partial x_{n_2}}\sum_{k_2=1}^{K}\int_{t_0}^{t} C_{\Xi_{k_1}\Xi_{k_2}}(t,s)\,\mathbb{E}^{\theta}\left[\delta(\boldsymbol{x}-\boldsymbol{X}(t;\theta))\,V_{n_2k_2}^{\boldsymbol{\Xi}(s)}(t;\theta)\right]ds\right), \tag{4.34}$$

where $\boldsymbol{V}^{\boldsymbol{X}^0}(t;\theta)$ and $\boldsymbol{V}^{\boldsymbol{\Xi}(s)}(t;\theta)$ are the ***matrices of the variational derivatives*** of the response with respect to initial value and excitation respectively, defined as

$$V_{nm}^{\boldsymbol{X}^0}(t;\theta) = \frac{\partial X_n[\boldsymbol{X}^0(\theta);\boldsymbol{\Xi}(\bullet|_{t_0}^{t};\theta)]}{\partial X_m^0(\theta)},\quad n,m=1,\dots,N, \tag{4.35a}$$

$$V_{nk}^{\boldsymbol{\Xi}(s)}(t;\theta) = \frac{\delta X_n[\boldsymbol{X}^0(\theta);\boldsymbol{\Xi}(\bullet|_{t_0}^{t};\theta)]}{\delta\Xi_k(s;\theta)},\quad n=1,\dots,N,\ k=1,\dots,K. \tag{4.35b}$$

As in the scalar case of the previous section, the drift coefficient of the final SLE (4.34) is augmented, compared to the drift of the original SLE (4.31), by $\sum_{k=1}^{K} q_{nk}(\boldsymbol{x})\,m_{\Xi_k}(t)$, which models the effect of the average of the excitation to response pdf evolution. Note also that SLE (4.34), for the special case of zero-mean value excitation, $\boldsymbol{m}_{\boldsymbol{\Xi}}(t)=\boldsymbol{0}$, uncorrelated to the initial value, $\boldsymbol{C}_{\boldsymbol{X}^0\boldsymbol{\Xi}}(t)=\boldsymbol{0}$, reduces to the SLE derived in (Cetto et al., 1984; Dekker, 1982; Hernandez-Machado et al., 1983; San Miguel & Sancho, 1980).

### 4.2.1 Formulation and solution of the variational equations

Since $\boldsymbol{X}^0(\theta)$ and $\Xi(s;\theta)$, $s\in[t_0,t]$ are parameters of $\boldsymbol{X}[\boldsymbol{X}^0(\theta);\Xi(\bullet|_{t_0}^{t};\theta)]$, we may calculate the derivatives of the response with respect initial value and excitation by formulating and solving the corresponding variational equations (Amann, 1990, Chapter II, section 9; Anosov & Arnold, 1987, sec.2.7; Grigorian, 2008, sec. 2.10) along ***each path of solution***, i.e. for every value of the stochastic argument $\theta$ separately. This pathwise consideration equals, on a practical level, to discarding the stochastic argument $\theta$, and working with the deterministic initial value problem:

$$\dot{X}_n(t) = h_n\big(\boldsymbol{X}(t)\big) + \sum_{k=1}^{K} q_{nk}\big(\boldsymbol{X}(t)\big)\,\Xi_k(t), \quad X_n(t_0) = X_n^0, \quad n=1,\dots,N. \qquad (4.36a,b)$$

**(a)** ***Variational IVP for the initial value, and its solution***

By applying the differential operator $\partial\bullet/\partial X_\nu^0$, $\nu\in\{1,\dots,N\}$ at both sides of Eqs. (4.32a,b), and since a given excitation function $\Xi(\bullet)$ is not functionally dependent on $X_\nu^0$, we obtain:

$$\dot{V}_{n\nu}^{\boldsymbol{X}^0}(t) = \sum_{m=1}^{N}\left[\frac{\partial h_n\big(\boldsymbol{X}(t)\big)}{\partial X_m(t)} + \sum_{k=1}^{K}\frac{\partial q_{nk}\big(\boldsymbol{X}(t)\big)}{\partial X_m(t)}\,\Xi_k(t)\right]V_{m\nu}^{\boldsymbol{X}^0}(t), \qquad (4.37a)$$

$$V_{n\nu}^{\boldsymbol{X}^0}(t_0) = \delta_{n\nu}, \qquad n,\nu=1,\dots,N, \qquad (4.37b)$$

where $\delta_{n\nu}$ is Kronecker's delta. Since Eqs. (4.37a,b) constitute a linear multidimensional ordinary differential equation, its the solution can be expressed in closed form by utilizing the corresponding ***state-transition matrix*** $\boldsymbol{\Phi}(t,t_0)$ (Brockett, 1970), sec. 1.3:

$$V_{n\nu}^{\boldsymbol{X}^0}(t) = \sum_{m=1}^{N}\Phi_{nm}(t,t_0)\,V_{m\nu}^{\boldsymbol{X}^0}(t_0) = \sum_{m=1}^{N}\Phi_{nm}(t,t_0)\,\delta_{m\nu} = \Phi_{n\nu}(t,t_0). \qquad (4.38)$$

State-transition matrix $\boldsymbol{\Phi}(t,t_0)$, whose calculation is described at length in the previous reference, is the generalization of the exponential solution of a scalar ODE, to the case of multidimensional systems of linear ODEs with time-varying coefficients.

By returning now from the pathwise treatment, Eqs. (4.36a,b), to the random system (4.32a,b), the variational derivative with respect to initial value is expressed as

$$V_{n\nu}^{\boldsymbol{X}^0}(t;\theta) = \Phi_{n\nu}(t,t_0;\theta). \qquad (4.39)$$

**(b)** ***Variational IVP for the excitation, and its solution***

In this case, a procedure similar to the one applied for $V_{n\nu}^{\boldsymbol{X}^0}(t)$ is employed. For this, we assume that Volterra derivative $\delta\bullet/\delta\Xi_\ell(s)$, $\ell\in\{1,\dots,K\}$, $s\in[t_0,t]$, has similar properties to the usual partial derivative $\partial\bullet/\partial X_\nu^0$, i.e. is linear, can be interchanged with the temporal derivative and obeys the usual chain rule of differentiation. Thus, by applying $\delta\bullet/\delta\Xi_\ell(s)$ at both sides of Eq. (4.32a), using the aforementioned properties of Volterra de-rivative, and since for a given path function $\boldsymbol{\Xi}(\bullet)$, the value $\Xi_\ell(s)$ is not functionally dependent on $\boldsymbol{X}^0$, as well as on $\Xi_k(t)$'s for $s \neq t$ and $\ell \neq k$, we obtain

$$\dot{V}_{n\ell}^{\Xi(s)}(t) = \sum_{m=1}^{N}\left[\frac{\partial h_n(\boldsymbol{X}(t))}{\partial X_m(t)} + \sum_{k=1}^{K}\frac{\partial q_{nk}(\boldsymbol{X}(t))}{\partial X_m(t)}\,\Xi_k(t)\right] V_{m\ell}^{\Xi(s)}(t) + q_{n\ell}(\boldsymbol{X}(t))\,\delta(t-s), \quad (4.40a)$$

$$V_{n\ell}^{\Xi(s)}(t_0) = 0, \qquad n = 1, \dots, N, \qquad \ell = 1, \dots, K, \quad (4.40b)$$

where $\delta(t-s)$ is Dirac's delta function. Since, by causality (Cetto et al., 1984; Ueda, 2016), any perturbation $\delta\boldsymbol{\Xi}(s)$ in excitation at time $s$, cannot result in a perturbation $\delta\boldsymbol{X}(t)$ in the response for $t < s$: $\boldsymbol{V}^{\Xi(s)}(t) = \boldsymbol{0}$ for $t < s$. By integrating Eq. (4.40a) over $[s-\varepsilon, t]$, for small $\varepsilon > 0$, and taking the limit $\varepsilon \to 0$, we obtain

$$V_{n\ell}^{\Xi(s)}(t) = \sum_{m=1}^{N}\int_{s}^{t}\left[\frac{\partial h_n(\boldsymbol{X}(u))}{\partial X_m(u)} + \sum_{k=1}^{K}\frac{\partial q_{nk}(\boldsymbol{X}(u))}{\partial X_m(u)}\,\Xi_k(u)\right] V_{m\ell}^{\Xi(s)}(u)\,du + q_{n\ell}(\boldsymbol{X}(s)),$$
$$n = 1, \dots, N, \quad \ell = 1, \dots, K, \quad (4.41)$$

which is a Volterra integral equation of the second kind, equivalent to the linear IVP (Polyanin & Manzhirov, 2008)

$$\dot{V}_{n\ell}^{\Xi(s)}(t) = \sum_{m=1}^{N}\left[\frac{\partial h_n(\boldsymbol{X}(t))}{\partial X_m(t)} + \sum_{k=1}^{K}\frac{\partial q_{nk}(\boldsymbol{X}(t))}{\partial X_m(t)}\,\Xi_k(t)\right] V_{m\ell}^{\Xi(s)}(t), \quad (4.42a)$$

$$V_{n\ell}^{\Xi(s)}(s) = q_{n\ell}(\boldsymbol{X}(s)), \qquad n = 1, \dots, N, \qquad \ell = 1, \dots, K. \quad (4.42b)$$

Since linear IVP (4.42a,b) has the same matrix with IVP (4.37a,b), its solution can be expressed using the same state-transition matrix, albeit with different initial time value, as

$$V_{n\ell}^{\Xi(s)}(t) = \sum_{m=1}^{N}\Phi_{nm}(t,s)\,V_{m\ell}^{\Xi(s)}(s) = \sum_{m=1}^{N}\Phi_{nm}(t,s)\,q_{m\ell}(\boldsymbol{X}(s)). \quad (4.43)$$

As previously, by returning to the random problem (4.32a.b), Eq. (4.43) is expressed as

$$V_{n\ell}^{\Xi(s)}(t;\theta) = \sum_{m=1}^{N}\Phi_{nm}(t,s;\theta)\,q_{m\ell}(\boldsymbol{X}(s;\theta)). \quad (4.44)$$

**Identification of non-local terms in SLE (4.34).** As in the scalar case, the variational derivatives $\boldsymbol{V}^{X^0}(t;\theta)$, $\boldsymbol{V}^{\Xi(s)}(t;\theta)$ appearing in the multidimensional SLE (4.34) are again functionals of response and excitation. A straightforward way of observing this, is by considering the ***Peano-Baker series expansion*** (Brockett, 1970), sec. 1.3, for the state-transition matrix $\boldsymbol{\Phi}(t,s;\theta)$. By denoting the matrix of linear RDEs (4.37a), (4.42a) as $\boldsymbol{A}(\boldsymbol{X}(t;\theta), \boldsymbol{\Xi}(t;\theta))$,

$$A_{nm}(\boldsymbol{X}(t;\theta)\,;\,\boldsymbol{\Xi}(t;\theta)) = \frac{\partial h_n(\boldsymbol{X}(t;\theta))}{\partial X_m(t;\theta)} + \sum_{k=1}^{K}\frac{\partial q_{nk}(\boldsymbol{X}(t;\theta))}{\partial X_m(t;\theta)}\,\Xi_k(t;\theta), \quad (4.45)$$

state-transition matrix is expanded in series as

$$\boldsymbol{\Phi}(t,s;\theta) = \mathbf{I} + \int_s^t \boldsymbol{A}\big(\boldsymbol{X}(u;\theta);\Xi(u;\theta)\big)\,du + \\ + \int_s^t \boldsymbol{A}\big(\boldsymbol{X}(u_1;\theta);\Xi(u_1;\theta)\big)\int_s^{u_1} \boldsymbol{A}\big(\boldsymbol{X}(u_2;\theta);\Xi(u_2;\theta)\big)\,du_2\,du_1 + \ldots \tag{4.46}$$

where $\mathbf{I}$ is the $N\times N$ unit matrix. Due to the integrals in the right-hand side of Eq. (4.46), state-transition matrix is easily identified as a functional of the response and excitation, $\boldsymbol{\Phi}(t,s;\theta) = \boldsymbol{\Phi}[\boldsymbol{X}(\bullet|_s^t;\theta);\Xi(\bullet|_s^t;\theta)]$. Via Eqs. (4.39), (4.44), this functional character is inherited by the matrices of the variational derivatives, denoted thus equivalently as

$$\boldsymbol{V}^{X^0}(t;\theta) = \boldsymbol{V}^{X^0}[\boldsymbol{X}(\bullet|_{t_0}^t;\theta);\Xi(\bullet|_{t_0}^t;\theta)],\ \ \boldsymbol{V}^{\Xi(s)}(t;\theta) = \boldsymbol{V}^{\Xi(s)}[\boldsymbol{X}(\bullet|_s^t;\theta);\Xi(\bullet|_s^t;\theta)].$$

### 4.2.2 Rederivation of the multidimensional Fokker-Planck-Kolmogorov equation

As a first application of the multidimensional SLE (4.34), we consider the nonlinear RDE, Eq. (4.32a), under $K-$dimensional ***zero-mean white noise excitation*** whose intensity is the symmetric, positive definite matrix $\boldsymbol{D}(t)$, and for uncorrelated initial value to excitation

$$\boldsymbol{m}_\Xi(t) = \mathbf{0},\quad \boldsymbol{C}_{\Xi\Xi}^{\mathrm{WN}}(t,s) = 2\boldsymbol{D}(t)\,\delta(t-s),\quad \boldsymbol{C}_{X^0\Xi}(t) = \mathbf{0}.$$

As explained in Section 4.1.1, the singular autocovariance $\boldsymbol{C}_{\Xi\Xi}^{\mathrm{WN}}(t,s)$ is approximated by an weighted delta family, $\boldsymbol{C}_{\Xi\Xi}^{(\varepsilon)}(t,s) = 2\boldsymbol{D}(t)\,\delta_\varepsilon(t-s)$, whose smooth functions $\delta_\varepsilon(t-s)$ possess the property: $\lim_{\varepsilon\downarrow 0}\int_{t_0}^t \delta_\varepsilon(t-s)\,g(s)\,ds = (1/2)\,g(t)$ for any continuous function $g(\bullet)$, Eq. (4.14). Thus, since $V_{n_2k_2}^{\Xi(s)}(t;\theta)$ is a continuous function with respect to $s$, see Eq. (4.42), the integrals in SLE (4.34) are calculated as

$$\int_{t_0}^t C_{\Xi_{k_1}\Xi_{k_2}}(t,s)\,V_{n_2k_2}^{\Xi(s)}(t;\theta)\,ds \equiv \lim_{\varepsilon\downarrow 0}\int_{t_0}^t C_{\Xi_{k_1}\Xi_{k_2}}^{(\varepsilon)}(t,s)\,V_{n_2k_2}^{\Xi(s)}(t;\theta)\,ds = \\ = 2D_{k_1k_2}(t)\lim_{\varepsilon\downarrow 0}\int_{t_0}^t \delta_\varepsilon(t-s)\,V_{n_2k_2}^{\Xi(s)}(t;\theta)\,ds = D_{k_1k_2}(t)\,V_{n_2k_2}^{\Xi(t)}(t;\theta). \tag{4.47}$$

Since, now $\Phi_{nm}(t,t;\theta) = \delta_{nm}$ (Brockett, 1970), $V_{n_2k_2}^{\Xi(t)}(t;\theta)$ is expressed via Eq. (4.44) as

$$V_{n_2k_2}^{\Xi(t)}(t;\theta) = \sum_{m=1}^N \Phi_{n_2m}(t,t;\theta)\,q_{mk_2}\big(\boldsymbol{X}(t;\theta)\big) = q_{n_2k_2}\big(\boldsymbol{X}(t;\theta)\big). \tag{4.48}$$

By substituting Eq. (4.48) into Eq. (4.47) we finally obtain

$$\int_{t_0}^t C_{\Xi_{k_1}\Xi_{k_2}}(t,s)\,V_{n_2k_2}^{\Xi(s)}(t;\theta)\,ds = D_{k_1k_2}(t)\,q_{n_2k_2}\big(\boldsymbol{X}(t;\theta)\big). \tag{4.49}$$

Substitution of Eq. (4.49) into SLE (4.34) for $\boldsymbol{m}_{\Xi}(t) = \mathbf{0}$, $\boldsymbol{C}_{X^0\Xi}(t) = \mathbf{0}$, and use of delta projection, results into

$$\frac{\partial f_{X(t)}(\boldsymbol{x})}{\partial t} + \sum_{n=1}^{N} \frac{\partial}{\partial x_n}\left(h_n(\boldsymbol{x})\, f_{X(t)}(\boldsymbol{x})\right) =$$
$$= \sum_{n_1=1}^{N} \frac{\partial}{\partial x_{n_1}}\left[\sum_{k_1=1}^{K} q_{n_1k_1}(\boldsymbol{x}) \sum_{k_2=1}^{K} D_{k_1k_2}(t) \sum_{n_2=1}^{N} \frac{\partial}{\partial x_{n_2}}\left(q_{n_2k_2}(\boldsymbol{x})\, f_{X(t)}(\boldsymbol{x})\right)\right]. \tag{4.50}$$

Eq. (4.50) is the multidimensional FPK equation under the Stratonovich interpretation (Öttinger, 1996) equation (3.107). Note that, for additively excited RDE (4.32a), i.e. $q_{nk}(\boldsymbol{x}) = \kappa_{nk} =$ constants, Stratonovich FPK (4.50) coincides with the Itō one, and is expressed in the usual drift-diffusion form of the FPK equation, see e.g. (Gardiner, 2004) Eq. (3.5.6):

$$\frac{\partial f_{X(t)}(\boldsymbol{x})}{\partial t} + \sum_{n=1}^{N} \frac{\partial}{\partial x_n}\left(h_n(\boldsymbol{x})\, f_{X(t)}(\boldsymbol{x})\right) = \sum_{n_1=1}^{N}\sum_{n_2=1}^{N} B_{n_1n_2}^{\mathrm{WN}}(t)\, \frac{\partial^2 f_{X(t)}(\boldsymbol{x})}{\partial x_{n_1}\partial x_{n_2}}, \tag{4.51}$$

where diffusion is defined as the symmetric matrix

$$B_{n_1n_2}^{\mathrm{WN}}(t) = \sum_{k_1=1}^{K}\sum_{k_2=1}^{K} \kappa_{n_1k_1} D_{k_1k_2}(t)\, \kappa_{n_2k_2}\,. \tag{4.52}$$

### 4.2.3 The case of linear dynamical system under additive Gaussian noise

If we consider now $h_n(\boldsymbol{x}) = \sum_{m=1}^{N} \eta_{nm} x_m$, $q_{nk}(\boldsymbol{x}) = \kappa_{nk} =$ constants, multidimensional RIVP (4.32a,b) becomes the linear additively excited dynamical system

$$\dot{X}_n(t;\theta) = \sum_{m=1}^{N} \eta_{nm} X_m(t;\theta) + \sum_{k=1}^{K} \kappa_{nk}\, \Xi_k(t;\theta), \tag{4.53a}$$

$$X_n(t_0;\theta) = X_n^0(\theta), \qquad n = 1, \dots, N. \tag{4.53b}$$

For the case of RIVP (4.53a,b), matrix $\boldsymbol{A}\left(\boldsymbol{X}(t;\theta), \boldsymbol{\Xi}(t;\theta)\right)$ of the corresponding variational equations is specified, via Eq. (4.45), into

$$A_{nm}\left(\boldsymbol{X}(t;\theta);\boldsymbol{\Xi}(t;\theta)\right) = \eta_{nm}. \tag{4.54}$$

Thus, in the linear-additive case, the matrix of the variational equation does not depend on the response and excitation, and, a fortiori, is a time-independent constant matrix. In this case, state-transition matrix $\boldsymbol{\Phi}(t,s;\theta)$ is expressed as the ***matrix exponential*** (Brockett, 1970; Perko, 2001)

$$\boldsymbol{\Phi}(t,s;\theta) = e^{\boldsymbol{\eta}(t-s)}. \tag{4.55}$$

**Remark 4.5: On the calculation of the elements of matrix exponential.** The components of matrix exponential $e^{\boldsymbol{\eta}(t-s)}$ can be determined using the Jordan canonical form $\boldsymbol{K}$ of matrix $\boldsymbol{\eta}$; $\boldsymbol{\eta} = \boldsymbol{P}\,\boldsymbol{K}\,\boldsymbol{P}^{-1}$, where $\boldsymbol{P}$ is an appropriate invertible matrix defined via the (generalized) eigenvectors of $\boldsymbol{\eta}$ (Perko, 2001), whose inverse is denoted by $\boldsymbol{P}^{-1}$. Then, $e^{\boldsymbol{\eta}(t-s)} = \boldsymbol{P}\,e^{\boldsymbol{K}(t-s)}\,\boldsymbol{P}^{-1}$, with the matrix exponential $e^{\boldsymbol{K}(t-s)}$ of a Jordan canonical form being easily determined, see e.g. (Perko, 2001), sec. 1.8.

By substituting Eq. (4.55) into Eqs. (4.38), (4.44), the variational derivatives of the linear dynamical system (4.53a,b) are determined into

$$V_{n\nu}^{X^0}(t;\theta) = \left(e^{\boldsymbol{\eta}(t-t_0)}\right)_{n\nu}, \qquad V_{n\ell}^{\Xi(s)}(t;\theta) = \sum_{m=1}^{N} \kappa_{m\ell}\left(e^{\boldsymbol{\eta}(t-s)}\right)_{nm}. \tag{4.56a,b}$$

Then, by substituting Eqs. (4.56a,b) into SLE (4.34), we obtain the *exact* response pdf evolution equation

$$\begin{aligned}\frac{\partial f_{\boldsymbol{X}(t)}(\boldsymbol{x})}{\partial t} + \sum_{n=1}^{N}\frac{\partial}{\partial x_n}\left[\left(\sum_{m=1}^{N}\eta_{nm}\,x_m + \sum_{k=1}^{K}\kappa_{nk}\,m_{\Xi_k}(t)\right) f_{\boldsymbol{X}(t)}(\boldsymbol{x})\right] &= \\ = \sum_{n_1=1}^{N}\sum_{n_2=1}^{N}\frac{\partial^2}{\partial x_{n_1}\partial x_{n_2}}&\left(B_{n_1n_2}^{\text{eff}}(t)\, f_{\boldsymbol{X}(t)}(\boldsymbol{x})\right),\end{aligned} \tag{4.57}$$

in which the ***effective diffusion matrix*** $\boldsymbol{B}^{\text{eff}}(t)$ is defined as

$$\begin{aligned}B_{n_1n_2}^{\text{eff}}(t) = &\sum_{n_3=1}^{N}\sum_{k=1}^{K}\kappa_{n_1k}\,C_{X_{n_3}^0\Xi_k}(t)\left(e^{\boldsymbol{\eta}(t-t_0)}\right)_{n_2n_3} + \\ &+\sum_{n_3=1}^{N}\sum_{k_1=1}^{K}\sum_{k_2=1}^{K}\kappa_{n_1k_1}\,\kappa_{n_3k_2}\int_{t_0}^{t} C_{\Xi_{k_1}\Xi_{k_2}}(t,s)\left(e^{\boldsymbol{\eta}(t-s)}\right)_{n_2n_3} ds.\end{aligned} \tag{4.58}$$

In order to solve Eq. (4.57), we perform the $N-$dimensional Fourier transform; $\varphi_{\boldsymbol{X}(t)}(\boldsymbol{y}) = \int_{\mathbb{R}^N} e^{i\boldsymbol{y}^{\mathrm{T}}\boldsymbol{x}} f_{\boldsymbol{X}(t)}(\boldsymbol{x})\,d\boldsymbol{x}$, which gives rise to the following linear equation of first partial derivatives for the characteristic function $\varphi_{\boldsymbol{X}(t)}(\boldsymbol{y})$:

$$\begin{aligned}\frac{\partial\varphi_{\boldsymbol{X}(t)}(\boldsymbol{y})}{\partial t} = &\sum_{m=1}^{N}\left(\sum_{n=1}^{N}\eta_{nm}\,y_n\right)\frac{\partial\varphi_{\boldsymbol{X}(t)}(\boldsymbol{y})}{\partial y_m} + \\ &+\left[i\sum_{n=1}^{N}\left(\sum_{k=1}^{K}\kappa_{nk}\,m_{\Xi_k}(t)\right)y_n - \sum_{n_1=1}^{N}\sum_{n_2=1}^{N}B_{n_1n_2}^{\text{eff}}(t)\,y_{n_1}y_{n_2}\right]\varphi_{\boldsymbol{X}(t)}(\boldsymbol{y}).\end{aligned} \tag{4.59}$$

Eq. (4.59) is supplemented by the Gaussian initial condition, which, in the Fourier domain, is expressed as

$$\varphi_{\boldsymbol{X}_0}(\boldsymbol{y}) = \exp\left(i\sum_{n=1}^{N}m_{X_n^0}\,y_n - \frac{1}{2}\sum_{n_1=1}^{N}\sum_{n_2=1}^{N}C_{X_{n_1}^0X_{n_2}^0}\,y_{n_1}y_{n_2}\right). \tag{4.60}$$

Following (Polyanin et al., 2001), we first study the homogeneous variant of Eq. (4.59)

$$\frac{\partial \varphi_{\boldsymbol{X}(t)}(\boldsymbol{y})}{\partial t} - \sum_{m=1}^{N} \left( \sum_{n=1}^{N} \eta_{nm} \, y_n \right) \frac{\partial \varphi_{\boldsymbol{X}(t)}(\boldsymbol{y})}{\partial y_m} = 0 . \tag{4.61}$$

The characteristic equation corresponding to homogeneous Eq. (4.61) is

$$\frac{dt}{1} = - \frac{dy_1}{\sum_{n=1}^{N} \eta_{n1} \, y_n} = \cdots = - \frac{dy_N}{\sum_{n=1}^{N} \eta_{nN} \, y_n}$$

which is expressed equivalently as the linear ODE, $\dot{\boldsymbol{y}} = - \boldsymbol{\eta}^{\mathrm{T}} \boldsymbol{y}$, where superscript T denotes the transpose of the matrix. Via the solution of the said ODE, the following change in variables is determined; $\boldsymbol{y} = e^{-\boldsymbol{\eta}^{\mathrm{T}} t} \boldsymbol{\upsilon} \Rightarrow \boldsymbol{\upsilon}(\boldsymbol{y}, t) = e^{\boldsymbol{\eta}^{\mathrm{T}} t} \boldsymbol{y}$. It is easy to see that, under this change of variables, Eq. (4.59) is simplified into the linear ODE with respect to time:

$$\begin{aligned} \frac{\partial \varphi_{\boldsymbol{X}(t)}(\boldsymbol{\upsilon})}{\partial t} = \Bigg[ & i \sum_{n=1}^{N} \sum_{k=1}^{K} \sum_{m=1}^{N} \kappa_{nk} \, \upsilon_m \, m_{\Xi_k}(t) \, (e^{-\boldsymbol{\eta}^{\mathrm{T}} t})_{nm} - \\ & - \sum_{n_1=1}^{N} \sum_{n_2=1}^{N} \sum_{m_1=1}^{N} \sum_{m_2=1}^{N} \upsilon_{m_1} \, \upsilon_{m_2} \, (e^{-\boldsymbol{\eta}^{\mathrm{T}} t})_{n_1 m_1} \, B^{\mathrm{eff}}_{n_1 n_2}(t) \, (e^{-\boldsymbol{\eta}^{\mathrm{T}} t})_{n_2 m_2} \Bigg] \varphi_{\boldsymbol{X}(t)}(\boldsymbol{\upsilon}), \end{aligned} \tag{4.62}$$

while the initial condition (4.60) is transformed into

$$\begin{aligned} \varphi_{X_0}(\boldsymbol{\upsilon}) = {} & \exp\left( i \sum_{n=1}^{N} \sum_{m=1}^{N} m_{X_n^0} (e^{-\boldsymbol{\eta}^{\mathrm{T}} t_0})_{nm} \, \upsilon_m \right) \times \\ & \times \exp\left( - \frac{1}{2} \sum_{n_1=1}^{N} \sum_{n_2=1}^{N} \sum_{m_1=1}^{N} \sum_{m_2=1}^{N} \upsilon_{m_1} \upsilon_{m_2} (e^{-\boldsymbol{\eta}^{\mathrm{T}} t_0})_{n_1 m_1} C_{X_{n_1}^0 X_{n_2}^0} \, (e^{-\boldsymbol{\eta}^{\mathrm{T}} t_0})_{n_2 m_2} \right). \end{aligned} \tag{4.63}$$

Solution to IVP (4.62)-(4.63) is easily determined to

$$\begin{aligned} \varphi_{\boldsymbol{X}(t)}(\boldsymbol{\upsilon}) = {} & \exp\left[ i \sum_{m=1}^{N} \upsilon_m \left( \sum_{n=1}^{N} m_{X_n^0} (e^{-\boldsymbol{\eta}^{\mathrm{T}} t_0})_{nm} + \sum_{k=1}^{K} \kappa_{nk} \int_{t_0}^{t} m_{\Xi_k}(u) \, (e^{-\boldsymbol{\eta}^{\mathrm{T}} u})_{nm} \, du \right) \right] \times \\ & \times \exp\Bigg[ - \frac{1}{2} \sum_{m_1=1}^{N} \sum_{m_2=1}^{N} \upsilon_{m_1} \upsilon_{m_2} \sum_{n_1=1}^{N} \sum_{n_2=1}^{N} \Bigg( (e^{-\boldsymbol{\eta}^{\mathrm{T}} t_0})_{n_1 m_1} C_{X_{n_1}^0 X_{n_2}^0} \, (e^{-\boldsymbol{\eta}^{\mathrm{T}} t_0})_{n_2 m_2} + \\ & \qquad + 2 \int_{t_0}^{t} (e^{-\boldsymbol{\eta}^{\mathrm{T}} u})_{n_1 m_1} \, B^{\mathrm{eff}}_{n_1 n_2}(u) \, (e^{-\boldsymbol{\eta}^{\mathrm{T}} u})_{n_2 m_2} \, du \Bigg) \Bigg]. \end{aligned} \tag{4.64}$$

By returning now to the original variables, and employing the identities of exponential matrices; $\sum_{m=1}^{N} (e^{-\boldsymbol{\eta}^{\mathrm{T}} u})_{nm} (e^{\boldsymbol{\eta}^{\mathrm{T}} t})_{m\nu} = (e^{\boldsymbol{\eta}^{\mathrm{T}} (t-u)})_{n\nu}$, and $(e^{\boldsymbol{\eta}^{\mathrm{T}} (t-u)})_{n\nu} = (e^{\boldsymbol{\eta} (t-u)})_{\nu n}$, solution (4.64) is written as

$$\varphi_{X(t)}(\boldsymbol{y}) = \exp\left[ i \sum_{\nu=1}^{N} y_\nu \left( \sum_{n=1}^{N} (e^{\boldsymbol{\eta}(t-t_0)})_{\nu n} \, m_{X_n^0} + \int_{t_0}^{t} \sum_{n=1}^{N} (e^{\boldsymbol{\eta}(t-u)})_{\nu n} \left( \sum_{k=1}^{K} \kappa_{nk} \, m_{\Xi_k}(u) \right) du \right) \right] \times$$
$$\times \exp\left[ -\frac{1}{2} \sum_{\nu_1=1}^{N} \sum_{\nu_2=1}^{N} y_{\nu_1} y_{\nu_2} \left( \sum_{n_1=1}^{N} \sum_{n_1=1}^{N} (e^{\boldsymbol{\eta}(t-t_0)})_{\nu_1 n_1} C_{X_{n_1}^0 X_{n_2}^0} (e^{\boldsymbol{\eta}(t-t_0)})_{\nu_2 n_2} + \right.\right.$$
$$\left.\left. + 2 \int_{t_0}^{t} \sum_{n_1=1}^{N} \sum_{n_2=1}^{N} (e^{\boldsymbol{\eta}(t-u)})_{\nu_1 n_1} B_{n_1 n_2}^{\text{eff}}(u) \, (e^{\boldsymbol{\eta}(t-u)})_{\nu_2 n_2} \, du \right) \right]. \tag{4.65}$$

Eq. (4.65) is expressed equivalently in matrix notation as

$$\varphi_{X(t)}(\boldsymbol{y}) = \exp\left[ i \, \boldsymbol{y}^{\mathrm{T}} \left( e^{\boldsymbol{\eta}(t-t_0)} \boldsymbol{m}_{X^0} + \int_{t_0}^{t} e^{\boldsymbol{\eta}(t-u)} \boldsymbol{\kappa} \, \boldsymbol{m}_{\Xi}(u) \, du \right) - \frac{1}{2} \, \boldsymbol{y}^{\mathrm{T}} \boldsymbol{Q}(t) \, \boldsymbol{y} \right], \tag{4.66}$$

with the square matrix $\boldsymbol{Q}(t)$ of the quadratic form defined as

$$\boldsymbol{Q}(t) = e^{\boldsymbol{\eta}(t-t_0)} \boldsymbol{C}_{X^0 X^0} \, e^{\boldsymbol{\eta}^{\mathrm{T}}(t-t_0)} + 2 \int_{t_0}^{t} e^{\boldsymbol{\eta}(t-u)} \boldsymbol{B}^{\text{eff}}(u) \, e^{\boldsymbol{\eta}^{\mathrm{T}}(t-u)} \, du \,,$$

and by substituting effective diffusion $\boldsymbol{B}^{\text{eff}}(u)$ from its definition relation, Eq. (4.58)

$$\boldsymbol{Q}(t) = e^{\boldsymbol{\eta}(t-t_0)} \boldsymbol{C}_{X^0 X^0} \, e^{\boldsymbol{\eta}^{\mathrm{T}}(t-t_0)} + 2 \left( \int_{t_0}^{t} e^{\boldsymbol{\eta}(t-u)} \boldsymbol{\kappa} \, \boldsymbol{C}_{X^0 \Xi}^{\mathrm{T}}(u) \, du \right) e^{\boldsymbol{\eta}^{\mathrm{T}}(t-t_0)} +$$
$$+ 2 \int_{t_0}^{t} \int_{t_0}^{u} e^{\boldsymbol{\eta}(t-u)} \boldsymbol{\kappa} \, \boldsymbol{C}_{\Xi\Xi}(u,s) \, \boldsymbol{\kappa}^{\mathrm{T}} e^{\boldsymbol{\eta}^{\mathrm{T}}(t-s)} \, ds \, du \,. \tag{4.67}$$

In Eq. (4.67), the integrand of the double integral is symmetric, permitting us to write

$$2 \int_{t_0}^{t} \int_{t_0}^{u} e^{\boldsymbol{\eta}(t-u)} \boldsymbol{\kappa} \, \boldsymbol{C}_{\Xi\Xi}(u,s) \, \boldsymbol{\kappa}^{\mathrm{T}} e^{\boldsymbol{\eta}^{\mathrm{T}}(t-s)} \, ds \, du = \int_{t_0}^{t} \int_{t_0}^{t} e^{\boldsymbol{\eta}(t-u)} \boldsymbol{\kappa} \, \boldsymbol{C}_{\Xi\Xi}(u,s) \, \boldsymbol{\kappa}^{\mathrm{T}} e^{\boldsymbol{\eta}^{\mathrm{T}}(t-s)} \, ds \, du \,.$$

Furthermore, we observe that matrix $\boldsymbol{Q}(t)$ is not symmetric, due to the presence of the term containing the single integral. Thus, in solution (4.67), matrix $\boldsymbol{Q}(t)$ can be substituted by its symmetric part $\boldsymbol{Q}^{\text{sym}}(t)$:

$$\boldsymbol{Q}^{\text{sym}}(t) = \frac{1}{2} \left( \boldsymbol{Q}(t) + \boldsymbol{Q}^{\mathrm{T}}(t) \right) = e^{\boldsymbol{\eta}(t-t_0)} \boldsymbol{C}_{X^0 X^0} \, e^{\boldsymbol{\eta}^{\mathrm{T}}(t-t_0)} +$$
$$+ \left( \int_{t_0}^{t} e^{\boldsymbol{\eta}(t-u)} \boldsymbol{\kappa} \, \boldsymbol{C}_{X^0 \Xi}^{\mathrm{T}}(u) \, du \right) e^{\boldsymbol{\eta}^{\mathrm{T}}(t-t_0)} + e^{\boldsymbol{\eta}(t-t_0)} \left( \int_{t_0}^{t} \boldsymbol{C}_{X^0 \Xi}(u) \boldsymbol{\kappa}^{\mathrm{T}} e^{\boldsymbol{\eta}^{\mathrm{T}}(t-u)} \, du \right) + \tag{4.68}$$
$$+ \int_{t_0}^{t} \int_{t_0}^{t} e^{\boldsymbol{\eta}(t-u)} \boldsymbol{\kappa} \, \boldsymbol{C}_{\Xi\Xi}(u,s) \, \boldsymbol{\kappa}^{\mathrm{T}} e^{\boldsymbol{\eta}^{\mathrm{T}}(t-s)} \, ds \, du \,,$$

since a non-symmetric square matrix $\boldsymbol{Q}(t)$ can be decomposed into a symmetric and a skew-symmetric part; $\boldsymbol{Q}(t) = \boldsymbol{Q}^{\text{sym}}(t) + \boldsymbol{Q}^{\text{skew}}(t)$, $\boldsymbol{Q}^{\text{sym}}(t) = (1/2)\left(\boldsymbol{Q}(t) + \boldsymbol{Q}^{\text{T}}(t)\right)$, $\boldsymbol{Q}^{\text{skew}}(t) = (1/2)\left(\boldsymbol{Q}(t) - \boldsymbol{Q}^{\text{T}}(t)\right)$, and the quadratic form with a skew-symmetric matrix equals to zero (Kolter & Do, 2015). Thus, solution (4.68) is updated to

$$\varphi_{X(t)}(\boldsymbol{y}) = \exp\left[ i\,\boldsymbol{y}^{\text{T}} \left( e^{\boldsymbol{\eta}(t-t_0)} \boldsymbol{m}_{X^0} + \int_{t_0}^{t} e^{\boldsymbol{\eta}(t-u)} \boldsymbol{\kappa}\, \boldsymbol{m}_{\Xi}(u)\, du \right) - \frac{1}{2} \boldsymbol{y}^{\text{T}} \boldsymbol{Q}^{\text{sym}}(t)\, \boldsymbol{y} \right], \tag{4.69}$$

The form of characteristic function $\varphi_{X(t)}(\boldsymbol{y})$, as given by Eq. (4.69), is identified as Gaussian, with the $N\times 1$ moment vector $\boldsymbol{m}_X(t)$ and $N\times N$ covariance matrix $\boldsymbol{C}_{XX}(t,t)$ determined to

$$\boldsymbol{m}_X(t) = e^{\boldsymbol{\eta}(t-t_0)} \boldsymbol{m}_{X^0} + \int_{t_0}^{t} e^{\boldsymbol{\eta}(t-u)} \boldsymbol{\kappa}\, \boldsymbol{m}_{\Xi}(u)\, du, \qquad \text{and} \tag{4.70}$$

$$\begin{aligned}\boldsymbol{C}_{XX}(t,t) &= \boldsymbol{Q}^{\text{sym}}(t) = e^{\boldsymbol{\eta}(t-t_0)} \boldsymbol{C}_{X^0X^0}\, e^{\boldsymbol{\eta}^{\text{T}}(t-t_0)} + \\ &+ \left( \int_{t_0}^{t} e^{\boldsymbol{\eta}(t-u)} \boldsymbol{\kappa}\, \boldsymbol{C}^{\text{T}}_{X^0\Xi}(u)\, du \right) e^{\boldsymbol{\eta}^{\text{T}}(t-t_0)} + e^{\boldsymbol{\eta}(t-t_0)} \left( \int_{t_0}^{t} \boldsymbol{C}_{X^0\Xi}(u) \boldsymbol{\kappa}^{\text{T}} e^{\boldsymbol{\eta}^{\text{T}}(t-u)}\, du \right) + \\ &+ \int_{t_0}^{t}\int_{t_0}^{t} e^{\boldsymbol{\eta}(t-u)} \boldsymbol{\kappa}\, \boldsymbol{C}_{\Xi\Xi}(u,s)\, \boldsymbol{\kappa}^{\text{T}} e^{\boldsymbol{\eta}^{\text{T}}(t-s)}\, ds\, du.\end{aligned} \tag{4.71}$$

**Verification of the moments.** Formulae (4.70), (4.71) for the mean value vector and auto-covariance matrix of the response can be verified by their direct calculation from RIVP (4.53a,b), since it is linear. First, for convenience, we express RIVP (4.53a,b) in matrix form

$$\dot{\boldsymbol{X}}(t;\theta) = \boldsymbol{\eta}\, \boldsymbol{X}(t;\theta) + \boldsymbol{\kappa}\, \Xi(t;\theta), \boldsymbol{X}(t_0;\theta) = \boldsymbol{X}^0(\theta). \tag{4.72}$$

Again, its solution is expressed in closed form by employing the matrix exponential $e^{\boldsymbol{\eta}(t-u)}$ as

$$\boldsymbol{X}(t;\theta) = e^{\boldsymbol{\eta}(t-t_0)} \boldsymbol{X}^0(\theta) + \int_{t_0}^{t} e^{\boldsymbol{\eta}(t-u)} \boldsymbol{\kappa}\, \Xi(u;\theta)\, du. \tag{4.73}$$

By applying the average operator $\mathbb{E}^{\theta}[\bullet]$ at both sides of Eq. (4.73), formula (4.70) for the mean value is readily obtained. Let us now calculate the quantity $\boldsymbol{X}(t;\theta)\,\boldsymbol{X}^{\text{T}}(t;\theta)$ from Eq. (4.73)

$$\begin{aligned}&\boldsymbol{X}(t;\theta)\,\boldsymbol{X}^{\text{T}}(t;\theta) = \\ &= e^{\boldsymbol{\eta}(t-t_0)} \boldsymbol{X}^0(\theta) \left(\boldsymbol{X}^0(\theta)\right)^{\text{T}} e^{\boldsymbol{\eta}^{\text{T}}(t-t_0)} + e^{\boldsymbol{\eta}(t-t_0)} \int_{t_0}^{t} \boldsymbol{X}^0(\theta)\, \Xi^{\text{T}}(u;\theta) \boldsymbol{\kappa}^{\text{T}} e^{\boldsymbol{\eta}^{\text{T}}(t-u)}\, du + \\ &+ \left( \int_{t_0}^{t} e^{\boldsymbol{\eta}(t-u)} \boldsymbol{\kappa}\, \Xi(u;\theta) \left(\boldsymbol{X}^0(\theta)\right)^{\text{T}} du \right) e^{\boldsymbol{\eta}^{\text{T}}(t-t_0)} + \int_{t_0}^{t}\int_{t_0}^{t} e^{\boldsymbol{\eta}(t-u)} \boldsymbol{\kappa}\, \Xi(u;\theta) \Xi^{\text{T}}(s;\theta) \boldsymbol{\kappa}^{\text{T}} e^{\boldsymbol{\eta}^{\text{T}}(t-s)}\, ds\, du.\end{aligned} \tag{4.74}$$

We also calculate the quantity $\boldsymbol{m}_X(t)\,\boldsymbol{m}_X^{\mathrm{T}}(t)$, by employing the verified formula (4.70):

$$\boldsymbol{m}_X(t)\,\boldsymbol{m}_X^{\mathrm{T}}(t) = e^{\boldsymbol{\eta}(t-t_0)}\,\boldsymbol{m}_{X^0}\,\boldsymbol{m}_{X^0}^{\mathrm{T}}\,e^{\boldsymbol{\eta}^{\mathrm{T}}(t-t_0)} + e^{\boldsymbol{\eta}(t-t_0)}\int_{t_0}^{t}\boldsymbol{m}_{X^0}\,\boldsymbol{m}_{\Xi}^{\mathrm{T}}(u)\,\boldsymbol{\kappa}^{\mathrm{T}}\,e^{\boldsymbol{\eta}^{\mathrm{T}}(t-u)}\,du + \left(\int_{t_0}^{t} e^{\boldsymbol{\eta}(t-u)}\,\boldsymbol{\kappa}\,\boldsymbol{m}_{\Xi}(u)\,\boldsymbol{m}_{X^0}^{\mathrm{T}}\,du\right)e^{\boldsymbol{\eta}^{\mathrm{T}}(t-t_0)} + \int_{t_0}^{t}\int_{t_0}^{t} e^{\boldsymbol{\eta}(t-u)}\,\boldsymbol{\kappa}\,\boldsymbol{m}_{\Xi}(u)\,\boldsymbol{m}_{\Xi}^{\mathrm{T}}(s)\,\boldsymbol{\kappa}^{\mathrm{T}}\,e^{\boldsymbol{\eta}^{\mathrm{T}}(t-s)}\,ds\,du. \tag{4.75}$$

By using Eqs. (4.74), (4.76), and the definition relations of the matrices of second-order central moments, e.g. $\boldsymbol{C}_{XX}(t) = \mathbb{E}^{\theta}\left[\boldsymbol{X}(t;\theta)\,\boldsymbol{X}^{\mathrm{T}}(t;\theta)\right] - \boldsymbol{m}_X(t)\,\boldsymbol{m}_X^{\mathrm{T}}(t)$, we obtain

$$\boldsymbol{C}_{XX}(t,t) = \mathbb{E}^{\theta}\left[\boldsymbol{X}(t;\theta)\,\boldsymbol{X}^{\mathrm{T}}(t;\theta) - \boldsymbol{m}_X(t)\,\boldsymbol{m}_X^{\mathrm{T}}(t)\right] =$$
$$= e^{\boldsymbol{\eta}(t-t_0)}\,\boldsymbol{C}_{X^0X^0}\,e^{\boldsymbol{\eta}^{\mathrm{T}}(t-t_0)} + \left(\int_{t_0}^{t} e^{\boldsymbol{\eta}(t-u)}\,\boldsymbol{\kappa}\,\boldsymbol{C}_{X^0\Xi}^{\mathrm{T}}(u)\,du\right)e^{\boldsymbol{\eta}^{\mathrm{T}}(t-t_0)} + e^{\boldsymbol{\eta}(t-t_0)}\left(\int_{t_0}^{t}\boldsymbol{C}_{X^0\Xi}(u)\,\boldsymbol{\kappa}^{\mathrm{T}}\,e^{\boldsymbol{\eta}^{\mathrm{T}}(t-u)}\,du\right) + \int_{t_0}^{t}\int_{t_0}^{t} e^{\boldsymbol{\eta}(t-u)}\,\boldsymbol{\kappa}\,\boldsymbol{C}_{\Xi\Xi}(u,s)\,\boldsymbol{\kappa}^{\mathrm{T}}\,e^{\boldsymbol{\eta}^{\mathrm{T}}(t-s)}\,ds\,du. \tag{4.76}$$

which coincides with expression (4.71) obtained from the solution of the pdf evolution equation.

**Remark 4.6: Connection between effective diffusion matrix and cross-covariance.** By following the same procedure as for the calculation of $\boldsymbol{C}_{XX}(t,t)$ from RIVP solution (4.73), we obtain the following expression for the one-time, excitation-response cross-covariance matrix:

$$\boldsymbol{C}_{\Xi X}(t,t) = \boldsymbol{C}_{X^0\Xi}^{\mathrm{T}}(t)\,e^{\boldsymbol{\eta}^{\mathrm{T}}(t-t_0)} + \int_{t_0}^{t}\boldsymbol{C}_{\Xi\Xi}(t,u)\,\boldsymbol{\kappa}^{\mathrm{T}}\,e^{\boldsymbol{\eta}^{\mathrm{T}}(t-u)}\,du. \tag{4.77}$$

By comparing the above expression to the definition relation of effective diffusion matrix, Eq. (4.58), we easily observe that

$$\boldsymbol{B}^{\mathrm{eff}}(t) = \boldsymbol{\kappa}\,\boldsymbol{C}_{\Xi X}(t,t). \tag{4.78}$$

Furthermore, as we have seen in the solution procedure for pdf Eq. (4.57), the non-symmetric square matrix $\boldsymbol{B}^{\mathrm{eff}}(t)$ can be symmetrized, resulting into

$$\boldsymbol{B}_{\mathrm{sym}}^{\mathrm{eff}}(t) = \frac{1}{2}\left(\boldsymbol{\kappa}\,\boldsymbol{C}_{\Xi X}(t,t) + \boldsymbol{C}_{X\Xi}(t,t)\,\boldsymbol{\kappa}^{\mathrm{T}}\right). \tag{4.79}$$

The symmetrisation of the diffusion matrix is a topic also discussed in (Cetto et al., 1984).

As in the scalar case, the ***uniqueness of Gaussian solution*** to the exact pdf evolution Eq. (4.57) for multidimensional linear systems is guaranteed by the injectivity of Fourier transform and the uniqueness of solution for the transformed problem (4.59), (4.60) for $\varphi_{X(t)}(\boldsymbol{y})$ (Polyanin et al., 2001), sec. 10.1.2. Furthermore, the verification of solution constitutes a first validation for the multidimensional extension of the NF theorem, Eq. (3.60), whose proof was not presented at length in the previous Chapter. The solution of the multidimensional evolution Eq. (4.57) via Fourier transform was presented here in a detailed way, since it serves as the prototype of the solution of the other multidimensional evolution equations corresponding to linear systems, that will studied subsequently in this Chapter, namely the two-dimensional equations for the one-time response-excitation, response-initial value and the two-time response pdfs.

## 4.3 The final form of one-time response-excitation stochastic Liouville equation

Moving to SLEs for pdfs of higher order, we first consider the one-time response-excitation SLE (2.32):

$$\frac{\partial f_{X(t)\Xi(t)}(x,u)}{\partial t} + \frac{\partial}{\partial x}\Big[\big(h(x)+q(x)u\big)\, f_{X(t)\Xi(t)}(x,u)\Big] = \\ = -\frac{\partial}{\partial u}\,\mathbb{E}^{\theta}\Big[\delta(x-X(t;\theta))\,\delta(u-\Xi(t;\theta))\,\dot{\Xi}(t;\theta)\Big], \tag{4.80}$$

corresponding to scalar RIVP (4.2a,b). As in the previous Sections, for jointly Gaussian initial value and excitation, the average in the right-hand side of SLE (4.80) is evaluated further by employing the adequate extension of the NF theorem, Eq. (3.69), which, for its FFℓ being the product of the random delta functions of the response and excitation, reads

$$\mathbb{E}^{\theta}\Big[\dot{\Xi}(t;\theta)\,\delta(x-X(t;\theta))\,\delta(u-\Xi(t;\theta))\Big] = \dot{m}_{\Xi}(t)\, f_{X(t)\Xi(t)}(x,u) + \\ + \dot{C}_{X_0\Xi}(t)\,\mathbb{E}^{\theta}\left[\frac{\partial\Big(\delta(x-X[X_0(\theta);\Xi(\bullet|_{t_0}^{t};\theta)])\,\delta(u-\Xi(t;\theta))\Big)}{\partial X_0(\theta)}\right] + \\ + \int_{t_0}^{t}\partial_t C_{\Xi\Xi}(t,s)\,\mathbb{E}^{\theta}\left[\frac{\delta\Big(\delta(x-X[X_0(\theta);\Xi(\bullet|_{t_0}^{t};\theta)])\,\delta(u-\Xi(t;\theta))\Big)}{\delta\Xi(s;\theta)}\right]ds. \tag{4.81}$$

By taking into consideration that the paths of excitation are functionally independent from the initial value, and that $\delta\Xi(t;\theta)/\delta\Xi(s;\theta)=\delta(t-s)$, the application of product rule for the derivatives in the right-hand side of Eq. (4.81) reads

$$\mathbb{E}^{\theta}\Big[\dot{\Xi}(t;\theta)\,\delta(x-X(t;\theta))\,\delta(u-\Xi(t;\theta))\Big] = \dot{m}_{\Xi}(t)\, f_{X(t)\Xi(t)}(x,u) + \\ + \dot{C}_{X_0\Xi}(t)\,\mathbb{E}^{\theta}\left[\frac{\partial\delta(x-X[X_0(\theta);\Xi(\bullet|_{t_0}^{t};\theta)])}{\partial X_0(\theta)}\,\delta(u-\Xi(t;\theta))\right] + \\ + \int_{t_0}^{t}\partial_t C_{\Xi\Xi}(t,s)\,\mathbb{E}^{\theta}\left[\frac{\delta\delta(x-X[X_0(\theta);\Xi(\bullet|_{t_0}^{t};\theta)])}{\delta\Xi(s;\theta)}\,\delta(u-\Xi(t;\theta))\right]ds + \\ + \int_{t_0}^{t}\partial_t C_{\Xi\Xi}(t,s)\,\mathbb{E}^{\theta}\left[\delta(x-X(t;\theta))\,\frac{\partial\delta(u-\Xi(t;\theta))}{\partial\Xi(t;\theta)}\,\delta(t-s)\right]ds.$$

and by employing the identity of $\delta(t-s)$

$$\begin{aligned}
&\mathbb{E}^{\theta}\Big[\dot{\Xi}(t;\theta)\,\delta(x-X(t;\theta))\,\delta(u-\Xi(t;\theta))\Big] = \dot{m}_{\Xi}(t)\; f_{X(t)\Xi(t)}(x,u) + \\
&\quad + \dot{C}_{X_0\Xi}(t)\;\mathbb{E}^{\theta}\left[\frac{\partial\delta(x-X[X_0(\theta);\Xi(\bullet|_{t_0}^{t};\theta)])}{\partial X_0(\theta)}\;\delta(u-\Xi(t;\theta))\right] + \\
&\quad + \int_{t_0}^{t}\partial_t C_{\Xi\Xi}(t,s)\;\mathbb{E}^{\theta}\left[\frac{\delta\delta(x-X[X_0(\theta);\Xi(\bullet|_{t_0}^{t};\theta)])}{\delta\Xi(s;\theta)}\;\delta(u-\Xi(t;\theta))\right]ds + \\
&\quad + \partial_t C_{\Xi\Xi}(t,s)\Big|_{s=t}\;\mathbb{E}^{\theta}\left[\delta(x-X(t;\theta))\;\frac{\partial\delta(u-\Xi(t;\theta))}{\partial\Xi(t;\theta)}\right].
\end{aligned}\tag{4.82}$$

Finally, by employing the chain rule for the derivatives of random delta function of the response, as well as the delta projection formulae of Section 2.2, Eq. (4.82) is written as

$$\begin{aligned}
&\mathbb{E}^{\theta}\Big[\dot{\Xi}(t;\theta)\,\delta(x-X(t;\theta))\,\delta(u-\Xi(t;\theta))\Big] = \dot{m}_{\Xi}(t)\; f_{X(t)\Xi(t)}(x,u) - \\
&- \dot{C}_{X_0\Xi}(t)\,\frac{\partial}{\partial x}\,\mathbb{E}^{\theta}\Big[\delta(x-X(t;\theta))\,\delta(u-\Xi(t;\theta))\,V_{X_0}[X(\bullet|_{t_0}^{t};\theta);\Xi(\bullet|_{t_0}^{t};\theta)]\Big] - \\
&- \frac{\partial}{\partial x}\int_{t_0}^{t}\partial_t C_{\Xi\Xi}(t,s)\;\mathbb{E}^{\theta}\Big[\delta(x-X(t;\theta)\,\delta(u-\Xi(t;\theta))\,V_{\Xi(s)}[X(\bullet|_{s}^{t};\theta);\Xi(\bullet|_{s}^{t};\theta)]\Big]\,ds - \\
&\quad - \partial_t C_{\Xi\Xi}(t,s)\Big|_{s=t}\;\frac{\partial f_{X(t)\Xi(t)}(x,u)}{\partial u},
\end{aligned}\tag{4.83}$$

where $V_{X_0}[X(\bullet|_{t_0}^{t};\theta);\Xi(\bullet|_{t_0}^{t};\theta)]$, $V_{\Xi(s)}[X(\bullet|_{s}^{t};\theta);\Xi(\bullet|_{s}^{t};\theta)]$ are the variational derivatives of the response with respect to initial value and excitation, defined by Eqs. (4.8), (4.9). Substitution of NF theorem, Eq. (4.83), into SLE (4.80), results in the ***final form of one-time, response-excitation SLE***:

$$\begin{aligned}
&\frac{\partial f_{X(t)\Xi(t)}(x,u)}{\partial t} + \frac{\partial}{\partial x}\Big[\big(h(x)+q(x)u\big)\,f_{X(t)\Xi(t)}(x,u)\Big] + \dot{m}_{\Xi}(t)\,\frac{\partial f_{X(t)\Xi(t)}(x,u)}{\partial u} = \\
&= \dot{C}_{X_0\Xi}(t)\,\frac{\partial^2}{\partial x\partial u}\,\mathbb{E}^{\theta}\Big[\delta(x-X(t;\theta))\,\delta(u-\Xi(t;\theta))\,V_{X_0}[X(\bullet|_{t_0}^{t};\theta);\Xi(\bullet|_{t_0}^{t};\theta)]\Big] + \\
&+ \frac{\partial^2}{\partial x\partial u}\int_{t_0}^{t}\partial_t C_{\Xi\Xi}(t,s)\;\mathbb{E}^{\theta}\Big[\delta(x-X(t;\theta)\,\delta(u-\Xi(t;\theta))\,V_{\Xi(s)}[X(\bullet|_{s}^{t};\theta);\Xi(\bullet|_{s}^{t};\theta)]\Big]\,ds + \\
&\quad + \frac{1}{2}\,\dot{\sigma}^2_{\Xi}(t)\,\frac{\partial^2 f_{X(t)\Xi(t)}(x,u)}{\partial u^2}.
\end{aligned}\tag{4.84}$$

For the derivation of SLE (4.84), the following straightforward relation was employed

$$\partial_t C_{\Xi\Xi}(t,s)\Big|_{s=t} = C_{\dot{\Xi}\Xi}(t,t) = \mathbb{E}^{\theta}[\dot{\Xi}(t;\theta)\,\Xi(t;\theta)] = \frac{1}{2}\frac{\partial}{\partial t}\mathbb{E}^{\theta}[\Xi^2(t;\theta)] = \frac{1}{2}\,\dot{\sigma}^2_{\Xi}(t).$$

As the SLEs derived in the previous Sections, response-excitation SLE (4.84) contains non-local terms that depend on the history of the response and excitation, which are identified as the variational derivatives of the response.

**The case of linear, additively excited RDE.** In the linear-additive case, i.e. $h(x)=\eta x$ and $q(x)=\kappa$, the variational derivatives lose their non-local character, and are specified, by Eqs. (4.19a,b), into $V_{X_0}(t;\theta)=e^{\eta(t-t_0)}$, $V_{\Xi(s)}(t;\theta)=\kappa\, e^{\eta(t-s)}$. By substituting these expressions into SLE (4.84), an exact response-excitation pdf evolution equation is obtained, in closed form:

$$\begin{aligned}\frac{\partial f_{X(t)\Xi(t)}(x,u)}{\partial t}&+\frac{\partial}{\partial x}\Big[\big(\eta x+\kappa u\big)\, f_{X(t)\Xi(t)}(x,u)\Big]+\dot{m}_{\Xi}(t)\,\frac{\partial f_{X(t)\Xi(t)}(x,u)}{\partial u}=\\&=G(t)\,\frac{\partial^2 f_{X(t)\Xi(t)}(x,u)}{\partial x\partial u}+\frac{1}{2}\,\dot{\sigma}_{\Xi}^2(t)\,\frac{\partial^2 f_{X(t)\Xi(t)}(x,u)}{\partial u^2}.\end{aligned} \tag{4.85}$$

where

$$G(t)=\dot{C}_{X_0\Xi}(t)\,e^{\eta(t-t_0)}+\kappa\int_{t_0}^{t}\partial_t C_{\Xi\Xi}(t,s)\,e^{\eta(t-s)}\,ds. \tag{4.86}$$

In Eq. (4.85), we observe that, the ***drift-diffusion differential operator is degenerate***, since the second derivative with respect to $x$ is missing. However, as we will see subsequently, this does not pose any problem in solving Eq. (4.85).

**Remark 4.7: Connection between $G(t)$ and $D^{\text{eff}}(t)$.** By using Eq. (4.86), as well as the definition relation (4.21) for the effective noise intensity $D^{\text{eff}}(t)$ that appears in the exact response pdf evolution Eq. (4.20), it is easily derived that

$$\dot{D}^{\text{eff}}(t)=\eta D^{\text{eff}}(t)+\kappa^2\sigma_{\Xi}^2(t)+\kappa G(t),\quad D^{\text{eff}}(t_0)=\kappa C_{X_0\Xi}(t_0), \tag{4.87a,b}$$

and by solving the above IVP

$$D^{\text{eff}}(t)=\kappa\, C_{X_0\Xi}(t_0)\,e^{\eta(t-t_0)}+\int_{t_0}^{t}\Big(\kappa^2\sigma_{\Xi}^2(\tau)+\kappa G(\tau)\Big)\,e^{\eta(t-\tau)}\,d\tau. \tag{4.88}$$

Relation (4.88) will be proven quite useful in validating the moments obtained from the solution of Eq. (4.85).

**Solution of Eq. (4.85) using Fourier transform.** As performed, in paragraph 4.2.3, for the $N-$dimensional Eq. (4.57), response-excitation evolution Eq. (4.85) is solved by utilizing the two-dimensional Fourier transform; $\varphi_{X(t)\Xi(t)}(\mathbf{y})=\int_{\mathbb{R}^2}e^{i(y_1x+y_2u)}\,f_{X(t)\Xi(t)}(x,u)\,dxdu$, resulting in the equation

$$\begin{aligned}\frac{\partial\varphi_{X(t)\Xi(t)}(\mathbf{y})}{\partial t}&=\eta\, y_1\frac{\partial\varphi_{X(t)\Xi(t)}(\mathbf{y})}{\partial y_1}+\kappa\, y_1\frac{\partial\varphi_{X(t)\Xi(t)}(\mathbf{y})}{\partial y_2}+\\&+\left(i\dot{m}_{\Xi}(t)\,y_2-G(t)\,y_1y_2-\frac{1}{2}\dot{\sigma}_{\Xi}^2(t)\,y_2^2\right)\varphi_{X(t)\Xi(t)}(\mathbf{y}),\end{aligned} \tag{4.89}$$

supplemented by the Gaussian initial condition

$$\varphi_{X(t_0)\Xi(t_0)}(\boldsymbol{y}) = \exp\left( i m_{X_0} y_1 + i m_{\Xi}(t_0) y_2 - \frac{1}{2}\sigma^2_{X_0} y_1^2 - \frac{1}{2}\sigma^2_{\Xi}(t_0)^2 y_2^2 - C_{X_0\Xi}(t_0) y_1 y_2 \right). \tag{4.90}$$

First, the solution to the homogeneous variant of Eq. (4.89) determines the change of variables, $\upsilon_1 = y_1 e^{\eta t}$, $\upsilon_2 = \kappa y_1 - \eta y_2$, under which Eq. (4.89) is transformed into a linear ODE with respect time. By solving the said equation and then returning to the original variables $\boldsymbol{y}$, we determine the solution of IVP (4.89)-(4.90) to the Gaussian characteristic function

$$\begin{aligned}\varphi_{X(t)\Xi(t)}(\boldsymbol{y}) &= \exp\left[ i y_1 \left( m_{X_0} e^{\eta(t-t_0)} + \kappa \int_{t_0}^{t} m_{\Xi}(\tau)\, e^{\eta(t-\tau)}\, d\tau \right) + i y_2 m_{\Xi}(t) - \frac{1}{2} y_2^2\, \sigma^2_{\Xi}(t) \right] \times \\ &\times \exp\left\{ -\frac{1}{2} y_1^2 \left[ \sigma^2_{X_0} e^{2\eta(t-t_0)} + \frac{2\kappa}{\eta} \left( C_{X_0\Xi}(t_0)(e^{2\eta(t-t_0)} - e^{\eta(t-t_0)}) + \right.\right.\right. \\ &\left.\left.\left. + \int_{t_0}^{t} \left( \kappa\sigma^2_{\Xi}(\tau) + G(\tau) \right) (e^{2\eta(t-\tau)} - e^{\eta(t-\tau)})\, d\tau \right) \right] \right\} \times \\ &\times \exp\left[ -y_1 y_2 \left( C_{X_0\Xi}(t_0) e^{\eta(t-t_0)} + \int_{t_0}^{t} \left( \kappa\sigma^2_{\Xi}(\tau) + G(\tau) \right) e^{\eta(t-\tau)}\, d\tau \right) \right].\end{aligned} \tag{4.91}$$

Detailed derivation of solution (4.91) can be found in (Mavromatis, 2020). From Eq. (4.91), we can identify the first and second moments of $X(t;\theta)$, $\Xi(t;\theta)$. First, since the moments of the excitation are data of the problem, Eq. (4.91) returns the trivial relations $m_{\Xi}(t) = m_{\Xi}(t)$, $\sigma^2_{\Xi}(t) = \sigma^2_{\Xi}(t)$. Moving now to the moments of response, as well as the response-excitation covariance, we retrieve, from Eq. (4.91), the relations

$$m_X(t) = m_{X_0} e^{\eta(t-t_0)} + \kappa \int_{t_0}^{t} m_{\Xi}(\tau)\, e^{\eta(t-\tau)}\, d\tau, \tag{4.92a}$$

$$\begin{aligned}\sigma^2_X(t) &= \sigma^2_{X_0} e^{2\eta(t-t_0)} + \\ &+ \frac{2\kappa}{\eta} \left( C_{X_0\Xi}(t_0)(e^{2\eta(t-t_0)} - e^{\eta(t-t_0)}) + \int_{t_0}^{t} \left( \kappa\sigma^2_{\Xi}(\tau) + G(\tau) \right) (e^{2\eta(t-\tau)} - e^{\eta(t-\tau)})\, d\tau \right),\end{aligned} \tag{4.92b}$$

$$C_{X\Xi}(t,t) = C_{X_0\Xi}(t_0) e^{\eta(t-t_0)} + \int_{t_0}^{t} \left( \kappa\sigma^2_{\Xi}(\tau) + G(\tau) \right) e^{\eta(t-\tau)}\, d\tau. \tag{4.92c}$$

Eq. (4.92a) is validated as the solution for the mean value of the response, see Eq. (D.3) of Appendix D, while, by employing Eq. (4.88) and after some algebraic manipulations, Eqs. (4.92a,c) are expressed equivalently as

$$\sigma_X^2(t) = \sigma_{X_0}^2 e^{2\eta(t-t_0)} + 2\int_{t_0}^{t} D^{\text{eff}}(\tau)\, e^{2\eta(t-\tau)}\, d\tau, \tag{4.92b′}$$

$$C_{X\Xi}(t,t) = \frac{1}{\kappa} D^{\text{eff}}(t). \tag{4.92c′}$$

Eqs. (4.92b′,c′) are correct, since they coincide with the validated relations (4.26), (4.27) for $C_{X\Xi}(t,t)$, $\sigma_X^2(t)$. Thus, by solving Eq. (4.85), the expected Gaussian solution was obtained. Additionally, we have to note that response-excitation pdf evolution Eq. (4.85) is solved alternatively in Appendix F, without resorting to Fourier transform, by employing a generalization of the ansatz of (Shtelen & Stogny, 1989). While this second solution method does not guarantee the uniqueness of solution, as does the method employing the Fourier transform, it does reveal however a connection between the two-dimensional, degenerate drift-diffusion Eq. (4.85) and a system of two one-dimensional heat equations.

### 4.4 The final form of response-initial value stochastic Liouville equation

Similarly to the formulation of response-excitation SLE in its final form, Eq. (4.84), we formulate the final, response-initial value SLE corresponding to RIVP (4.2a,b). In this case, we begin from the response-initial value SLE (2.45), derived in paragraph 2.5.3

$$\begin{aligned}\frac{\partial f_{X_0 X(t)}(x_0, x_1)}{\partial t} &+ \frac{\partial}{\partial x_1}\Big(h(x_1)\, f_{X_0 X(t)}(x_0, x_1)\Big) = \\ &= -\frac{\partial}{\partial x_1}\Big(q(x_1)\, \mathbb{E}^{\theta}\Big[\delta(x_0 - X_0(\theta))\, \delta(x_1 - X(t;\theta))\, \Xi(t;\theta)\Big]\Big).\end{aligned} \tag{4.93}$$

The average in the right-hand side of Eq. (4.93) is elaborated further by using the extended NF theorem, Eq. (3.32), which for FFℓ $\mathcal{F}[\cdots] = \delta(x_0 - X_0(\theta))\delta(x_1 - X[X_0(\theta); \Xi(\bullet|_{t_0}^{t}; \theta)])$, and after the application of product and chain rules reads

$$\begin{aligned}&\mathbb{E}^{\theta}\Big[\delta(x_0 - X_0(\theta))\, \delta(x_1 - X(t;\theta))\, \Xi(t;\theta)\Big] = \\ &= m_{\Xi}(t)\, f_{X_0 X(t)}(x_0, x_1) + C_{X_0\Xi}(t)\, \mathbb{E}^{\theta}\left[\frac{\partial \delta(x_0 - X_0(\theta))}{\partial X_0(\theta)}\, \delta(x_1 - X(t;\theta))\right] + \\ &+ C_{X_0\Xi}(t)\, \mathbb{E}^{\theta}\left[\delta(x_0 - X_0(\theta))\, \frac{\partial \delta(x_1 - X(t;\theta))}{\partial X(t;\theta)}\, V_{X_0}[X(\bullet|_{t_0}^{t};\theta); \Xi(\bullet|_{t_0}^{t};\theta)]\right] + \\ &+ \int_{t_0}^{t} C_{\Xi\Xi}(t,s)\, \mathbb{E}^{\theta}\left[\delta(x_0 - X_0(\theta))\, \frac{\partial \delta(x_1 - X(t;\theta))}{\partial X(t;\theta)}\, V_{\Xi(s)}[X(\bullet|_{s}^{t};\theta); \Xi(\bullet|_{s}^{t};\theta)]\right] ds.\end{aligned} \tag{4.94}$$

Substitution of Eq. (4.94) into Eq. (4.93) and use of delta projection formulae results into the ***final response-initial value SLE***

$$\frac{\partial f_{X_0 X(t)}(x_0, x_1)}{\partial t} + \frac{\partial}{\partial x_1}\left[\left(h(x_1) + q(x_1)\, m_\Xi(t)\right) f_{X_0 X(t)}(x_0, x_1)\right] =$$

$$= C_{X_0 \Xi}(t)\, \frac{\partial}{\partial x_1}\left( q(x_1)\, \frac{\partial}{\partial x_0}\, f_{X_0 X(t)}(x_0, x_1) \right) +$$

$$+ C_{X_0 \Xi}(t)\, \frac{\partial}{\partial x_1}\left( q(x_1)\, \frac{\partial}{\partial x_1}\, \mathbb{E}^\theta\left[ \delta(x_0, x_1; X_0(\theta), X(t;\theta))\, V_{X_0}[X(\bullet|_{t_0}^{t};\theta);\Xi(\bullet|_{t_0}^{t};\theta)]\right]\right) +$$

$$+ \frac{\partial}{\partial x_1}\left( q(x_1)\, \frac{\partial}{\partial x_1} \int_{t_0}^{t} C_{\Xi\Xi}(t,s)\, \mathbb{E}^\theta\left[\delta(x_0, x_1; X_0(\theta), X(t;\theta))\, V_{\Xi(s)}[X(\bullet|_s^t;\theta);\Xi(\bullet|_s^t;\theta)]\right] ds \right), \tag{4.95}$$

where $\delta(x_0, x_1; X_0(\theta), X(t;\theta))$ is a shorthand for the product of the random delta functions $\delta(x_0 - X_0(\theta))\, \delta(x_1 - X(t;\theta))$. As expected, the final SLE (4.95) is a non-closed pdf evolution equation, with its nonlocal terms being identified as the variational derivatives of the response.

**The case of linear, additively excited RDE.** For the linear-additive case, $h(x) = \eta x$, $q(x) = \kappa$, the variational derivatives of the response are specified into $V_{X_0}(t;\theta) = e^{\eta(t-t_0)}$, $V_{\Xi(s)}(t;\theta) = \kappa\, e^{\eta(t-s)}$, see Eqs. (4.19a,b). Thus, their substitution into the SLE (4.95), results into the exact response-initial value pdf evolution equation

$$\frac{\partial f_{X_0 X(t)}(x_0, x_1)}{\partial t} + \frac{\partial}{\partial x_1}\left[\left(h(x_1) + \kappa\, m_\Xi(t)\right) f_{X_0 X(t)}(x_0, x_1)\right] = \tag{4.96}$$

$$= \kappa\, C_{X_0 \Xi}(t)\, \frac{\partial^2 f_{X_0 X(t)}(x_0, x_1)}{\partial x_0\, \partial x_1} + D^{\text{eff}}(t)\, \frac{\partial^2 f_{X_0 X(t)}(x_0, x_1)}{\partial x_1^2},$$

where coefficient $D^{\text{eff}}(t)$ is defined by Eq. (4.21), as in the case of the exact response pdf evolution Eq. (4.20). Similarly to the exact response-excitation pdf Eq. (4.85), Eq. (4.96) is also a degenerate drift-diffusion equation, since the second order $x_0$ – derivative is missing from its right-hand side.

Recalling the discussion in paragraph 2.5.3, evolution Eq. (4.96) is supplemented with the initial condition (2.47)

$$f_{X_0 X(t_0)}(x_0, x_1) = f_{X_0}(x_0)\, \delta(x_0 - x_1), \tag{4.97a}$$

where the initial value pdf $f_{X_0}(x_0)$ is considered Gaussian

$$f_{X_0}(x) = \frac{1}{\sqrt{2\pi\sigma_{X_0}^2}} \exp\left[ -\frac{1}{2}\, \frac{(x - m_{X_0})^2}{\sigma_{X_0}^2} \right]. \tag{4.97b}$$

**Solution of Eq. (4.96) using Fourier transform.** As usual, we employ the two-dimensional Fourier transform $\varphi_{X_0 X(t)}(y_0, y_1) = \int_{\mathbb{R}^2} e^{i(y_0 x_0 + y_1 x_1)} f_{X_0 X(t)}(x_0, x_1)\, dx_1 dx_0$, which results in the transformed Eq. (4.96)

$$\frac{\partial \varphi_{X_0 X(t)}(y_0, y_1)}{\partial t} = \eta y_1 \frac{\partial \varphi_{X_0 X(t)}(y_0, y_1)}{\partial y_1} + \\ + \left( i\kappa m_{\Xi}(t)\, y_1 - \kappa\, C_{X_0 \Xi}(t)\, y_0 y_1 - D^{\text{eff}}(t)\, y_1^2 \right) \varphi_{X_0 X(t)}(y_0, y_1). \tag{4.98}$$

Under the above Fourier transform, initial condition (4.97a) is expressed as

$$\varphi_{X_0 X(t_0)}(y_0, y_1) = \int_{\mathbb{R}} e^{i y_0 x_0} f_{X_0}(x_0) \left( \int_{\mathbb{R}} e^{i y_1 x_1} \delta(x_0 - x_1)\, dx_1 \right) dx_0 .$$

By employing the identity of delta function, and the definition of the characteristic function of initial value; $\varphi_{X_0}(y) = \int_{\mathbb{R}} e^{iyx} f_{X_0}(x)\, dx$, we obtain

$$\varphi_{X_0 X(t_0)}(y_0, y_1) = \int_{\mathbb{R}} e^{i(y_0 + y_1) x_0} f_{X_0}(x_0)\, dx_0 = \varphi_{X_0}(y_0 + y_1). \tag{4.99}$$

Since the Fourier transform of the Gaussian pdf (4.97b) is

$$\varphi_{X_0}(y) = \exp\left( i m_{X_0}\, y - \frac{1}{2} \sigma_{X_0}^2\, y^2 \right),$$

the transformed initial condition (4.99) is expressed in its final form as

$$\varphi_{X_0 X(t_0)}(y_0, y_1) = \exp\left( i m_{X_0} (y_0 + y_1) - \frac{1}{2} \sigma_{X_0}^2 (y_0 + y_1)^2 \right) = \\ = \exp\left( i m_{X_0} y_0 + i m_{X_0} y_1 - \frac{1}{2} \sigma_{X_0}^2 y_0^2 - \frac{1}{2} \sigma_{X_0}^2 y_1^2 - \sigma_{X_0}^2 y_0 y_1 \right). \tag{4.100}$$

By first examining the homogenous variant of Eq. (4.98), the characteristic curve $\upsilon(y_1, t) = y_1 e^{\eta t}$ is determined, and thus under the variable change $(y_0, y_1) \to (y_0, \upsilon)$, Eq. (4.98) is transformed into a linear ODE with respect to $t$. By solving the aforementioned ODE and returning to the original variables, we obtain the solution

$$\varphi_{X_0 X(t)}(y_0, y_1) = \exp\left[ i m_{X_0} y_0 + i y_1 \left( m_{X_0} e^{\eta(t - t_0)} + \kappa \int_{t_0}^{t} m_{\Xi}(\tau)\, e^{\eta(t-\tau)}\, d\tau \right) \right] \times \\ \times \exp\left[ -\frac{1}{2} \sigma_{X_0}^2 y_0^2 - \frac{1}{2} y_1^2 \left( \sigma_{X_0}^2 e^{2\eta(t - t_0)} + 2 \int_{t_0}^{t} D^{\text{eff}}(\tau)\, e^{2\eta(t-\tau)}\, d\tau \right) \right] \times \\ \times \exp\left[ - y_0 y_1 \left( \sigma_{X_0}^2 e^{\eta(t - t_0)} + \kappa \int_{t_0}^{t} C_{X_0 \Xi}(\tau)\, e^{\eta(t-\tau)}\, d\tau \right) \right]. \tag{4.101}$$

By employing now the formulae (D.3), (D.23), derived in Appendix D, for response mean value $m_X(t)$ and cross-covariance $C_{X_0X}(t)$ respectively, as well as the verified formula (4.27) relating the variance of the response $\sigma_X^2(t)$ to $D^{\text{eff}}(t)$, solution (4.101) is expressed equivalently as

$$\varphi_{X_0X(t)}(y_0, y_1) = \exp\left( i m_{X_0} y_0 + i m_X(t) y_1 - \frac{1}{2}\sigma_{X_0}^2 y_0^2 - \frac{1}{2}\sigma_X^2(t) y_1^2 - C_{X_0X}(t) y_0 y_1 \right). \tag{4.102}$$

Eq. (4.102) gives the expected Gaussian form for $\varphi_{X_0X(t)}(y_0, y_1)$, verifying thus the validity of response-initial value pdf evolution Eq. (4.96).

## 4.5 The final form of two-time response stochastic Liouville equation

Let us now move on to SLE (2.39), for the two-time response pdf $f_{X(t)X(s)}(x_1, x_2)$, corresponding to RIVP (4.2a,b):

$$\begin{aligned} \frac{\partial f_{X(t)X(s)}(x_1, x_2)}{\partial t} + \frac{\partial}{\partial x_1}\left( h(x_1) f_{X(t)X(s)}(x_1, x_2) \right) &= \\ = -\frac{\partial}{\partial x_1}\left( q(x_1)\, \mathbb{E}^{\theta}\left[ \delta(x_1 - X(t;\theta))\, \delta(x_2 - X(s;\theta))\, \Xi(t;\theta) \right] \right)&. \end{aligned} \tag{4.103}$$

As we have discussed in paragraph 2.5.2, where the above SLE (4.103) was derived, the two time instances $t, s$ are different, but no ordering between them was assumed. This is of particular importance in the evaluation of the SLE average via the NF theorem, Eq. (3.32), in which, the FFℓ is specified to

$$\delta(x_1 - X[X_0(\theta); \Xi(\bullet|_{t_0}^{t}; \theta)])\, \delta(x_2 - X[X_0(\theta); \Xi(\bullet|_{t_0}^{s}; \theta)]).$$

We have to note that, for any ordering of $t, s$ ($t < s$ or $s < t$) the above FFℓ is a functional on excitation $\Xi(\bullet;\theta)$ over the interval $[t_0, t_1]$, with $t_1 = \max\{t, s\}$. Thus, NF theorem, Eq. (3.32), reads

$$\begin{aligned} \mathbb{E}^{\theta}\left[ \delta(x_1 - X(t;\theta))\, \delta(x_2 - X(s;\theta))\, \Xi(t;\theta) \right] &= m_{\Xi}(t)\, f_{X(t)X(s)}(x_1, x_2) + \\ &+ C_{X_0\Xi}(t)\, \mathbb{E}^{\theta}\left[ \frac{\partial\left( \delta(x_1 - X(t;\theta))\, \delta(x_2 - X(s;\theta)) \right)}{\partial X_0(\theta)} \right] + \\ &+ \int_{t_0}^{t_1} C_{\Xi\Xi}(t,\tau)\, \mathbb{E}^{\theta}\left[ \frac{\delta\left( \delta(x_1 - X(t;\theta))\, \delta(x_2 - X(s;\theta)) \right)}{\delta \Xi(\tau;\theta)} \right] d\tau, \end{aligned}$$

and after the application of product and chain rules, the variational derivatives of the response appear

$$\begin{aligned}\mathbb{E}^{\theta}\Big[\delta(x_1-X(t;\theta))\,\delta(x_2-X(s;\theta))\,\Xi(t;\theta)\Big] &= m_{\Xi}(t)\; f_{X(t)X(s)}(x_1,x_2)+\\ &+C_{X_0\Xi}(t)\,\mathbb{E}^{\theta}\left[\frac{\partial\delta(x_1-X(t;\theta))}{\partial X(t;\theta)}\,\delta(x_2-X(s;\theta))\,V_{X_0}(t;\theta)\right]+\\ &+C_{X_0\Xi}(t)\,\mathbb{E}^{\theta}\left[\delta(x_1-X(t;\theta))\,\frac{\partial\delta(x_2-X(s;\theta))}{\partial X(s;\theta)}\,V_{X_0}(s;\theta)\right]+\\ &+\int_{t_0}^{t_1}C_{\Xi\Xi}(t,\tau)\,\mathbb{E}^{\theta}\left[\frac{\partial\delta(x_1-X(t;\theta))}{\partial X(t;\theta)}\,\delta(x_2-X(s;\theta))\,V_{\Xi(\tau)}(t;\theta)\right]d\tau+\\ &+\int_{t_0}^{t_1}C_{\Xi\Xi}(t,\tau)\,\mathbb{E}^{\theta}\left[\delta(x_1-X(t;\theta))\,\frac{\partial\delta(x_2-X(s;\theta))}{\partial X(s;\theta)}\,V_{\Xi(\tau)}(s;\theta)\right]d\tau.\end{aligned}\tag{4.104}$$

Furthermore, due to ***causality***: $V_{\Xi(\tau)}(t;\theta)=\delta X(t;\theta)/\delta\Xi(\tau;\theta)=0$ for $\tau>t$, since a variation in excitation, $\delta\Xi(\tau;\theta)$, at a certain time instance, cannot result in a variation in response, $\delta X(t;\theta)$, at a previous time. Thus, the upper limits of the integrals in Eq. (4.104) are adjusted accordingly to

$$\begin{aligned}\mathbb{E}^{\theta}\Big[\delta(x_1-X(t;\theta))\,\delta(x_2-X(s;\theta))\,\Xi(t;\theta)\Big] &= m_{\Xi}(t)\; f_{X(t)X(s)}(x_1,x_2)+\\ &+C_{X_0\Xi}(t)\,\mathbb{E}^{\theta}\left[\frac{\partial\delta(x_1-X(t;\theta))}{\partial X(t;\theta)}\,\delta(x_2-X(s;\theta))\,V_{X_0}(t;\theta)\right]+\\ &+C_{X_0\Xi}(t)\,\mathbb{E}^{\theta}\left[\delta(x_1-X(t;\theta))\,\frac{\partial\delta(x_2-X(s;\theta))}{\partial X(s;\theta)}\,V_{X_0}(s;\theta)\right]+\\ &+\int_{t_0}^{t}C_{\Xi\Xi}(t,\tau)\,\mathbb{E}^{\theta}\left[\frac{\partial\delta(x_1-X(t;\theta))}{\partial X(t;\theta)}\,\delta(x_2-X(s;\theta))\,V_{\Xi(\tau)}(t;\theta)\right]d\tau+\\ &+\int_{t_0}^{s}C_{\Xi\Xi}(t,\tau)\,\mathbb{E}^{\theta}\left[\delta(x_1-X(t;\theta))\,\frac{\partial\delta(x_2-X(s;\theta))}{\partial X(s;\theta)}\,V_{\Xi(\tau)}(s;\theta)\right]d\tau.\end{aligned}\tag{4.105}$$

Substitution of Eq. (4.105) into (4.103) results in the ***final form of two-time response SLE***:

$$\begin{aligned}&\frac{\partial f_{X(t)X(s)}(x_1,x_2)}{\partial t}+\frac{\partial}{\partial x_1}\Big[\big(h(x_1)+q(x_1)\,m_{\Xi}(t)\big)\,f_{X(t)X(s)}(x_1,x_2)\Big]=\\ &=C_{X_0\Xi}(t)\frac{\partial}{\partial x_1}\left(q(x_1)\frac{\partial}{\partial x_1}\mathbb{E}^{\theta}\Big[\delta(x_1,x_2;X(t;\theta),X(s;\theta))V_{X_0}[X(\bullet|_{t_0}^{t};\theta);\Xi(\bullet|_{t_0}^{t};\theta)]\Big]\right)+\\ &+C_{X_0\Xi}(t)\frac{\partial}{\partial x_1}\left(q(x_1)\frac{\partial}{\partial x_2}\mathbb{E}^{\theta}\Big[\delta(x_1,x_2;X(t;\theta),X(s;\theta))V_{X_0}[X(\bullet|_{t_0}^{s};\theta);\Xi(\bullet|_{t_0}^{s};\theta)]\Big]\right)+\\ &+\frac{\partial}{\partial x_1}\left(q(x_1)\frac{\partial}{\partial x_1}\int_{t_0}^{t}C_{\Xi\Xi}(t,\tau)\mathbb{E}^{\theta}\Big[\delta(x_1,x_2;X(t;\theta),X(s;\theta))V_{\Xi(\tau)}[X(\bullet|_{\tau}^{t};\theta);\Xi(\bullet|_{\tau}^{t};\theta)]\Big]d\tau\right)+\\ &+\frac{\partial}{\partial x_1}\left(q(x_1)\frac{\partial}{\partial x_2}\int_{t_0}^{s}C_{\Xi\Xi}(t,\tau)\mathbb{E}^{\theta}\Big[\delta(x_1,x_2;X(t;\theta),X(s;\theta))V_{\Xi(\tau)}[X(\bullet|_{\tau}^{s};\theta);\Xi(\bullet|_{\tau}^{s};\theta)]\Big]d\tau\right),\end{aligned}\tag{4.106}$$

where $\delta(x_1, x_2\,; X(t\,;\theta), X(s\,;\theta))$ is a shorthand for the product of the random delta functions $\delta(x_1 - X(t\,;\theta))\,\delta(x_2 - X(s\,;\theta))$.

**The case of linear, additively excited RDE.** For the linear-additive case, i.e. $h(x) = \eta x$, $q(x) = \kappa$, the variational derivatives appearing in Eq. (4.106) are calculated to $V_{X_0}(t\,;\theta) = e^{\eta(t-t_0)}$, $V_{\Xi(s)}(t\,;\theta) = \kappa e^{\eta(t-s)}$. Thus, SLE (4.106), is specified into the following exact, two-time response pdf evolution equation:

$$\frac{\partial f_{X(t)X(s)}(x_1, x_2)}{\partial t} + \frac{\partial}{\partial x_1}\left[\left(h(x_1) + \kappa m_\Xi(t)\right) f_{X(t)X(s)}(x_1, x_2)\right] = \\ = D^{\text{eff}}(t)\,\frac{\partial^2 f_{X(t)X(s)}(x_1, x_2)}{\partial x_1^2} + D(t, s)\,\frac{\partial^2 f_{X(t)X(s)}(x_1, x_2)}{\partial x_1 \partial x_2}, \tag{4.107}$$

where $D^{\text{eff}}(t)$ is defined by Eq. (4.21) and

$$D(t, s) = \kappa\, C_{X_0\Xi}(t)\, e^{\eta(s-t_0)} + \kappa^2 \int_{t_0}^{s} C_{\Xi\Xi}(t, \tau)\, e^{\eta(s-\tau)}\, d\tau\,. \tag{4.108}$$

Note that, comparison of Eq. (4.21) with Eq. (4.108) yields that $D^{\text{eff}}(t) = D(t, t)$. As discussed in paragraph 2.5.2, the initial condition of Eq. (4.107) is $f_{X(t_0)X(s)}(x_1, x_2) = f_{X_0 X(s)}(x_1, x_2)$, with its right-hand side being determined by the solution of response-initial value pdf evolution Eq. (4.96).

**Solution of Eq. (4.107) using Fourier transform.** As for the previous exact pdf equations, we employ the Fourier transform $\varphi_{X(t)X(s)}(\boldsymbol{y}) = \int_{\mathbb{R}^2} e^{i(y_1x_1 + y_2x_2)}\, f_{X(t)X(s)}(x_1, x_2)\, dx_1 dx_2$, which results in the transformed equation

$$\frac{\partial \varphi_{X(t)X(s)}(\boldsymbol{y})}{\partial t} = \eta\, y_1 \frac{\partial \varphi_{X(t)X(s)}(\boldsymbol{y})}{\partial y_1} + \\ + \left(i\kappa m_\Xi(t) y_1 - D(t, s) y_1 y_2 - \kappa D^{\text{eff}}(t) y_1^2\right) \varphi_{X(t)X(s)}(\boldsymbol{y}). \tag{4.109}$$

Initial condition for Eq. (4.109) is determined via the solution (4.102) of the response-initial value pdf evolution equation for linear RDE

$$\varphi_{X_0 X(s)}(\boldsymbol{y}) = \exp\left(i m_{X_0} y_1 + i m_X(s) y_2 - \frac{1}{2}\sigma_{X_0}^2 y_1^2 - \frac{1}{2}\sigma_X^2(s) y_2^2 - C_{X_0X}(s) y_1 y_2\right). \tag{4.110}$$

We observe that Eq. (4.109) is of the same form as the respective Eq. (4.98) for the response-initial value pdf. Thus, it is solved by employing the change of variables $(y_1, y_2) \to (\upsilon, y_2)$, $\upsilon = y_1 e^{\eta t}$. After calculations, presented in detailed in (Mavromatis, 2020), its solution in the original variables reads

$$\varphi_{X(t)\,X(s)}(y) = \exp\left[ i\,y_1 \left( m_{X_0}\, e^{\eta(t-t_0)} + \kappa \int_{t_0}^{t} m_{\Xi}(\tau)\, e^{\eta(t-\tau)}\, d\tau \right) + i\,y_2\, m_X(s) \right] \times$$
$$\times \exp\left[ -\frac{1}{2} y_1^2 \left( \sigma_{X_0}^2\, e^{2\eta(t-t_0)} + 2 \int_{t_0}^{t} D^{\text{eff}}(t)\, e^{2\eta(t-\tau)}\, d\tau \right) - \frac{1}{2} y_2^2\, \sigma_X^2(s) \right] \times \qquad (4.111)$$
$$\times \exp\left[ -\,y_1\, y_2 \left( C_{X_0 X}(s)\, e^{\eta(t-t_0)} + \int_{t_0}^{t} D(\tau, s)\, e^{\eta(t-\tau)}\, d\tau \right) \right].$$

By utilizing Eq. (D.3) for $m_X(t)$, the verified Eq. (4.27) for $\sigma_X^2(t)$, as well as Eq. (D.24) for $C_{XX}(t, s)$ in conjunction with the definition relation (4.108) for $D(t, s)$, solution (4.111) is written equivalently as

$$\varphi_{X(t)X(s)}(y) = \exp\left( i m_X(t) y_1 + i m_X(s) y_2 - \frac{1}{2} \sigma_X^2(t) y_1^2 - \frac{1}{2} \sigma_X^2(s) y_2^2 - C_{XX}(t, s) y_1 y_2 \right). \qquad (4.112)$$

Eq. (4.112) is the expected Gaussian form of the two-time response pdf, yielding both the one-time; $m_X(t)$, $m_X(s)$, $\sigma_X^2(t)$, $\sigma_X^2(s)$, and the two-time, $C_{XX}(t, s)$, moments correctly. This positive result for the linear case, as well as the other similar results obtained in previous Sections, constitute a fact supporting the conjecture that the SLEs, under an appropriate closure scheme, could also be well-posed and yield satisfactory results for the case of non-linear random dynamical systems.

### 4.6 Towards non-Gaussian excitations: SLEs for systems under polynomially Gaussian noise

In Chapter 2, we also derived the SLE (2.21)

$$\frac{\partial f_{X(t)}(x)}{\partial t} + \frac{\partial}{\partial x}\left( h(x)\, f_{X(t)}(x) \right) = -\frac{\partial}{\partial x} \sum_{n=1}^{N} q_n(x)\, \mathbb{E}^{\theta}\left[ \delta(x - X(t;\theta))\, \Xi^n(t;\theta) \right], \qquad (4.113)$$

corresponding to the RDE excited by a polynomial of the noise $\Xi(t;\theta)$

$$\dot{X}(t;\theta) = h(X(t;\theta)) + \sum_{n=1}^{N} q_n(X(t;\theta))\, \Xi^n(t;\theta), \qquad X(t_0;\theta) = X_0(\theta). \qquad (4.114\text{a,b})$$

In this Section, noise $\Xi(t;\theta)$ will be considered Gaussian. Considering RDE (4.114a) has the following two advantages. On the one hand, $\sum_{n=1}^{N} q_n(X(t;\theta))\, \Xi^n(t;\theta)$ is a more general excitation, called polynomially Gaussian henceforth. On the other hand, since $\Xi(t;\theta)$ per se *is* Gaussian, the averages in the right-hand side of SLE (4.113) can still be elaborated by using the novel generalizations of NF theorem derived in Section 3.5.

### 4.6.1 Response SLE for a nonlinear RDE under additive quadratic Gaussian noise

As a first example of formulating response SLEs for RDEs under polynomially Gaussian excitation, we consider a nonlinear RDE excited by additive quadratic Gaussian noise $\Xi^2(t;\theta)$:

$$\dot{X}(t;\theta) = h\big(X(t;\theta)\big) + \kappa\, \Xi^2(t;\theta). \tag{4.115}$$

For the sake of simplicity, we consider $\Xi(t;\theta)$ to be a zero-mean valued, and the initial value of response to be deterministic; $X(t_0;\theta) = x_0$. Thus, for RDE (4.115), SLE (4.113) is specified into

$$\frac{\partial f_{X(t)}(x)}{\partial t} + \frac{\partial}{\partial x}\big(h(x)\, f_{X(t)}(x)\big) = -\kappa \frac{\partial}{\partial x}\mathbb{E}^{\theta}[\delta(x - X(t;\theta))\, \Xi^2(t;\theta)]. \tag{4.116}$$

In this case, where initial value is deterministic, random delta function is only a functional on excitation; $\delta(x - X(t;\theta)) = \delta(x - X[\Xi(\bullet|_{t_0}^{t};\theta)])$, and thus the average in the right-hand side of SLE (4.116) can be further elaborated using the generalization of NF theorem, Eq. (3.75), for $\mathcal{F}[\Xi(\bullet|_{t_0}^{t};\theta)] = \delta(x - X[\Xi(\bullet|_{t_0}^{t};\theta)])$, $n = 2$, and $m_{\Xi}(t) = C_{X_0\Xi}(t) = 0$:

$$\begin{aligned}\mathbb{E}^{\theta}\Big[\delta(x - X[\Xi(\bullet|_{t_0}^{t};\theta)])\, \Xi^2(t;\theta)\Big] &= C_{\Xi\Xi}(t,t)\, f_{X(t)}(x) + \\ &+ \int_{t_0}^{t}\int_{t_0}^{t} C_{\Xi\Xi}(t,s_1)\, C_{\Xi\Xi}(t,s_2)\, \mathbb{E}^{\theta}\left[\frac{\delta^2 \delta(x - X[\Xi(\bullet|_{t_0}^{t};\theta)])}{\delta\Xi(s_1;\theta)\,\delta\Xi(s_2;\theta)}\right] ds_1\, ds_2.\end{aligned} \tag{4.117}$$

The second order Volterra derivative of the random delta function, appearing in the right-hand side of Eq. (4.117) is further evaluated using the multivariate version of Faà di Bruno formula (Hardy, 2006):

$$\begin{aligned}\frac{\delta^2 \delta(x - X[\Xi(\bullet|_{t_0}^{t};\theta)])}{\delta\Xi(s_1;\theta)\,\delta\Xi(s_2;\theta)} &= \frac{\partial \delta(x - X(t;\theta))}{\partial X(t;\theta)}\, \frac{\delta^2 X[\Xi(\bullet|_{t_0}^{t};\theta)]}{\delta\Xi(s_1;\theta)\,\delta\Xi(s_2;\theta)} + \\ &+ \frac{\partial^2 \delta(x - X(t;\theta))}{\partial X^2(t;\theta)}\, \frac{\delta X[\Xi(\bullet|_{t_0}^{t};\theta)]}{\delta\Xi(s_1;\theta)}\, \frac{\delta X[\Xi(\bullet|_{t_0}^{t};\theta)]}{\delta\Xi(s_2;\theta)}.\end{aligned} \tag{4.118}$$

As usual, for determining the first order Volterra derivative of the response with respect to excitation; $V^{(1)}_{\Xi(s)}(t;\theta) = \delta X[\Xi(\bullet|_{t_0}^{t};\theta)]/\delta\Xi(s;\theta)$, an RDE can be formulated by applying the Volterra derivative operator $\delta X \bullet/\delta\Xi(s;\theta)$ at both sides of RDE (4.115):

$$\frac{\partial V^{(1)}_{\Xi(s)}(t;\theta)}{\partial t} = h'\big(X(t;\theta)\big)V^{(1)}_{\Xi(s)}(t;\theta) + 2\kappa\, \Xi(t;\theta)\, \delta(t-s). \tag{4.119}$$

Following paragraph 4.2.1, the solution to Eq. (4.119) is determined to

$$V^{(1)}_{\Xi(s)}(t;\theta) = 2\kappa\, \Xi(s;\theta)\, \mathcal{E}[X(\bullet|_{s}^{t};\theta)], \qquad s \le t, \qquad \text{with} \tag{4.120}$$

$$\mathcal{E}[X(\bullet|_{s}^{t};\theta)] = \exp\left(\int_{s}^{t} h'\big(X(u;\theta)\big)\, du\right). \tag{4.121}$$

Note that, Volterra derivative $V^{(1)}_{\Xi(s)}(t;\theta)$ is defined for $s \le t$ since, by causality, any perturbation $\delta\Xi(s;\theta)$, acting at time $s$, cannot result in a perturbation $\delta X(t;\theta)$ for $t < s$; thus $V^{(1)}_{\Xi(s)}(t;\theta) = 0$ for $t < s$, giving rise to the convention that $\mathcal{E}[X(\bullet|^t_s;\theta)] = 0$ for $t < s$. Having the first order Volterra derivative in closed form, Eq. (4.120), we differentiate it to obtain second order derivative $V^{(2)}_{\Xi(s_1)\Xi(s_2)}(t;\theta) = \delta^2 X[\Xi(\bullet|^t_{t_0};\theta)]/\big(\delta\Xi(s_1;\theta)\,\delta\Xi(s_2;\theta)\big)$:

$$\begin{aligned} V^{(2)}_{\Xi(s_1)\Xi(s_2)}(t;\theta) = {} & 2\kappa\,\delta(s_2 - s_1)\,\mathcal{E}[X(\bullet|^t_{s_1};\theta)] + \\ & + 4\kappa^2\,\Xi(s_1;\theta)\,\Xi(s_2;\theta)\,\mathcal{E}[X(\bullet|^t_{s_1};\theta)]\,\mathcal{I}[X(\bullet|^t_{s_2};\theta)], \quad s_1 \le s_2 \le t, \end{aligned} \tag{4.122}$$

with

$$\mathcal{I}[X(\bullet|^t_{s_2};\theta)] = \int_{s_2}^{t} h''\big(X(u;\theta)\big)\,\mathcal{E}[X(\bullet|^u_{s_2};\theta)]\,du. \tag{4.123}$$

As in the case of first order Volterra derivative, $V^{(2)}_{\Xi(s_1)\Xi(s_2)}(t;\theta)$ is non-zero for $s_1 \le s_2 \le t$ due to causality. Also, by Eq. (4.123), $\mathcal{I}[X(\bullet|^t_{s_2};\theta)] = 0$ for $t < s_2$. By substituting Eqs. (4.120), (4.122) into Eq. (4.118), we obtain the expression

$$\begin{aligned} \frac{\delta^2\,\delta(x - X[\Xi(\bullet|^t_{t_0};\theta)])}{\delta\Xi(s_1;\theta)\,\delta\Xi(s_2;\theta)} = {} & 2\kappa\,\delta(s_2 - s_1)\,\frac{\partial\delta(x - X(t;\theta))}{\partial X(t;\theta)}\,\mathcal{E}[X(\bullet|^t_{s_1};\theta)] + \\ & + 4\kappa^2\,\Xi(s_1;\theta)\,\Xi(s_2;\theta)\,\mathcal{G}_1[X(\bullet|^t_{s_*};\theta)], \end{aligned} \tag{4.124}$$

with

$$\begin{aligned} \mathcal{G}_1[X(\bullet|^t_{s_*};\theta)] = {} & \frac{\partial\delta(x - X(t;\theta))}{\partial X(t;\theta)}\,\mathcal{E}[X(\bullet|^t_{s_1};\theta)]\,\mathcal{I}[X(\bullet|^t_{s_2};\theta)] + \\ & + \frac{\partial^2\delta(x - X(t;\theta))}{\partial X^2(t;\theta)}\,\mathcal{E}[X(\bullet|^t_{s_1};\theta)]\,\mathcal{E}[X(\bullet|^t_{s_2};\theta)], \end{aligned} \tag{4.125}$$

where $s_* = \min\{s_1, s_2\}$. By now employing Eq. (4.124), NF theorem (4.117), as well as the delta projection formulae, SLE (4.116) is specified into

$$\begin{aligned} & \frac{\partial f_{X(t)}(x)}{\partial t} + \frac{\partial}{\partial x}\Big[\big(h(x) + \kappa\,C_{\Xi\Xi}(t,t)\big)\,f_{X(t)}(x)\Big] = \\ & = 2\kappa^2\,\frac{\partial^2}{\partial x^2}\int_{t_0}^{t} C^2_{\Xi\Xi}(t,s)\,\mathbb{E}^{\theta}\Big[\delta(x - X(t;\theta))\,\mathcal{E}[X(\bullet|^t_s;\theta)]\Big]\,ds - \\ & - 4\kappa^3\,\frac{\partial}{\partial x}\int_{t_0}^{t}\!\!\int_{t_0}^{t} C_{\Xi\Xi}(t,s_1)\,C_{\Xi\Xi}(s_2,t)\,\mathbb{E}^{\theta}\Big[\Xi(s_1;\theta)\,\Xi(s_2;\theta)\,\mathcal{G}_1[X(\bullet|^t_{s_*};\theta)]\Big]\,ds_1 ds_2. \end{aligned} \tag{4.126}$$

**Remark 4.8:** As SLE (4.12) for linear Gaussian excitation, SLE (4.126) is a non-closed evolution equation, whose non-local terms depend on the history of response and excitation. However, in the last term of the right-hand side of SLE (4.126), the average is in a form adequate for applying generalized NF theorem anew, since $\mathcal{G}_1[X(\bullet|_{s_*}^{t};\theta)] = \mathcal{G}_1[X[\Xi(\bullet|_{t_0}^{t};\theta)]]$. This feature, permitting of ***multiple applications of NF theorem*** was not present in Eq. (4.12).

Thus, the aforementioned average in SLE (4.126) is evaluated further using the generalized NF theorem, Eq. (3.76), for $n=2$ and $\mathcal{F}[\Xi(\bullet|_{t_0}^{t};\theta)] = \mathcal{G}_1[X[\Xi(\bullet|_{t_0}^{t};\theta)]]$:

$$\mathbb{E}^{\theta}\Big[\Xi(s_1;\theta)\,\Xi(s_2;\theta)\,\mathcal{G}_1[X(\bullet|_{s_*}^{t};\theta)]\Big] = C_{\Xi\Xi}(s_1,s_2)\,\mathbb{E}^{\theta}\Big[\mathcal{G}_1[X(\bullet|_{s_*}^{t};\theta)]\Big] + \\ + \int_{t_0}^{t}\int_{t_0}^{t} C_{\Xi\Xi}(s_1,\tau_1)\,C_{\Xi\Xi}(s_2,\tau_2)\,\mathbb{E}^{\theta}\left[\frac{\delta^2\mathcal{G}_1[X[\Xi(\bullet|_{t_0}^{t};\theta)]]}{\delta\Xi(\tau_1;\theta)\,\delta\Xi(\tau_2;\theta)}\right] d\tau_1 d\tau_2. \tag{4.127}$$

Similarly to the procedure for determining the second-order Volterra derivative of random delta function, Eqs. (4.118)-(4.125), derivative $\delta^2\mathcal{G}_1[X[\Xi(\bullet|_{t_0}^{t};\theta)]]/\big(\delta\Xi(\tau_1;\theta)\,\delta\Xi(\tau_2;\theta)\big)$ is calculated to

$$\frac{\delta^2\mathcal{G}_1[X[\Xi(\bullet|_{t_0}^{t};\theta)]]}{\delta\Xi(\tau_1;\theta)\,\delta\Xi(\tau_2;\theta)} = \int_{t_0}^{t}\frac{\delta\mathcal{G}_1[X(\bullet|_{s_*}^{t};\theta)]}{\delta X(u;\theta)}\,V^{(2)}_{\Xi(\tau_1)\Xi(\tau_2)}(u;\theta)\,du + \\ + \int_{t_0}^{t}\int_{t_0}^{t}\frac{\delta^2\mathcal{G}_1[X(\bullet|_{s_*}^{t};\theta)]}{\delta X(u_2;\theta)\,\delta X(u_1;\theta)}\,V^{(1)}_{\Xi(\tau_2)}(u_2;\theta)\,V^{(1)}_{\Xi(\tau_1)}(u_1;\theta)\,du_1du_2, \tag{4.128}$$

and by substituting Eqs. (4.120), (4.122):

$$\frac{\delta^2\mathcal{G}_1[X[\Xi(\bullet|_{t_0}^{t};\theta)]]}{\delta\Xi(\tau_1;\theta)\,\delta\Xi(\tau_2;\theta)} = 2\kappa\,\delta(\tau_2-\tau_1)\int_{t_0}^{t}\frac{\delta\mathcal{G}_1[X(\bullet|_{s_*}^{t};\theta)]}{\delta X(u;\theta)}\,\mathcal{I}[X(\bullet|_{\tau_1}^{u};\theta)]\,du + \\ + 4\kappa^2\,\Xi(\tau_1;\theta)\,\Xi(\tau_2;\theta)\,\mathcal{G}_2[X(\bullet|_{\tau_*}^{t};\theta)], \tag{4.129}$$

with

$$\mathcal{G}_2[X(\bullet|_{\tau_*}^{t};\theta)] = \int_{t_0}^{t}\frac{\delta\mathcal{G}_1[X(\bullet|_{s_*}^{t};\theta)]}{\delta X(u;\theta)}\,\mathcal{I}[X(\bullet|_{\tau_1}^{u};\theta)]\,\mathcal{I}[X(\bullet|_{\tau_2}^{u};\theta)]\,du + \\ + \int_{t_0}^{t}\int_{t_0}^{t}\frac{\delta^2\mathcal{G}_1[X(\bullet|_{s_*}^{t};\theta)]}{\delta X(u_2;\theta)\,\delta X(u_1;\theta)}\,\mathcal{I}[X(\bullet|_{\tau_2}^{u_2};\theta)]\,\mathcal{I}[X(\bullet|_{\tau_1}^{u_1};\theta)]\,du_1du_2, \tag{4.130}$$

where $\tau_* = \min\{\tau_1,\tau_2\}$. By substituting NF theorem, Eq. (4.127), and Eqs. (4.129) into SLE (4.126), we have

$$\frac{\partial f_{X(t)}(x)}{\partial t} + \frac{\partial}{\partial x}\left[\left(h(x) + \kappa\, C_{\Xi\Xi}(t,t)\right) f_{X(t)}(x)\right] =$$

$$+\,2\kappa^2 \frac{\partial^2}{\partial x^2}\int_{t_0}^{t} C_{\Xi\Xi}^2(t,s)\, \mathbb{E}^{\theta}\left[\delta(x - X(t;\theta))\, \mathcal{E}[X(\bullet|_s^t;\theta)]\right] ds\, -$$

$$-\,4\kappa^3 \frac{\partial}{\partial x}\int_{t_0}^{t}\int_{t_0}^{t} C_{\Xi\Xi}(t,s_1)\, C_{\Xi\Xi}(s_1,s_2)\, C_{\Xi\Xi}(s_2,t)\, \mathbb{E}^{\theta}\left[\mathcal{G}_1[X(\bullet|_{s_*}^t;\theta)]\right] ds_1 ds_2\, -$$

$$-\,8\kappa^4 \frac{\partial}{\partial x}\int_{t_0}^{t}\int_{t_0}^{t}\int_{t_0}^{t} C_{\Xi\Xi}(t,s_1)\, C_{\Xi\Xi}(s_1,\tau)\, C_{\Xi\Xi}(\tau,s_2)\, C_{\Xi\Xi}(s_2,t)\, \times$$

$$\times\, \mathbb{E}^{\theta}\left[\int_{t_0}^{t} \frac{\delta \mathcal{G}_1[X(\bullet|_{s_*}^t;\theta)]}{\delta X(u;\theta)}\, \mathcal{E}[X(\bullet|_\tau^u;\theta)]\, du\right] d\tau\, ds_1 ds_2\, -$$

$$-\,16\kappa^5 \frac{\partial}{\partial x}\int_{t_0}^{t}\int_{t_0}^{t}\int_{t_0}^{t}\int_{t_0}^{t} C_{\Xi\Xi}(t,s_1)\, C_{\Xi\Xi}(s_1,\tau_1)\, C_{\Xi\Xi}(\tau_2,s_2)\, C_{\Xi\Xi}(s_2,t)\, \times$$

$$\times\, \mathbb{E}^{\theta}\left[\Xi(\tau_1;\theta)\,\Xi(\tau_2;\theta)\, \mathcal{G}_2[X(\bullet|_{\tau_*}^t;\theta)]\right] d\tau_1 d\tau_2 ds_1 ds_2\,. \qquad (4.131)$$

By comparing SLE in the form of Eq. (4.131) to SLE (4.126), we observe that the additional application of the NF theorem gave rise to two new terms (the second and third term in the right-hand side of Eq. (4.131)) whose integrands depend on the history of the response, while $\mathbb{E}^{\theta}\left[\Xi(\tau_1;\theta)\,\Xi(\tau_2;\theta)\, \mathcal{G}_2[X(\bullet|_{\tau_*}^t;\theta)]\right]$, appearing in the last term in the right-hand side of Eq. (4.131) can be again further evaluated by NF theorem, Eq. (4.127). Since the second Volterra derivative of $\mathcal{G}_2$ with respect to excitation is of the form given by Eq. (4.129) for $\mathcal{G}_1 = \mathcal{G}_2$, another application of the NF theorem will result in nonlocal terms depending on $X(\bullet;\theta)$, and a term needing the application same NF theorem, Eq. (4.127) again. This procedure can be performed repeatedly; however, we observe that the terms in the right-hand side of Eq. (4.131) are ordered in increasing powers of $\kappa$. By neglecting the term with $\kappa^5$ in Eq. (4.131), substituting the definition relation, Eq. (4.125), for $\mathcal{G}_1[X(\bullet|_{s_*}^t;\theta)]$, and after calculations presented at length in Appendix G, we have the ***approximate SLE***:

$$\frac{\partial f_{X(t)}(x)}{\partial t} + \frac{\partial}{\partial x}\left[\left(h(x) + \kappa\, C_{\Xi\Xi}(t,t)\right) f_{X(t)}(x)\right] =$$

$$= \frac{\partial^2}{\partial x^2}\,\mathbb{E}^{\theta}\left[\delta(x - X(t;\theta))\left(2\kappa^2 \mathcal{N}_{2,1} + 4\kappa^3 \mathcal{N}_{2,2} + 8\kappa^4 \mathcal{N}_{2,3}\right)\right] -$$

$$-\,\frac{\partial^3}{\partial x^3}\,\mathbb{E}^{\theta}\left[\delta(x - X(t;\theta))\left(4\kappa^3 \mathcal{N}_{3,1} + 8\kappa^4 \mathcal{N}_{3,2}\right)\right] + \qquad (4.132)$$

$$+\,\frac{\partial^4}{\partial x^4}\,\mathbb{E}^{\theta}\left[\delta(x - X(t;\theta))\left(8\kappa^4 \mathcal{N}_{4,1}\right)\right].$$

In SLE (4.132), all $\mathcal{N}-$terms are functionals depending on the history of the response, and their exact forms is presented in Appendix G, Eqs. (G.14)-(G.19), due to their length.

**Remark 4.9:** While the case of SLE (4.132) is not studied further in the present thesis, all $\mathcal{N}-$ terms can be approximated using the novel current-time approximation closure proposed in par. 5.1.2 of the following Chapter, resulting thus in a pdf evolution equation in closed form, whose numerical solution and evaluations will be the topic of a future work. The aim of the present section was indeed to show the possibility of moving to polynomially Gaussian excitations in a systematic way, as well as the adequacy of the tools we have at our disposal, namely the delta projection and the generalized NF theorem for multiple times, Eq. (3.70)

### 4.6.2 The case of linear RDE under additive quadratic Gaussian noise

As for all SLEs derived in the previous Sections, we shall also consider the case of linear RDE, $h(x) = \eta x$, under additive quadratic Gaussian excitation. In this case, Eqs. (4.121), (4.123) are specified into

$$\mathcal{E}[X(\bullet|_s^t;\theta)] = e^{\eta(t-s)}, \qquad \mathcal{I}[X(\bullet|_s^t;\theta)] = 0, \tag{4.133a,b}$$

that is, $\mathcal{E}$ is just a function of time, independent from $X(\bullet;\theta)$, and $\mathcal{I}$ vanishes since it depends on the second derivative of $h(x)$. Under Eqs. (4.133a,b), SLE (4.126) after the first application of NF theorem, reads

$$\frac{\partial f_{X(t)}(x)}{\partial t} + \frac{\partial}{\partial x}\left(\eta x f_{X(t)}(x)\right) = -\kappa A_1(t)\frac{\partial f_{X(t)}(x)}{\partial x} + 2\kappa^2 A_2(t)\frac{\partial^2 f_{X(t)}(x)}{\partial x^2} -$$
$$- 4\kappa^3 \frac{\partial^3}{\partial x^3}\int_{t_0}^{t}\int_{t_0}^{t} C_{\Xi\Xi}(t,s_1)\, C_{\Xi\Xi}(s_2,t)\, e^{\eta(t-s_1)} e^{\eta(t-s_2)} \times \tag{4.134}$$
$$\times \mathbb{E}^{\theta}\left[\Xi(s_1;\theta)\,\Xi(s_2;\theta)\,\delta(x-X(t;\theta))\right] ds_1 ds_2,$$

with

$$A_1(t) = C_{\Xi\Xi}(t,t), \qquad A_2(t) = \int_{t_0}^{t} C_{\Xi\Xi}^2(t,s)\, e^{\eta(t-s)}\, ds. \tag{4.135a,b}$$

In SLE (4.134), the average $\mathbb{E}^{\theta}\left[\Xi(s_1;\theta)\,\Xi(s_2;\theta)\,\delta(x-X(t;\theta))\right]$ can be further evaluated by using NF theorem, Eq. (4.127), for $\mathcal{G}_1[X[\Xi(\bullet|_{t_0}^t;\theta)]] = \delta(x-X[\Xi(\bullet|_{t_0}^t;\theta)])$:

$$\mathbb{E}^{\theta}\left[\Xi(s_1;\theta)\,\Xi(s_2;\theta)\,\delta(x-X(t;\theta))\right] = C_{\Xi\Xi}(s_1,s_2)\, f_{X(t)}(x) +$$
$$+ \int_{t_0}^{t}\int_{t_0}^{t} C_{\Xi\Xi}(s_1,\tau_1)\, C_{\Xi\Xi}(s_2,\tau_2)\, \mathbb{E}^{\theta}\left[\frac{\delta^2 \delta(x-X[\Xi(\bullet|_{t_0}^t;\theta)])}{\delta\Xi(\tau_1;\theta)\,\delta\Xi(\tau_2;\theta)}\right] d\tau_1 d\tau_2. \tag{4.136}$$

Since the second order Volterra derivative of random delta function, given by Eq. (4.118) is specified, for the linear case, into

$$\frac{\delta^2 \delta(x-X(t;\theta))}{\delta\Xi(\tau_1;\theta)\,\delta\Xi(\tau_2;\theta)} = 2\kappa e^{\eta(t-\tau_1)}\,\delta(\tau_2-\tau_1)\frac{\partial\delta(x-X(t;\theta))}{\partial X(t;\theta)} +$$
$$+ 4\kappa^2 e^{\eta(t-\tau_1)} e^{\eta(t-\tau_2)}\,\Xi(\tau_1;\theta)\,\Xi(\tau_2;\theta)\frac{\partial^2\delta(x-X(t;\theta))}{\partial X^2(t;\theta)}, \tag{4.137}$$

NF theorem, Eq. (4.136), reads

$$\mathbb{E}^{\theta}\Big[\Xi(s_1;\theta)\,\Xi(s_2;\theta)\,\delta(x-X(t;\theta))\Big] = C_{\Xi\Xi}(s_1,s_2)\,f_{X(t)}(x) - $$
$$- 2\kappa\int_{t_0}^{t} C_{\Xi\Xi}(s_1,\tau)\,C_{\Xi\Xi}(\tau,s_2)\,e^{\eta(t-\tau)}\,d\tau\,\frac{\partial f_{X(t)}(x)}{\partial x} + $$
$$+ 4\kappa^2\,\frac{\partial^2}{\partial x^2}\int_{t_0}^{t}\int_{t_0}^{t} C_{\Xi\Xi}(s_1,\tau_1)\,C_{\Xi\Xi}(s_2,\tau_2)\,e^{\eta(t-\tau_1)}e^{\eta(t-\tau_2)}\times \tag{4.138}$$
$$\times\,\mathbb{E}^{\theta}\Big[\Xi(\tau_1;\theta)\,\Xi(\tau_2;\theta)\,\delta(x-X(t;\theta))\Big]\,d\tau_1 d\tau_2.$$

Substitution of NF theorem, Eq. (4.138), into SLE (4.134) results into

$$\frac{\partial f_{X(t)}(x)}{\partial t} + \frac{\partial}{\partial x}\big(\eta x\, f_{X(t)}(x)\big) = -\kappa A_1(t)\,\frac{\partial f_{X(t)}(x)}{\partial x} + 2\kappa^2 A_2(t)\,\frac{\partial^2 f_{X(t)}(x)}{\partial x^2} - $$
$$- 4\kappa^3 A_3(t)\,\frac{\partial^3 f_{X(t)}(x)}{\partial x^3} + 8\kappa^4 A_4(t)\,\frac{\partial^4 f_{X(t)}(x)}{\partial x^4} - $$
$$- 16\kappa^5\,\frac{\partial^5}{\partial x^5}\int_{t_0}^{t}\int_{t_0}^{t}\int_{t_0}^{t}\int_{t_0}^{t} C_{\Xi\Xi}(t,s_1)\,C_{\Xi\Xi}(s_1,\tau_1)\,C_{\Xi\Xi}(\tau_2,s_2)\,C_{\Xi\Xi}(s_2,t)\times \tag{4.139}$$
$$\times\, e^{\eta(t-s_1)}e^{\eta(t-s_2)}e^{\eta(t-\tau_1)}e^{\eta(t-\tau_2)}\,\mathbb{E}^{\theta}\Big[\Xi(\tau_1;\theta)\,\Xi(\tau_2;\theta)\,\delta(x-X(t;\theta))\Big]\,d\tau_1 d\tau_2\,ds_1 ds_2,$$

with

$$A_3(t) = \int_{t_0}^{t}\int_{t_0}^{t} C_{\Xi\Xi}(t,s_1)\,C_{\Xi\Xi}(s_1,s_2)\,C_{\Xi\Xi}(s_2,t)\,e^{\eta(t-s_1)}e^{\eta(t-s_2)}\,ds_1 ds_2, \tag{4.140}$$

$$A_4(t) = \int_{t_0}^{t}\int_{t_0}^{t}\int_{t_0}^{t} C_{\Xi\Xi}(t,s_1)\,C_{\Xi\Xi}(s_1,\tau)\,C_{\Xi\Xi}(\tau,s_2)\,C_{\Xi\Xi}(s_2,t)\,e^{\eta(t-s_1)}e^{\eta(t-s_2)}e^{\eta(t-\tau)}\,d\tau\,ds_1 ds_2. \tag{4.141}$$

As for the nonlinear RDE, in paragraph 4.6.1, term $\mathbb{E}^{\theta}\Big[\Xi(\tau_1;\theta)\,\Xi(\tau_2;\theta)\,\delta(x-X(t;\theta))\Big]$ can be elaborated further by applying again the NF theorem, Eq. (4.136). However, the forms in the linear case are simpler, since the functional inside the average, at which the NF theorem is applied in every step, is always the random delta function of the response. Thus, the structure of $A_n(t)$'s, resulting from the repeated applications of NF theorem, Eq. (4.136), is easily identified. This results in the SLE taking the following, ***Kramers-Moyal-like expansion form*** (van Kampen, 2007) Eq. (VIII.2.6):

$$\frac{\partial f_{X(t)}(x)}{\partial t} + \frac{\partial}{\partial x}\big(\eta x\, f_{X(t)}(x)\big) = \sum_{n=1}^{\infty}(-1)^n\,2^{n-1}\kappa^n A_n(t)\,\frac{\partial^n f_{X(t)}(x)}{\partial x^n}, \tag{4.142}$$

with

$$A_n(t) = \int_{t_0}^{t}\overset{(n-1)}{\cdots}\int_{t_0}^{t}\prod_{\substack{k=1\\ s_0\equiv t}}^{n-1} C_{\Xi\Xi}(s_{k-1},s_k)\,e^{\eta(t-s_k)}\,C_{\Xi\Xi}(s_{n-1},t)\,ds_1\cdots ds_{n-1}. \tag{4.143}$$

The finite expansion in Eq. (4.142) is formal, since we do not know if it converges. However, under the assumption that the said series is convergent, we can perform the usual Fourier transform on Eq. (4.142), obtaining thus the following evolution equation for the characteristic function of the response, $\varphi_{X(t)}(y) = \int_{\mathbb{R}} e^{iyx} f_{X(t)}(x)\,dx$:

$$\frac{\partial \varphi_{X(t)}(y)}{\partial t} - \eta y \frac{\partial \varphi_{X(t)}(y)}{\partial y} = \left( \sum_{n=1}^{\infty} i^n y^n 2^{n-1} \kappa^n A_n(t) \right) \varphi_{X(t)}(y). \tag{4.144}$$

Eq. (4.144) is supplemented with the initial condition; $\varphi_{X(t_0)}(y) = e^{iyx_0}$, which is the Fourier transform of the deterministic initial condition for response pdf, $f_{X(t)}(x) = \delta(x - x_0)$, since $X(t_0;\theta) = x_0$. Following the method of paragraph 4.1.2, Eq. (4.144) can be easily solved, resulting into

$$\varphi_{X(t)}(y) =$$

$$= \exp\left[ iy \left( x_0 e^{\eta(t-t_0)} + \kappa \int_{t_0}^{t} C_{\Xi\Xi}(s,s)\, e^{\eta(t-s)}\, ds \right) + \sum_{n=2}^{\infty} i^n y^n 2^{n-1} \kappa^n \int_{t_0}^{t} A_n(s)\, e^{n\eta(t-s)}\, ds \right]. \tag{4.145}$$

Under definition relation (4.143) for $A_n(t)$'s, the integrals inside the series in the right-hand side of Eq. (4.145) are expressed as

$$\int_{t_0}^{t} A_n(s)\, e^{n\eta(t-s)}\, ds = \int_{t_0}^{t}\int_{t_0}^{s} \overset{(n-1)}{\cdots} \int_{t_0}^{s} \prod_{\substack{k=1 \\ s_0 = s_n \equiv s}}^{n} C_{\Xi\Xi}(s_{k-1}, s_k)\, e^{\eta(t-s_k)}\, ds_1 \cdots ds_{n-1}\, ds. \tag{4.146}$$

**Lemma 4.1.** It holds true that

$$\int_{t_0}^{t} A_n(s)\, e^{n\eta(t-s)}\, ds = \frac{1}{n} \int_{t_0}^{t} \overset{(n)}{\cdots} \int_{t_0}^{t} \prod_{\substack{k=1 \\ s_0 = s_n}}^{n} C_{\Xi\Xi}(s_{k-1}, s_k)\, e^{\eta(t-s_k)}\, ds_1 \cdots ds_n. \tag{4.147}$$

**Proof.** The multiple integral of $\prod_{\substack{k=1 \\ s_0 = s_n}}^{n} C_{\Xi\Xi}(s_{k-1}, s_k)\, e^{\eta(t-s_k)}$ over the *n*-dimensional hyper-cube $[t_0, t]^n$ can be partitioned into a sum of *n* multiple integrals, each one defined over one *n*-dimensional hyperpyramid:

$$\int_{t_0}^{t} \overset{(n)}{\cdots} \int_{t_0}^{t} \prod_{\substack{k=1 \\ s_0 = s_n}}^{n} C_{\Xi\Xi}(s_{k-1}, s_k)\, e^{\eta(t-s_k)}\, ds_1 \cdots ds_n =$$

$$= \sum_{m=1}^{n} \int_{t_0}^{t} \left( \int_{t_0}^{s_m} \overset{(n-1)}{\cdots} \int_{t_0}^{s_m} \prod_{\substack{k=1 \\ s_0 = s_n}}^{n} C_{\Xi\Xi}(s_{k-1}, s_k)\, e^{\eta(t-s_k)} \prod_{\substack{\ell=1 \\ \ell \neq m}}^{n} ds_\ell \right) ds_m.$$

In each multiple integral of the *m*-sum, we rename $s_m = s$. The rest integration arguments, $s_1, \dots, s_{m-1}, s_{m+1}, \dots, s_n$ -whose range is the same, $[t_0, s]$- are renamed as follows; for $\ell = 1, \dots, m-1$: $s_\ell = s_{n-m+\ell}$, and for $\ell = m+1, \dots, n$: $s_\ell = s_{\ell-m}$. Under this renaming, it is easy to see that all terms of *m*-sum are equal, resulting into

$$\int_{t_0}^{t} \overset{(n)}{\cdots} \int_{t_0}^{t} \prod_{\substack{k=1 \\ s_0 = s_n}}^{n} C_{\Xi\Xi}(s_{k-1}, s_k)\, e^{\eta(t-s_k)}\, ds_1 \cdots ds_n =$$

$$= n \int_{t_0}^{t} \int_{t_0}^{s} \overset{(n-1)}{\cdots} \int_{t_0}^{s} \prod_{\substack{k=1 \\ s_0 = s_n \equiv s}}^{n} C_{\Xi\Xi}(s_{k-1}, s_k)\, e^{\eta(t-s_k)}\, ds_1 \cdots ds_{n-1}\, ds. \quad \blacksquare$$

Substitution of Eq. (4.147) in Eq. (4.145) yields the characteristic function $\varphi_{X(t)}(y)$ in the form:

$$\varphi_{X(t)}(y) = \exp\left[ \sum_{n=1}^{\infty} \frac{i^n}{n!} \varkappa_X^{(n)}(t)\, y^n \right], \qquad \text{with} \tag{4.148}$$

$$\varkappa_X^{(1)}(t) = x_0 e^{\eta(t-t_0)} + \kappa \int_{t_0}^{t} C_{\Xi\Xi}(s, s)\, e^{\eta(t-s)}\, ds, \tag{4.149}$$

and for $n = 2, 3, \ldots$:

$$\varkappa_X^{(n)}(t) = 2^{n-1}(n-1)!\, \kappa^n \int_{t_0}^{t} \overset{(n)}{\cdots} \int_{t_0}^{t} \prod_{\substack{k=1 \\ s_0 \equiv s_n}}^{n} C_{\Xi\Xi}(s_{k-1}, s_k)\, e^{\eta(t-s_k)}\, ds_1 \cdots ds_n\,. \tag{4.150}$$

Since characteristic function $\varphi_{X(t)}(y)$ is expressed as the exponential of this particular infinite series, each quantity $\varkappa_X^{(n)}(t)$ is identified as the $n^{\text{th}}$-order cumulant of the response (Abramowitz & Stegun, 1964), formula 26.1.12. Thus, by formally solving Eq. (4.142), we were able to determine the whole hierarchy of response cumulants.

**Validation of the formulae for response cumulants.** As in the previous Sections, and since the RDE is linear, cumulant formulae (4.149), (4.150) can be verified by their direct derivation from the closed-form solution of RDE (4.115) for $h(x) = \eta x$, which reads

$$X(t;\theta) = x_0 e^{\eta(t-t_0)} + \kappa I(t;\theta), \quad \text{with} \quad I(t;\theta) = \int_{t_0}^{t} \Xi^2(s;\theta)\, e^{\eta(t-s)}\, ds\,. \tag{4.151a,b}$$

First, by applying the mean value operator $\mathbb{E}^{\theta}[\bullet]$ on both sides of Eq. (4.151a), and remembering that excitation was assumed as zero-mean, Eq. (4.149) for the first cumulant, that is, the mean value is obtained. On the other hand, for the validation of Eq. (4.150) for higher-order cumulants, we observe that, in Eq. (4.151a), the terms $x_0 e^{\eta(t-t_0)}$, $\kappa$ are deterministic, and thus, by employing the semi-invariance property of $\varkappa_X^{(n)}(t)$, $n = 2, 3, \ldots$ (Cramér, 1946) sec. 15.10, we obtain

$$\varkappa_X^{(n)}(t) = \kappa^n \varkappa_I^{(n)}(t), \qquad n = 2, 3, \ldots, \tag{4.152}$$

where $\varkappa_I^{(n)}(t)$ are the cumulants of integral $I(t;\theta)$. Thus, the problem is now transformed into the calculation of the cumulants of $I(t;\theta)$. To this end, we shall invoke Volterra's concept of passing from the discrete to continuous, see Appendix A, and approximate integral $I(t;\theta)$ with its discrete analogue

$$S(t;\theta) = \sum_{n=1}^{N} \Xi_n^2(\theta)\, u_n(t), \tag{4.153}$$

where the $N-$dimensional vectors $\boldsymbol{\Xi}(\theta) = \{\Xi_n(\theta)\}_{n=1}^{N}$, $\boldsymbol{u}(t) = \{u_n(t)\}_{n=1}^{N}$, are defined by the sampling continuous functions $\Xi(s;\theta)$, $u(t,s) = \kappa e^{\eta(t-s)}$ over a partition of the interval $[t_0, t]$:

$\Xi_n(\theta) = \Xi(s_n;\theta)$, $u_n(t) = u(t, s_n) = \kappa e^{\eta(t-s_n)}$, $t_0 \le s_1 < s_2 < \cdots < s_n < \cdots < s_N \le t$.

By Eq. (4.153), sum $S(t;\theta)$ is identified as a quadratic form of the $N-$dimensional Gaussian random vector $\boldsymbol{\Xi}(\theta)$ with zero mean value and autocovariance $\boldsymbol{C}_{\boldsymbol{\Xi\Xi}}$, with the matrix of the quadratic form being the diagonal $N \times N$ $\boldsymbol{A}(t) = \{u_n(t)\,\delta_{nm}\}_{n,m=1}^{N}$. The formulae for cumulants of a quadratic form of Gaussian vector with diagonal matrix $\boldsymbol{A}(t)$ have been determined in the literature, see e.g. (Rodhe, Urquhart, & Searle, 1968) Eq. (10), (Searle, Urquhart, & Evans, 1970) corollary 2.1, (Scarowsky, 1973) corollary 2.1, into

$$\varkappa_S^{(n)}(t) = 2^{n-1}(n-1)!\, \mathrm{tr}\left[\left(\boldsymbol{C}_{\boldsymbol{\Xi\Xi}}\boldsymbol{A}(t)\right)^n\right], \tag{4.154}$$

where $\mathrm{tr}[\bullet]$ denotes the trace of the matrix. By denoting now the elements of the $n^{\mathrm{th}}$ power of $\left(\boldsymbol{C}_{\boldsymbol{\Xi\Xi}}\boldsymbol{A}(t)\right)_{nm} = \sum_{\nu=1}^{N} C_{\Xi_n\Xi_\nu}\, u_\nu(t)\,\delta_{\nu m} = C_{\Xi_n\Xi_m}\, u_m(t)$ as

$$\left[\left(\boldsymbol{C}_{\boldsymbol{\Xi\Xi}}\boldsymbol{A}(t)\right)^n\right]_{k\ell} = \sum_{m_1=1}^{N} \overset{(n-1)}{\cdots} \sum_{m_{n-1}=1}^{N} C_{\Xi_k\Xi_{m_1}} u_{m_1}(t)\, C_{\Xi_{m_1}\Xi_{m_2}} u_{m_2}(t) \cdots C_{\Xi_{m_{n-1}}\Xi_\ell} u_\ell(t),$$

which can be expressed in the more contracted form

$$\left[\left(\boldsymbol{C}_{\boldsymbol{\Xi\Xi}}\boldsymbol{A}(t)\right)^n\right]_{k\ell} = \sum_{m_1=1}^{N} \overset{(n-1)}{\cdots} \sum_{m_{n-1}=1}^{N} \prod_{\substack{p=1\\ m_0\equiv k,\, m_n\equiv\ell}}^{n} C_{\Xi_{m_{p-1}}\Xi_p} u_p(t),$$

the trace appearing in Eq. (4.154) is written as

$$\mathrm{tr}\left[\left(\boldsymbol{C}_{\boldsymbol{\Xi\Xi}}\boldsymbol{A}(t)\right)^n\right] = \sum_{k=1}^{N}\left[\left(\boldsymbol{C}_{\boldsymbol{\Xi\Xi}}\boldsymbol{A}(t)\right)^n\right]_{kk} = \sum_{k=1}^{N}\sum_{m_1=1}^{N} \overset{(n-1)}{\cdots} \sum_{m_{n-1}=1}^{N} \prod_{\substack{p=1\\ m_0\equiv m_n\equiv k}}^{n} C_{\Xi_{m_{p-1}}\Xi_p} u_p(t) =$$

$$= \sum_{m_1=1}^{N} \overset{(n)}{\cdots} \sum_{m_n=1}^{N} \prod_{\substack{p=1\\ m_0\equiv m_n}}^{n} C_{\Xi_{m_{p-1}}\Xi_p} u_p(t). \tag{4.155}$$

Substitution of Eq. (4.155) into Eq. (4.154) results in the expression for cumulants of $S(t;\theta)$

$$\varkappa_S^{(n)}(t) = 2^{n-1}(n-1)! \sum_{m_1=1}^{N} \overset{(n)}{\cdots} \sum_{m_n=1}^{N} \prod_{\substack{p=1\\ m_0\equiv m_n}}^{n} C_{\Xi_{m_{p-1}}\Xi_{m_p}}\, u_{m_p}. \tag{4.156}$$

By considering now the Volterra limit, $N \to \infty$, of Eq. (4.16), a formula for the cumulants of integral $I(t;\theta)$ is obtained

$$\varkappa_I^{(n)}(t) = 2^{n-1}(n-1)! \int_{t_0}^{t} \overset{(n)}{\cdots} \int_{t_0}^{t} \prod_{\substack{p=1 \\ s_0 \equiv s_n}}^{n} C_{\Xi\Xi}(s_{p-1}, s_p)\, e^{\eta(t-s_p)}\, ds_1 \cdots ds_n \,. \tag{4.157}$$

Substitution of Eq. (4.157) into Eq. (4.152) yields the relation (4.150) for cumulants obtained by solving the response pdf evolution Eq. (4.142). Note also that cumulant formulae (4.149), (4.150) could be also verified by following the more complicated, yet more mathematically rigorous way of utilizing the results of (Krée & Soize, 1986; Christian Soize, 1980) with regard to the infinite-dimensional probability measure of the response to a linear system under quadratic Gaussian excitation.

**Remark 4.10:** The correct calculation of all response cumulants via the formal solution of Eq. (4.142) is a first positive result in the direction of non-Gaussian excitations, which has to be followed by numerical solution and investigation of the said equation. Furthermore, we have to note that, while linear RDEs under quadratic Gaussian noise have been considered before, see e.g. (Łuczka, 1986a, 1986b, 1987; Łuczka, Hänggi, & Gadomski, 1995; Sagués et al., 1984; San Miguel & Sancho, 1981), the above result has not been obtained.

# Chapter 5: Novel approximation closure and formulation of pdf evolution equations for non-linear systems

**Summary.** In the first Section of the present Chapter, we present the approximation closures for the averaged terms of SLEs corresponding to a nonlinear, additively excited, scalar RDE. Apart from discussing the existing ones, we also introduce a novel closure that retains a tractable, and thus easily computable, amount of nonlocality of the averaged SLE terms, since it employs the history of appropriate moments of the response. Substitution of the novel closure in the one-time response SLE results into response pdf evolution equations of the general form

$$\frac{\partial f_{X(t)}(x)}{\partial t} + \frac{\partial}{\partial x}\Big[\big(h(x) + \kappa m_{\Xi}(t)\big)\, f_{X(t)}(x)\Big] = \frac{\partial^2}{\partial x^2}\Big[\mathcal{B}[\, f_X \,;x,t]\; f_{X(t)}(x)\Big].$$

The diffusion coefficient $\mathcal{B}[\, f_X \,;x,t]$ of this novel evolution equation, defined in Eqs. (5.24), (5.25) below, apart from being a function of state and time variables, is also a functional on the unknown response pdf. This feature, which differentiates the above equation from the classical FPK, is indeed desired, since it reflects the non-Markovian character of the response. In Section 5.2, the same novel closure is used for the derivation of evolution equations for the pdf of higher order, namely the response-initial value, and the two-time response pdf. In Section 5.3, the approximation schemes are generalized in order to be applied for the closure of multidimensional SLEs. Last, in Section 5.4, we discuss the case of RDEs excited by multiplicative noise. After presenting Fox's trick, which is the usual way of treating the multiplicative case in the existing literature, we introduce a new, alternative way of tackling the problem, by moving, from the one-time response pdf evolution equation, to considering the one-time response-excitation pdf equation.

### 5.1 One-time response pdf evolution equations for scalar, additively excited RDEs

As a first case, we consider the scalar nonlinear RDE under additive Gaussian excitation

$$\dot{X}(t;\theta) = h(X(t;\theta)) + \kappa\,\Xi(t;\theta), \qquad X(t_0;\theta) = X_0(\theta). \tag{5.1a,b}$$

For RIVP (5.1a,b), the variational derivatives of the response, defined by Eqs. (4.8), (4.9), are specified into

$$V_{X_0}(t;\theta) = \exp\left(\int_{t_0}^{t} h'(X(u;\theta))\,du\right), \quad V_{\Xi(s)}(t;\theta) = \kappa\exp\left(\int_{s}^{t} h'(X(u;\theta))\,du\right). \tag{5.2a,b}$$

Thus, in the additive case, variational derivatives are functionals only of the response and not of the excitation. This results into SLE (4.12) taking the form

$$\frac{\partial f_{X(t)}(x)}{\partial t} + \frac{\partial}{\partial x}\left[\left(h(x) + \kappa\, m_{\Xi}(t)\right) f_{X(t)}(x)\right] =$$

$$= \kappa\, C_{X_0\Xi}(t)\, \frac{\partial^2}{\partial x^2}\, \Xi^{\theta}\left[\delta(x - X(t;\theta)) \exp\left(\int_{t_0}^{t} h'(X(u;\theta))\, du\right)\right] + \tag{5.3}$$

$$+ \kappa^2 \frac{\partial^2}{\partial x^2} \int_{t_0}^{t} C_{\Xi\Xi}(t, s)\, \Xi^{\theta}\left[\delta(x - X(t;\theta)) \exp\left(\int_{s}^{t} h'(X(u;\theta))\, du\right)\right] ds .$$

In the averages appearing in the right-hand side of SLE (5.3), random delta function $\delta(x - X(t;\theta))$ of the response at current time $t$ is multiplied by nonlocal terms of the form $\exp\left(\int_s^t h'(X(u;\theta))\, du\right)$ that depend on the history of the response, and not only on response at current time. Thus, in order the application of delta projection to lead to a pdf evolution equation in closed form, all closure schemes for SLE (5.3) consist, in fact, of a current-time approximation for the said nonlocal term.

### 5.1.1 Existing approximation closures

**Small correlation time approximation.** The first, straightforward closure is to treat the nonlocal term $\exp\left(\int_s^t h'(X(u;\theta))\, du\right)$ as a function of $s$, and thus to approximate it by a first-order Taylor series around current time $t$:

$$\exp\left(\int_{s}^{t} h'(X(u;\theta))\, du\right) \cong 1 + h'(X(t;\theta))\,(t - s) . \tag{5.4}$$

Eq. (5.4) introduces the small correlation time (SCT) approximation, since it is valid for small correlation times of excitation and small cross-correlation times between excitation and initial value, in which the main effects of $C_{\Xi\Xi}(t, s)$ and $C_{X_0\Xi}(t)$ in SLE (5.3) are concentrated around current time $t$ and initial time $t_0$ respectively. Substitution of Eq. (5.4) into SLE (5.3), and use of delta projection formula (2.8), yields the SCT response pdf evolution equation

$$\frac{\partial f_{X(t)}(x)}{\partial t} + \frac{\partial}{\partial x}\left[\left(h(x) + \kappa m_{\Xi}(t)\right) f_{X(t)}(x)\right] = \frac{\partial^2}{\partial x^2}\left[\left(D_0^{\mathrm{SCT}}(t) + D_1^{\mathrm{SCT}}(t) h'(x)\right) f_{X(t)}(x)\right], \tag{5.5}$$

in which the coefficients $D_0(t)$, $D_1(t)$ are given by the relation

$$D_n^{\mathrm{SCT}}(t) = \kappa\, C_{X_0\Xi}(t)\, (t - t_0)^n + \kappa^2 \int_{t_0}^{t} C_{\Xi\Xi}(t, s)\, (t - s)^n\, ds , \qquad \text{for } n = 0, 1. \tag{5.6}$$

Eq. (5.5) extends the time-dependent SCT response pdf evolution equation derived in the existing literature, see e.g. (Hänggi & Jung, 1995) Eq. (5.5), to the case of correlated initial value and nonzero mean Gaussian excitation having general (not only Ornstein-Uhlenbeck) autocorrelation function. However, the main drawback of SCT Eq. (5.5) is that its diffusion coefficient in Eq. (5.5) may also take negative values.

In most of the existing literature, see e.g. (Hänggi & Jung, 1995; Sancho et al., 1982) to name two seminal works, the stationary, $t \to \infty$, pdf evolution Eq. (5.5) is specified for deterministic initial value, and zero-mean Ornstein-Uhlenbeck (OU) excitation ($m_\Xi(t) = 0$, $C_{\Xi\Xi}(t, s) = De^{-|t-s|/\tau}/\tau$, $D > 0$). In this case, SCT Eq. (5.5) takes the form

$$\frac{\partial}{\partial x}\left(h(x)\, f_X(x)\right) = D\kappa^2 \frac{\partial^2}{\partial x^2}\left[\left(1 + \tau h'(x)\right) f_X(x)\right], \tag{5.7}$$

which is also called the *standard SCT FPK-like equation*[23], see (Hänggi & Jung, 1995) Eq. 5.6.

**Remark 5.1:** In the linear case, $h(x) = \eta x$, diffusion coefficient of SCT Eq. (5.5) is specified into

$$D_0^{\mathrm{SCT}}(t) + D_1^{\mathrm{SCT}}(t)\, h'(x) = \kappa\, C_{X_0\Xi}(t)\left(1 + \eta(t - t_0)\right) + \kappa^2 \int_{t_0}^{t} C_{\Xi\Xi}(t, s)\left(1 + \eta(t - s)\right) ds\,. \tag{5.8}$$

By comparing Eq. (5.8) to the definition relation (4.21) of the diffusion coefficient $D^{\mathrm{eff}}(t)$ in the exact response pdf evolution Eq. (4.20) for the linear case

$$D^{\mathrm{eff}}(t) = \kappa\, e^{\eta(t-t_0)}\, C_{X_0\Xi}(t) + \kappa^2 \int_{t_0}^{t} e^{\eta(t-s)}\, C_{\Xi\Xi}(t, s)\, ds, \tag{5.9}$$

we observe that SCT diffusion coefficient is obtained from Eq. (5.9) under first order Taylor expansions of term $e^{\eta(t-t_0)}$ around $t_0$ with respect to $t$, and $e^{\eta(t-s)}$ around $t$ with respect to $s$. Thus, SCT Eq. (5.5) fails to yield the exact response pdf evolution equation in the linear case.

**Fox's approximation.** Another closure scheme has been proposed in (Fox, 1986a), under which, in the nonlocal terms of SLE (5.3), only the integral inside the exponential is approximated using a first-order Taylor series around current time $t$:

$$\exp\left(\int_s^t h'(X(u;\theta))\, du\right) \cong \exp\left[h'\left(X(t;\theta)\right)(t - s)\right]. \tag{5.10}$$

By substituting approximation (5.10) into SLE (5.3) and employing the delta projection, we obtain Fox's response pdf evolution equation

$$\frac{\partial f_{X(t)}(x)}{\partial t} + \frac{\partial}{\partial x}\left[\left(h(x) + \kappa m_\Xi(t)\right) f_{X(t)}(x)\right] = \frac{\partial^2}{\partial x^2}\left(D^{\mathrm{Fox}}(x,t)\, f_{X(t)}(x)\right), \tag{5.11}$$

with diffusion coefficient

[23] In Eq. (5.7), and in all stationary pdf equations subsequently, steady-state pdf $f_{X(t)}(x)$ is denoted by $f_X(x)$.

$$D^{\mathrm{Fox}}(x,t) = \kappa\, C_{X_0\Xi}(t) \exp\big(h'(x)\,(t-t_0)\big) + \kappa^2 \int_{t_0}^{t} C_{\Xi\Xi}(t,\,s) \exp\big(h'(x)\,(t-s)\big)\, ds\,. \quad (5.12)$$

We can easily see that, contrary to SCT Eq. (5.5), Fox's response pdf evolution Eq. (5.11) is exact in the linear case. Furthermore, its diffusion coefficient $D^{\mathrm{Fox}}(x,t)$ is always positive, as required physical (interpretation) and mathematical (well-posedness) reasons. However, Fox's diffusion coefficient is plagued by another problem. By observing Eq. (5.12), we can easily deduce that the stationary $D^{\mathrm{Fox}}(x,\infty)$ can take infinite values for a range of $x$. For illustration purposes, we consider again the usual case of deterministic initial value and OU excitation, where $D^{\mathrm{Fox}}(x,t)$ is calculated to

$$D^{\mathrm{Fox}}(x,t) = \kappa^2 \int_{t_0}^{t} C_{\Xi\Xi}(t,\,s) \exp\big(h'(x)\,(t-s)\big)\, ds = \frac{\kappa^2 D}{\tau} \int_{t_0}^{t} \exp\left[ -\left( \frac{1-\tau\, h'(x)}{\tau} \right) (t-s) \right] ds =$$

$$= \frac{\kappa^2 D}{1-\tau\, h'(x)} \left\{ 1 - \exp\left[ -\left( \frac{1-\tau\, h'(x)}{\tau} \right) (t-t_0) \right] \right\}. \quad (5.13)$$

Thus, in the stationary case, $t \to \infty$, $D^{\mathrm{Fox}}(x,\infty) = \kappa^2 D/(1-\tau\, h'(x))$ for $\tau\, h'(x) < 1$ and $D^{\mathrm{Fox}}(x,\infty) = +\infty$ for $\tau\, h'(x) \geq 1$. As discussed also in (Faetti et al., 1988; Tsironis & Grigolini, 1988a), the infinite value of stationary diffusion for a range of $x$ means that, in this region, the probability mass is vanishing, i.e. the stationary pdf there has to be zero. Thus, the validity of Fox's stationary FPK-like equation[24]

$$\frac{\partial}{\partial x}\big(h(x)\; f_X(x)\big) = D\kappa^2\, \frac{\partial^2}{\partial x^2}\left[ \frac{1}{1-\tau\, h'(x)}\; f_X(x) \right], \quad (5.14)$$

derived in (Fox, 1986a), is restricted to $\tau\, h'(x) < 1$, which is indeed a small correlation time condition. Note also that, by formally expanding the term $1/\big(1-\tau\, h'(x)\big)$ of Eq. (5.14) in terms of $\tau$ and keeping only up to the linear term, $1/\big(1-\tau\, h'(x)\big) \cong 1+\tau\, h'(x)$, the standard SCT FPK-like Eq. (5.7) is retrieved.

**Hänggi's ansatz.** Last, another popular approximation scheme is the so-called Hänggi's ansatz, first introduced in (Hänggi et al., 1985) and elaborated further in (Hänggi & Jung, 1995), where it was termed *decoupling approximation* by the authors. Under Hänggi's ansatz, the random quantity inside the nonlocal terms of the SLE (5.3) is approximated by its mean value

$$\exp\left( \int_{s}^{t} h'(X(u;\theta))\, du \right) \cong \exp\left( \int_{s}^{t} R_{h'}(u)\, du \right), \qquad \text{where} \quad (5.15)$$

$$R_{h'}(u) = \mathbb{E}^{\theta}\big[ h'(X(u;\theta)) \big]. \quad (5.16)$$

After approximation (5.15), the nonlocal terms in SLE become deterministic, albeit depending on a particular moment of the response, leading thus to Hänggi's response pdf evolution equation

[24] also termed the effective Fokker-Planck equation, see also Eq. (1.19) in the Introduction.

$$\frac{\partial f_{X(t)}(x)}{\partial t} + \frac{\partial}{\partial x}\left[\left(h(x) + \kappa\, m_{\Xi}(t)\right) f_{X(t)}(x)\right] = D^{\text{Han}}[R_{h'}(\bullet|_{t_0}^{t}) \,;\, t]\, \frac{\partial^2 f_{X(t)}(x)}{\partial x^2}, \tag{5.17}$$

with its diffusion coefficient $D^{\text{Han}}[R_{h'}(\bullet|_{t_0}^{t}) \,;\, t]$ defined as

$$D^{\text{Han}}[R_{h'}(\bullet|_{t_0}^{t}) \,;\, t] = \kappa \exp\left(\int_{t_0}^{t} R_{h'}(u)\, du\right) C_{X_0\Xi}(t) + \kappa^2 \int_{t_0}^{t} \exp\left(\int_{s}^{t} R_{h'}(u)\, du\right) C_{\Xi\Xi}(t, s)\, ds. \tag{5.18}$$

As for the previous cases, the stationary variant of pdf evolution Eq. (5.18), for the case of OU excitation, is easily determined into

$$\frac{\partial}{\partial x}\left(h(x)\, f_X(x)\right) = \frac{D}{1 - \tau R_{h'}(\infty)} \frac{\partial^2 f_X(x)}{\partial x^2}, \tag{5.19}$$

where $R_{h'}(\infty)$ is the steady-state value of the moment $R_{h'}(t)$. Stationary Eq. (5.19) is valid under the condition $\tau R_{h'}(\infty) < 1$, which, as stated in (Hänggi & Jung, 1995), is valid for all globally stable physical systems, for which $R_{h'}(\infty) < 0$ holds true. Thus, Hänggi's Eq. (5.17) is a legitimate FPK-like evolution equation with bounded-valued diffusion coefficient, which can be solved up to the stationary regime, ***without being restricted to small correlation times***, as SCT and Fox's equations.

**Remark 5.2:** Hänggi's evolution Eq. (5.17) is inherently different than the exact response pdf evolution equations for the linear case derived in the previous chapter, as well as than SCT and Fox's evolution Eqs. (5.5), (5.11) respectively. While all the aforementioned evolution equations are of drift-diffusion type, resembling thus the classical FPK equation, the diffusion coefficient in Hänggi's case depends on the whole history of the unknown pdf $f_{X(t)}(x)$ through its dependence on the time integral of a particular moment of the response. Equations of FPK type whose coefficients depend on the unknown pdf are commonly called ***nonlinear FPK equations*** (Frank, 2005), and are encountered in many fields of physics, e.g. in quantum systems (Carrillo, Rosado, & Salvarani, 2008; Sakhnovich & Sakhnovich, 2012), in problems with anomalous diffusion (Curado & Nobre, 2003), in non-equilibrium thermodynamics (Frank, 2002) or in systems with long-range interactions (Zaikin & Schimansky-Geier, 1998).

### 5.1.2 A novel approximation closure of nonlocal character[25]

Having presented the most widely-used closure techniques for SLEs, we introduce now a novel approximation, that combines some of the advantages of the existing ones. First, we begin with the ***decomposition*** of the effect of nonlinearity $h'(X(u;\theta))$ into its ***mean effect*** $R_{h'}(u)$, defined by Eq. (5.16), and ***fluctuation*** $\varphi_{h'}$ around it

$$\varphi_{h'}\left(X(u;\theta); R_{h'}(u)\right) = h'(X(u;\theta)) - R_{h'}(u). \tag{5.20}$$

[25] This closure was first derived in our works (Athanassoulis, Kapelonis, & Mamis, 2018; Mamis et al., 2019) in a more convoluted way, in which the stochastic Volterra-Taylor functional expansion around an appropriate (adjustable) deterministic functional was employed. Due to this fact, in conjunction with the use of decoupling between mean value and fluctuations, our closure in the said works was termed VADA: Volterra adjustable decoupling approximation. The use of term VADA is discontinued in the present work, opting for the simpler terms "novel approximation closure", resulting into the "novel pdf evolution equations".

Thus, the nonlocal SLE terms are expressed equivalently as

$$\exp\left(\int_s^t h'(X(u;\theta))\,du\right) = \exp\left(\int_s^t R_{h'}(u)\,du\right)\cdot\exp\left(\int_s^t \varphi_{h'}\big(X(u;\theta);R_{h'}(u)\big)\,du\right). \tag{5.21}$$

Under the assumption that ***fluctuations are small***, we first employ a linear current-time approximation for the integral of $\varphi_{h'}$

$$\exp\left(\int_s^t h'(X(u;\theta))\,du\right) \cong \exp\left(\int_s^t R_{h'}(u)\,du\right)\cdot\exp\Big(\varphi_{h'}\big(X(t;\theta);R_{h'}(t)\big)(t-s)\Big), \tag{5.22}$$

and also perform a Taylor expansion for the exponential containing the fluctuations, which is truncated at its $M^{\text{th}}$ term

$$\exp\left(\int_s^t h'(X(u;\theta))\,du\right) \cong \exp\left(\int_s^t R_{h'}(u)\,du\right)\cdot\sum_{m=0}^{M}\frac{1}{m!}\,\varphi_{h'}^{m}\big(X(t;\theta);R_{h'}(t)\big)(t-s)^m\,. \tag{5.23}$$

Eq. (5.23) constitutes our novel approximation. It contains, in its right-hand side, two factors; the first, being nonlinear and nonlocal, encapsulates the mean effect of nonlinearity *without any approximation*. The approximation is performed only on the second part, i.e. the effect of fluctuations around the mean value. Note that the importance of a mean value-fluctuations decomposition has been also recognized in (Terwiel, 1974; van Kampen, 2007) for linear RDEs, where it was performed at the level of the RDE itself.

By substituting approximation (5.23) into SLE (5.3), and employing the delta projection, we obtain the family of ***novel response pdf evolution equations***

$$\begin{aligned}\frac{\partial f_{X(t)}(x)}{\partial t} &+ \frac{\partial}{\partial x}\Big[\big(h(x)+\kappa m_{\Xi}(t)\big)\,f_{X(t)}(x)\Big] = \\ &= \frac{\partial^2}{\partial x^2}\left\{\left[\sum_{m=0}^{M}\frac{1}{m!}\,D_m[R_{h'}(\bullet|_{t_0}^{t});t]\,\varphi_{h'}^{m}\big(x;R_{h'}(t)\big)\right]f_{X(t)}(x)\right\},\end{aligned} \tag{5.24}$$

in which the coefficients $D_m$, called the ***generalized effective noise intensities***, are given by

$$\begin{aligned}D_m[R_{h'}(\bullet|_{t_0}^{t});t] &= \kappa\exp\left(\int_{t_0}^{t} R_{h'}(u)\,du\right)C_{X_0\Xi}(t)\,(t-t_0)^m + \\ &+\kappa^2\int_{t_0}^{t}\exp\left(\int_s^t R_{h'}(u)\,du\right)C_{\Xi\Xi}(t,s)\,(t-s)^m\,ds.\end{aligned} \tag{5.25}$$

We observe that, under approximation (5.23), coefficients $D_m$, retain a tractable amount of ***probabilistic nonlocality***, since the term $\exp\left(\int_s^t R_{h'}(u)\,du\right)$ depends on the time history of the unknown response pdf $f_{X(t)}(x)$. Furthermore, the dependence of $\varphi_{h'}$ on $R_{h'}(t)$ introduces a new type of local ***probabilistic nonlinearity***, since $R_{h'}(t)$ depends on the unknown $f_{X(t)}(x)$ at current time $t$. Thus, response pdf evolution Eq. (5.24) is identified as a ***nonlinear FPK equation***, as discussed in Remark 5.2.

**Remark 5.3: Retrieving Hänggi's pdf evolution equation.** For $M=0$, the novel pdf evolution Eq. (5.24) coincides with Hänggi's Eq. (5.17). Thus, our approach generalizes Hänggi's ansatz, providing also a more systematic way of derivation than the one presented in (Hänggi & Jung, 1995; Hänggi et al., 1985). We have to underline that such a generalization of Hänggi's ansatz, without increasing the order of $x-$derivatives in the pdf evolution equation, was deemed not possible by Hänggi himself, see (Hänggi & Jung, 1995) p.273.

**Remark 5.4: On the solution scheme for Eq. (5.24).** While the numerical solution of Eq. (5.24) is the topic of Section 6.1 of the following Chapter, we have state, at this point, that it is a self-standing procedure, since the calculation of response moments $R_{h'}(u)$ needed in the solution are also calculated from Eq. (5.24) per se, by employing an iterative scheme.

## 5.2 Application of novel closure to derive evolution equations for higher-order response pdfs

Novel approximation (5.32), introduced in the previous paragraph, is also applicable for the closure of SLE for higher-order response pdfs. First, we specify, for the additively excited case, the response-initial value SLE (4.95), and two-time response SLE (4.106):

$$\frac{\partial f_{X_0X(t)}(x_0,x_1)}{\partial t}+\frac{\partial}{\partial x_1}\Big[\big(h(x_1)+\kappa m_\Xi(t)\big)f_{X_0X(t)}(x_0,x_1)\Big]=\kappa C_{X_0\Xi}(t)\,\frac{\partial^2 f_{X_0X(t)}(x_0,x_1)}{\partial x_0\,\partial x_1}+$$
$$+\kappa C_{X_0\Xi}(t)\,\frac{\partial^2}{\partial x_1^2}\,\mathbb{E}^\theta\left[\delta(x_0-X_0(\theta))\,\delta(x_1-X(t;\theta))\,\exp\left(\int_{t_0}^{t}h'(X(u;\theta))\,du\right)\right]+$$
$$+\kappa^2\frac{\partial^2}{\partial x_1^2}\int_{t_0}^{t}C_{\Xi\Xi}(t,s)\,\mathbb{E}^\theta\left[\delta(x_0-X_0(\theta))\,\delta(x_1-X(t;\theta))\,\exp\left(\int_{s}^{t}h'(X(u;\theta))\,du\right)\right]ds, \tag{5.26}$$

and

$$\frac{\partial f_{X(t)X(s)}(x_1,x_2)}{\partial t}+\frac{\partial}{\partial x_1}\Big[\big(h(x_1)+\kappa m_\Xi(t)\big)\,f_{X(t)X(s)}(x_1,x_2)\Big]=$$
$$=\kappa C_{X_0\Xi}(t)\frac{\partial^2}{\partial x_1^2}\,\mathbb{E}^\theta\left[\delta(x_1-X(t;\theta))\,\delta(x_2-X(s;\theta))\exp\left(\int_{t_0}^{t}h'(X(u;\theta))\,du\right)\right]+$$
$$+\kappa^2\frac{\partial^2}{\partial x_1^2}\int_{t_0}^{t}C_{\Xi\Xi}(t,\tau)\,\mathbb{E}^\theta\left[\delta(x_1-X(t;\theta))\,\delta(x_2-X(s;\theta))\exp\left(\int_{\tau}^{t}h'(X(u;\theta))\,du\right)\right]d\tau+$$
$$+\kappa C_{X_0\Xi}(t)\frac{\partial^2}{\partial x_1\,\partial x_2}\,\mathbb{E}^\theta\left[\delta(x_1-X(t;\theta))\,\delta(x_2-X(s;\theta))\exp\left(\int_{t_0}^{s}h'(X(u;\theta))\,du\right)\right]+$$
$$+\kappa^2\frac{\partial^2}{\partial x_1\,\partial x_2}\int_{t_0}^{s}C_{\Xi\Xi}(t,\tau)\,\mathbb{E}^\theta\left[\delta(x_1-X(t;\theta))\,\delta(x_2-X(s;\theta))\exp\left(\int_{\tau}^{s}h'(X(u;\theta))\,du\right)\right]d\tau. \tag{5.27}$$

Use of novel approximation closure (5.23) in SLE (5.26) results into the closed, yet approximate response-initial value pdf evolution equation

$$\frac{\partial f_{X_0 X(t)}(x_0, x_1)}{\partial t} + \frac{\partial}{\partial x_1}\Big[\big(h(x_1) + \kappa m_{\Xi}(t)\big) f_{X_0 X(t)}(x_0, x_1)\Big] = \kappa C_{X_0 \Xi}(t) \frac{\partial^2 f_{X_0 X(t)}(x_0, x_1)}{\partial x_0 \partial x_1} + \\ + \frac{\partial^2}{\partial x_1^2}\left\{\left[\sum_{m=0}^{M} \frac{1}{m!} D_m[R_{h'}(\bullet|_{t_0}^{t}); t]\, \varphi_{h'}^m\big(x_1; R_{h'}(t)\big)\right] f_{X_0 X(t)}(x_0, x_1)\right\}, \tag{5.28}$$

where $D_m$'s and fluctuation $\varphi_{h'}$ are defined as in the one-time response pdf evolution Eq. (5.24). Similarly, by substituting closure (5.23) into SLE (5.27) we obtain the two-time response pdf equation

$$\frac{\partial f_{X(t) X(s)}(x_1, x_2)}{\partial t} + \frac{\partial}{\partial x_1}\Big[\big(h(x_1) + \kappa m_{\Xi}(t)\big) f_{X(t) X(s)}(x_1, x_2)\Big] = \\ = \frac{\partial^2}{\partial x_1^2}\left\{\left[\sum_{m=0}^{M} \frac{1}{m!} D_m[R_{h'}(\bullet|_{t_0}^{t}); t]\, \varphi_{h'}^m\big(x_1; R_{h'}(t)\big)\right] f_{X(t) X(s)}(x_1, x_2)\right\} + \\ + \frac{\partial^2}{\partial x_1 \partial x_2}\left\{\left[\sum_{m=0}^{M} \frac{1}{m!} D_m[R_{h'}(\bullet|_{t_0}^{s}); t, s]\, \varphi_{h'}^m\big(x_2; R_{h'}(s)\big)\right] f_{X(t) X(s)}(x_1, x_2)\right\}, \tag{5.29}$$

where the ***two-time generalized noise intensities*** $D_m[R_{h'}(\bullet|_{t_0}^{s}); t, s]$ are defined as

$$D_m[R_{h'}(\bullet|_{t_0}^{s}); t, s] = \kappa \exp\left(\int_{t_0}^{s} R_{h'}(u)\, du\right) C_{X_0 \Xi}(t)\, (s - t_0)^m + \\ + \kappa^2 \int_{t_0}^{s} \exp\left(\int_{\tau}^{s} R_{h'}(u)\, du\right) C_{\Xi\Xi}(t, \tau)\, (s - \tau)^m\, d\tau. \tag{5.30}$$

In Eq. (5.29), one-time noise intensity $D_m[R_{h'}(\bullet|_{t_0}^{t}); t]$ is defined by Eq. (5.25), as for the one-time response pdf Eq. (5.24). Also, by comparing Eq. (5.24) to (5.30), we observe that $D_m[R_{h'}(\bullet|_{t_0}^{t}); t] = D_m[R_{h'}(\bullet|_{t_0}^{t}); t, t]$.

**Remark 5.5: Compatibility with the one-time response pdf evolution equation.** Note that, by integrating both sides of response-initial value pdf Eq. (5.28) with respect to $x_0$, and under the plausible assumption that $f_{X_0 X(t)}(\pm\infty, x_1) = 0$, the one-time response pdf Eq. (5.24) is readily obtained. Eq. (5.24) is also obtained by integrating two-time response pdf Eq. (5.29) with respect to $x_2$, under $\varphi_{h'}^m\big(\pm\infty; R_{h'}(s)\big) f_{X(t) X(s)}(x_1, \pm\infty) = 0$, which, at least for the case of polynomial nonlinearities, is equivalent to the assumption that the pdf, at infinity, goes to zero more rapidly than polynomial. Both the above assumptions are deduced from the need for normalization (finite integral) of the pdfs, which are also considered well-behaved. i.e. they do not oscillate infinitely rapidly as the state variables go to infinity (Gardiner, 2004) par. 5.2.1.f. The compatibility of these two evolution equations for higher order pdfs with the one-time pdf equation is a first positive result in the direction that, by employing closure (5.32), a legitimate hierarchy of approximate, yet closed pdf evolution equations regarding the response of an RDE can be formulated.

## 5.3 One-time response pdf evolution equations for multidimensional, additively excited systems

SCT, Fox's, Hänggi's, as well as our novel approximation scheme, presented in Section 5.1 for scalar RDEs, can be generalized for the multidimensional, nonlinear, additively excited dynamical systems

$$\dot{X}_n(t;\theta) = h_n\big(\boldsymbol{X}(t;\theta)\big) + \sum_{k=1}^{K} \kappa_{nk}\, \Xi_k(t;\theta), \tag{5.31a}$$

$$X_n(t_0;\theta) = X_n^0(\theta), \qquad n = 1, \dots, N, \tag{5.31b}$$

for which multidimensional SLE (4.34) is specified into

$$\frac{\partial f_{\boldsymbol{X}(t)}(\boldsymbol{x})}{\partial t} + \sum_{n=1}^{N} \frac{\partial}{\partial x_n}\left[\left(h_n(\boldsymbol{x}) + \sum_{k=1}^{K} \kappa_{nk}\, m_{\Xi_k}(t)\right) f_{\boldsymbol{X}(t)}(\boldsymbol{x})\right] =$$
$$= \sum_{n_1=1}^{N}\sum_{n_2=1}^{N} \frac{\partial^2}{\partial x_{n_1}\partial x_{n_2}} \sum_{n_3=1}^{N}\sum_{k=1}^{K} \kappa_{n_1 k}\, C_{X^0_{n_3}\Xi_k}(t)\, \mathbb{E}^{\theta}\Big[\delta(\boldsymbol{x}-\boldsymbol{X}(t;\theta))\Phi_{n_2 n_3}[\boldsymbol{X}(\bullet|_{t_0}^{t};\theta)]\Big] +$$
$$+ \sum_{n_1=1}^{N}\sum_{n_1=1}^{K} \frac{\partial^2}{\partial x_{n_1}\partial x_{n_2}} \sum_{n_3=1}^{N}\sum_{k_1=1}^{K}\sum_{k_2=1}^{K} \kappa_{n_1 k_1}\kappa_{n_3 k_2} \int_{t_0}^{t} C_{\Xi_{k_1}\Xi_{k_2}}(t,s)\, \mathbb{E}^{\theta}\Big[\delta(\boldsymbol{x}-\boldsymbol{X}(t;\theta))\Phi_{n_2 n_3}[\boldsymbol{X}(\bullet|_{s}^{t};\theta)]\Big] ds. \tag{5.32}$$

Via the specification, for the additive case, of Eq. (4.45), state-transition matrix $\boldsymbol{\Phi}[\boldsymbol{X}(\bullet|_s^t;\theta)]$ in SLE (5.32) is determined to correspond to a linear ODE system with matrix

$$A_{nm}\big(\boldsymbol{X}(t;\theta)\big) = \frac{\partial h_n\big(\boldsymbol{X}(t;\theta)\big)}{\partial X_m(t;\theta)}, \quad n, m = 1, \dots, N, \tag{5.33}$$

which is identified as the Jacobian matrix of vector function $\boldsymbol{h}(\boldsymbol{x})$. Thus, the multidimensional counterpart of $\exp\left(\int_s^t h'(X(u;\theta))\, du\right)$, needing a current-time approximation, is the state-transition matrix $\boldsymbol{\Phi}[\boldsymbol{X}(\bullet|_s^t;\theta)]$, which is also denoted as $\boldsymbol{\Phi}[\boldsymbol{A}(\boldsymbol{X}(\bullet|_s^t;\theta))]$ whenever the dependence on matrix $\boldsymbol{A}\big(\boldsymbol{X}(t;\theta)\big)$ has to be explicitly mentioned.

As discussed in Section 4.2, matrix $\boldsymbol{\Phi}[\boldsymbol{X}(\bullet|_s^t;\theta)]$ can be expanded in ***Peano-Baker series***, Eq. (4.46). By keeping only its first two terms, the said expansion reads

$$\boldsymbol{\Phi}[\boldsymbol{X}(\bullet|_s^t;\theta)] \cong \mathbf{I} + \int_s^t \boldsymbol{A}\big(\boldsymbol{X}(u;\theta)\big)\, du. \tag{5.34}$$

By performing now a current-time approximation of the integral, we obtain

$$\boldsymbol{\Phi}[\boldsymbol{X}(\bullet|_s^t;\theta)] \cong \mathbf{I} + \boldsymbol{A}\big(\boldsymbol{X}(t;\theta)\big)(t-s). \tag{5.35}$$

By comparing Eq. (5.35) to SCT approximation (5.4), Eq. (5.35) is identified as the generalization of ***SCT approximate closure for the multidimensional case***. Substitution of Eq. (5.35) into SLE (5.32) yields the SCT pdf evolution equation

$$\begin{aligned}\frac{\partial f_{\boldsymbol{X}(t)}(\boldsymbol{x})}{\partial t} + \sum_{n=1}^{N} \frac{\partial}{\partial x_n}\left[\left(h_n(\boldsymbol{x}) + \sum_{k=1}^{K} \kappa_{nk}\, m_{\Xi_k}(t)\right) f_{\boldsymbol{X}(t)}(\boldsymbol{x})\right] = \\ = \sum_{n_1=1}^{N}\sum_{n_2=1}^{N} \frac{\partial^2}{\partial x_{n_1}\,\partial x_{n_2}}\left[\left(D^{(0)}_{n_1 n_2}(t) + \sum_{n_3=1}^{N} D^{(1)}_{n_1 n_3}(t)\, A_{n_2 n_3}(\boldsymbol{x})\right) f_{\boldsymbol{X}(t)}(\boldsymbol{x})\right].\end{aligned} \tag{5.36}$$

where

$$D^{(i)}_{n_1 n_2}(t) = \sum_{k=1}^{K} \kappa_{n_1 k}\, C_{X^0_{n_2}\Xi_k}(t)\,(t-t_0)^i + \sum_{k_1=1}^{K}\sum_{k_2=1}^{K} \kappa_{n_1 k_1}\kappa_{n_2 k_2} \int_{t_0}^{t} C_{\Xi_{k_1}\Xi_{k_2}}(t,s)\,(t-s)^i\, ds,\ i = 0, 1. \tag{5.37}$$

For the case of zero mean value excitation, $\boldsymbol{m}_{\boldsymbol{\Xi}}(t) = \mathbf{0}$, and deterministic initial condition, $\boldsymbol{C}_{\boldsymbol{X}^0\boldsymbol{\Xi}}(t) = \mathbf{0}$, Eq. (5.36) coincides with the usual FPK-like equation derived, in the existing literature and in a much more laborious way, by employing the correlation time expansion (Dekker, 1982; Hernandez-Machado et al., 1983; San Miguel & Sancho, 1980), or the ordered cumulant expansion (Fox, 1983; Garrido & Sancho, 1982). However, similarly to its counterpart for scalar RDEs, it can be easily observed that SCT pdf Eq. (5.36) fails to yield the exact response pdf evolution Eq. (4.57) for the case of a linear multidimensional random system.

In order to obtain a multidimensional response pdf evolution equation that is exact in the linear case, we can approximate the transition matrix using ***Magnus expansion*** (Blanes, Casas, Oteo, & Ros, 2009):

$$\boldsymbol{\Phi}[\boldsymbol{X}(\bullet|_s^t;\theta)] = \exp\left(\sum_{m=1}^{\infty} \boldsymbol{\Omega}_m[\boldsymbol{X}(\bullet|_s^t;\theta)]\right), \tag{5.38}$$

in which the first two terms of the series inside the exponential read

$$\boldsymbol{\Omega}_1[\boldsymbol{X}(\bullet|_s^t;\theta)] = \int_s^t \boldsymbol{A}\big(\boldsymbol{X}(u;\theta)\big)\, du,$$

$$\boldsymbol{\Omega}_2[\boldsymbol{X}(\bullet|_s^t;\theta)] = \frac{1}{2}\int_s^t \int_s^{u_1} \Big[\boldsymbol{A}\big(\boldsymbol{X}(u_1;\theta)\big), \boldsymbol{A}\big(\boldsymbol{X}(u_2;\theta)\big)\Big]\, du_2\, du_1,$$

with $[\boldsymbol{A},\boldsymbol{B}] = \boldsymbol{A}\boldsymbol{B} - \boldsymbol{B}\boldsymbol{A}$ being the matrix commutator. Keeping only the first term $\boldsymbol{\Omega}_1$ in Magnus expansion (5.38), and considering a linear current-time approximation for its temporal integral, results in

$$\boldsymbol{\Phi}[\boldsymbol{X}(\bullet|_s^t;\theta)] \cong \exp\Big[\boldsymbol{A}\big(\boldsymbol{X}(t;\theta)\big)(t-s)\Big]. \tag{5.39}$$

Eq. (5.39), by which state-transition matrix is expressed as a matrix exponential, is exact for the case where matrix $\boldsymbol{A}$ is time-independent; see e.g. (Brockett, 1970) and the linear multi-dimensional case in paragraph 4.2.3. Thus, Eq. (5.39) is a good approximation in cases where response $\boldsymbol{X}(t;\theta)$ does not vary much in time, and, a fortiori, in cases where $\boldsymbol{X}(t;\theta)$ is small. Since, for the scalar case, Eq. (5.39) yields Fox approximation (5.10), Eq. (5.39) constitutes the generalization of ***Fox's approximation for the multidimensional case.*** By substituting approximation (5.39) into SLE (5.32), we obtain

$$\frac{\partial f_{\boldsymbol{X}(t)}(\boldsymbol{x})}{\partial t}+\sum_{n=1}^{N}\frac{\partial}{\partial x_n}\left[\left(h_n(\boldsymbol{x})+\sum_{k=1}^{K}\kappa_{nk}\, m_{\Xi_k}(t)\right) f_{\boldsymbol{X}(t)}(\boldsymbol{x})\right]=\sum_{n_1=1}^{N}\sum_{n_2=1}^{N}\frac{\partial^2\left(B^{\text{Fox}}_{n_1n_2}(\boldsymbol{x},t)\, f_{\boldsymbol{X}(t)}(\boldsymbol{x})\right)}{\partial x_{n_1}\partial x_{n_2}}, \tag{5.40}$$

where the diffusion matrix is defined as

$$\begin{aligned} B^{\text{Fox}}_{n_1n_2}(\boldsymbol{x},t) = {} & \sum_{n_3=1}^{N}\sum_{k=1}^{K}\kappa_{n_1k}\, C_{X^0_{n_3}\Xi_k}(t)\left(\exp\left[\boldsymbol{A}(\boldsymbol{x})\,(t-t_0)\right]\right)_{n_2n_3}+ \\ & +\sum_{n_3=1}^{N}\sum_{k_1=1}^{K}\sum_{k_2=1}^{K}\kappa_{n_1k_1}\kappa_{n_3k_2}\int_{t_0}^{t} C_{\Xi_{k_1}\Xi_{k_2}}(t,s)\left(\exp\left[\boldsymbol{A}(\boldsymbol{x})\,(t-s)\right]\right)_{n_2n_3}ds. \end{aligned} \tag{5.41}$$

To the best of our knowledge, response pdf evolution Eq. (5.40) has not been presented before. Furthermore, since approximation (5.39) is exact when $\boldsymbol{A}$ is a constant matrix, Eq. (5.40) is exact in the linear case.

What is more, and in spite of not being proposed before, ***Hänggi's ansatz for the multidimensional case*** can be also easily formulated, by substituting $\boldsymbol{\Phi}[\boldsymbol{A}(\boldsymbol{X}(\bullet|_s^t;\theta))]$ with the deterministic state-transition matrix that corresponds to a linear ODE with the mean value of $\boldsymbol{A}(\boldsymbol{X}(t;\theta))$ as its matrix:

$$\boldsymbol{\Phi}[\boldsymbol{A}(\boldsymbol{X}(\bullet|_s^t;\theta))] \cong \boldsymbol{\Phi}[\boldsymbol{R}_{\boldsymbol{A}}(\bullet|_s^t)], \qquad \text{where} \tag{5.42}$$

$$\boldsymbol{R}_{\boldsymbol{A}}(u) = \mathbb{E}^{\theta}\left[\boldsymbol{A}(\boldsymbol{X}(u;\theta))\right]. \tag{5.43}$$

Substitution of Eq. (5.42) into SLE (5.32) results in the response pdf evolution equation

$$\begin{aligned} \frac{\partial f_{\boldsymbol{X}(t)}(\boldsymbol{x})}{\partial t}+\sum_{n=1}^{N}\frac{\partial}{\partial x_n}\left[\left(h_n(\boldsymbol{x})+\sum_{k=1}^{K}\kappa_{nk}\, m_{\Xi_k}(t)\right) f_{\boldsymbol{X}(t)}(\boldsymbol{x})\right] = {} & \\ = \sum_{n_1=1}^{N}\sum_{n_2=1}^{N} B^{\text{Han}}_{n_1n_2}[\boldsymbol{R}_{\boldsymbol{A}}(\bullet|_{t_0}^t);t]\,\frac{\partial^2 f_{\boldsymbol{X}(t)}(\boldsymbol{x})}{\partial x_{n_1}\partial x_{n_2}}, & \end{aligned} \tag{5.44}$$

with diffusion matrix

$$\begin{aligned} B^{\text{Han}}_{n_1n_2}[\boldsymbol{R}_{\boldsymbol{A}}(\bullet|_{t_0}^t);t] = {} & \sum_{n_3=1}^{N}\sum_{k=1}^{K}\kappa_{n_1k}\, C_{X^0_{n_3}\Xi_k}(t)\,\Phi_{n_2n_3}[\boldsymbol{R}_{\boldsymbol{A}}(\bullet|_{t_0}^t)]+ \\ & +\sum_{n_3=1}^{N}\sum_{k_1=1}^{K}\sum_{k_2=1}^{K}\kappa_{n_1k_1}\kappa_{n_3k_2}\int_{t_0}^{t} C_{\Xi_{k_1}\Xi_{k_2}}(t,s)\,\Phi_{n_2n_3}[\boldsymbol{R}_{\boldsymbol{A}}(\bullet|_s^t)]\,ds. \end{aligned} \tag{5.45}$$

**The novel approximation closure.** Having formulated the multidimensional generalizations of the usual closure schemes for SLEs, we shall now derive the multidimensional analogue of our closure (5.23). First, we perform a decomposition of random matrix $\boldsymbol{A}\big(\boldsymbol{X}(t;\theta)\big)$ into its mean value $\boldsymbol{R}_A(t)$ and fluctuation

$$\varphi_A\big(\boldsymbol{X}(t;\theta);\boldsymbol{R}_A(t)\big) = \boldsymbol{A}\big(\boldsymbol{X}(t;\theta)\big) - \boldsymbol{R}_A(t), \tag{5.46}$$

around it. As in the scalar case, the central idea behind the novel closure is to encapsulate the mean effect of nonlinearities without any approximation. Thus, we search for a representation of the state-transition matrix $\boldsymbol{\Phi}[\boldsymbol{A}(\boldsymbol{X}(\bullet|_s^t;\theta))]$ in terms of state-transition matrix $\boldsymbol{\Phi}[\boldsymbol{R}_A(\bullet|_s^t)]$ of the mean problem, which is given by the following theorem.

**Theorem 5.1: Decomposition of state-transition matrix.** Matrix $\boldsymbol{\Phi}[\boldsymbol{A}(\boldsymbol{X}(\bullet|_s^t;\theta))]$ is expressed as

$$\boldsymbol{\Phi}[\boldsymbol{A}(\boldsymbol{X}(\bullet|_s^t;\theta))] = \boldsymbol{\Phi}[\boldsymbol{R}_A(\bullet|_s^t)]\,\boldsymbol{\Phi}[\boldsymbol{B}(\bullet|_s^t,s;\theta)], \qquad \text{where} \tag{5.47}$$

$$\boldsymbol{B}(t,s;\theta) = \boldsymbol{\Phi}^{-1}[\boldsymbol{R}_A(\bullet|_s^t)]\,\varphi_A\big(\boldsymbol{X}(t;\theta);\boldsymbol{R}_A(t)\big)\,\boldsymbol{\Phi}[\boldsymbol{R}_A(\bullet|_s^t)]. \tag{5.48}$$

**Proof.** We seek a representation of matrix $\boldsymbol{\Phi}[\boldsymbol{A}(\boldsymbol{X}(\bullet|_s^t;\theta))]$ in the form

$$\boldsymbol{\Phi}[\boldsymbol{A}(\boldsymbol{X}(\bullet|_s^t;\theta))] = \boldsymbol{\Phi}[\boldsymbol{R}_A(\bullet|_s^t)]\,\hat{\boldsymbol{\Phi}}(t,s;\theta), \tag{5.49}$$

where matrix $\hat{\boldsymbol{\Phi}}(t,s;\theta)$ has to be determined. First, we recall that state-transition matrices $\boldsymbol{\Phi}[\boldsymbol{A}(\boldsymbol{X}(\bullet|_s^t;\theta))]$ and $\boldsymbol{\Phi}[\boldsymbol{R}_A(\bullet|_s^t)]$ satisfy the following matrix ODEs, see theorem 1 in (Brockett, 1970, sec. 1.3):

$$\frac{\partial \boldsymbol{\Phi}[\boldsymbol{A}(\boldsymbol{X}(\bullet|_s^t;\theta))]}{\partial t} = \boldsymbol{A}(\boldsymbol{X}(t;\theta))\,\boldsymbol{\Phi}[\boldsymbol{A}(\boldsymbol{X}(\bullet|_s^t;\theta))], \quad \boldsymbol{\Phi}[\boldsymbol{A}(\boldsymbol{X}(\bullet|_s^s;\theta))] = \mathbf{I}, \tag{5.50a,b}$$

and

$$\frac{\partial \boldsymbol{\Phi}[\boldsymbol{R}_A(\bullet|_s^t)]}{\partial t} = \boldsymbol{R}_A(t)\,\boldsymbol{\Phi}[\boldsymbol{R}_A(\bullet|_s^t)], \qquad \boldsymbol{\Phi}[\boldsymbol{R}_A(\bullet|_s^s)] = \mathbf{I}. \tag{5.51a,b}$$

Now, by differentiating both sides of Eq. (5.49) with respect to $t$, we have

$$\frac{\partial \boldsymbol{\Phi}[\boldsymbol{A}(\boldsymbol{X}(\bullet|_s^t;\theta))]}{\partial t} = \frac{\partial \boldsymbol{\Phi}[\boldsymbol{R}_A(\bullet|_s^t)]}{\partial t}\,\hat{\boldsymbol{\Phi}}(t,s;\theta) + \boldsymbol{\Phi}[\boldsymbol{R}_A(\bullet|_s^t)]\,\frac{\partial \hat{\boldsymbol{\Phi}}(t,s;\theta)}{\partial t}. \tag{5.52}$$

In Eq. (5.52), the left-hand side is substituted by Eq. (5.50a), while the derivative appearing in the first term of the right-hand side is substituted by Eq. (5.51a), resulting into

$$\boldsymbol{A}(\boldsymbol{X}(t;\theta))\, \boldsymbol{\Phi}[\boldsymbol{A}(\boldsymbol{X}(\bullet|_s^t;\theta))] = \boldsymbol{R}_A(t)\, \boldsymbol{\Phi}[\boldsymbol{R}_A(\bullet|_s^t)]\, \hat{\boldsymbol{\Phi}}(t,s;\theta) + \boldsymbol{\Phi}[\boldsymbol{R}_A(\bullet|_s^t)]\, \frac{\partial \hat{\boldsymbol{\Phi}}(t,s;\theta)}{\partial t}.$$

In the above relation, $\boldsymbol{\Phi}[\boldsymbol{A}(\boldsymbol{X}(\bullet|_s^t;\theta))]$ is substituted by Eq. (5.49), and after some algebraic manipulations, we obtain the equation

$$\frac{\partial \hat{\boldsymbol{\Phi}}(t,s;\theta)}{\partial t} = \boldsymbol{\Phi}^{-1}[\boldsymbol{R}_A(\bullet|_s^t)]\left(\boldsymbol{A}(\boldsymbol{X}(t;\theta)) - \boldsymbol{R}_A(t)\right) \boldsymbol{\Phi}[\boldsymbol{R}_A(\bullet|_s^t)]\, \hat{\boldsymbol{\Phi}}(t,s;\theta)$$

and by employing fluctuation definition relation, Eq. (5.46):

$$\frac{\partial \hat{\boldsymbol{\Phi}}(t,s;\theta)}{\partial t} = \boldsymbol{\Phi}^{-1}[\boldsymbol{R}_A(\bullet|_s^t)]\, \boldsymbol{\varphi}_A\left(\boldsymbol{X}(t;\theta);\boldsymbol{R}_A(t)\right) \boldsymbol{\Phi}[\boldsymbol{R}_A(\bullet|_s^t)]\, \hat{\boldsymbol{\Phi}}(t,s;\theta). \tag{5.53}$$

ODE (5.53) is supplemented with the initial condition $\hat{\boldsymbol{\Phi}}(s,s;\theta) = \mathbf{I}$, which is derived from Eq. (5.49) for $t = s$ and employing Eqs. (5.50b), (5.51b). By employing the definition relation (5.48) for $\boldsymbol{B}(t,s;\theta)$, matrix $\hat{\boldsymbol{\Phi}}(t,s;\theta)$, since it is the solution of matrix ODE (5.53), is identified as the state-transition matrix $\boldsymbol{\Phi}[\boldsymbol{B}(\bullet|_s^t, s;\theta)]$, completing thus the proof of Eq. (5.47). ■

At this point, we introduce the assumption that fluctuation $\boldsymbol{\varphi}_A$ is small. Thus, the state-transition matrix $\boldsymbol{\Phi}[\boldsymbol{B}(\bullet|_s^t, s;\theta)]$ containing the fluctuation is approximated by the respective matrix exponential

$$\boldsymbol{\Phi}[\boldsymbol{B}(\bullet|_s^t, s;\theta)] \cong \exp\left[\boldsymbol{B}(t,s;\theta)\,(t-s)\right] = \\ = \exp\left[\boldsymbol{\Phi}^{-1}[\boldsymbol{R}_A(\bullet|_s^t)]\, \boldsymbol{\varphi}_A\left(\boldsymbol{X}(t;\theta);\boldsymbol{R}_A(t)\right) \boldsymbol{\Phi}[\boldsymbol{R}_A(\bullet|_s^t)]\,(t-s)\right], \tag{5.54}$$

which constitutes a current-time approximation for $\boldsymbol{\Phi}[\boldsymbol{B}(\bullet|_s^t, s;\theta)]$. By using the properties of matrix exponential, see theorem 1 in (Brockett, 1970, sec. 1.5), Eq. (5.54) is written equivalently as

$$\boldsymbol{\Phi}[\boldsymbol{B}(\bullet|_s^t, s;\theta)] \cong \boldsymbol{\Phi}^{-1}[\boldsymbol{R}_A(\bullet|_s^t)] \exp\left[\boldsymbol{\varphi}_A\left(\boldsymbol{X}(t;\theta);\boldsymbol{R}_A(t)\right)(t-s)\right] \boldsymbol{\Phi}[\boldsymbol{R}_A(\bullet|_s^t)]. \tag{5.55}$$

Substitution of Eq. (5.55) into representation (5.47) results into

$$\boldsymbol{\Phi}[\boldsymbol{A}(\boldsymbol{X}(\bullet|_s^t;\theta))] \cong \exp\left[\boldsymbol{\varphi}_A\left(\boldsymbol{X}(t;\theta);\boldsymbol{R}_A(t)\right)(t-s)\right] \boldsymbol{\Phi}[\boldsymbol{R}_A(\bullet|_s^t)]. \tag{5.56}$$

Last, as in the scalar case, exponential in the right-hand side of Eq. (5.56) is expanded in Taylor series, truncated at the term of $M^{\text{th}}$ order:

$$\boldsymbol{\Phi}[\boldsymbol{A}(\boldsymbol{X}(\bullet|_s^t;\theta))] \cong \left[\sum_{m=0}^{M} \frac{1}{m!}\, \boldsymbol{\varphi}_A^m\left(\boldsymbol{X}(t;\theta);\boldsymbol{R}_A(t)\right)(t-s)^m\right] \boldsymbol{\Phi}[\boldsymbol{R}_A(\bullet|_s^t)]. \tag{5.57}$$

Eq. (5.57) constitutes our ***novel approximation for the multidimensional case***. By substituting it in the SLE (5.32), we obtain the response pdf evolution equation in closed form

$$\frac{\partial f_{\boldsymbol{X}(t)}(\boldsymbol{x})}{\partial t} + \sum_{n=1}^{N} \frac{\partial}{\partial x_n} \left[ \left( h_n(\boldsymbol{x}) + \sum_{k=1}^{K} \kappa_{nk}\, m_{\Xi_k}(t) \right) f_{\boldsymbol{X}(t)}(\boldsymbol{x}) \right] =$$
$$= \sum_{n_1=1}^{N} \sum_{n_2=1}^{N} \frac{\partial^2}{\partial x_{n_1} \partial x_{n_2}} \left[ \left( \sum_{m=0}^{M} \frac{1}{m!} \sum_{n_3=1}^{N} D^{(m)}_{n_1 n_3}[\boldsymbol{R}_{\boldsymbol{A}}(\bullet|_{t_0}^{t}); t]\, \boldsymbol{\varphi}^{m}_{n_2 n_3}(\boldsymbol{x}; \boldsymbol{R}_{\boldsymbol{A}}(t)) \right) f_{\boldsymbol{X}(t)}(\boldsymbol{x}) \right], \tag{5.58}$$

in which $\boldsymbol{\varphi} = \boldsymbol{\varphi}_{\boldsymbol{A}}$ and

$$D^{(m)}_{n_1 n_3}[\boldsymbol{R}_{\boldsymbol{A}}(\bullet|_{t_0}^{t}); t] = \sum_{n_4=1}^{N} \sum_{k=1}^{K} \kappa_{n_1 k}\, \Phi_{n_3 n_4}[\boldsymbol{R}_{\boldsymbol{A}}(\bullet|_{t_0}^{t})]\, C_{X^0_{n_4} \Xi_k}(t)\,(t-t_0)^m +$$
$$+ \sum_{n_4=1}^{N} \sum_{k_1=1}^{K} \sum_{k_2=1}^{K} \kappa_{n_1 k_1} \kappa_{n_4 k_2} \int_{t_0}^{t} \Phi_{n_3 n_4}[\boldsymbol{R}_{\boldsymbol{A}}(\bullet|_{s}^{t})]\, C_{\Xi_{k_1}\Xi_{k_2}}(t,s)\,(t-s)^m\, ds. \tag{5.59}$$

Note that, as in the scalar case, the novel pdf evolution Eq. (5.58) for $M = 0$ coincides with Hänggi's multidimensional pdf Eq. (5.44), since, by comparing Eq. (5.45) with (5.59); $B^{\mathrm{Han}}_{n_1 n_2}[\boldsymbol{R}_{\boldsymbol{A}}(\bullet|_{t_0}^{t}); t] = D^{(0)}_{n_1 n_3}[\boldsymbol{R}_{\boldsymbol{A}}(\bullet|_{t_0}^{t}); t]$.

## 5.4 The case of multiplicative excitation

Having derived pdf evolution equations in closed form for scalar and multidimensional systems under additive coloured excitation, we shall now proceed with examining the more general case of multiplicative excitation. We thus consider the SLE (4.12):

$$\frac{\partial f_{X(t)}(x)}{\partial t} + \frac{\partial}{\partial x}\left[ \left( h(x) + q(x)\, m_{\Xi}(t) \right) f_{X(t)}(x) \right] =$$
$$= C_{X_0 \Xi}(t) \frac{\partial}{\partial x}\left( q(x) \frac{\partial}{\partial x} \mathbb{E}^{\theta}\left[ \delta(x - X(t;\theta)) \exp\left( \mathcal{I}[X(\bullet|_{t_0}^{t};\theta); \Xi(\bullet|_{t_0}^{t};\theta)] \right) \right] \right) +$$
$$+ \frac{\partial}{\partial x}\left( q(x) \frac{\partial}{\partial x} \int_{t_0}^{t} C_{\Xi\Xi}(t,s)\, \mathbb{E}^{\theta}\left[ \delta(x - X(t;\theta))\, q(X(s;\theta)) \exp\left( \mathcal{I}[X(\bullet|_{s}^{t};\theta); \Xi(\bullet|_{s}^{t};\theta)] \right) \right] ds \right), \tag{5.60}$$

with

$$\mathcal{I}[X(\bullet|_{s}^{t};\theta); \Xi(\bullet|_{s}^{t};\theta)] = \int_{s}^{t} \left[ h'(X(u;\theta)) + q'(X(u;\theta))\, \Xi(u;\theta) \right] du, \tag{5.61}$$

corresponding to the scalar, multiplicatively excited RDE:

$$\dot{X}(t;\theta) = h(X(t;\theta)) + q(X(t;\theta))\, \Xi(t;\theta), \qquad X(t_0;\theta) = X_0(\theta). \tag{5.62a,b}$$

**Main difficulties in the multiplicative case.** By comparing the nonlinear terms in SLE (5.60) with the respective terms in SLE (5.3) for the additive case, we observe the following. First, inside the second average in the right-hand side of SLE (5.60), the term $q(X(s;\theta))$ appears, on which, in order to be treated by the delta projection method, a current-time approximation should be applied. However, the most important complicacy in the multiplicative case is that the integrand of $\mathcal{I}$, inside the exponential, depends not only on the response but also on the excitation. This means, that even after a current-time approximation, term $\exp(\mathcal{I})$ would depend not only $X(t;\theta)$ but also on excitation at current time $\Xi(t;\theta)$; see, for example, Fox's approximation for the multiplicative case:

$$\exp\Big(\mathcal{I}[X(\bullet|_s^t;\theta);\Xi(\bullet|_s^t;\theta)]\Big) \cong \exp\Big(\big[h'(X(t;\theta))+q'(X(t;\theta))\,\Xi(t;\theta)\big](t-s)\Big). \quad (5.63)$$

The problem in Eq. (5.63) is that the term containing $\Xi(t;\theta)$ cannot be treated by delta projection, since only the random delta function $\delta(x-X(t;\theta))$, and not $\delta(x-\Xi(t;\theta))$ is present inside the averages in SLE (5.60). Thus, as mentioned also in (Fox, 1986a, sec. V), our only choice is to neglect term $q'(X(t;\theta))\,\Xi(t;\theta)$ in Eq. (5.63), thus practically limiting our approximation to the small excitation regime.

### 5.4.1 Fox's trick

In the literature, there exists a method of recasting the multiplicative SLE (5.60) into the form of the additive one, Eq. (5.3). This method was proposed in (Fox, 1986b) and thus has been subsequently called Fox's trick, see e.g. (Hänggi, 1989). Fox's trick consists of substituting excitation $\Xi(u;\theta)$ appearing in Eq. (5.61), by solving RDE (5.62a) with respect to excitation:

$$\Xi(u;\theta) = \frac{\dot{X}(u;\theta)}{q(X(u;\theta))} - \frac{h(X(u;\theta))}{q(X(u;\theta))}. \quad (5.64)$$

However, in order expression (5.64) to be correct, $q(X(u;\theta)) \neq 0$ for $u \in [s,t] \subseteq [t_0,t]$. This is a ***rather restrictive condition***, since, for the usual cases where $q(\bullet)$ is a power or a polynomial, it means that we have to a priori ascertain that the unknown response $X(u;\theta)$ does not cross zero. Interestingly enough, condition $q(X(u;\theta)) \neq 0$, despite being obvious, has not been discussed before, see e.g. (Fox, 1986b; Hänggi & Jung, 1995; Peacock-López et al., 1988).

Substitution of Eq. (5.64) into Eq. (5.61) results in integral $\mathcal{I}$ explicitly depending only on the history of response, as in the additive case

$$\mathcal{I}[X(\bullet|_s^t;\theta);\Xi(\bullet|_s^t;\theta)] = \mathcal{I}[X(\bullet|_s^t;\theta)] =$$

$$= \int_s^t \frac{q'(X(u;\theta))\,\dot{X}(u;\theta)}{q(X(u;\theta))}\,du + \int_s^t \left[h'(X(u;\theta)) - q'(X(u;\theta))\,\frac{h(X(u;\theta))}{q(X(u;\theta))}\right] du =$$

$$= \int_s^t \frac{\partial q(X(u;\theta))/\partial u}{q(X(u;\theta))}\,du + \int_s^t q(X(u;\theta))\left(\frac{h(X(u;\theta))}{q(X(u;\theta))}\right)' du,$$

*and since* $q(X(u;\theta))$ *does not change sign as it is assumed continuous and non-zero* for $u \in [s,t] \subseteq [t_0, t]$,

$$\mathcal{I}[X(\bullet|_s^t;\theta)] = \ln\left(\frac{q(X(t;\theta))}{q(X(s;\theta))}\right) + \int_s^t q(X(u;\theta))\left(\frac{h(X(u;\theta))}{q(X(u;\theta))}\right)' du. \tag{5.65}$$

Use of Eq. (5.65) in SLE (5.60) results into

$$\frac{\partial f_{X(t)}(x)}{\partial t} + \frac{\partial}{\partial x}\Big[\big(h(x) + q(x)\, m_\Xi(t)\big) f_{X(t)}(x)\Big] =$$

$$= C_{X_0\Xi}(t)\,\frac{\partial}{\partial x}\left(q(x)\,\frac{\partial}{\partial x}\,\mathbb{E}^\theta\left[\delta(x - X(t;\theta))\,\frac{q(X(t;\theta))}{q(X(t_0;\theta))}\exp\left(\int_{t_0}^t \zeta(X(u;\theta))\,du\right)\right]\right) +$$

$$+ \frac{\partial}{\partial x}\left(q(x)\frac{\partial}{\partial x}\int_{t_0}^t C_{\Xi\Xi}(t,s)\,\mathbb{E}^\theta\left[\delta(x - X(t;\theta))\, q(X(t;\theta))\exp\left(\int_s^t \zeta(X(u;\theta))\,du\right)\right] ds\right), \tag{5.66}$$

where

$$\zeta(X(u;\theta)) = q(X(u;\theta))\left(\frac{h(X(u;\theta))}{q(X(u;\theta))}\right)', \tag{5.67}$$

and after an application of the delta projection

$$\frac{\partial f_{X(t)}(x)}{\partial t} + \frac{\partial}{\partial x}\Big[\big(h(x) + q(x)\, m_\Xi(t)\big) f_{X(t)}(x)\Big] =$$

$$= C_{X_0\Xi}(t)\,\frac{\partial}{\partial x}\left\{q(x)\,\frac{\partial}{\partial x}\left(q(x)\,\mathbb{E}^\theta\left[\frac{\delta(x - X(t;\theta))}{q(X(t_0;\theta))}\exp\left(\int_{t_0}^t \zeta(X(u;\theta))\,du\right)\right]\right)\right\} +$$

$$+ \frac{\partial}{\partial x}\left\{q(x)\,\frac{\partial}{\partial x}\left(q(x)\int_{t_0}^t C_{\Xi\Xi}(t,s)\,\mathbb{E}^\theta\left[\delta(x - X(t;\theta))\exp\left(\int_s^t \zeta(X(u;\theta))\,du\right)\right] ds\right)\right\}, \tag{5.68}$$

We observe that, apart form the elimination of explicit dependence of the nonlocal terms on $\Xi(\bullet;\theta)$, Fox's trick resulted also in the elimination of $q(X(s;\theta))$ from the second average in the right-hand side of the SLE. On the other hand, Fox's trick made $q(X(t_0;\theta))$ to appear inside the first SLE average, which, since it is not at $t$, needs a current-time approximation. Note that the appearance of $q(X(t_0;\theta))$, and the complicacy it induces, was not mentioned in the aforementioned literature, since, there, SLEs for cases with correlated initial value with excitation were not considered. Thus, for $C_{X_0\Xi}(t) = 0$, the nonlocal term in SLE (5.68) can be straightforwardly approximated using SCT, Fox's Hänggi's or our new closure scheme, leading thus to a closed response pdf evolution equation.

### 5.4.2 The one-time response-excitation pdf evolution equation

An alternative method, in which the multiplicative case can be treated without the restriction imposed by Fox's trick, is to consider, instead of one-time response SLE, the one-time response-excitation SLE (4.84), which for the case of RDE (5.62a,b)[26], reads

$$\frac{\partial f_{X(t)\Xi(t)}(x,u)}{\partial t} + \frac{\partial}{\partial x}\Big[\big(h(x)+q(x)u\big)\, f_{X(t)\Xi(t)}(x,u)\Big] + \dot{m}_{\Xi}(t)\,\frac{\partial f_{X(t)\Xi(t)}(x,u)}{\partial u} =$$

$$= \dot{C}_{X_0\Xi}(t)\,\frac{\partial^2}{\partial x\partial u}\,\mathbb{E}^{\theta}\Big[\delta(x,u\,;X,\Xi)\exp\Big(\mathcal{I}[X(\bullet|_{t_0}^{t};\theta);\Xi(\bullet|_{t_0}^{t};\theta)]\Big)\Big] +$$

$$+\frac{\partial^2}{\partial x\partial u}\int_{t_0}^{t}\partial_t C_{\Xi\Xi}(t,s)\,\mathbb{E}^{\theta}\Big[\delta(x,u\,;X,\Xi)\,q(X(s;\theta))\exp\Big(\mathcal{I}[X(\bullet|_{s}^{t};\theta);\Xi(\bullet|_{s}^{t};\theta)]\Big)\Big]\,ds +$$

$$+\frac{1}{2}\,\dot{\sigma}_{\Xi}^{2}(t)\,\frac{\partial^2 f_{X(t)\Xi(t)}(x,u)}{\partial u^2}, \tag{5.69}$$

where $\delta(x,u\,;X,\Xi)$ is a shorthand for $\delta(x-X(t;\theta))\,\delta(u-\Xi(t;\theta)$ Thus, since both random delta functions of response and excitation at current time are present inside the averages of SLE (4.69), we can apply our novel current-time approximation closure, as presented in paragraph 5.1.2, to the nonlocal terms $\exp(\mathcal{I})$ of SLE (5.69). First, we decompose the integrand of $\mathcal{I}$ into its mean value

$$R(u) = R_{h'}(u) + R_{q'\Xi}(u) = \mathbb{E}^{\theta}\big[h'(X(u;\theta))\big] + \mathbb{E}^{\theta}\big[q'(X(u;\theta))\,\Xi(u;\theta)\big], \tag{5.70}$$

and fluctuations

$$\varphi\big(X(u;\theta),\,\Xi(u;\theta);R(u)\big) = h'(X(u;\theta)) + q'(X(u;\theta))\,\Xi(u;\theta) - R(u)\,. \tag{5.71}$$

Under this decomposition, nonlocal term $\exp(\mathcal{I})$ is expressed equivalently as

$$\exp\Big(\mathcal{I}[X(\bullet|_{s}^{t};\theta);\Xi(\bullet|_{s}^{t};\theta)]\Big) = \exp\left(\int_s^t R(u)\,du\right)\exp\left(\int_s^t \varphi\big(X(u;\theta),\,\Xi(u;\theta);R(u)\big)\,du\right). \tag{5.72}$$

Then, by assuming that fluctuations are small, we perform a current-time approximation for their integral

$$\exp\Big(\mathcal{I}[X(\bullet|_{s}^{t};\theta);\Xi(\bullet|_{s}^{t};\theta)]\Big) \cong \exp\left(\int_s^t R(u)\,du\right)\exp\Big(\varphi\big(X(t;\theta),\,\Xi(t;\theta);R(t)\big)\,(t-s)\Big),$$

and a Taylor expansion of the exponential containing the fluctuations

$$\exp\Big(\mathcal{I}[X(\bullet|_{s}^{t};\theta);\Xi(\bullet|_{s}^{t};\theta)]\Big) \cong \exp\left(\int_s^t R(u)\,du\right)\sum_{m=0}^{M}\frac{(t-s)^m}{m!}\,\varphi^m\big(X(t;\theta),\,\Xi(t;\theta);R(t)\big). \tag{5.73}$$

[26] We remind that the only restriction for moving on to the response-excitation SLE formulation is that the paths of excitation should be differentiable, as discussed in paragraph 2.5.1.

Eq. (5.73) constitutes the novel approximation for the multiplicative case. Furthermore, as we have already discussed in the beginning of the present section, a current-time approximation should be also performed on the term $q(X(s;\theta)$ appearing inside the second average in SLE (5.69). For this term, we perform a linear Taylor expansion around current time

$$\begin{aligned} q(X(s;\theta)) &\cong q(X(t;\theta)) - q'(X(t;\theta))\dot{X}(t;\theta)\,(t-s) = \\ &= q(X(t;\theta)) - q'(X(t;\theta))\big(h(X(t;\theta)) + q(X(t;\theta))\,\Xi(t;\theta)\big)\,(t-s). \end{aligned} \tag{4.74}$$

Substitution of approximations (5.73) and (5.74) into SLE (5.69) results in the one-time response-excitation pdf evolution equation in closed form

$$\begin{aligned} &\frac{\partial f_{X(t)\Xi(t)}(x,u)}{\partial t} + \frac{\partial}{\partial x}\Big[\big(h(x)+q(x)u\big)\, f_{X(t)\Xi(t)}(x,u)\Big] + \dot{m}_{\Xi}(t)\,\frac{\partial f_{X(t)\Xi(t)}(x,u)}{\partial u} = \\ &= \frac{\partial^2}{\partial x \partial u}\Bigg\{\Bigg[\sum_{m=0}^{M}\frac{1}{m!}\Big(D_m^{(X_0\Xi)}[R(\bullet|_{t_0}^{t});t] + D_m^{(1)}[R(\bullet|_{t_0}^{t});t]\,q(x) - \\ &\qquad - D_m^{(2)}[R(\bullet|_{t_0}^{t});t]\,q'(x)\,\big(h(x)+q(x)u\big)\Big)\,\varphi^m\big(x,u;R(t)\big)\Bigg]\,f_{X(t)\Xi(t)}(x,u)\Bigg\} + \\ &\qquad + \frac{1}{2}\,\dot{\sigma}_{\Xi}^2(t)\,\frac{\partial^2 f_{X(t)\Xi(t)}(x,u)}{\partial u^2}, \end{aligned} \tag{5.74}$$

where

$$D_m^{(X_0\Xi)}[R(\bullet|_{t_0}^{t});t] = \exp\left(\int_{t_0}^{t} R(u)\,du\right)\dot{C}_{X_0\Xi}(t)\,(t-t_0)^m, \tag{5.75a}$$

$$D_m^{(1)}[R(\bullet|_{t_0}^{t});t] = \int_{t_0}^{t}\exp\left(\int_{s}^{t} R(u)\,du\right)\partial_t C_{\Xi\Xi}(t,s)\,(t-s)^m\,ds, \tag{5.75b}$$

$$D_m^{(2)}[R(\bullet|_{t_0}^{t});t] = \int_{t_0}^{t}\exp\left(\int_{s}^{t} R(u)\,du\right)\partial_t C_{\Xi\Xi}(t,s)\,(t-s)^{m+1}\,ds. \tag{5.75c}$$

**Remark 5.6: Marginal equation for one-time response pdf.** By integrating both sides of one-time response-excitation Eq. (5.74) with respect to the state variable $u$ corresponding to excitation, and under the assumption that $f_{X(t)\Xi(t)}(x,u)$ and its derivatives go to zero fast enough for $u \to \pm\infty$ (see also the analogous discussion in Remark 5.5), we easily obtain the equation with respect to marginal pdf $f_{X(t)}(x)$:

$$\frac{\partial f_{X(t)}(x)}{\partial t} + \frac{\partial}{\partial x}\int_{\mathbb{R}}\big(h(x)+q(x)u\big)\,f_{X(t)\Xi(t)}(x,u)\,du = 0,$$

which is easily recast to

$$\frac{\partial f_{X(t)}(x)}{\partial t} + \frac{\partial}{\partial x}\big(h(x)\,f_{X(t)}(x)\big) = -\frac{\partial}{\partial x}\left(q(x)\int_{\mathbb{R}} u\,f_{X(t)\Xi(t)}(x,u)\,du\right). \tag{5.76}$$

We observe that Eq. (5.76) coincides with Eq. (2.21), which is an alternative form of the ***exact stochastic Liouville equation*** for the response, before the application of any approximation scheme[27]. Thus, we have to note that, while the above result verifies the compatibility between the derivations of response and response-excitation pdf evolution equations, its main importance lies elsewhere: The fact that the marginal equation for $f_{X(t)}(x)$ resulted in the exact SLE and not in an approximate pdf evolution equation, supports the conjunction that, by solving the response-excitation Eq. (5.74) we obtain a better approximation for $f_{X(t)}(x)$ than the one obtained from solving an approximate response pdf evolution equation, even for the case of an additive RDE.

**Remark 5.7: Marginal equation for the one-time excitation pdf.** Analogously to the derivation of Eq. (5.76), and under the assumption that $f_{X(t)\Xi(t)}(x,u)$ and its derivatives also go to zero fast enough for $x \to \pm\infty$, we obtain the following pdf evolution equation for $f_{\Xi(t)}(u)$, by integrating both sides of Eq. (5.74) with respect to $x$:

$$\frac{\partial f_{\Xi(t)}(u)}{\partial t} + \dot{m}_{\Xi}(t)\,\frac{\partial f_{\Xi(t)}(u)}{\partial u} = \frac{1}{2}\,\dot{\sigma}^2_{\Xi}(t)\,\frac{\partial^2 f_{\Xi(t)}(u)}{\partial u^2}. \tag{5.77}$$

Eq. (5.77) can be solved by employing the Fourier transform; $\varphi_{\Xi(t)}(y) = \int_{\mathbb{R}} e^{iyu} f_{\Xi(t)}(u)\,du$, as presented in paragraph 4.1.2. By following this solution procedure, the unique solution of Eq. (5.77) is identified to the Gaussian pdf

$$f_{\Xi(t)}(u) = \frac{1}{\sqrt{2\pi\sigma^2_{\Xi}(t)}}\exp\left[-\frac{1}{2}\frac{\left(u - m_{\Xi}(t)\right)^2}{\sigma^2_{\Xi}(t)}\right]. \tag{5.78}$$

Eq. (5.78) gives the expected form for $f_{\Xi(t)}(u)$, since, from the start, excitation was considered Gaussian with mean value $m_{\Xi}(t)$ and variance $\sigma^2_{\Xi}(t)$. The derivation of Eq. (5.77) from Eq. (5.74), and the validation of the correctness of its solution, constitute a compatibility check for response-excitation pdf evolution Eq. (5.74), since it ensures that the said equation encapsulates the evolution of *the given* excitation pdf correctly.

**Response-excitation pdf evolution equation for the additive case.** Last, and for the sake of completeness, we also specify response-excitation pdf evolution Eq. (5.74) for the case of additive excitation, $q(x) = \kappa$:

$$\frac{\partial f_{X(t)\Xi(t)}(x,u)}{\partial t} + \frac{\partial}{\partial x}\Big[\big(h(x) + \kappa u\big)\, f_{X(t)\Xi(t)}(x,u)\Big] + \dot{m}_{\Xi}(t)\,\frac{\partial f_{X(t)\Xi(t)}(x,u)}{\partial u} =$$
$$= \frac{\partial^2}{\partial x \partial u}\left\{\left[\sum_{m=0}^{M}\frac{1}{m!}G_m[R_{h'}(\bullet|_{t_0}^{t});t]\,\varphi^m_{h'}\big(x;R_{h'}(t)\big)\right] f_{X(t)\Xi(t)}(x,u)\right\} + \tag{5.79}$$
$$+\frac{1}{2}\,\dot{\sigma}^2_{\Xi}(t)\,\frac{\partial^2 f_{X(t)\Xi(t)}(x,u)}{\partial u^2},$$

[27] We remind that, as discussed in Section 2.3, Eq. (5.76) cannot be solved per se, since it contains, as unknowns, both the one-time response pdf $f_{X(t)}(x)$, and the one-time response-excitation pdf $f_{X(t)\Xi(t)}(x,u)$.

where

$$G_m[R_{h'}(\bullet|_{t_0}^{t});t] = \exp\left(\int_{t_0}^{t} R_{h'}(\tau)\,d\tau\right)\dot{C}_{X_0\Xi}(t)\,(t-t_0)^m + \\ + \kappa\int_{t_0}^{t}\exp\left(\int_{s}^{t} R_{h'}(\tau)\,d\tau\right)\partial_t C_{\Xi\Xi}(t,s)\,(t-s)^m\,ds. \tag{5.80}$$

Note that in Eqs. (5.79), (5.80), the mean value $R_{h'}(t)$ and fluctuation $\varphi_{h'}\left(x;R_{h'}(t)\right)$ of the nonlinearities of the additive case appear, defined by Eqs. (5.16), (5.20) respectively.

# Chapter 6: Numerical solution scheme & results

**Summary.** In order to quantify the range of validity and assess the accuracy of the novel pdf evolution equations derived in the previous Chapter, a number of numerical simulations has been performed in the present Chapter[28]. The numerical method, presented briefly in Sec. 6.1, used for solving the various versions of pdf evolution equations, employs: **i)** a partition of unity finite element method for the discretization of the response pdf in the state space, **ii)** a Bubnov-Galerkin technique for deriving ODEs governing the evolution of the response pdf, and **iii)** a Crank-Nicolson scheme for solving the said ODEs in the time domain. A feature peculiar to our novel pdf evolution equations, calling for special numerical treatment, is their nonlinear/nonlocal character. This peculiarity is treated by a self-contained, iterative scheme as follows: the current-time values of the response moments, needed for the calculation of their diffusion coefficient, are estimated by extrapolation based on the two previous time steps, and then are improved by iterations at the current time. Usually one or two iterations suffice. The final values of these moments, for each time instant, are stored and used for the calculation of the nonlocal terms (time integrals). Finally, in Sec. 6.2, the above numerical scheme is implemented for the solution of a benchmark bistable case, showing the accuracy of our novel pdf evolution equations even in the regime of both high noise intensities and large correlation times of the excitation.

## 6.1 Numerical scheme for the solution of scalar pdf evolution equations[29]

The Partition of Unity Finite Element Method (PUFEM), introduced by (Melenk & Babuska, 1996), is a numerical method that generalizes the standard Finite Element Method (FEM) by utilizing the Partition of Unity (PU) concept. PUFEM has been already used for the numerical solution of the classical FPK equations, both in the stationary case (M. Kumar, Chakravorty, Singla, & Junkins, 2009) and in the transient case (M. Kumar, Chakravorty, & Junkins, 2010). In this work, a PUFEM numerical scheme is adapted and implemented for the numerical solution of the scalar response pdf evolution equations derived in the previous Chapter, including their non-linear, nonlocal variants.

### 6.1.1 Partition of Unity and PU approximation spaces

Let $\Omega \subset \mathbb{R}^n$ be an open domain, and $\{\Omega_k\}$, $k = 1, \dots, K$, $K \in \mathbb{N}$ be an open cover of $\Omega$ which satisfies *the pointwise overlap condition*:

$$\left(\exists M \in \mathbb{N}\right)\left(\forall \boldsymbol{x} \in \Omega\right): \ \mathrm{Card}\left(\{k_j\} \mid \boldsymbol{x} \in \Omega_{k_j}\right) \le M .$$ [30]

[28] The numerical solution scheme has been developed by colleague Z. G. Kapelonis in (Kapelonis, 2020)

[29] Earlier version of Section 6.1 constitutes Appendix C of (Mamis et al., 2019).

[30] Note that, for any $\boldsymbol{x} \in \Omega$, $\mathrm{Card}\left(\{k_j\} \mid \boldsymbol{x} \in \Omega_{k_j}\right) \ge 1$, since $\{\Omega_k\}$ is a cover of $\Omega$. In practice, the method is implemented so that $\mathrm{Card}\left(\{k_j\} \mid \boldsymbol{x} \in \Omega_{k_j}\right) \ge 2$ (except for a boundary "layer" near the physical boundary of the computational domain $\Omega$, where the said cardinality may be 1).

Then, for any given $s \in \mathbb{N}$, there exists a set of functions $\left\{ \varphi_k(\boldsymbol{x}) \right\}$, each one associated with an open domain $\Omega_k$, such that, for all $k = 1(1)K$ :

1. $\varphi_k(\bullet) \in C^s\left(\mathbb{R}^n \to \mathbb{R}\right)$,
2. $\operatorname{supp}(\varphi_k(\bullet)) \subseteq \bar{\Omega}_k$, [31]
3. $\left(\forall\, \boldsymbol{x} \in \Omega\right)\ 0 \le \varphi_k(\boldsymbol{x}) \le 1$, and
4. $\left(\forall\, \boldsymbol{x} \in \Omega\right)\ \sum_{k=1}^{K} \varphi_k(\boldsymbol{x}) = 1$.

We say that the set of functions $\{\varphi_k(\boldsymbol{x})\}$ forms a $C^s$**−*partition of unity*** ($C^s$−PU) associated with the cover $\{\Omega_k\}$. Functions $\varphi_k(\boldsymbol{x})$ are called partition of unity functions (PUF).
For each subdomain $\Omega_k$ we consider an approximate function basis

$$\left\{ b_\mu^k(\bullet) \in C^\ell\left(\Omega_k \to \mathbb{R}\right),\ \mu = 1, 2, \cdots, \mathrm{M}(k) \right\},$$

where the necessary order of smoothness $\ell$ is dictated by the specific problem. The set $\left\{ b_\mu^k(\bullet) \right\}$ is called the *local basis associated with* $\Omega_k$ (or the *local basis of* $\Omega_k$), and may contain different number of elements, $\mathrm{M}(k)$, for different subdomains $\Omega_k$. In each $\Omega_k$, the local basis defines the *local approximation space*

$$V^{\mathrm{M}_k}(\Omega_k) = \operatorname{span}\left(\{b_\mu^k(\boldsymbol{x})\}\right). \tag{6.1}$$

In this work, local approximation spaces will be constructed by using Legendre orthogonal polynomials and, thus, they will be dense in $C(\Omega_k)$, $C^1(\Omega_k)$ and $H^1(\Omega_k)$.
The *approximate basis in the global domain* $\Omega$ is then constructed by means of the functions

$$u_\mu^k(\boldsymbol{x}) = \varphi_k(\boldsymbol{x})\, b_\mu^k(\boldsymbol{x}), \quad \boldsymbol{x} \in \Omega, \quad \begin{cases} k = 1, 2, \cdots, K, \\ \mu = 1, 2, \cdots, \mathrm{M}(k). \end{cases} \tag{6.2}$$

Following the FEM tradition, we shall call $u_\mu^k(\boldsymbol{x})$ *shape functions*. The span of shape functions defines the *global approximation space*, called also *partition of unity approximation space*, or *partition of unity space*, for short:

$$V^{\mathrm{PU}}(\Omega) = \operatorname{span}\left( \{u_\mu^k(\boldsymbol{x}) = \varphi_k(\boldsymbol{x})\, b_\mu^k(\boldsymbol{x}),\ k = 1, 2, \cdots, K,\ \mu = 1, 2, \cdots, \mathrm{M}(k)\} \right). \tag{6.3}$$

Given that the local approximation spaces $V^{\mathrm{M}_k}(\Omega_k)$ have the necessary approximation properties, the global approximation space $V^{\mathrm{PU}}(\Omega)$ is also dense in $C(\Omega)$, $C^1(\Omega)$ and $H^1(\Omega)$, see Theorem 2.1 in (Melenk & Babuska, 1996), providing us with an efficient approximation of functions $F(\bullet)$ from the aforementioned spaces:

$$F(\boldsymbol{x}) \cong \hat{F}(\boldsymbol{x}) = \sum_{k=1}^{K} \sum_{\mu=1}^{\mathrm{M}(k)} w_\mu^k\, u_\mu^k(\boldsymbol{x}), \quad \boldsymbol{x} \in \Omega, \tag{6.4}$$

where $\hat{F}(\boldsymbol{x})$ denotes the PU approximation of $F(\boldsymbol{x})$ and $w_\mu^k$ are scalar weights.

[31] From Conditions 1, 2, it is easy to infer that $\operatorname{supp}(\varphi_k'(\xi)) = \operatorname{supp}(\varphi_k''(\xi)) = \operatorname{supp}(\partial^s \varphi_k(\xi) / \partial \xi^s) = \bar{\Omega}_k$.

The ***main advantage*** PUFEM has over the classical FEM, is that, by introducing the PUF $\varphi_k(\bullet)$, the continuity between the overlapping elements $\Omega_k$ is satisfied by construction, making the consideration of matching conditions redundant (Melenk & Babuska, 1996). This PUFEM property is especially important for moving towards solutions of pdf evolution equations that correspond to systems with many (more than three) state variables, where the matching coditions for classical FEM would be particulary cumbersome.

### 6.1.2 Construction of cover, PU functions, and local basis functions for 1D domains

We shall now present a specific implementation of the above concepts and constructions for one-dimensional (1D) domains. In our implementation $\mathrm{Card}\left(\{k_j\} \mid \boldsymbol{x} \in \Omega_{k_j}\right) = 2$ for almost all $\boldsymbol{x} \in \Omega$, with the exception of the "outer half" of the end subdomains ($\Omega_1$ and $\Omega_K$) where $\mathrm{Card}\left(\{k_j\} \mid \boldsymbol{x} \in \Omega_{k_j}\right) = 1$.

Let $\Omega = [\omega_{\text{g-min}}, \omega_{\text{g-max}}] \subset \mathbb{R}$ be a given domain (*the computational domain*) on the real axis, and $K$ be the number of subdomains of the cover $\{\Omega_k\}$ of $\Omega$. Then, each subdomain (interval) $\Omega_k = [\omega^k_{\min}, \omega^k_{\max}]$ is defined by means of the equations

$$\begin{aligned} \omega^k_{\min} &= \omega_{\text{g-min}} + (k-1)\,h, \\ \omega^k_{\max} &= \omega_{\text{g-min}} + (k+1)\,h = \omega^k_{\min} + 2h, \end{aligned} \tag{6.5}$$

where $h = (\omega_{\text{g-max}} - \omega_{\text{g-min}})/(K+1)$. For this cover, the length of each subdomain is equal to $2h$, and the overlapping between adjacent subdomains is $h$. The layout of such an 1D cover is shown in Figure 1.

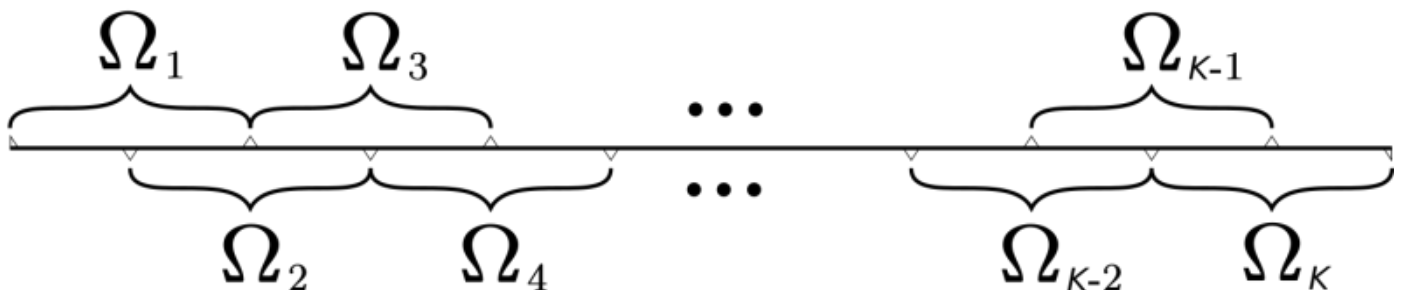

**Figure 1.** Layout of cover $\{\Omega_k\}$ of the computational domain $\Omega = \bigcup_{k=1}^{K} \Omega_k$.

For the specification of PU functions and the local basis functions, it is more convenient to work in a *reference domain*, say $\Omega_{\text{ref}} = [-1, +1]$, and then transform the results into the physical domain. All calculations are performed in $\Omega_{\text{ref}}$ (the reference view) and converted to arbitrary $\Omega_k$ (the absolute view) by an affine map, defined by

$$\begin{aligned} &\xi_k(\omega) : [\omega^k_{\min}, \omega^k_{\max}] \to [-1, +1] \\ &\xi_k(\omega) = \frac{2\omega - \omega^k_{\min} - \omega^k_{\max}}{\omega^k_{\max} - \omega^k_{\min}}, \quad \omega \in [\omega^k_{\min}, \omega^k_{\max}] \end{aligned} \tag{6.6}$$

with inverse map (from reference to absolute coordinates):

$$\begin{aligned} &\omega_k(\xi) : [-1, +1] \to [\omega^k_{\min}, \omega^k_{\max}] \\ &\omega_k(\xi) = \frac{(\omega^k_{\max} - \omega^k_{\min})\xi + \omega^k_{\max} + \omega^k_{\min}}{\omega^k_{\max} - \omega^k_{\min}}, \quad \xi \in [-1, 1]. \end{aligned} \tag{6.7}$$

We shall now construct the mother $C^s$−PU function in the reference domain $\Omega_{\text{ref}}$, which is denoted by $\tilde{\varphi}^{(s)}(\xi)$.[32] This function is defined as a piecewise polynomial, by

$$\tilde{\varphi}^{(s)}(\xi) = \begin{cases} g_s(y_L(\xi)), & \text{for } \xi \in [-1, 0] \\ g_s(y_R(\xi)), & \text{for } \xi \in [0, 1] \\ 0 \quad , & \text{otherwise} \end{cases} \tag{6.8}$$

where:

$$y_L(\xi) = 2\xi + 1 , \qquad y_R(\xi) = -2\xi + 1, \tag{6.9}$$

and $g_s(\bullet)$ is a polynomial satisfying the following conditions[33]

$$g_s(\xi) + g_s(-\xi) = 1, \quad \forall \xi \in \Re,$$
$$g_s(1) = 1,$$
$$\frac{d^{\ell} g_s(1)}{d\xi^{\ell}} = 0, \quad \text{for } \ell = 1, \dots, s.$$

It is only a matter of algebraic manipulations to show that such a polynomial exists, contains a constant term and only odd powers of the independent variable, and is of degree $2s-1$. Thus, its general form is

$$g_s(z) = a_0 + \sum_{i=1}^{i=s} a_i z^{2i-1}, \tag{6.10}$$

with coefficients $a_i$ dependent on $s$ (values for cases $s = 1, 2, 3$ are shown in Table 1).

| **Continuity** | **Coefficients** |
|---|---|
| $C^1\ (s=1)$ | $a_0 = \frac{1}{2},\quad a_1 = \frac{1}{2},\quad a_n = 0 \text{ for } n \geq 2$ |
| $C^2\ (s=2)$ | $a_0 = \frac{1}{2},\quad a_1 = \frac{3}{4},\quad a_2 = -\frac{1}{4},\quad a_n = 0 \text{ for } n \geq 3$ |
| $C^3\ (s=3)$ | $a_0 = \frac{1}{2},\quad a_1 = \frac{15}{16},\quad a_2 = -\frac{5}{8},\quad a_3 = \frac{3}{16},\quad a_n = 0 \text{ for } n \geq 4$ |

**Table 1.**: Polynomial coefficients for the PU function branch function (C10) with $s = 1, 2, 3$.

For the local basis functions, the Legendre polynomials are used, usually defined by Rodrigues' formula, $P_n(\xi) = \frac{1}{2^n n!} \frac{d^n}{dx^n} (\xi^2 - 1)^n$. They are also expressed by the explicit formula

[32] Note that the superscript ($s$) in $\tilde{\varphi}^{(s)}(\xi)$ indicates the order of smoothness of the PU functions (not to be confused with differentiation).
[33] Sufficient conditions in order that Conditions 1-4 stated above are satisfied.

$$P_n(\xi) = 2^n \sum_{i=0}^{n} \xi^i \binom{n}{i} \binom{\frac{n+i-1}{2}}{n}, \qquad \xi \in \mathbb{R}. \tag{6.11}$$

The local basis functions in $\Omega_{\text{ref}}$, denoted by $\tilde{b}_\mu(\xi)$, are defined by $\tilde{b}_\mu(\xi) = P_{\mu-1}(\xi)$. The first five of them are listed below

$$\tilde{b}_1(\xi) = P_0(\xi) = 1, \quad \tilde{b}_2(\xi) = P_1(\xi) = \xi, \quad \tilde{b}_3(\xi) = P_2(\xi) = \frac{1}{2}(3\xi^2 - 1),$$

$$\tilde{b}_4(\xi) = P_3(\xi) = \frac{1}{2}(5\xi^3 - 3\xi), \quad \tilde{b}_5(\xi) = P_4(\xi) = \frac{1}{8}(35\xi^4 - 30\xi^2 + 3).$$

Now, Eq. (C2) for the reference shape functions reads as follows

$$\tilde{u}_\mu(\xi) = \tilde{\varphi}^{(s)}(\xi)\, \tilde{b}_\mu(\xi). \tag{6.12}$$

Figure 2 shows both the local basis functions and the corresponding shape functions (in the reference domain $\Omega_{\text{ref}}$) for smoothness of order $s = 2$. Note that $\tilde{u}_1(\xi) = \tilde{\varphi}^{(s)}(\xi)$.

Now, by using the affine transformation (6.6), (6.7), we are able to pass to the physical domain, obtaining the PU functions $\varphi_k^{(s)}(x)$, the local basis functions $b_\mu^k(x)$, and the shape functions $u_\mu^k(x)$, for any subdomain $\Omega_k$:

$$\varphi_k^{(s)}(x) = \tilde{\varphi}^{(s)}(\xi_k(x)), \qquad b_\mu^k(x) = \tilde{b}_\mu(\xi_k(x)) \tag{6.13a,b}$$

$$u_\mu^k(x) = \varphi_k^{(s)}(x)\, b_\mu^k(x) = \tilde{u}_\mu(\xi_k(x)). \tag{6.13c}$$

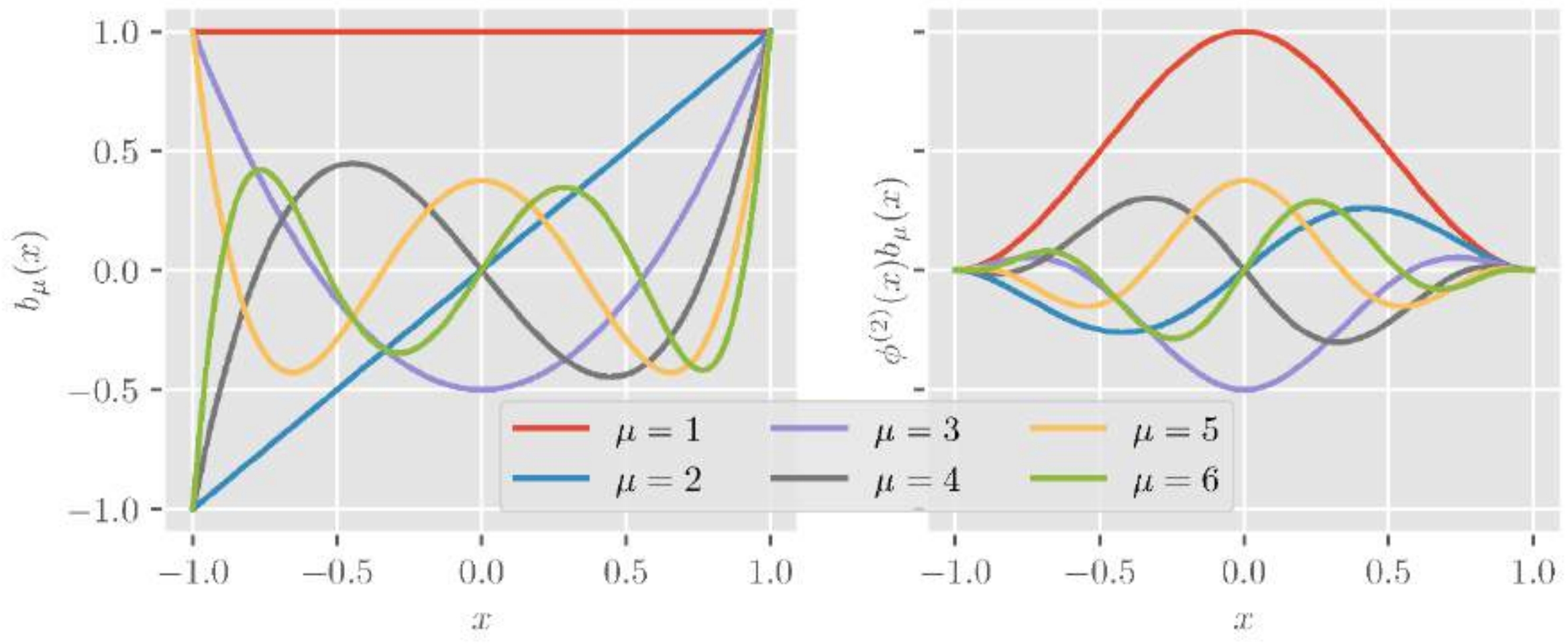

**Figure 2.** Legendre polynomial local basis functions $\tilde{b}_\mu(\xi)$ (left panel) and corresponding shape functions $\tilde{\varphi}^{(2)}(\xi)\cdot\tilde{b}_\mu(\xi)$ (right panel), for different values of $\mu$.

### 6.1.3 PUFEM discretization and numerical solution of pdf evolution equations

Since the scalar response pdf evolution equations, presented in Section 5.1 of the previous Chapter, can attain a variety of forms, the following generic form is adapted for developing the numerical scheme

$$\frac{\partial f(x,t)}{\partial t} = -\frac{\partial}{\partial x}\left[q(x,t)\,f(x,t)\right] + \frac{\partial^2 \mathcal{B}[f;x,t]\,f(x,t)}{\partial x^2}, \tag{6.14}$$

where now the unknown pdf is denoted by $f(x,t)$ instead of $f_{X(t)}(x)$. In Eq. (6.14) $q(x,t)$ is the *drift coefficient*, and $\mathcal{B}[f;x,t]$ is the *diffusion coefficient*. Recall that the diffusion coefficient in response pdf equations (5.24) under the novel closure (5.23) depends on the unknown pdf $f(x,t)$, rendering Eq. (6.14) nonlinear and nonlocal. Nevertheless, in this subsection, the simple notation $B(x,t)$ will be used for the diffusion coefficient, with notation $\mathcal{B}[f_X;x,t]$ being employed in the next subsection, where the treatment of nonlinearity/nonlocality will be explained.

Eq. (C14) is supplemented by the following conditions:

| | | |
|---|---|---|
| $f(x,t_0) = f_0(x) = \text{known}$ , | (initial condition) | (6.15a) |
| $f(x,t) \geq 0, \quad \forall\, x \in \Re$, | (positivity condition) | (6.15b) |
| $f(x,t)$ decays sufficiently fast as $\lvert x\rvert \to \infty$ .[34] | (decay condition) | (6.15c) |

In addition, since it is a pdf, the unknown function $f(x,t)$ should satisfy the condition

$$\int_{-\infty}^{+\infty} f(x,t)\,dx = 1 \tag{6.16}$$

A weak form of Eq. (6.14) is obtained by projecting both members on the elements $v_j(x)$ of a test function space, $V^{\text{Test}}(\Re)$:

$$\frac{\partial}{\partial t}\int_{S(v_j)} f(x,t)\cdot v_j(x)dx = -\int_{S(v_j)} \frac{\partial}{\partial x}\left[q(x,t)\,f(x,t)\right]\cdot v_j(x)dx \; + \\ + \int_{S(v_j)} \frac{\partial^2 B(x,t)\,f(x,t)}{\partial x^2}\cdot v_j(x)dx, \tag{6.17}$$

where $S(v_j) = \text{supp}(v_j(\bullet))$. Integration by parts on both terms in the right-hand-side of Eq. (6.17) leads to the following equation, containing only first derivatives with respect to the unknown function $f(x,t)$:

$$\frac{\partial}{\partial t}\int_{S(v_j)} f(x,t)v_j(x)dx = \left(\frac{\partial B(x,t)\,f(x,t)}{\partial x}v_j(x) - q(x,t)\,f(x,t)v_j(x)\right)\Bigg|_{\partial S(v_j)} + \\ + \int_{S(v_j)}\left(q(x,t) - \frac{\partial B(x,t)}{\partial x}\right) f(x,t)\frac{\partial v_j(x)}{\partial x}\,dx - \int_{S(v_j)} B(x,t)\frac{\partial f(x,t)}{\partial x}\frac{\partial v_j(x)}{\partial x}dx. \tag{6.18}$$

To derive a discrete system of equations, the unknown solution $f(x,t)$ is approximated by using Eq. (6.4):

$$f(x,t) \cong \hat{f}(x,t) = \sum_{k=1}^{K}\sum_{\mu=1}^{\mathrm{M}(k)} w_\mu^k(t)u_\mu^k(x) = \sum_{m=1}^{M} w_m(t)u_m(x), \tag{6.19}$$

[34] The appropriate decay rate at infinity needs for ensuring the existence of the integrals over the $\Re$, which appear in the weak form of Eq. (6.10).

where $w_\mu^k(t) = w_m(t)$ are time-dependent weights, to be determined numerically, and the *global index* $m = m(k, \mu)$ is defined by:

$$m = m(k, \mu) = \sum_{i=1}^{k-1} \mathrm{M}(i) + \mu \le \sum_{i=1}^{K} \mathrm{M}(i) \equiv M . \qquad (6.20)$$

(Recall that $\mathrm{M}(i)$ denotes the number of basis functions in subdomain $\Omega_i$). Eq. (6.20) can be inverted, to derive $k, \mu$ from $m$, by means of the formulae

$$k = k(m) = \max\left\{ j \in \mathbb{N} : \sum_{i=1}^{i=j} \mathrm{M}(i) \le m \right\},$$
$$\mu = \mu(m,k) = m - \sum_{i=1}^{i=k-1} \mathrm{M}(i), \qquad \left( \sum_{i=1}^{i=0} \mathrm{M}(i) \equiv 0 \right).$$

The integer $M = \sum_{i=1}^{i=K} \mathrm{M}(i)$ counts the total number of shape functions used in the discretization, and defines the *global Degree of Freedom* (DOF) of the scheme. To achieve better accuracy, the total DOF can be increased either by refining the cover $\{\Omega_k\}$ (which corresponds to $h-$FE refinement), or by increasing the number $\mathrm{M}(k)$ of local basis functions in some (or all) subdomains $\Omega_k$ (which corresponds to $p-$FE refinement), or both ($hp-$FE refinement).

In all applications presented herein, the test space $V^{\mathrm{Test}}(\mathfrak{R})$ is chosen to be the same as the approximation space $V^{\mathrm{PU}}(\Omega)$ (i.e. $u_m(x) = v_m(x), \forall\, m$), i.e. Bubnov-Galerkin approach (Hughes, 2000), leading to the following linear system:

$$\boldsymbol{C}\,\dot{\boldsymbol{w}}(t) = \boldsymbol{A}(t)\,\boldsymbol{w}(t), \qquad (6.21)$$

where:

$$\boldsymbol{C} \in \mathfrak{R}^{M\times M} : c_{j,m} = \int_{S(u_m,u_j)} u_m(x)\, u_j(x)\, dx \qquad (6.22a)$$

$$\boldsymbol{A}(t): t \to \mathfrak{R}^{M\times M} : a_{j,m} = \int_{S(u_m,u_j)} \left[ q(x,t) - \frac{\partial B(x,t)}{\partial x} \right] u_m(x) \frac{\partial u_j(x)}{\partial x}\, dx - \\ - \int_{S(u_m,u_j)} B(x,t) \frac{\partial u_m}{\partial x} \frac{\partial u_j(x)}{\partial x} dx, \qquad (6.22b)$$

where $S(u_m,u_j) = \mathrm{supp}(u_m(\cdot)) \cap \mathrm{supp}(u_j(\cdot))$. Note that the boundary terms of Eq. (6.18) do not appear in Eq. (6.22b), since they are all zero because of the properties of PU functions. Non-trivial integrals appearing in elements of matrix $\boldsymbol{C}$, Eq. (6.22a), are computed using analytically calculated forms for the implemented representation, and integrals in matrix $\boldsymbol{A}(t)$ Eq. (6.22b), are also computed analytically in the case of pdf evolution equations with polynomial drift and diffusion coefficients.

Approximating the time derivative in Eq. (6.21) by means of a Crank-Nicolson scheme and grouping the terms with respect to their temporal argument, we obtain:

$$\left( \boldsymbol{C} - \frac{\Delta t}{2} \boldsymbol{A}(t + \Delta t) \right) \boldsymbol{w}(t + \Delta t) = \left( \frac{\Delta t}{2} \boldsymbol{A}(t) + \boldsymbol{C} \right) \boldsymbol{w}(t) \tag{6.23}$$

where $\Delta t$ is the timestep. For each $t$, Eq. (6.23) leads to an algebraic system of equations which can be solved to obtain $\boldsymbol{w}(t + \Delta t)$ in terms of $\boldsymbol{w}(t)$. Initialization of the numerical scheme requires knowledge of initial values of the weights $w(t_0) = w_0$, which are obtained by fitting the PU representation $\hat{f}(x, t_0)$ to the known initial data $f_0(x)$:

$$\boldsymbol{C}\, \boldsymbol{w}_0 = \boldsymbol{f} \qquad \text{where} \tag{6.24}$$

$$\boldsymbol{f} \in \mathbb{R}^M : \; f_j = \int_{S(u_j)} f_0(x)\, u_j(x)\, dx . \tag{6.25}$$

Numerical solution of the linear systems (6.23), (6.24) is performed via the SciPy (Jones, Oliphant, Peterson, & Al., 2001) interface to LAPACK's (Anderson et al., 1999) routine GESV. For solving the problem $\boldsymbol{A}\boldsymbol{X} = \boldsymbol{B}$, GESV performs an LU-decomposition with partial pivoting and row interchanges to matrix $\boldsymbol{A}$, namely $\boldsymbol{A} = \boldsymbol{P}\boldsymbol{L}\boldsymbol{U}$, where $\boldsymbol{P}$ is a permutation matrix, $\boldsymbol{L}$ is unit lower triangular, and $\boldsymbol{U}$ is upper triangular; the factored form of matrix $\boldsymbol{A}$ is used to solve the original system.

### 6.1.4 Validation of the proposed scheme for the linear-additive case

Having described the numerical scheme for the solution of scalar pdf evolution equations, we shall now present a first application to the simplest case that corresponds to the linear RDE under additive coloured noise excitation:

$$\dot{X}(t;\theta) = \eta X(t;\theta) + \kappa \Xi(t;\theta), \; X(t_0;\theta) = X_0(\theta). \tag{6.28a,b}$$

As it is derived in paragraph 4.1.2, the response pdf of RDE (6.28) satisfies *exactly* the following pdf evolution equation

$$\frac{\partial f_{X(t)}(x)}{\partial t} + \frac{\partial}{\partial x}\left[ \left( \eta x + \kappa m_{\Xi}(t) \right) f_{X(t)}(x) \right] = D^{\text{eff}}(t) \frac{\partial^2 f_{X(t)}(x)}{\partial x^2}, \tag{6.29}$$

with effective noise intensity

$$D^{\text{eff}}(t) = \kappa\, e^{\eta(t - t_0)} C_{X_0 \Xi}(t) + \kappa^2 \int_{t_0}^{t} e^{\eta(t-s)} C_{\Xi\Xi}(t, s)\, ds . \tag{6.30}$$

As proved in the aforementioned paragraph, exact pdf evolution equation (6.29) has the Gaussian pdf

$$f_{X(t)}(x) = \frac{1}{\sqrt{2\pi \sigma_X^2(t)}} \exp\left[ -\frac{1}{2} \frac{\left( x - m_X(t) \right)^2}{\sigma_X^2(t)} \right] \tag{6.31}$$

as its unique solution, with $m_X(t)$, $\sigma_X^2(t)$ being expressed in closed form via Eqs. (4.23), (4.27):

$$m_X(t) = m_{X_0} e^{\eta(t - t_0)} + \kappa \int_{t_0}^{t} m_{\Xi}(\tau)\, e^{\eta(t-\tau)}\, d\tau , \tag{6.32}$$

and

$$\sigma_X^2(t) = \sigma_{X_0}^2 e^{2\eta(t-t_0)} + 2\int_{t_0}^{t} D^{\text{eff}}(\tau)\, e^{2\eta(t-\tau)}\, d\tau\,. \tag{6.33}$$

The numerical solution to pdf evolution Eq. (6.29), obtained by using PUFEM with 200 degrees of freedom (50 PU functions, each with 4 basis functions), is compared to the analytic solution (6.31) in Fig. 3 (left panels, a and c). Besides, in the right panels of Fig. 3, the approximate pdf $f_{X(t)}(x)$, obtained by direct Monte Carlo (MC) simulations of RIVP (6.28a,b), is compared to the exact solution (6.31). For the MC simulations, $5{\cdot}10^4$ realizations were solved and the pdf was approximated using a kernel density estimator (D. W. Scott, 2015). As can be seen from Fig.3, numerical results indicate that both the PUFEM numerical scheme and the MC simulations are able to accurately approximate the analytical solution. These results provide a first validation of the proposed new pdf equation, as well as the implementation of the PUFEM numerical scheme for its solution.

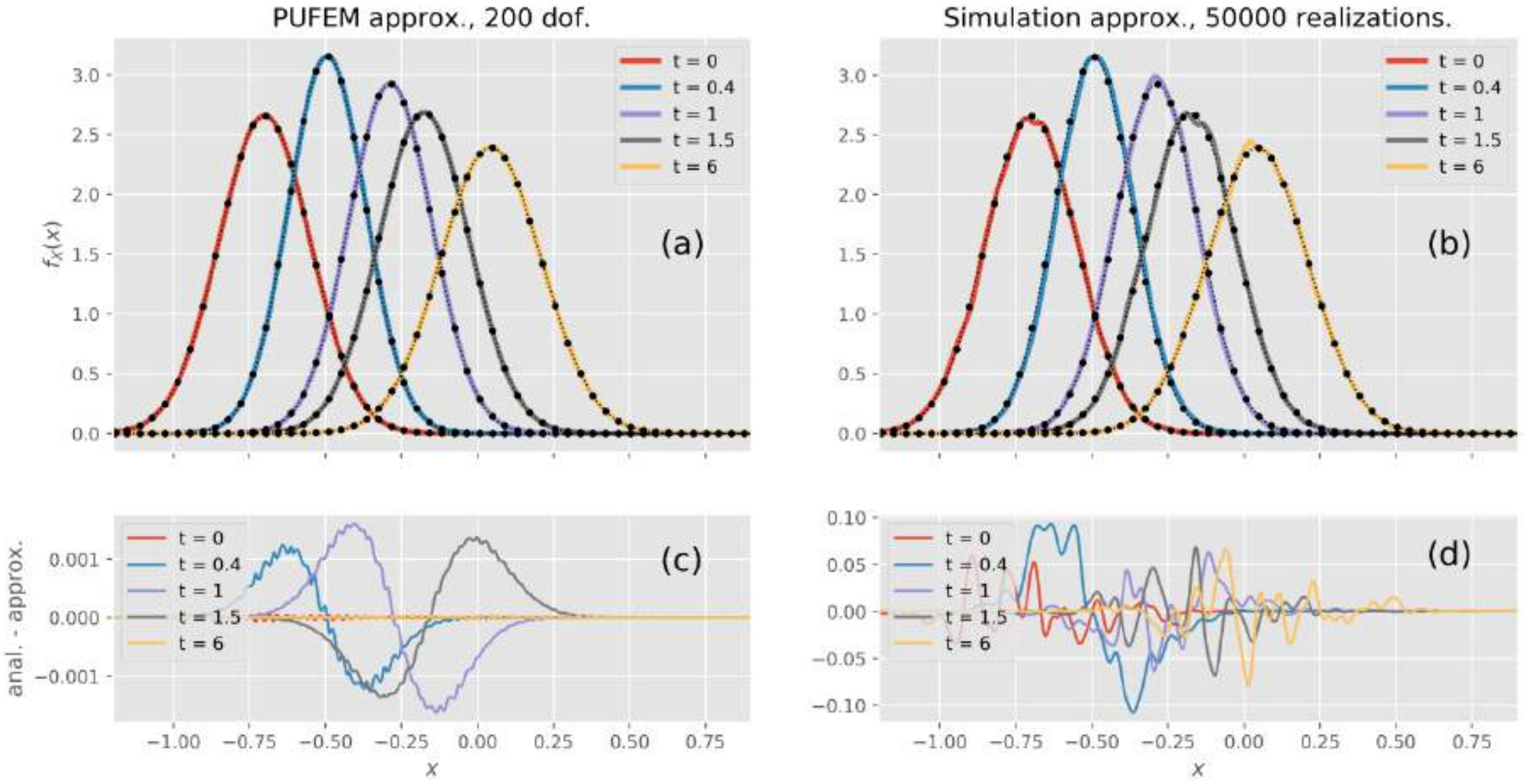

**Figure 3.** Evolution of response pdf for the linear RIVP (6.28a,b): $\eta = -0.8$, $\kappa = 0.2$, under non-zero mean OU excitation ($m_\Xi = 0.2$, $D = 1$, $\tau = 1$) and Gaussian initial data ($m_{X_0} = -0.7$, $\sigma_{X_0} = 0.15$). Analytical solutions (round markers in panels (a) and (b)) are compared to the PUFEM solution (panel (a)) and MC simulation approximation (panel (b)); errors of PUFEM and MC against the analytical solution are shown in panels (c) and (d), respectively.

### 6.1.5 Treatment of nonlinear, nonlocal terms. Algorithm of numerical solution scheme

As discussed above, a feature peculiar to novel pdf evolution equations, corresponding to nonlinear RDEs, is their nonlocality and nonlinearity, coming from the dependence of the diffusion coefficient on the unknown pdf through a specific moment. The exact form of diffusion coefficient reads now as follows, see also Eqs. (5.24), (5.25) of the previous Chapter:

$$\mathcal{B}[f;x,t] = \sum_{m=0}^{M} \frac{1}{m!} D_m[R_{h'}(\bullet|_{t_0}^{t});t]\, \varphi_{h'}^{m}\big(x;R_{h'}(t)\big), \tag{6.34}$$

where $R_{h'}(s) = \Xi^{\theta}\left[h'\left(X(s;\theta)\right)\right]$. As can be seen in Eq. (6.34), $\mathcal{B}[f;x,t]$ depends on the current-time value of moment $R_{h'}(t)$, as well as on the whole time history of the same moment $R_{h'}(\bullet)$, via coefficients $D_m$, see Eq. (5.25):

$$D_m[R_{h'}(\bullet|_{t_0}^{t}),t] = \kappa \exp\left(\int_{t_0}^{t} R_{h'}(u)\,du\right) C_{X_0\Xi}(t)\,(t-t_0)^m + \\ + \kappa^2 \int_{t_0}^{t} \exp\left(\int_{s}^{t} R_{h'}(u)\,du\right) C_{\Xi\Xi}(t,s)\,(t-s)^m\,ds. \tag{6.35}$$

Thus, in each time step, we have to calculate the instantaneous values of the moment $R_{h'}(t)$, as well as the integral $\int_s^t R_{h'}(u)\,du$. This is done by means of a prediction-correction iterative scheme, which is incorporated in the following algorithmic presentation of the whole scheme for the numerical solutions of pdf evolution equations.

**<u>Initial data problem</u>** $t = t_0$**:**

**Calculate** matrix $\boldsymbol{C}$ from Eq. (6.22a).

**Solve the initial data problem** (6.24) to obtain initial weights $\boldsymbol{w}_0$.

**<u>First time step</u>** $i = 1$**:**

**Calculate** $R_{h'}(t_0)$ from the given initial pdf

**Set** $R_{h'}(t_1) = R_{h'}(t_0)$

(*) **Set** $\int R_{h'}(u)\,du = R_{h'}(t_1)\,\Delta t$, where $\Delta t$ is the time step

**Calculate** matrices $\boldsymbol{A}(t_0)$ and $\boldsymbol{A}(t_1)$ from Eq. 6.22b.

**Solve system** (6.23) to obtain weights $\boldsymbol{w}(t_1)$

**Calculate** $f(x,t_1)$ from representation (6.19)

**Update** $R_{h'}(t_1)$ to a new value $R_{h'}^{\text{upd}}(t_1)$, by using $f(x,t_1)$

**Check convergence condition** $\left|R_{h'}(t_1) - R_{h'}^{\text{upd}}(t_1)\right| \le \varepsilon_{\text{tol}}$, for a given error $\varepsilon_{\text{tol}}$

**If** convergence condition is **not satisfied**

**Set** $R_{h'}(t_1) = R_{h'}^{\text{upd}}(t_1)$ and **return** to line (*)

**<u>Time marching</u>** $i = 2, 3, \ldots, I$**:**

**Calculate** $R_{h'}(t_i)$ via linear extrapolation of $R_{h'}(t_{i-1})$ and $R_{h'}(t_{i-2})$

(†) **Calculate** $\int R_{h'}(u)\,du$ using current and previous values (stored) $R_{h'}(t_k)$, $k = 0(1)i$

**Calculate** matrices $\boldsymbol{A}(t_{i-1})$ and $\boldsymbol{A}(t_i)$ from Eq. (6.22b).

**Solve system** (6.23) to obtain weights $\boldsymbol{w}(t_i)$

**Calculate** $f(x,t_i)$ from representation (6.19)

**Update** $R_{h'}(t_i)$ to a new value $R_{h'}^{\text{upd}}(t_i)$, by using $f(x,t_i)$

**Check convergence condition** $\left|R_{h'}(t_i) - R_{h'}^{\text{upd}}(t_i)\right| \le \varepsilon_{\text{tol}}$, for a given error $\varepsilon_{\text{tol}}$

**If** convergence condition is **not satisfied**
**Set** $R_{h'}(t_i) = R_{h'}^{\text{upd}}(t_i)$ and **return** to line (†)

In every time step, the iterative scheme typically converges within only a few (1-2) iterations.

## 6.2 Results for the benchmark bistable case under Ornstein-Uhlenbeck excitation[35]

As a first case to formulate and solve numerically the novel equations for the response pdf, we consider the nonlinear, bistable RDE under additive, zero-mean OU noise:

$$\dot{X}(t;\theta) = \eta_1 X(t;\theta) + \eta_3 X^3(t;\theta) + \kappa \Xi(t;\theta), \quad X(t_0;\theta) = X_0(\theta), \quad \text{(6.36a,b)}$$

with $\eta_1 > 0$, $\eta_3 < 0$. In this case, initial value $X_0(\theta)$ is taken uncorrelated to the excitation. The bistable RIVP (6.36a,b) is the usual benchmark case considered in most of the relevant works up to now, see e.g. (Fox, 1986a; Grigolini, 1986; Hänggi, 1989; Hänggi & Jung, 1995, sec. VI; Hänggi et al., 1984, 1985; Ridolfi et al., 2011, sec. 2.5; Sancho & San Miguel, 1989).

**Normalization of the RDE.** Following (Hänggi & Jung, 1995, sec. VI), we introduce the dimensionless variables $\tilde{t} = \eta_1 t$, $\tilde{X} = X\sqrt{|\eta_3|/\eta_1}$, $\tilde{X}_0 = X_0\sqrt{|\eta_3|/\eta_1}$, $\tilde{\Xi} = \kappa \Xi \sqrt{|\eta_3|/\eta_1^3}$. Thus, RIVP (6.36a,b) is rescaled to the dimensionless (normalized) form

$$\dot{\tilde{X}}(\tilde{t};\theta) = \tilde{X}(\tilde{t};\theta) - \tilde{X}^3(\tilde{t};\theta) + \tilde{\Xi}(\tilde{t};\theta), \qquad \tilde{X}(\tilde{t}_0;\theta) = \tilde{X}_0(\theta). \quad \text{(6.37a,b)}$$

The determination of autocorrelation function of the normalized $\tilde{\Xi}(\tilde{t};\theta)$ calls for the normalization of the correlation time $\tau_{\text{cor}}$ and intensity $D_{\text{OU}}$ of OU noise excitation. For this purpose, the relaxation time $\tau_{\text{rel}}$ of the homogeneous variant ($\Xi(t;\theta) = 0$) of Eq. (6.36a) is chosen as reference time. Since homogeneous Eq. (6.36a) is a Bernoulli equation, its relaxation time is found to be $\tau_{\text{rel}} = 1/(2\eta_1)$ (by studying the long-time behaviour of analytic solution). Thus, the dimensionless correlation time, called also *relative correlation time*, is defined as[36]

$$\tilde{\tau} = \tau_{\text{cor}} / \tau_{\text{rel}} = 2\eta_1 \tau_{\text{cor}}. \quad \text{(6.38)}$$

On the basis of Eq. (6.38) and the definition of the normalized excitation $\tilde{\Xi}(\tilde{t};\theta)$, the autocorrelation function of the latter is expressed as

$$C_{\tilde{\Xi}\tilde{\Xi}}(\tilde{t}, \tilde{s}) = \frac{\tilde{D}}{\tilde{\tau}} \exp\left( -\frac{2|\tilde{t}-\tilde{s}|}{\tilde{\tau}} \right), \quad \text{(6.39)}$$

with dimensionless intensity

$$\tilde{D} = 2\kappa^2 D_{\text{OU}} \frac{|\eta_3|}{\eta_1^2}. \quad \text{(6.40)}$$

---

[35] Results of Section 6.2 were presented in Sec. 4 and Appendix D of (Mamis et al., 2019).

[36] Note that, in (Hänggi & Jung, 1995, sec. VI), as well as in most works in the existing literature, the normalization of $\tau_{\text{cor}}$ is not performed by using $\tau_{\text{rel}} = 1/(2\eta_1)$, opting instead for the use of the same normalization as for time variable; $\tilde{\tau} = \eta_1 \tau_{\text{cor}}$, which also results in $\tilde{D} = \kappa^2 D_{\text{OU}} |\eta_3|/\eta_1^2$. While this discrepancy is not significant, we feel that our normalization choice is better justified from a physics point of view.

Now, as a first example, we consider the dimensionless RIVP (6.37a,b) with $\tilde{D}=1$, and $\tilde{\tau}$ taking values in the range $0.1-3.0$. That is, we study a *strongly nonlinear, bistable* case, under *strong random excitation* outside of the small-noise intensity regime (Hänggi & Jung, 1995; Ridolfi et al., 2011), over a *wide range of relative correlation times*. Initial pdf is taken to be Gaussian with zero mean value and variance $\sigma = 0.6$.

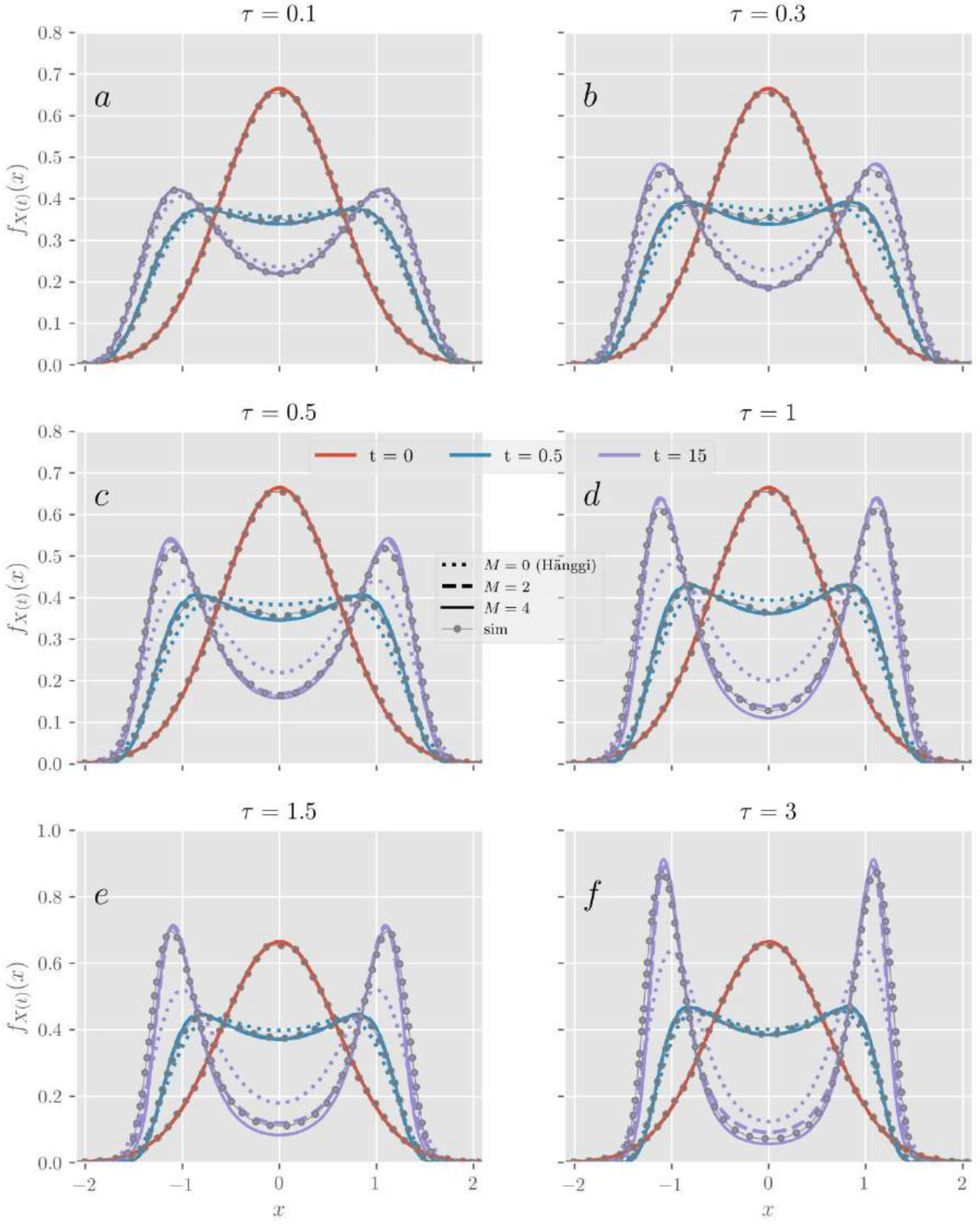

**Figure 4.** Evolution of response pdf for the RIVP (6.37a,b) excited by a zero-mean OU process with $D=1$ and $\tau = 0.1, 0.3, 0.5, 1.0, 1.5, 3.0$. Initial pdf is zero-mean Gaussian with $\sigma = 0.6$. Results from various methods are presented along with MC simulations.

In Figure 4 we present the evolution of the response pdf for increasing values of the ratio $\tilde{\tau} = \tau_{\text{cor}} / \tau_{\text{rel}}$, as calculated by solving the:

0th-order response pdf evolution Eq. (5.24) with $M = 0$ (Hänggi's equation),
2nd-order response pdf evolution Eq. (5.24) with $M = 2$ and
4th-order response pdf evolution Eq. (5.24) with $M = 4$.

In the same figure, results obtained by Monte Carlo (MC) simulations are plotted, denoted by sim in the legends, for comparison purposes. The choice of novel pdf evolution equations of even order ($M = 0, 2, 4$) is made to ensure the global positivity of the corresponding diffusion coefficients. Note that the final time instant in all plots is chosen to be in the long-time stationary regime, in order to check the validity of pdf evolution equations in both the transient and the stationary regimes. Increasing $M$ induces no significant increase in the ***computational effort*** for solving the pdf evolution equations. However, for $M = 4$, the emergence of higher-order polynomials in the weak form of the equation induces complicacies in the numerical implementation that are discussed at length in (Kapelonis, 2020). This, in conjunction with the fact that the resulting pdf for $M = 4$ is only slightly better than the respective pdf for $M = 2$, prevented us from considering pdf evolution equations of higher order, e.g. $M = 6$.

As shown in Figure 4, for small values of $\tau$ (Fig. 4a, $\tau = \tau_{\text{cor}} / \tau_{\text{rel}} = 0.1$)[37], all pdf evolution equations predict a time evolution of response pdf in agreement with MC simulations, both in the transient and the steady state regime. Only Hänggi's equation slightly underestimates the peak values of the stationary pdf. As $\tau$ increases (Fig. 4b, $\tau = 0.3$) SCT Hänggi's underestimates the pdf peak values more, while both 2nd and 4th-order pdf equations are in almost full agreement with MC simulations. This picture is practically the same in Fig. 4c ($\tau = 0.5$), with Hänggi's being even worse. For larger values of $\tau$ (Fig. 4d, $\tau = 1$), both novel pdf evolution equations are fairly accurate; however, they are a bit off at the peak values of the stationary pdf. For even larger values of $\tau$ (Fig. 4e, $\tau = 1.5$ and Fig. 4f, $\tau = 3$) Hänggi's fails totally, while $M = 2$ and $M = 4$ equations provide fairly accurate approximations, except for a minor failure at predicting the pdf peak values. Besides, the abscissae of the peak values predicted by the novel pdf equations are somewhat shifted closer to zero, in comparison with the ones of the MC results.

Having examined a case where pdf evolution equations were solved in both the transient and the stationary regime, we move on to a ***more detailed investigation*** of the range $(D, \tau)$ in which pdf evolution equations with $M = 0, 2, 4$ yield accurate results, by comparing their stationary pdf to MC simulations, shown in Figures 5a,b. For the MC simulations, the end of the transient state, $t_{\text{st}}$, is estimated by monitoring the stationarity (time invariance) of the first two moments of the response. From this time instant $t_{\text{st}}$ forward, we keep all simulation samples, with time step equal to the double of correlation time $\tau$ of excitation. This choice of time step ensures that we obtain uncorrelated samples. On the other hand, the time-dependent pdf evolution equations with $M = 0, 2, 4$ are solved from the initial time up to their stationary regime. The end of the transient state of pdf evolution equations is taken equal to $t_{\text{st}}$ of MC simulations; then we check that their solutions remain unchanged for time equal to $2t_{\text{st}}$, to ensure that they are indeed stationary. By observing Figures 5a,b, we come up with the following three remarks:

[37] From now on (in the text and in figures' captions) we omit the tilde from the non-dimensional quantities.

**Remark 6.1: On the range of validity of Hänggi's ansatz.** For $D = 0.2$ and $1.0$ (Fig. 4a), Hänggi's stationary pdfs are good approximations for $\tau = 0.1$ and $1.0$, while for larger noise intensity ($D = 2.0$ in Fig. 4b), the good performance of Hänggi's method is restricted to the small correlation time regime ($\tau = 0.1$). For even larger noise intensities ($D = 5.0$ in Fig. 4b), Hänggi's equation fails even for $\tau = 0.1$. In general, Hänggi's method fails to capture the phenomenon of higher and narrower pdf peaks as the parameter $D\tau$ increases, resulting thus in pdfs that are more diffusive than the ones obtained by MC simulations. Roughly speaking, Hänggi's equation becomes unreliable when $D\tau$ becomes larger than 1.

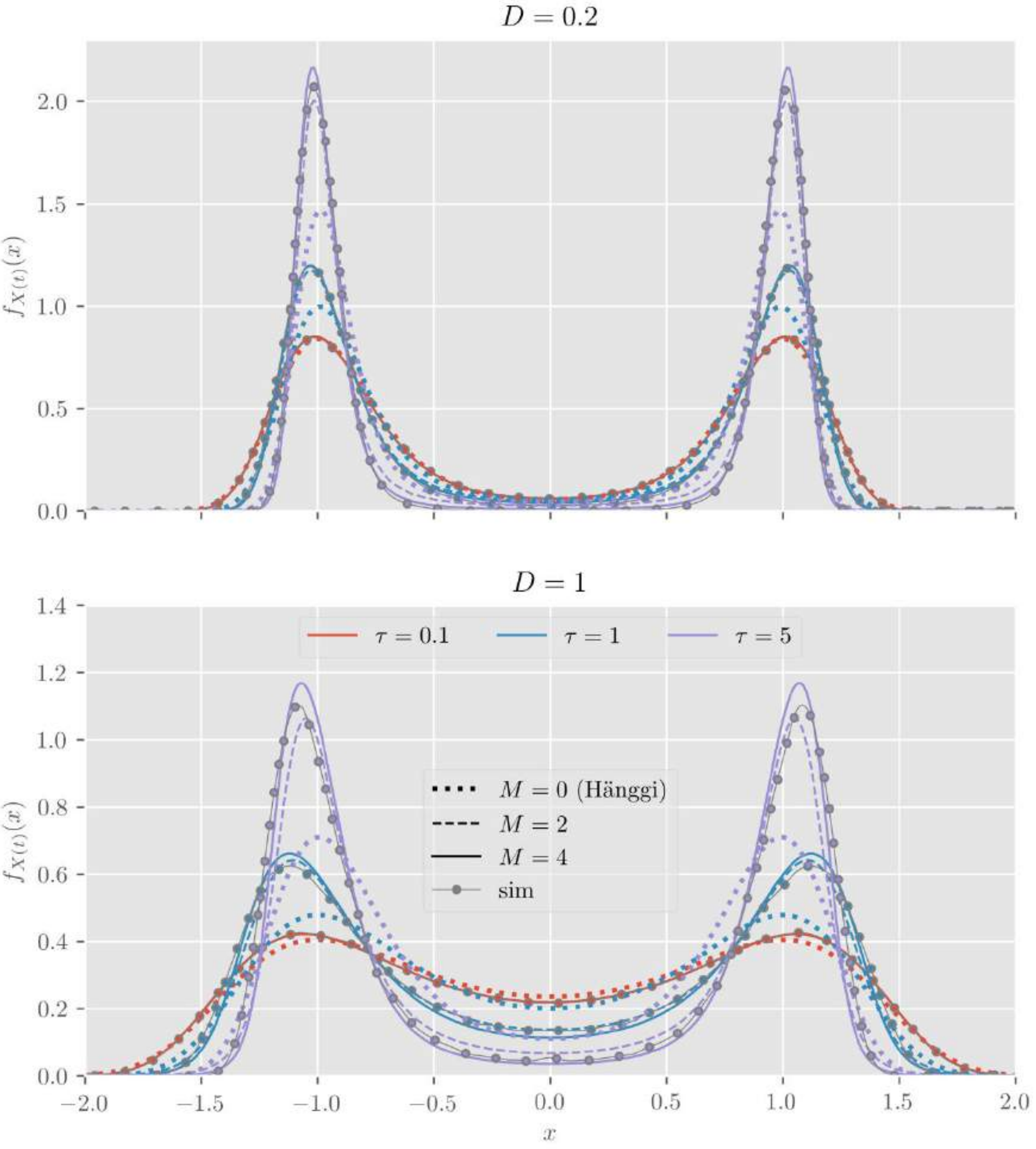

**Figure 5a.** Stationary response pdf for RIVP (6.37a,b) excited by a zero-mean OU process with $D = 0.2$ (upper panel) and $D = 1$ (lower panel), for $\tau = 0.1, 1.0, 5.0$. Results from pdf evolution equations with $M = 0$ ( Hänggi's), and $M = 2$, $4$ are presented along with MC simulations.

**Remark 6.2: On the range of validity of novel pdf evolution equations.** In Figs. 5a,b, we observe that the stationary pdfs obtained by solving the novel pdf evolution equations with $M = 2,\ 4$ are in good agreement with pdfs obtained by MC simulations, even in cases where both $D$ and $\tau$ attain large values. Note that, even for $D = 5.0$ and $\tau = 5.0$, the novel pdf evolution equations continue to give acceptable approximations of the stationary pdf. A general trend is that, as $D$ and $\tau$ increase, their stationary pdfs begin to fail at predicting the exact values and abscissae of the response pdf peaks. Roughly speaking, the range of validity of the novel equations seems to be limited by the condition $D\tau < 25$, providing a substantial improvement in comparison to Hänggi's which is limited by the condition $D\tau < 1$.

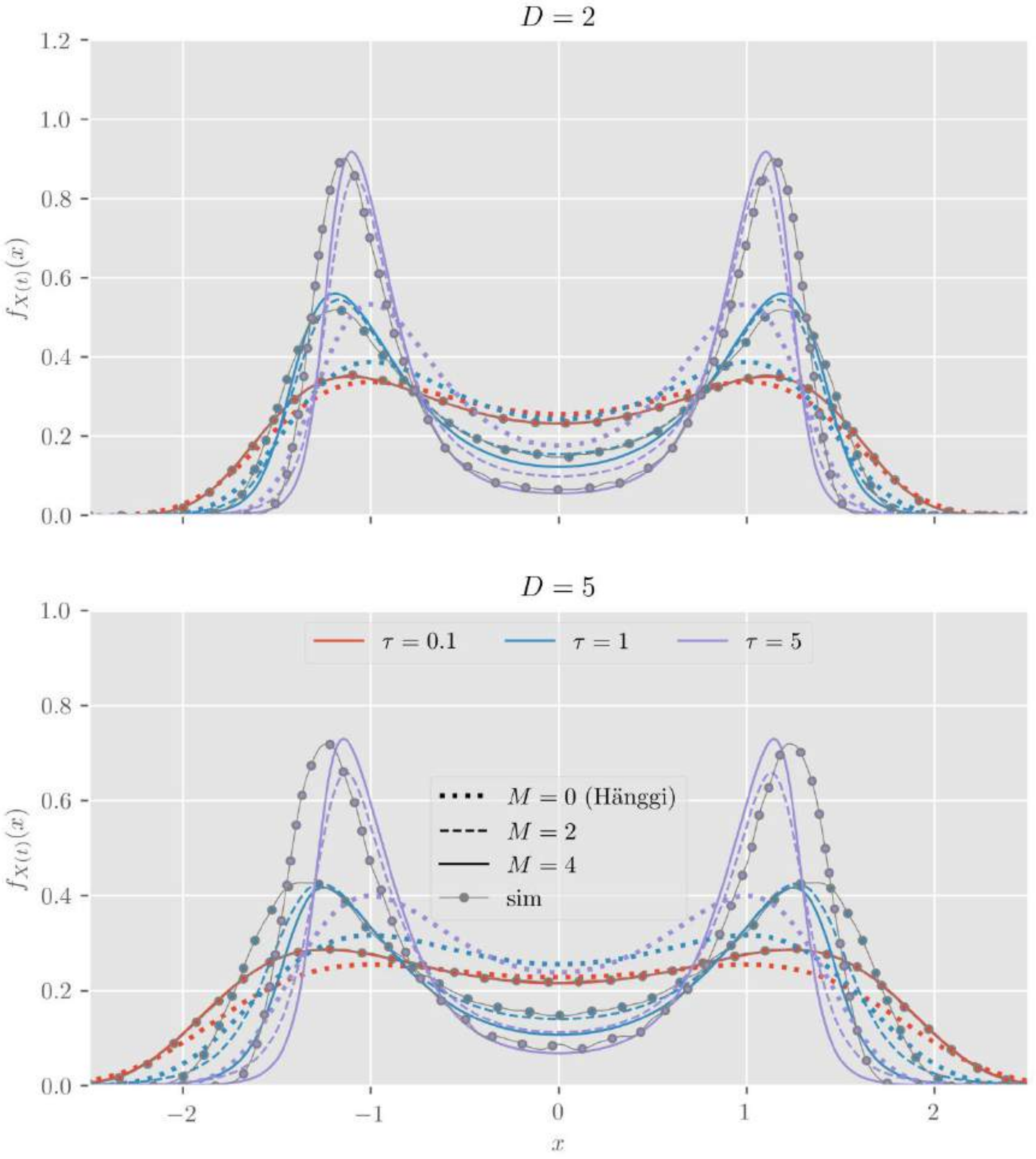

**Figure 5b.** Stationary response pdf for RIVP (6.37a,b) excited by a zero-mean OU process with $D = 2$ (upper panel) and $D = 5$ (lower panel), for $\tau = 0.1, 1.0, 5.0$. Results from pdf evolution equations with $M = 0$ ( Hänggi's), and $M = 2,\ 4$ are presented along with MC simulations.

As $D$ and $\tau$ increase further, the problem becomes more difficult to solve numerically; the numerical scheme, in its present form, exhibits instabilities or divergence towards the end of the transient state for various cases where $D\tau > 25$. The numerical solution of equation with $M = 4$ fails earlier than the one with $M = 2$ (and Hänggi's). This is reasonable, since, in 4$^{\text{th}}$-order pdf equation, higher-order polynomial terms are included, amplifying thus any instabilities in the numerical scheme.

**Remark 6.3: On the prediction of peak value drift.** In Figs. 5a,b, we observe that, for large values of $D$ and $\tau$, the pdfs obtained by MC simulations exhibit peak values at abscissae larger than 1 in absolute value. This peak value drift, which has been documented before, see e.g. (Hänggi & Jung, 1995) p. 294, is predicted quite accurately by the novel equations with $M = 2,\ 4$ in the regime of $(D,\ \tau) = [0,\ 5.0] \times [0,\ 5.0]$, with the 4$^{\text{th}}$-order pdf equation being consistently more accurate than the 2$^{\text{nd}}$-order one. Note that this phenomenon is not captured at all by Hänggi's equation ($M = 0$), which predicts pdfs with peaks always fixed at $\pm 1$.

**Hänggi's peaks.** The above finding, regarding the peaks of Hänggi's response pdf, can also be retrieved analytically. First, we remind that the solution to a stationary, FPK-like equation of the form

$$\frac{\partial}{\partial x}\Big(h(x)\ f_X(x)\Big) = \frac{\partial^2}{\partial x^2}\Big(B(x)\ f_X(x)\Big), \tag{6.41}$$

is determined to (Gardiner, 2004, para. 5.2.2; Mamis, 2015, sec. 4.2)

$$f_X(x) = \frac{C}{B(x)} \exp\left(\int \frac{h(x)}{B(x)}\, dx\right), \tag{6.42}$$

where $C$ is the normalizing constant and $\int \bullet\, dx$ denotes the antiderivative. For Hänggi's stationary Eq. (5.19), the diffusion coefficient is specified into the $x$-independent value $B^{\text{Han}} \equiv B^{\text{Han}}[\,f_X(\bullet)] = D/\big(2 - \tau R_{h'}(\infty)\big)$, where $R_{h'}(\infty)$ is the stationary value of an appropriate moment of the response defined in the aforementioned equation[38]. Thus, solution (6.42) is specified into

$$f_X^{\text{Han}}(x) = C \exp\left(\frac{1}{B^{\text{Han}}} \int h(x)\, dx\right). \tag{6.43}$$

By calculating its first derivative, the stationary points of Hänggi's pdf (6.43) are easily identified as the roots of the function $h(x)$. For our case where $h(x) = x - x^3$, its roots are 0, $\pm 1$, and by checking the sign of the second derivative of $f_X^{\text{Han}}(x)$, 0 is identified as a point of local minimum, and $\pm 1$ as points of local maxima.

Thus, while Hänggi's diffusion is of nonlocal character, its independence from $x$ results in stationary points of the pdf that do not depend on neither the intensity or correlation time of the excitation, nor the moments of the response. This is remedied by considering the novel pdf evolution equations for $M = 2,\ 4$.

---

[38] In Eq. (5.19), stationary diffusion coefficient is $D/\big(1 - \tau R_{h'}(\infty)\big)$. The discrepancy with the diffusion coefficient as stated above, $D/\big(2 - \tau R_{h'}(\infty)\big)$, is due to the fact that, here, we employ the dimensionless variants of $D$, $\tau$, as defined by Eqs. (6.38), (6.40).

# Chapter 7: Conclusions and future directions

**Recapitulation.** In the present thesis, the methodology for deriving evolution equations for pdfs that describe the probabilistic structure of the response to a dynamical system under Gaussian coloured noise has been revisited and extended. More specifically, we rederived and generalized the main steps of the methodology, which are the formulation of *stochastic Liouville equations via delta projection formalism* (Chapter 2), the *Novikov-Furutsu theorem* (Chapter 3), the calculation of *variational derivatives of the response* (Chapter 4), and last, the *current-time closure schemes for the nonlocal terms* of the stochastic Liouville equations (Chapter 5), in order to obtain pdf evolution equations in closed form. This layout served also the first goal of this thesis, which is the presentation of the existing practices in a unified way. In Chapter 5, we also proposed a *novel current-time scheme* that retained a tractable amount of the original nonlocality of the stochastic Liouville equation, by employing the history of an appropriate moment of the response (or a joint moment of response and excitation in the case of the response-excitation SLE). Application of this closure scheme resulted in a family of new pdf evolution equations belonging to the class of nonlinear FPKs (Frank, 2005), that also contain the widely-used Hänggi's equation (Hänggi & Jung, 1995, sec. V.B) as a special case. In Chapter 6, we considered the benchmark case of a bistable RDE under Ornstein-Uhlenbeck excitation. For this case, the new, one-time response pdf evolution equations were solved and compared to results from Monte Carlo simulations. The correct prediction of the response pdf to this benchmark case was important, since it is a bimodal pdf, and thus essentially different than the unimodal Gaussian paradigm. The numerical solution scheme, presented briefly in Sec. 6.1, as well as the source code for the Monte Carlo simulations, were developed and implemented by colleague Z.G. Kapelonis. As shown in the figures of Sec. 6.2, the new pdf evolution equations are *valid even in the regime of high noise intensity and large correlation times of the excitation*, i.e. away from the white noise limit. What is more, the computational effort for solving the new response pdf evolution equations is comparable to the effort required for solving the respective classical FPK equation with one state variable.

**Numerical solution of 2D pdf evolution equations.** The first, straightforward direction for future works is the extension of the numerical solver for the case of multidimensional pdf evolution equations. More specifically, the solver for two-dimensional equations, which is currently under development by our scientific group, will allow for the solution and testing of the other pdf evolution equations derived in Chapter 5, namely the equations for the joint, one-time response-excitation pdf, the joint two-time response pdf, as well as the one-time response pdf for systems with two state variables, such as random oscillators. The study of random oscillators is clearly more interesting from an engineering point of view, and can also lead to applications in the field of our School, such as ***ship motions under random sea wave excitation.*** As a first example, we can consider the roll motion regarding the angular displacement of ship around its fore-and-aft axis. A typical roll model is an oscillator whose restoring and damping are nonlinear (Oh, Nayfeh, & Mook, 2000; Taylan, 2000). Note that the problem of roll motion under irregular waves, i.e. coloured random excitation, has been treated before in (Francescutto & Naito, 2004), via the filtering method and the resulting FPK equation for the augmented system.

**First-passage problems.** Furthermore, the methodology presented herein may be applied for the formulation of evolution equations for other pdfs of significance. In this direction, the equation for the joint, two-time pdf of the response and its first temporal derivative was considered in the diploma thesis of I. Mavromatis, a member of our group. The aforementioned pdf is useful in answering to first-passage problems (Verechtchaguina, Sokolov, & Schimansky-Geier, 2006).

**Turbulence.** We have to also note that representing the desired pdf as the average of an appropriate random delta function (the method we called "delta projection" herein), in order to derive evolution equations for it, has been already used extensively in turbulence (Pope, 1985), where it is termed the "pdf method". In particular, in a series of works (Bragg, Swailes, & Skartlien, 2012; Hyland et al., 1999; Jaume Masoliver & Wang, 1995; Reeks, 1991, 2005; Reeks, Simonin, & Fede, 2016; Swailes & Darbyshire, 1997, 1999; van Dijk & Swailes, 2010, 2012) the problem of *particle transport in random media or in turbulent flows* has been approached by the derivation of FPK-like equations via the delta projection method. The main difference between the present thesid and the aforementioned works is that, in turbulence and random media, excitation is a random field, exhibiting thus correlations in both time and space. In this case, the Novikov-Furutsu theorem is appropriately generalized, see e.g. (Swailes & Darbyshire, 1997). Similarly to our case, the evolution equation for the joint pdf of the displacement and velocity of the particle is non-closed. Thus, one could consider adjusting the novel closure proposed herein for the closure of the said equation.

**Random PDEs.** Last, a broad direction for future works, that is worth to be mentioned, is the generalization of the methodology to PDEs excited by random fields. Similarly to the derivation of the classical FPK equation corresponding to ODEs under white noise, the derivation of the functional FPK equation for PDEs under random fields that are delta-correlated in time, is also standard, see e.g. (Gardiner, 2004, para. 8.1.1). Interestingly enough, in the aforementioned work, the functional FPK equation was derived as the Volterra's continuous limit of the multidimensional classical FPK equation, a practice that we also favoured in the present thesis, and was presented in Appendix A. By extending the delta projection method to represent the probability density functional of the random field of the response to a PDE, we can commence the derivation of *functional FPK-like equations corresponding to PDEs* under the excitation of random fields that are correlated both spatially and temporally.

**Beyond the restriction of Gaussian excitation.** The main restriction of our methodology is the assumption that the coloured excitation is a Gaussian process. The step in which this assuption is made is the application of the Novikov-Furutsu theorem, which has been proven for mappings with Gaussian arguments. Thus, for the restriction to be lifted, appropriate generalizations of the NF theorem have to be performed. Such an attempt was performed in Sec. 4.6, where response pdf evolution equations for systems under polynomially Gaussian excitation were derived, making use of the generalized NF theorem of Sec. 3.5. A more promising direction is to represent the non-Gaussian excitation pdfs as *mixtures of Gaussian densities*. The appropriate generalization of NF theorem needed for this approach is under consideration.

# Appendix A: Volterra's principle of passing from discrete to continuous. Volterra calculus[39]

A naïve, yet useful, way to describe Volterra's principle of passing from the discrete to continuous is the following. Consider the discrete quantities $\Xi_n$, $u_n$, $n = 1(1)N$, obtained by sampling the continuous functions $\Xi(t)$, $u(t)$ over the interval $[a, b]$:

$$\Xi_n = \Xi(t_n), \qquad u_n = u(t_n), \qquad t_n \in [a, b].$$

Assume further that the points $t_n$ are ordered, $a \le t_1 < t_2 < \cdots < t_n < \cdots < t_N \le b$, and, as $N \to \infty$, all increments $\Delta t_n = t_{n+1} - t_n$, $n = 1(1)(N-1)$, $\Delta t_0 = t_1 - a$ and $\Delta t_N = b - t_N$, tend to zero. Under this assumption, Volterra's principle of passing from the discrete to continuous consists of the replacement of the sums

$$\sum_{n=1}^{N} \Xi_n = \sum_{n=1}^{N} \Xi(t_n), \qquad \sum_{n=1}^{N} \Xi_n u_n = \sum_{n=1}^{N} \Xi(t_n)\, u(t_n)$$

by

$$\sum_{n=1}^{N} \Xi(s_n)\, \Delta t_n, \qquad \sum_{n=1}^{N} \Xi(s_n)\, u(s_n)\, \Delta t_n,$$ [40]

and their interpretation (in the limit, as $N \to \infty$) as Riemann integrals,

$$\int_a^b \Xi(s)\, ds, \qquad \int_a^b \Xi(s)\, u(s)\, ds.$$

This line of thought is especially useful for the probabilistic study of random functions (in continuous time), since the primitive probabilistic information associated with a random function $\Xi(t;\theta)$ is the joint distribution (or the joint characteristic function) of the random variables $\Xi_n(\theta) = \Xi(t_n;\theta)$, where the time instances $t_1 < t_2 < \cdots < t_n < \cdots < t_N$ are distributed over the (common) domain of definition of path functions. For example, considering the joint characteristic function of the random variables $X_0(\theta)$, $\Xi(s_1;\theta)$, $\Xi(s_2;\theta), \cdots$, $\Xi(s_N;\theta)$,

$$\varphi_{X_0 \Xi(s_1) \cdots \Xi(s_N)}\big(\upsilon; u(s_1), \ldots, u(s_N)\big) = \mathbb{E}^{\theta}\left[\exp\left(i X_0(\theta)\,\upsilon + i \sum_{n=1}^{N} \Xi(s_n;\theta)\, u(s_n)\, \Delta s\right)\right], \tag{A.1}$$

[39] The present Appendix was initially published in (Athanassoulis & Mamis, 2019).

[40] $s_n \in [t_n, t_{n+1}]$. Any choice of $s_n$ leads to the same results, since the functions $\Xi(t)$, $u(t)$ are assumed to be continuous.

and applying the above described Volterra's passing from the discrete to continuous, we obtain the joint characteristic function-functional (FFℓ) of $X_0(\theta)$, $\Xi(t;\theta)$:

$$\varphi_{X_0\Xi(\cdot)}[\upsilon\,;u(\bullet|_{t_0}^{t})] = \mathbb{E}^{\theta}\left[\exp\left(i\,X_0(\theta)\,\upsilon + i\int_{t_0}^{t}\Xi(s;\theta)\,u(s)\,ds\right)\right]. \tag{A.2}$$

Another interesting application of Volterra's passing from the discrete to continuous is the construction of the joint characteristic FFℓ of a jointly Gaussian scalar random variable and scalar random function, which is of fundamental importance for the proof of the extended NF theorem (Chapter 3). In order to find this joint characteristic FFℓ we begin from its discrete analogue, the joint characteristic function of a random variable $X_0(\theta)$ and a random vector $\boldsymbol{\Xi}(\theta)$. This is obtained by simple manipulations of a $(1+N)-$dimensional Gaussian characteristic function (see e.g. (Lukacs & Laha, 1964), Sec. 2.1), and reads as follows:

$$\begin{aligned}\varphi_{X_0\Xi}^{\text{Gauss}}(\upsilon;\boldsymbol{u}) = {} & \exp\left(i\sum_{n=1}^{N}\left(m_{\Xi}\right)_n u_n - \frac{1}{2}\sum_{n=1}^{N}\sum_{m=1}^{N}\left(C_{\Xi\Xi}\right)_{nm} u_n u_m\right)\times \\ & \times\exp\left(i m_{X_0}\,\upsilon - \frac{1}{2}\,C_{X_0X_0}\,\upsilon^2\right)\cdot\exp\left(-\,\upsilon\sum_{n=1}^{N}\left(C_{X_0\Xi}\right)_n u_n\right),\end{aligned} \tag{A.3}$$

where $m_{X_0}$, $C_{X_0X_0}$ are the mean value and the variance of the random variable $X_0(\theta)$, $\boldsymbol{m}_{\Xi}$, $\boldsymbol{C}_{\Xi\Xi}$ are the mean value and autocovariance of the random vector $\boldsymbol{\Xi}(\theta)$, and $\boldsymbol{C}_{X_0\Xi}$ denotes their cross-covariance. By setting $\Xi_n(\theta) = \Xi(t_n;\theta)$ and $u_n = u(t_n)$ and considering the Volterra's limiting process as described above, we obtain the Gaussian joint characteristic FFℓ

$$\begin{aligned}\varphi_{X_0\Xi(\cdot)}^{\text{Gauss}}[\upsilon\,;u(\bullet|_{t_0}^{t})] = {} & \\ = \exp & \left(i\int_{t_0}^{t} m_{\Xi}(s)u(s)\,ds - \frac{1}{2}\int_{t_0}^{t}\int_{t_0}^{t} C_{\Xi\Xi}(s_1,s_2)\,u(s_1)\,u(s_2)\,ds_1 ds_2\right)\times \\ & \times\exp\left(i m_{X_0}\,\upsilon - \frac{1}{2}\,C_{X_0X_0}\,\upsilon^2\right)\cdot\exp\left(-\,\upsilon\int_{t_0}^{t} C_{X_0\Xi}(s)\,u(s)\,ds\right).\end{aligned} \tag{A.4}$$

In Eq. (A.4), we observe that $\varphi_{X_0\Xi(\cdot)}^{\text{Gauss}}[\upsilon\,;u(\bullet|_{t_0}^{t})]$ is the product of the Gaussian characteristic functional of $\Xi(\bullet;\theta)$, the Gaussian characteristic function of $X_0(\theta)$, as well as a term encapsulating their probabilistic dependence via their cross-correlation function $C_{X_0\Xi}(s)$.

Passing from discrete to continuous is only a minor trick of Volterra's approach. The vigor of the latter is amply revealed only when consider the corresponding functional calculus. Volterra functional derivative $\delta\,/\delta u(s)$ is the continuous analogue of the (usual) partial derivative $\partial\,/\partial u_n$, and leads to a well-structured calculus on functionals (Averbukh & Smolyanov, 1967; Volterra, 1930; Volterra & Pérès, 1936). An example illustrating the rigorousness of Volterra calculus, as well as the relation between Volterra functional derivative and the more widely known Gâteaux derivative, is the following derivation of the of Taylor's theorem for functionals, called the Volterra-Taylor theorem:

Let us consider the functional $\mathcal{G}[u(\bullet|_{t_0}^{t})]$, assumed to be $M+1$ times Gâteaux differentiable, with the goal of expanding it around $u_0(\bullet)$. By assuming $u(\bullet)$, $u_0(\bullet)$ as fixed, we define the real-valued function of one scalar variable $g(\varepsilon)=\mathcal{G}\Big[u_0(\bullet|_{t_0}^{t})+\varepsilon\delta u(\bullet|_{t_0}^{t})\Big]$, where $\delta u(\bullet)=u(\bullet)-u_0(\bullet)$. Obviously, $g(\varepsilon)$ is well-defined and continuous in $[0,1]$, has the properties

$$g(0)=\mathcal{G}[u_0(\bullet|_{t_0}^{t})], \qquad \text{and} \qquad g(1)=\mathcal{G}[u(\bullet|_{t_0}^{t})],$$

and it is $M+1$ times differentiable, in the usual sense, in the interval $[0,1]$. Its Taylor expansion, for $g(1)$ around $g(0)$, reads as

$$g(1) = \sum_{m=0}^{M}\frac{1}{m!}\left.\frac{d^m g(\varepsilon)}{d\varepsilon^m}\right|_{\varepsilon=0} + \frac{1}{(M+1)!}\frac{d^{M+1}g(\theta)}{d\varepsilon^{M+1}}, \qquad 0<\theta<1. \tag{A.5}$$

Rewriting Eq. (A5) in terms of the functional $\mathcal{G}[u(\bullet|_{t_0}^{t})]$ we obtain

$$\begin{aligned}\mathcal{G}[u(\bullet|_{t_0}^{t})] = {}& \sum_{m=0}^{M}\frac{1}{m!}\frac{\partial^m}{\partial\varepsilon^m}\,\mathcal{G}\Big[u_0(\bullet|_{t_0}^{t})+\varepsilon\delta u(\bullet|_{t_0}^{t})\Big]\Big|_{\varepsilon=0} + \\ &+\frac{1}{(M+1)!}\frac{\partial^{M+1}}{\partial\varepsilon^{M+1}}\,\mathcal{G}\Big[u_0(\bullet|_{t_0}^{t})+\varepsilon\delta u(\bullet|_{t_0}^{t})\Big]\Big|_{\varepsilon=\theta}, \quad 0<\theta<1.\end{aligned} \tag{A.6}$$

The terms $\frac{\partial^m}{\partial\varepsilon^m}\,\mathcal{G}\Big[u_0(\bullet|_{t_0}^{t})+\varepsilon\delta u(\bullet|_{t_0}^{t})\Big]\Big|_{\varepsilon=0}$ are identified as the Gâteaux functional derivatives (Averbukh & Smolyanov, 1968), of $\mathcal{G}$ calculated at $u_0(\bullet)$ in the direction of $\delta u(\bullet)$ , denoted by $\delta^m\mathcal{G}\Big[u_0(\bullet|_{t_0}^{t})\,;\,\delta u(\bullet|_{t_0}^{t})\Big]$. Thus, Eq. (A.6) can be written as

$$\begin{aligned}\mathcal{G}[u(\bullet|_{t_0}^{t})] = {}& \sum_{m=0}^{M}\frac{1}{m!}\,\delta^m\mathcal{G}\Big[u_0(\bullet|_{t_0}^{t})\,;\,\delta u(\bullet|_{t_0}^{t})\Big] + \\ &+\frac{1}{(M+1)!}\frac{\partial^{M+1}}{\partial\varepsilon^{M+1}}\,\mathcal{G}\Big[u_0(\bullet|_{t_0}^{t})+\varepsilon\delta u(\bullet|_{t_0}^{t})\Big]\Big|_{\varepsilon=\theta}, \quad 0<\theta<1,\end{aligned} \tag{A.7}$$

Eq. (A.7) is the Gâteaux-Taylor series expansion for functionals. Under some technical assumptions (see (Hall, 1978; Hamilton, 1980)), it can be proved that Gâteaux derivatives are expressed in terms of Volterra derivatives by formulae

$$\delta\mathcal{G}\Big[u_0(\bullet|_{t_0}^{t})\,;\,\delta u(\bullet|_{t_0}^{t})\Big] = \int_{t_0}^{t}\frac{\delta\mathcal{G}[u_0(\bullet|_{t_0}^{t})]}{\delta u(s)}\,\delta u(s)\,ds,$$

for the first derivative and, in general,

$$\delta^m\mathcal{G}\Big[u_0(\bullet|_{t_0}^{t})\,;\,\delta u(\bullet|_{t_0}^{t})\Big] = \int_{t_0}^{t}\overset{(m)}{\cdots}\int_{t_0}^{t}\frac{\delta^m\mathcal{G}[u_0(\bullet|_{t_0}^{t})]}{\delta u(s_1)\cdots\delta u(s_m)}\,\delta u(s_1)\cdots\delta u(s_m)\,ds_1\cdots ds_m$$

for the $m^{\text{th}}$ derivative. Using the above equation, the Gâteaux-Taylor expansion, Eq. (A.7), is rewritten as a Volterra-Taylor expansion in the form

$$\mathcal{G}[u(\bullet|_{t_0}^{t})] = \sum_{m=0}^{M} \frac{1}{m!} \int_{t_0}^{t} \overset{(m)}{\cdots} \int_{t_0}^{t} \frac{\delta^m \mathcal{G}[u_0(\bullet|_{t_0}^{t})]}{\delta u(s_1)\cdots \delta u(s_m)} \delta u(s_1)\cdots \delta u(s_m)\, ds_1 \cdots ds_m + \\ + \big(\text{Residual term}\big). \tag{A.8}$$

By considering now that $\mathcal{G}[u(\bullet|_{t_0}^{t})]$ is infinitely Gâteaux-differentiable, and the functional series in the right-hand side of Eq. (A.8) converges, we obtain the infinite Volterra-Taylor series expansion of the functional,

$$\mathcal{G}[u(\bullet|_{t_0}^{t})] = \sum_{m=0}^{\infty} \frac{1}{m!} \int_{t_0}^{t} \overset{(m)}{\cdots} \int_{t_0}^{t} \frac{\delta^m \mathcal{G}[u_0(\bullet|_{t_0}^{t})]}{\delta u(s_1)\cdots \delta u(s_m)} \delta u(s_1) \cdots \delta u(s_m)\, ds_1 \cdots ds_m . \tag{A.9}$$

The structure of Eq. (A.9) is the same with the Taylor expansion for a function $G(\boldsymbol{u})$ of an $N-$dimensional vector $\boldsymbol{u}$ around $\boldsymbol{u}_0$:

$$G(\boldsymbol{u}) = \sum_{m=0}^{\infty} \frac{1}{m!} \sum_{n_1=1}^{N} \overset{(m)}{\cdots} \sum_{n_m=1}^{N} \frac{\partial^m G(\boldsymbol{u}_0)}{\partial u_{n_1} \cdots \partial u_{n_m}} \delta u_{n_1} \cdots \delta u_{n_m} , \tag{A.10}$$

where $\delta u_n = u_n - u_{0n}$. It is easy to see that Eq. (A.9) can be also obtained from Eq. (A.10) by applying the process of passing from the discrete to continuous, as described in the beginning of this section. The above analysis gives a clear indication of the validity of Volterra's principle (under some technical assumptions) and, in addition, establishes the fact the Volterra functional derivative is the continuous analogue of the (discrete) partial derivative.

# Appendix B: Constructive proof of the Novikov-Furutsu theorem for argument raised at any power

In Section 3.5 of Chapter 3, the following extension of the Novikov-Furutsu (NF) theorem for averages containing the Gaussian argument raised at any power, $\mathbb{E}^{\theta}\left[\Xi^{n}(s;\theta)\,\mathcal{F}[\cdots]\right] = \mathbb{E}^{\theta}\left[\,\Xi^{n}(s;\theta)\,\mathcal{F}[X_0(\theta)\,;\,\Xi(\bullet|_{t_0}^{t}\,;\theta)]\,\right]$, was derived, see Eq. (3.75):

$$\mathbb{E}^{\theta}\left[\,\Xi^{n}(s;\theta)\,\mathcal{F}[\cdots]\right] = \sum_{m_1+m_2+m_3=n} \binom{n}{m_1,\, m_2,\, m_3} m_{\Xi}^{m_1}(s)\, C_{X_0\Xi}^{m_2}(s) \sum_{k=0}^{\lfloor m_3/2 \rfloor} H_{m_3,k}\, C_{\Xi\Xi}^{k}(s,s) \times$$

$$\times \int_{t_0}^{t} \overset{(m_3-2k)}{\cdots} \int_{t_0}^{t} \prod_{i=1}^{m_3-2k} C_{\Xi\Xi}(s,\tau_i)\, \mathbb{E}^{\theta}\left[ \frac{\partial^{m_2}\, \delta^{m_3-2k}\, \mathcal{F}[\cdots]}{\partial X_0^{m_2}(\theta) \prod_{i=1}^{m_3-2k} \delta\Xi(\tau_i;\theta)} \right] \prod_{i=1}^{m_3-2k} d\tau_i . \qquad \text{(B.1)}$$

where $H_{n,k} = n!\big/\big(2^{k}\,k!\,(n-2k)!\big)$ are the absolute values of the coefficients of the $n-$th probabilist's Hermite polynomial, see Eqs. (3.72), (3.73). In Section 3.5, Eq. (B.1) was deduced from the more general NF theorem, Eq. (3.76). In the present Appendix, Eq. (B.1) will be proven independently, and for $s \in [t_0, t]$, by using the averaged shift operators $\bar{\mathcal{T}}^{(2)}$, defined in paragraph 3.4.1.

Thus, the present proof begins with the specification of Eq. (3.43) for $\mathbb{E}^{\theta}\left[\Xi^{n}(s;\theta)\,\mathcal{F}[\cdots]\right]$:

$$\mathbb{E}^{\theta}\left[\,\Xi^{n}(s;\theta)\,\mathcal{F}[X_0(\theta)\,;\,\Xi(\bullet|_{t_0}^{t}\,;\theta)]\right] =$$

$$= \left\{ \bar{\mathcal{T}}_{X_0\Xi}^{(0,2)}\, \bar{\mathcal{T}}_{X_0\Xi}^{(1,1)}\, \bar{\mathcal{T}}_{X_0\Xi}^{(2,0)} \left[ u^{n}(s)\, \mathcal{F}[\upsilon\,;\,u(\bullet|_{t_0}^{t})] \right] \right\}_{\substack{\upsilon = m_{X_0} \\ u(\bullet) = m_{\Xi}(\bullet)}} \qquad \text{(B.2)}$$

As in the proof of the extended NF theorem presented in paragraph 3.4.1, our goal is to determine the action of each operator $\bar{\mathcal{T}}^{(2)}$ on $u^{n}(s)\,\mathcal{F}[\upsilon\,;\,u(\bullet|_{t_0}^{t})]$. For operator $\bar{\mathcal{T}}_{X_0\Xi}^{(2,0)}\bullet$ this is easy, and is given by the following lemma. On the other hand, determining the actions of $\bar{\mathcal{T}}_{X_0\Xi}^{(1,1)}\bullet$, $\bar{\mathcal{T}}_{X_0\Xi}^{(0,2)}\bullet$ is more lengthy, and thus is the topic of the following two sections.

**Lemma B.1.** The action of operator $\bar{\mathcal{T}}^{(2,0)}_{X_0\Xi}$ on $u^n(s)\,\mathcal{F}[\upsilon\,;u(\bullet|_{t_0}^{t})]$ is given by

$$\bar{\mathcal{T}}^{(2,0)}_{X_0\Xi}\left[u^n(s)\,\mathcal{F}[\upsilon\,;u(\bullet|_{t_0}^{t})]\right] = u^n(s)\,\bar{\mathcal{T}}^{(2,0)}_{X_0\Xi}\left[\mathcal{F}[\upsilon\,;u(\bullet|_{t_0}^{t})]\right]. \tag{B.3}$$

**Proof.** By using $\bar{\mathcal{T}}^{(2,0)}_{X_0\Xi}\,\bullet$ in series form, Eq. (3.36a), the left-hand side of Eq. (B.3) is equal to

$$\bar{\mathcal{T}}^{(2,0)}_{X_0\Xi}\left[u^n(s)\,\mathcal{F}[\upsilon\,;u(\bullet|_{t_0}^{t})]\right] = \sum_{p=0}^{\infty}\frac{1}{p!}\,\frac{1}{2^p}\,C^{p}_{X_0X_0}\,\frac{\partial^{2p}\left[u^n(s)\,\mathcal{F}[\upsilon\,;u(\bullet|_{t_0}^{t})]\right]}{\partial\upsilon^{2p}},$$

and since $u^n(s)$ is $\upsilon-$independent:

$$\bar{\mathcal{T}}^{(2,0)}_{X_0\Xi}\left[u^n(t)\,\mathcal{F}[\upsilon\,;u(\bullet|_{t_0}^{t})]\right] = u^n(s)\sum_{p=0}^{\infty}\frac{1}{p!}\,\frac{1}{2^p}\,C^{p}_{X_0X_0}\,\frac{\partial^{2p}\,\mathcal{F}[\upsilon\,;u(\bullet|_{t_0}^{t})]}{\partial\upsilon^{2p}}. \tag{B.4}$$

Again, by virtue of Eq. (3.36a), the series in the right-hand side of Eq. (B.4) is identified as $\bar{\mathcal{T}}^{(2,0)}_{X_0\Xi}\left[\mathcal{F}[\upsilon\,;u(\bullet|_{t_0}^{t})]\right]$, completing thus the proof of Eq. (B.3). ■

Before proceeding to determine the actions of the other two operators $\bar{\mathcal{T}}^{(2)}$, we present two Lemmata that are used subsequently in both proofs.

**Lemma B.2: Product rule for higher-order Volterra derivatives.** For $q-$times Volterra differentiable FFℓs $\mathcal{F}_1[\upsilon\,;u(\bullet|_{t_0}^{t})]=\mathcal{F}_1[\cdots]$, $\mathcal{F}_2[\upsilon\,;u(\bullet|_{t_0}^{t})]=\mathcal{F}_2[\cdots]$, the $q^{\text{th}}$ Volterra derivative of their product is given by

$$\frac{\delta^q\left[\mathcal{F}_1[\cdots]\,\mathcal{F}_2[\cdots]\right]}{\delta u(\tau_1)\cdots\delta u(\tau_q)} = \sum_{m=0}^{q}\;\sum_{A\in\wp_m(\{\tau_1,\dots,\tau_q\})}\frac{\delta^m\,\mathcal{F}_1[\cdots]}{\prod_{\tau\in A}\delta u(\tau)}\,\frac{\delta^{q-m}\,\mathcal{F}_2[\cdots]}{\prod_{\tau\notin A}\delta u(\tau)}, \tag{B.5}$$

where $\wp_m\left(\{\tau_1,\cdots\tau_q\}\right)$ is the family of sets over $\{\tau_1,\cdots\tau_q\}$ whose elements $A$ have cardinality $m$. Lemma B.2 is proven by mathematical induction on index $q$, commencing from $q=1$. For $q=1$, $A\in\left\{\varnothing,\{\tau_1\}\right\}$, and thus Eq. (8) results in the product rule for first-order derivatives:

$$\frac{\delta\left[\mathcal{F}_1[\cdots]\,\mathcal{F}_2[\cdots]\right]}{\delta u(\tau_1)} = \frac{\delta\mathcal{F}_1[\cdots]}{\delta u(\tau_1)}\,\mathcal{F}_2[\cdots]+\mathcal{F}_1[\cdots]\,\frac{\delta\mathcal{F}_2[\cdots]}{\delta u(\tau_1)}.$$

Detailed proof of Lemma B.2 is given in the at the end of the present Appendix. ■

**Lemma B.3: Higher-order Volterra derivatives of $u^n(t)$.** The $m^{\text{th}}$ Volterra derivative of $u^n(t)$ is given by

$$\frac{\delta^m u^n(s)}{\delta u(\tau_1)\cdots\delta u(\tau_m)} = \begin{cases}\dfrac{n!}{(n-m)!}\,u^{n-m}(s)\,\prod\limits_{i=1}^{m}\delta(s-\tau_i) & \text{if } m\le n,\\ 0 & \text{otherwise,}\end{cases} \tag{B.6}$$

where $\delta(s-\tau_i)$ is Dirac's delta function. Eq. (B.6) is easily proven by mathematical induction on index $m$, commencing from the relation for $m=1$:

$$\frac{\delta u^n(s)}{\delta u(\tau_1)} = \begin{cases} n\, u^{n-1}(s)\, \delta(s-\tau_1) & \text{if } n \geq 1, \\ 0 & \text{if } n = 0. \end{cases}$$

which is true by virtue of the chain rule and the fact that $\delta u(s)/\delta u(\tau_1) = \delta(t-\tau_1)$. ■

**B1. The action of operator $\bar{\mathcal{T}}^{(1,1)}_{X_0\Xi}$ on $u^n(s)\,\mathcal{F}[\upsilon\,;u(\bullet|_{t_0}^{t})]$**

By using Eq. (3.36b), $\bar{\mathcal{T}}^{(1,1)}_{X_0\Xi}\left[u^n(s)\,\mathcal{F}[\upsilon\,;u(\bullet|_{t_0}^{t})]\right]$ is expressed in series form as

$$\bar{\mathcal{T}}^{(1,1)}_{X_0\Xi}\left[u^n(s)\,\mathcal{F}[\upsilon\,;u(\bullet|_{t_0}^{t})]\right] =$$

$$= \sum_{p=0}^{\infty}\frac{1}{p!}\int_{t_0}^{t}\overset{(p)}{\cdots}\int_{t_0}^{t} C_{X_0\Xi}(\tau^{(1)})\cdots C_{X_0\Xi}(\tau^{(p)})\,\frac{\partial^p\,\delta^p\left[u^n(s)\,\mathcal{F}[\upsilon\,;u(\bullet|_{t_0}^{t})]\right]}{\partial\upsilon^p\,\delta u(\tau^{(1)})\cdots\delta u(\tau^{(p)})}\,d\tau^{(1)}\cdots d\tau^{(p)}. \tag{B.7}$$

By using Lemma B.2, the derivative in the right-hand side of Eq. (B.7) is further evaluated as

$$\frac{\partial^p\,\delta^p\left[u^n(s)\,\mathcal{F}[\upsilon\,;u(\bullet|_{t_0}^{t})]\right]}{\partial\upsilon^p\,\delta u(\tau^{(1)})\cdots\delta u(\tau^{(p)})} = \frac{\delta^p}{\delta u(\tau^{(1)})\cdots\delta u(\tau^{(p)})}\left[u^n(s)\,\frac{\partial^p\mathcal{F}[\upsilon\,;u(\bullet|_{t_0}^{t})]}{\partial\upsilon^p}\right] =$$

$$= \sum_{m=0}^{p}\;\sum_{A\in\wp_m(S^{(p)})}\frac{\delta^m u^n(s)}{\prod\limits_{\tau\in A}\delta u(\tau)}\,\frac{\partial^p\,\delta^{p-m}\mathcal{F}[\upsilon\,;u(\bullet|_{t_0}^{t})]}{\partial\upsilon^p\prod\limits_{\tau\notin A}\delta u(\tau)}, \tag{B.8}$$

with $S^{(p)}=\{\tau^{(1)},\dots,\tau^{(p)}\}$. Using also Lemma B.3, Eq. (B.8) is expressed as

$$\frac{\partial^p\,\delta^p\left[u^n(s)\,\mathcal{F}[\upsilon\,;u(\bullet|_{t_0}^{t})]\right]}{\partial\upsilon^p\,\delta u(\tau^{(1)})\cdots\delta u(\tau^{(p)})} =$$

$$= \sum_{m=0}^{\min\{p,n\}}\frac{n!}{(n-m)!}\,u^{n-m}(s)\sum_{A\in\wp_m(S^{(p)})}\;\prod_{\tau\in A}\delta(s-\tau)\,\frac{\partial^p\,\delta^{p-m}\mathcal{F}[\upsilon\,;u(\bullet|_{t_0}^{t})]}{\partial\upsilon^p\prod\limits_{\tau\notin A}\delta u(\tau)}. \tag{B.9}$$

Note that the upper limit of the $m-$sum in the right-hand side of Eq. (B.9) is $\min\{p,n\}$, while in Eq. (B.8) was $2p$. This is due to the fact that $\delta^m u^n(s)/\prod\limits_{\tau\in A}\delta u(\tau)=0$ for $m>n$, see Eq. (B.6) of Lemma B.3. By substituting Eq. (B.9) into Eq. (B.7) we obtain

$$\bar{\mathcal{T}}^{(1,1)}_{X_0\Xi}\left[u^n(s)\,\mathcal{F}[\upsilon\,;u(\bullet|_{t_0}^{t})]\right] = \sum_{p=0}^{\infty}\frac{1}{p!}\sum_{m=0}^{\min\{p,n\}}\frac{n!}{(n-m)!}\,u^{n-m}(s)\sum_{A\in\wp_m(S^{(p)})}P_A\,, \tag{B.10}$$

with

$$P_A = \int_{t_0}^{t} \overset{(p)}{\cdots} \int_{t_0}^{t} C_{X_0\Xi}(\tau^{(1)}) \cdots C_{X_0\Xi}(\tau^{(p)}) \prod_{s \in A} \delta(s-\tau) \frac{\partial^p \, \delta^{p-m} \mathcal{F}[\upsilon ; u(\bullet|_{t_0}^{t})]}{\partial \upsilon^p \prod_{s \notin A} \delta u(\tau)} d\tau^{(1)} \cdots d\tau^{(p)}. \tag{B.11}$$

**Lemma B.4.** For all $A \in \wp_m(S^{(p)})$, $P_A$ is the same, and is equal to

$$P_m^{(p)} = C_{X_0\Xi}^m(s) \int_{t_0}^{t} \overset{(p-m)}{\cdots} \int_{t_0}^{t} \prod_{q=1}^{p-m} C_{X_0\Xi}(\tau^{(q)}) \frac{\partial^p \, \delta^{p-m} \mathcal{F}[\upsilon ; u(\bullet|_{t_0}^{t})]}{\partial \upsilon^p \prod_{q=1}^{p-m} \delta u(\tau^{(q)})} \prod_{q=1}^{p-m} d\tau^{(q)}. \tag{B.12}$$

**Proof.** For each subset $A \in \wp_m(S^{(p)})$, we define the index set $I = I(A)$ that contains all the $m$ indices of the $\tau^{(i)}$ elements that belong to $A$. Using this index set, Eq. (B.11) is expressed equivalently as

$$P_A = \int_{t_0}^{t} \overset{(p)}{\cdots} \int_{t_0}^{t} C_{X_0\Xi}(\tau^{(1)}) \cdots C_{X_0\Xi}(\tau^{(p)}) \prod_{i \in I} \delta(s-\tau^{(i)}) \frac{\partial^p \, \delta^{p-m} \mathcal{F}[\upsilon ; u(\bullet|_{t_0}^{t})]}{\partial \upsilon^p \prod_{\substack{i=1 \\ i \notin I}}^{p} \delta u(\tau^{(i)})} d\tau^{(1)} \cdots d\tau^{(p)},$$

and by employing the identities of $\prod_{i \in I} \delta(s-\tau^{(i)})$

$$P_A = C_{X_0\Xi}^m(s) \int_{t_0}^{t} \overset{(p)}{\cdots} \int_{t_0}^{t} \prod_{\substack{i=1 \\ i \notin I}}^{p} C_{X_0\Xi}(\tau^{(i)}) \frac{\partial^p \, \delta^{p-m} \mathcal{F}[\upsilon ; u(\bullet|_{t_0}^{t})]}{\partial \upsilon^p \prod_{\substack{i=1 \\ i \notin I}}^{p} \delta u(\tau^{(i)})} \prod_{\substack{i=1 \\ i \notin I}}^{p} d\tau^{(i)}.$$

By renaming the integration variables; $\tau^{(i)}$, $i = 1, \dots, p$, $i \notin I$ into $\tau^{(1)}, \dots, \tau^{(p-m)}$, we obtain Eq. (B.12). ■

By virtue of Lemma B.4, the right-hand side of Eq. (B.10) is expressed as

$$\bar{\mathcal{T}}_{X_0\Xi}^{(1,1)}\left[u^n(s) \mathcal{F}[\upsilon ; u(\bullet|_{t_0}^{t})]\right] = \sum_{p=0}^{\infty} \frac{1}{p!} \sum_{m=0}^{\min\{p,n\}} \frac{n!}{(n-m)!} u^{n-m}(s) |\wp_m(S^{(p)})| P_m^{(p)}, \tag{B.13}$$

where $|\bullet|$ denotes the cardinality of the set. Since $|\wp_m(S^{(p)})| = \binom{p}{m}$, for it is the number of subsets with $m$ elements chosen from a set of $p$ elements, Eq. (B.13) is written equivalently:

$$\bar{\mathcal{T}}_{X_0\Xi}^{(1,1)}\left[u^n(s) \mathcal{F}[\upsilon ; u(\bullet|_{t_0}^{t})]\right] = \sum_{p=0}^{\infty} \sum_{m=0}^{\min\{p,n\}} \binom{n}{m} u^{n-m}(s) \frac{P_m^{(p)}}{(p-m)!}. \tag{B.14}$$

In the right-hand side of Eq. (B.14), the order of summations can be interchanged, by employing the double summation rearrangement formula (B.43) proved at the end of the present Appendix. Thus, Eq. (B.14) is expressed as

$$\bar{\mathcal{T}}_{X_0\Xi}^{(1,1)}\left[u^n(s) \mathcal{F}[\upsilon ; u(\bullet|_{t_0}^{t})]\right] = \sum_{m=0}^{n} \binom{n}{m} u^{n-m}(s) \sum_{p=m}^{\infty} \frac{P_m^{(p)}}{(p-m)!}. \tag{B.15}$$

By using Eq. (B.12), the last sum in the right-hand side of Eq. (B.15) is written as

$$\sum_{p=m}^{\infty}\frac{P_m^{(p)}}{(p-m)!} = C_{X_0\Xi}^m(s)\sum_{p=m}^{\infty}\frac{1}{(p-m)!}\int_{t_0}^{t}\overset{(p-m)}{\cdots}\int_{t_0}^{t}\prod_{q=1}^{p-m}C_{X_0\Xi}(\tau^{(q)})\frac{\partial^p\,\delta^{p-m}\mathcal{F}[\upsilon\,;u(\bullet|_{t_0}^{t})]}{\partial\upsilon^p\prod_{q=1}^{p-m}\delta u(\tau^{(q)})}\prod_{q=1}^{p-m}d\tau^{(q)},$$

and after the change of index $\ell = p-m$

$$\sum_{p=m}^{\infty}\frac{P_m^{(p)}}{(p-m)!} = C_{X_0\Xi}^m(s)\sum_{\ell=0}^{\infty}\frac{1}{\ell!}\int_{t_0}^{t}\overset{(\ell)}{\cdots}\int_{t_0}^{t}\prod_{q=1}^{\ell}C_{X_0\Xi}(\tau^{(q)})\frac{\partial^\ell\,\delta^\ell}{\partial\upsilon^\ell\prod_{q=1}^{\ell}\delta u(\tau^{(q)})}\left[\frac{\partial^m\mathcal{F}[\upsilon\,;u(\bullet|_{t_0}^{t})]}{\partial\upsilon^m}\right]\prod_{q=1}^{\ell}d\tau^{(q)}.$$

(B.16)

By virtue of Eq. (3.36b), the right-hand side of Eq. (B.16) is identified as $\bar{\mathcal{T}}_{X_0\Xi}^{(1,1)}\left[\partial^m\mathcal{F}[\upsilon\,;u(\bullet|_{t_0}^{t})]/\partial\upsilon^m\right]$. Thus, Eq. (B.15) is expressed as

$$\bar{\mathcal{T}}_{X_0\Xi}^{(1,1)}\left[u^n(s)\,\mathcal{F}[\upsilon\,;u(\bullet|_{t_0}^{t})]\right] = \sum_{m=0}^{n}\binom{n}{m}u^{n-m}(s)\,C_{X_0\Xi}^m(s)\,\bar{\mathcal{T}}_{X_0\Xi}^{(1,1)}\left[\frac{\partial^m\mathcal{F}[\upsilon\,;u(\bullet|_{t_0}^{t})]}{\partial\upsilon^m}\right].$$

(B.17)

**B2. The action of operator $\bar{\mathcal{T}}_{X_0\Xi}^{(0,2)}$ on $u^n(s)\,\mathcal{F}[\upsilon\,;u(\bullet|_{t_0}^{t})]$**

By using Eq. (3.36c), $\bar{\mathcal{T}}_{X_0\Xi}^{(0,2)}\left[u^n(s)\,\mathcal{F}[\upsilon\,;u(\bullet|_{t_0}^{t})]\right]$ is expressed in series form as

$$\bar{\mathcal{T}}_{X_0\Xi}^{(0,2)}\left[u^n(s)\,\mathcal{F}[\upsilon\,;u(\bullet|_{t_0}^{t})]\right] = \sum_{p=0}^{\infty}\frac{1}{p!}\frac{1}{2^p}\int_{t_0}^{t}\overset{(2p)}{\cdots}\int_{t_0}^{t}\Big[C_{\Xi\Xi}(\tau_1^{(1)},\tau_2^{(1)})\cdots C_{\Xi\Xi}(\tau_1^{(p)},\tau_2^{(p)})\times$$

$$\times\frac{\delta^{2p}\left[u^n(s)\,\mathcal{F}[\upsilon\,;u(\bullet|_{t_0}^{t})]\right]}{\delta u(\tau_1^{(1)})\delta u(\tau_2^{(1)})\cdots\delta u(\tau_1^{(p)})\delta u(\tau_2^{(p)})}\,d\tau_1^{(1)}\,d\tau_2^{(1)}\cdots d\tau_1^{(p)}\,d\tau_2^{(p)}\Big].$$

(B.18)

Using Lemma B.2, the derivative in the right-hand side of Eq. (B.18) is expressed as

$$\frac{\delta^{2p}\left[u^n(s)\,\mathcal{F}[\upsilon\,;u(\bullet|_{t_0}^{t})]\right]}{\delta u(\tau_1^{(1)})\delta u(\tau_2^{(1)})\cdots\delta u(\tau_1^{(p)})\delta u(\tau_2^{(p)})} = \sum_{m=0}^{2p}\sum_{A\in\wp_m(S^{(2p)})}\frac{\delta^m u^n(s)}{\prod_{\tau\in A}\delta u(\tau)}\frac{\delta^{2p-m}\mathcal{F}[\upsilon\,;u(\bullet|_{t_0}^{t})]}{\prod_{\tau\notin A}\delta u(\tau)}$$

(B.19)

with $S^{(2p)} = \left\{\tau_1^{(1)},\tau_2^{(1)},\cdots\tau_1^{(p)},\tau_2^{(p)}\right\}$. Using Lemma B.3, the right-hand side of Eq. (B.19) is evaluated as

$$\frac{\delta^{2p}\left[u^n(s)\,\mathcal{F}[u(\bullet|_{t_0}^{t})]\right]}{\delta u(\tau_1^{(1)})\delta u(\tau_2^{(1)})\cdots\delta u(\tau_1^{(p)})\delta u(\tau_2^{(p)})} =$$

$$= \sum_{m=0}^{\min\{2p,n\}}\sum_{A\in\wp_m(S^{(2p)})}\frac{n!}{(n-m)!}u^{n-m}(s)\prod_{\tau\in A}\delta(s-\tau)\frac{\delta^{2p-m}\mathcal{F}[u(\bullet|_{t_0}^{t})]}{\prod_{\tau\notin A}\delta u(\tau)}.$$

(B.20)

By substituting Eq. (B.20) into Eq. (B.18) and defining $\alpha_m(s) = n!\, u^{n-m}(s)/(n-m)!$, we obtain

$$\bar{\mathcal{T}}^{(0,2)}_{X_0\Xi}\left[u^n(s)\,\mathcal{F}[\upsilon\,;u(\bullet|_{t_0}^{t})]\right] = \sum_{p=0}^{\infty}\frac{1}{p!}\frac{1}{2^p}\sum_{m=0}^{\min\{2p,\,n\}}\alpha_m(s)\times$$
$$\times\sum_{A\in\wp_m(S^{(2p)})}\int_{t_0}^{t}\overset{(2p)}{\cdots}\int_{t_0}^{t}\Bigg[C_{\Xi\Xi}(\tau_1^{(1)},\tau_2^{(1)})\cdots C_{\Xi\Xi}(\tau_1^{(p)},\tau_2^{(p)})\times \tag{B.21}$$
$$\times\prod_{\tau\in A}\delta(s-\tau)\,\frac{\delta^{2p-m}\mathcal{F}[\upsilon\,;u(\bullet|_{t_0}^{t})]}{\prod_{\tau\notin A}\delta u(\tau)}\,d\tau_1^{(1)}\,d\tau_2^{(1)}\cdots d\tau_1^{(p)}\,d\tau_2^{(p)}\Bigg].$$

**Remark B.1: Subsets with the same number of paired elements.** The family of all subsets $A\in\wp_m(S^{(2p)})$ with given cardinality $m\le 2p$ can be divided into classes $\wp_{m,k}(S^{(2p)})$, with regard to the number $k$ of pairs $\tau_1^{(i)},\tau_2^{(i)}$ (elements with the same upper index) contained in each $A$. If $A$ contains $k$ *pairs*, the rest $m-2k$ of its elements are *singletons*, i.e. their upper index is not repeated in any other element. The maximum number of pairs contained in an $A\in\wp_m(S^{(2p)})$ is $\lfloor m/2\rfloor$, with $\lfloor\bullet\rfloor$ being the floor function. For $m\le p$, the minimum number of pairs is 0, since we can choose all $m$ elements to be singletons. For $m>p$ this is not possible, since each one of the excessive $m-p$ elements should necessarily have the same upper index with another element, forming thus $m-p$ pairs. So, for the number of pairs $k$, it holds true that

$$k = m-\min\{p,m\},\; m-\min\{p,m\}+1,\;\dots,\;\lfloor m/2\rfloor.$$

By utilizing Remark B.1, Eq. (B.21) can be expressed as

$$\bar{\mathcal{T}}^{(0,2)}_{X_0\Xi}\left[u^n(s)\,\mathcal{F}[\upsilon\,;u(\bullet|_{t_0}^{t})]\right] = \sum_{p=0}^{\infty}\frac{1}{p!}\frac{1}{2^p}\sum_{m=0}^{\min\{2p,\,n\}}\alpha_m(s)\sum_{k=m-\min\{p,m\}}^{\lfloor m/2\rfloor}\;\sum_{A\in\wp_{m,k}(S^{(2p)})}P_A \tag{B.22}$$

with

$$P_A = \int_{t_0}^{t}\overset{(2p)}{\cdots}\int_{t_0}^{t}C_{\Xi\Xi}(\tau_1^{(1)},\tau_2^{(1)})\cdots C_{\Xi\Xi}(\tau_1^{(p)},\tau_2^{(p)})\times$$
$$\times\prod_{\tau\in A}\delta(s-\tau)\,\frac{\delta^{2p-m}\mathcal{F}[\upsilon\,;u(\bullet|_{t_0}^{t})]}{\prod_{\tau\notin A}\delta u(\tau)}\,d\tau_1^{(1)}\,d\tau_2^{(1)}\cdots d\tau_1^{(p)}\,d\tau_2^{(p)}. \tag{B.23}$$

**Lemma B.5.** For all $A\in\wp_{m,k}(S^{(2p)})$, $P_A$ is the same, and is equal to

$$P_{m,k}^{(p)} = C_{\Xi\Xi}^{k}(s,s)\int_{t_0}^{t}\overset{(m-2k)}{\cdots}\int_{t_0}^{t}\Bigg\{\prod_{\mu=1}^{m-2k}C_{\Xi\Xi}(s,\tau_\mu)\int_{t_0}^{t}\overset{2(p-(m-k))}{\cdots}\int_{t_0}^{t}\Bigg[\prod_{q=1}^{p-(m-k)}C_{\Xi\Xi}(\tau_1^{(q)},\tau_2^{(q)})\times$$
$$\times\frac{\delta^{2p-m}\mathcal{F}[\upsilon\,;u(\bullet|_{t_0}^{t})]}{\prod_{\mu=1}^{m-2k}\delta u(\tau_\mu)\prod_{q=1}^{p-(m-k)}\delta u(\tau_1^{(q)})\,\delta u(\tau_2^{(q)})}\Bigg]\prod_{q=1}^{p-(m-k)}d\tau_1^{(q)}\,d\tau_2^{(q)}\Bigg\}\prod_{\mu=1}^{m-2k}d\tau_\mu \tag{B.24}$$

**Proof.** For each subset $A \in \wp_{m,k}(S^{(2p)})$ two index sets can be defined; $I_{\text{pair}} = I_{\text{pair}}(A)$, containing the $k$ upper indices of the paired elements, and $I_{\text{singl}} = I_{\text{singl}}(A)$, containing the $m-2k$ upper indices of the singletons. In order to also keep the lower index of each singleton, we define, for each upper index $\mu \in I_{\text{singl}}$, another index $i_\mu$ that takes the appropriate value 1 or 2. Using the said index sets, Eq. (B.23) is written as

$$P_A = \int_{t_0}^{t} \overset{(2p)}{\cdots} \int_{t_0}^{t} C_{\Xi\Xi}(\tau_1^{(1)}, \tau_2^{(1)}) \cdots C_{\Xi\Xi}(\tau_1^{(p)}, \tau_2^{(p)}) \prod_{q \in I_{\text{pair}}} \delta(s-\tau_1^{(q)})\, \delta(s-\tau_2^{(q)}) \times$$
$$\times \prod_{\mu \in I_{\text{singl}}} \delta(s-\tau_{i_\mu}^{(\mu)}) \frac{\delta^{2p-m} \mathcal{F}[\upsilon; u(\bullet|_{t_0}^{t})]}{\prod_{\tau \notin A} \delta u(\tau)} d\tau_1^{(1)} d\tau_2^{(1)} \cdots d\tau_1^{(p)} d\tau_2^{(p)}.$$

First, we employ the identities of $\prod_{q \in I_{\text{pair}}} \delta(s-\tau_1^{(q)})\, \delta(s-\tau_2^{(q)})$, resulting in

$$P_A = C_{\Xi\Xi}^{k}(s,s) \int_{t_0}^{t} \overset{(2p-2k)}{\cdots} \int_{t_0}^{t} \prod_{\substack{q=1 \\ q \notin I_{\text{pair}}}}^{p} C_{\Xi\Xi}(\tau_1^{(q)}, \tau_2^{(q)}) \prod_{\mu \in I_{\text{singl}}} \delta(s-\tau_{i_\mu}^{(\mu)}) \times$$
$$\times \frac{\delta^{2p-m} \mathcal{F}[\upsilon; u(\bullet|_{t_0}^{t})]}{\prod_{\tau \notin A} \delta u(\tau)} \prod_{\substack{q=1 \\ q \notin I_{\text{pair}}}}^{p} d\tau_1^{(q)} d\tau_2^{(q)}.$$

Then, by employing the identities of $\prod_{\mu \in I_{\text{singl}}} \delta(s-\tau_{i_\mu}^{(\mu)})$ and the symmetry of $C_{\Xi\Xi}(\tau_1^{(q)}, \tau_2^{(q)})$,

$$P_A = C_{\Xi\Xi}^{k}(s,s) \int_{t_0}^{t} \overset{(2p-m)}{\cdots} \int_{t_0}^{t} \prod_{\substack{q=1 \\ q \notin I_{\text{pair}} \cup I_{\text{singl}}}}^{p} C_{\Xi\Xi}(\tau_1^{(q)}, \tau_2^{(q)}) \prod_{\mu \in I_{\text{singl}}} C_{\Xi\Xi}(s, \tau_{j_\mu}^{(\mu)}) \times$$
$$\times \frac{\delta^{2p-m} \mathcal{F}[\upsilon; u(\bullet|_{t_0}^{t})]}{\prod_{\mu \in I_{\text{singl}}} \delta u(\tau_{j_\mu}^{(\mu)}) \prod_{\substack{q=1 \\ q \notin I_{\text{pair}} \cup I_{\text{singl}}}}^{p} \delta u(\tau_1^{(q)})\, \delta u(\tau_1^{(q)})} \prod_{\substack{q=1 \\ q \notin I_{\text{pair}} \cup I_{\text{singl}}}}^{p} d\tau_1^{(q)} d\tau_2^{(q)} \prod_{\mu \in I_{\text{singl}}} d\tau_{j_\mu}^{(\mu)},$$

where $j_\mu = 2$ if $i_\mu = 1$ and vice versa. By renaming the integration variables; $\tau_{j_\mu}^{(\mu)}$, $\mu \in I_{\text{singl}}$ into $\tau_1, \tau_2, \ldots, \tau_{m-2k}$, and $s_i^{(q)}$, $i = 1, 2$, $q \notin I_{\text{pair}} \cup I_{\text{singl}}$ into $\tau_i^{(1)}, \tau_i^{(2)}, \ldots, \tau_i^{(p-m-k)}$, we obtain Eq. (27). ■

Using Lemma B.5, Eq. (B.22) is written as

$$\bar{\mathcal{T}}_{X_0\Xi}^{(0,2)}\left[u^n(s)\, \mathcal{F}[\upsilon; u(\bullet|_{t_0}^{t})]\right] = \sum_{p=0}^{\infty} \frac{1}{p!} \frac{1}{2^p} \sum_{m=0}^{\min\{2p,n\}} \alpha_m(s) \sum_{k=m-\min\{p,m\}}^{\lfloor m/2 \rfloor} |\wp_{m,k}(S^{(2p)})|\, P_{m,k}^{(p)}, \tag{B.25}$$

where $|\bullet|$ denotes the cardinality of the set.

**Lemma B.6: Calculating the number of subsets with cardinality $m$, and $k$ pairs.** For $m = 0, 1, \dots, \min\{2p, n\}$ and $k = m - \min\{p, m\}, m - \min\{p, m\} + 1, \dots, \lfloor m/2 \rfloor$, it holds true that

$$|\wp_{m,k}(S^{(2p)})| = \frac{2^{m-2k}\, p^{\underline{m-k}}}{k!\,(m-2k)!}, \tag{B.26}$$

where $p^{\underline{m-k}}$ is the falling factorial; $p^{\underline{m-k}} = p(p-1)\cdots(p-(m-k)+1)$.

**Proof.** The number of choosing $k$ pairs from the $p$ available pairs of $S^{(2p)} = \left\{\tau_1^{(1)}, \tau_2^{(1)}, \cdots \tau_1^{(p)}, \tau_2^{(p)}\right\}$ is given by the binomial coefficient $\binom{p}{k}$. These $k$ pairs contain $2k$ elements of $A$. The rest $m-2k$ elements of $A$ are singletons, and thus they must have different upper indices. The first singleton can be chosen from $(2p-2k)$ elements of $S^{(2p)}$, since $2k$ elements have been employed to form pairs. The second singleton can be chosen from $(2p-2k-2)$ elements, since we have excluded the $2k$ paired elements, the first singleton, as well as the element that has the same upper index with the first singleton. This argument can be followed for all $m-2k$ singletons, and since $A$ is an unordered set, we obtain the number of possible choices of singletons:

$$\frac{(2p-2k)\,(2p-2k-2)\,(2p-2k-4)\cdots(2p-2k-2(m-2k)+2)}{(m-2k)!} =$$

$$= \frac{2^{m-2k}\,(p-k)\,(p-k-1)\,(p-k-2)\cdots(p-(m-k)+1)}{(m-2k)!}.$$

Thus

$$|\wp_{m,k}(S^{(2p)})| = \binom{p}{k}\frac{2^{m-2k}\,(p-k)\,(p-k-1)\,(p-k-2)\cdots(p-(m-k)+1)}{(m-2k)!} =$$

$$= \frac{p!}{k!\,(p-k)!}\,\frac{2^{m-2k}\,(p-k)\,(p-k-1)\,(p-k-2)\cdots(p-(m-k)+1)}{(m-2k)!} =$$

$$= \frac{2^{m-2k}\;p\,(p-1)\,(p-k+1)\,(p-k)\,(p-k-1)\,(p-k-2)\cdots(p-(m-k)+1)}{k!\,(m-2k)!}.$$

By using the definition of falling factorial $p^{\underline{m-k}}$, Eq. (B.26) is obtained. ■

By substituting Eq. (B.26) into Eq. (B.25) we obtain, after algebraic manipulations

$$\bar{\mathcal{T}}^{(0,2)}_{X_0\Xi}\left[u^n(s)\,\mathcal{F}[\upsilon\,;u(\bullet|_{t_0}^{t})]\right] = \sum_{p=0}^{\infty}\sum_{m=0}^{\min\{2p,n\}}\alpha_m(s)\sum_{k=m-\min\{p,m\}}^{\lfloor m/2\rfloor}\frac{H_{m,k}}{m!}\,\frac{P^{(p)}_{m,k}}{2^{p-(m-k)}\,(p-(m-k))!}, \tag{B.27}$$

where $H_{m,k} = m!/[2^k\,k!\,(m-2k)!]$ are the absolute values of Hermite coefficients, Eq. (3.72). Now, by employing the double summation rearrangement formula (B.45) proved at the end of the Appendix, $p-$and $m-$sums are interchanged, resulting into

$$\bar{\mathcal{T}}^{(0,2)}_{X_0\Xi}\left[u^n(s)\,\mathcal{F}[\upsilon\,;u(\bullet|_{t_0}^{t})]\right] = \sum_{m=0}^{n}\alpha_m(s)\sum_{p=\lceil m/2\rceil}^{\infty}\;\sum_{k=m-\min\{p,m\}}^{\lfloor m/2\rfloor}\frac{H_{m,k}}{m!}\,\frac{P^{(p)}_{m,k}}{2^{p-(m-k)}\,(p-(m-k))!}. \tag{B.28}$$

where $\lceil \bullet \rceil$ denotes the ceiling function. Second, using the double summation rearrangement formula (B.49), $p-$and $k-$sums are also interchanged

$$\bar{\mathcal{T}}^{(0,2)}_{X_0\Xi}\left[u^n(s)\,\mathcal{F}[\upsilon\,;u(\bullet|_{t_0}^t)]\right] = \sum_{m=0}^{n}\alpha_m(s)\sum_{k=0}^{\lfloor m/2\rfloor}\frac{H_{m,k}}{m!}\sum_{p=m-k}^{\infty}\frac{P^{(p)}_{m,k}}{2^{p-(m-k)}\,(p-(m-k))!}. \quad \text{(B.29)}$$

By using the definition relation for $P^{(p)}_{m,k}$, Eq. (B.24), and after some algebraic manipulations, the last sum in the right-hand side of Eq. (B.29) is further evaluated as

$$\sum_{p=m-k}^{\infty}\frac{P^{(p)}_{m,k}}{2^{p-(m-k)}\,(p-(m-k))!} = C^k_{\Xi\Xi}(s,s)\int_{t_0}^{t}\overset{(m-2k)}{\cdots}\int_{t_0}^{t}\prod_{\mu=1}^{m-2k}C_{\Xi\Xi}(s,\tau_\mu)\,Q_{m,k}\prod_{\mu=1}^{m-2k}d\tau_\mu \quad \text{(B.30)}$$

with

$$Q_{m,k} = \sum_{p=m-k}^{\infty}\frac{1}{2^{p-(m-k)}\,(p-(m-k))!}\int_{t_0}^{t}\overset{2(p-(m-k))}{\cdots}\int_{t_0}^{t}\prod_{q=1}^{p-(m-k)}C_{\Xi\Xi}(\tau_1^{(q)},\tau_2^{(q)})\times$$
$$\times\frac{\delta^{2p-m}\mathcal{F}[\upsilon\,;u(\bullet|_{t_0}^t)]}{\prod_{\mu=1}^{m-2k}\delta u(\tau_\mu)\prod_{q=1}^{p-(m-k)}\delta u(\tau_1^{(q)})\,\delta u(\tau_2^{(q)})}\prod_{q=1}^{p-(m-k)}d\tau_1^{(q)}\,d\tau_2^{(q)}. \quad \text{(B.31)}$$

Under the change of index $r=p-(m-k)$, and by employing the series expansion of operator $\bar{\mathcal{T}}^{(0,2)}_{X_0\Xi}\bullet$, Eq. (3.36c), series $Q_{m,k}$ is identified as

$$Q_{m,k} = \bar{\mathcal{T}}^{(0,2)}_{X_0\Xi}\left[\frac{\delta^{m-2k}\mathcal{F}[\upsilon\,;u(\bullet|_{t_0}^t)]}{\prod_{\mu=1}^{m-2k}\delta u(\tau_\mu)}\right]. \quad \text{(B.32)}$$

By employing Eqs. (B.30), (B.32), as well as the definition relation for $\alpha_m(s)$, Eq. (B.29) is expressed as

$$\bar{\mathcal{T}}^{(0,2)}_{X_0\Xi}\left[u^n(s)\,\mathcal{F}[\upsilon\,;u(\bullet|_{t_0}^t)]\right] = \sum_{m=0}^{n}\binom{n}{m}u^{n-m}(s)\times$$
$$\times\sum_{k=0}^{\lfloor m/2\rfloor}H_{m,k}\,C^k_{\Xi\Xi}(s,s)\int_{t_0}^{t}\overset{(m-2k)}{\cdots}\int_{t_0}^{t}\prod_{\mu=1}^{m-2k}C_{\Xi\Xi}(s,\tau_\mu)\,\bar{\mathcal{T}}^{(0,2)}_{X_0\Xi}\left[\frac{\delta^{m-2k}\mathcal{F}[\upsilon\,;u(\bullet|_{t_0}^t)]}{\prod_{\mu=1}^{m-2k}\delta u(\tau_\mu)}\right]\prod_{\mu=1}^{m-2k}d\tau_\mu. \quad \text{(B.33)}$$

### B3. Finalization of the proof

Having now at our disposal the results of Lemma B.1 and sections B1, B2 regarding the actions of operators $\bar{\mathcal{T}}^{(2)}$, we start the final part of our proof by applying operator $\bar{\mathcal{T}}^{(1,1)}_{X_0\Xi}$ on both sides of Eq. (B.3), thus obtaining

$$\bar{\mathcal{T}}^{(1,1)}_{X_0\Xi}\bar{\mathcal{T}}^{(2,0)}_{X_0\Xi}\left[u^n(s)\,\mathcal{F}[\upsilon\,;u(\bullet|_{t_0}^t)]\right] = \bar{\mathcal{T}}^{(1,1)}_{X_0\Xi}\left[u^n(s)\,\mathcal{F}_1[\upsilon\,;u(\bullet|_{t_0}^t)]\right], \quad \text{(B.34)}$$

where $\mathcal{F}_1[\upsilon ; u(\bullet|_{t_0}^{t})] = \bar{\mathcal{T}}^{(2,0)}_{X_0\Xi}\left[\mathcal{F}[\upsilon ; u(\bullet|_{t_0}^{t})]\right]$. Under Eq. (B.17), the right-hand side of Eq. (B.34) is calculated as

$$\bar{\mathcal{T}}^{(1,1)}_{X_0\Xi}\bar{\mathcal{T}}^{(2,0)}_{X_0\Xi}\left[u^n(s)\,\mathcal{F}[\upsilon ; u(\bullet|_{t_0}^{t})]\right] = \sum_{m=0}^{n}\binom{n}{m}C^m_{X_0\Xi}(s)\,u^{n-m}(s)\,\mathcal{F}_2[\upsilon ; u(\bullet|_{t_0}^{t})], \quad \text{(B.35)}$$

where $\mathcal{F}_2[\upsilon ; u(\bullet|_{t_0}^{t})] = \bar{\mathcal{T}}_{\hat{X}_0\hat{\Xi}(\cdot)}\left[\partial^m\mathcal{F}_1[\upsilon ; u(\bullet|_{t_0}^{t})]/\partial\upsilon^m\right]$. By applying $\bar{\mathcal{T}}^{(0,2)}_{X_0\Xi}$ on both sides of Eq. (B.35), and employing the linearity of operators $\bar{\mathcal{T}}^{(2)}$ proven in paragraph 3.4.1, we obtain

$$\begin{aligned}\bar{\mathcal{T}}^{(0,2)}_{X_0\Xi}\bar{\mathcal{T}}^{(1,1)}_{X_0\Xi}\bar{\mathcal{T}}^{(2,0)}_{X_0\Xi}&\left[u^n(s)\,\mathcal{F}[\upsilon ; u(\bullet|_{t_0}^{t})]\right] = \\ &= \sum_{m=0}^{n}\binom{n}{m}C^m_{X_0\Xi}(s)\,\bar{\mathcal{T}}^{(0,2)}_{X_0\Xi}\left[u^{n-m}(s)\,\mathcal{F}_2[\upsilon ; u(\bullet|_{t_0}^{t})]\right].\end{aligned} \quad \text{(B.36)}$$

By virtue of Eq. (B.33) and some algebraic manipulations, Eq. (B.36) is evaluated to

$$\begin{aligned}\bar{\mathcal{T}}^{(0,2)}_{X_0\Xi}\bar{\mathcal{T}}^{(1,1)}_{X_0\Xi}\bar{\mathcal{T}}^{(2,0)}_{X_0\Xi}\left[u^n(s)\,\mathcal{F}[\upsilon ; u(\bullet|_{t_0}^{t})]\right] &= n!\sum_{m_1=0}^{n}\frac{C^{m_1}_{X_0\Xi}(s)}{m_1!}\sum_{m_2=0}^{n-m_1}\frac{u^{n-m_1-m_2}(s)}{(n-m_1-m_2)!}\times \\ \times\sum_{k=0}^{\lfloor m_2/2\rfloor}\frac{C^k_{\Xi\Xi}(s,s)}{2^k\,k!\,(m_2-2k)!}\int_{t_0}^{t}\overset{(m_2-2k)}{\cdots}\int_{t_0}^{t}\prod_{\mu=1}^{m_2-2k}C_{\Xi\Xi}(s,\tau_\mu)\,&\bar{\mathcal{T}}_{\hat{\Xi}(\cdot)\hat{\Xi}(\cdot)}\left[\frac{\delta^{m_2-2k}\mathcal{F}_2[\upsilon ; u(\bullet|_{t_0}^{t})]}{\prod_{\mu=1}^{m_2-2k}\delta u(\tau_\mu)}\right]\prod_{\mu=1}^{m_2-2k}d\tau_\mu\end{aligned}$$

and equivalently

$$\begin{aligned}\bar{\mathcal{T}}^{(0,2)}_{X_0\Xi}\bar{\mathcal{T}}^{(1,1)}_{X_0\Xi}\bar{\mathcal{T}}^{(2,0)}_{X_0\Xi}&\left[u^n(s)\,\mathcal{F}[\upsilon ; u(\bullet|_{t_0}^{t})]\right] = \\ &= \sum_{m_1+m_2+m_3=n}\binom{n}{m_1,\,m_2,\,m_3}u^{m_1}(s)\,C^{m_2}_{X_0\Xi}(s)\sum_{k=0}^{\lfloor m_3/2\rfloor}H_{m_3,k}\,C^k_{\Xi\Xi}(s,s)\times \\ &\quad\times\int_{t_0}^{t}\overset{(m_3-2k)}{\cdots}\int_{t_0}^{t}\prod_{\mu=1}^{m_3-2k}C_{\Xi\Xi}(s,\tau_\mu)\,\bar{\mathcal{T}}^{(0,2)}_{X_0\Xi}\left[\frac{\delta^{m_3-2k}\mathcal{F}_2[\upsilon ; u(\bullet|_{t_0}^{t})]}{\prod_{\mu=1}^{m_3-2k}\delta u(\tau_\mu)}\right]\prod_{\mu=1}^{m_3-2k}d\tau_\mu,\end{aligned} \quad \text{(B.37)}$$

where $\sum_{m_1+m_2=\ell}$ denotes the sum over all triplets $m_1$, $m_2$, $m_3$ with $m_1+m_2+m_3=n$, and $\binom{n}{m_1,\,m_2,\,m_3} = \dfrac{n!}{m_1!\,m_2!\,m_3!}$ is the multinomial coefficient. By employing the definitions of $\mathcal{F}_1$, $\mathcal{F}_2$ and the property that operators $\bar{\mathcal{T}}^{(2)}$ commute with each other, and with $\upsilon-$ and $u(s)-$ differentiations, see paragraph 3.4.1, Eq. (B.37) is expressed as

$$\bar{\mathcal{T}}^{(0,2)}_{X_0\Xi}\,\bar{\mathcal{T}}^{(1,1)}_{X_0\Xi}\,\bar{\mathcal{T}}^{(2,0)}_{X_0\Xi}\left[u^n(s)\,\mathcal{F}[\upsilon\,;u(\bullet|_{t_0}^{t})]\right]=$$

$$=\sum_{m_1+m_2+m_3=n}\binom{n}{m_1,\,m_2,\,m_3}u^{m_1}(s)\,C^{m_2}_{X_0\Xi}(s)\sum_{k=0}^{\lfloor m_3/2\rfloor}H_{m_3,k}\,C^{k}_{\Xi\Xi}(s,s)\times$$

$$\times\int_{t_0}^{t}\overset{(m_3-2k)}{\cdots}\int_{t_0}^{t}\prod_{\mu=1}^{m_3-2k}C_{\Xi\Xi}(s,\tau_\mu)\,\bar{\mathcal{T}}^{(0,2)}_{X_0\Xi}\,\bar{\mathcal{T}}^{(1,1)}_{X_0\Xi}\,\bar{\mathcal{T}}^{(2,0)}_{X_0\Xi}\left[\frac{\partial^{m_2}\delta^{m_3-2k}\mathcal{F}[\upsilon\,;u(\bullet|_{t_0}^{t})]}{\partial\upsilon^{m_2}\prod_{\mu=1}^{m_3-2k}\delta u(\tau_\mu)}\right]\prod_{\mu=1}^{m_3-2k}d\tau_\mu\,. \qquad \text{(B.38)}$$

By setting $\upsilon=m_{X_0}$ and $u(\bullet)=m_{\Xi}(\bullet)$ in Eq. (B.38), and applying Eq. (3.34) to each term of the form $\bar{\mathcal{T}}^{(0,2)}_{X_0\Xi}\,\bar{\mathcal{T}}^{(1,1)}_{X_0\Xi}\,\bar{\mathcal{T}}^{(2,0)}_{X_0\Xi}[\cdots]$ in both sides of Eq. (B.38), we obtain the generalization of NF theorem, Eq. (B.1).

At this point, we shall also present the proof of Lemma B.2 by mathematical induction, as well as the summation rearrangement formulae used in the above proof.

**Proof of Lemma B.2: Product rule for higher Volterra derivatives**. Eq. (B.5) is proven by mathematical induction on index $q$:

- For $q=1$, Eq. (B.5) hold true since $A\in\{\varnothing,\{\tau_1\}\}$, and thus is specified into

$$\frac{\delta\left[\mathcal{F}_1[\cdots]\mathcal{F}_2[\cdots]\right]}{\delta u(\tau_1)}=\frac{\delta\mathcal{F}_1[\cdots]}{\delta u(\tau_1)}\mathcal{F}_2[\cdots]+\mathcal{F}_1[\cdots]\frac{\delta\mathcal{F}_2[\cdots]}{\delta u(\tau_1)}$$

which is the product rule for first order derivatives.

- We assume that Eq. (B.5) holds true for one fixed $q$.
- Thus, for $q+1$:

$$\frac{\delta^{q+1}\left[\mathcal{F}_1[\cdots]\mathcal{F}_2[\cdots]\right]}{\delta u(\tau_1)\cdots\delta u(\tau_{q+1})}=\frac{\delta}{\delta u(\tau_{q+1})}\frac{\delta^{q}\left[\mathcal{F}_1[\cdots]\mathcal{F}_2[\cdots]\right]}{\delta u(\tau_1)\cdots\delta u(\tau_q)}=\ [\text{by employing Eq. (B.5) for } q]$$

$$=\frac{\delta}{\delta u(\tau_{q+1})}\sum_{m=0}^{q}\sum_{A\in\wp_m(\{\tau_1,\dots,\tau_q\})}\frac{\delta^m\mathcal{F}_1[\cdots]}{\prod_{\tau\in A}\delta u(\tau)}\frac{\delta^{q-m}\mathcal{F}_2[\cdots]}{\prod_{\tau\notin A}\delta u(\tau)}=$$

$$=\sum_{m=0}^{q}\sum_{A\in\wp_m(\{\tau_1,\dots,\tau_q\})}\frac{\delta}{\delta u(\tau_{q+1})}\left[\frac{\delta^m\mathcal{F}_1[\cdots]}{\prod_{\tau\in A}\delta u(\tau)}\frac{\delta^{q-m}\mathcal{F}_2[\cdots]}{\prod_{\tau\notin A}\delta u(\tau)}\right]=$$

$$=\sum_{m=0}^{q}\sum_{A\in\wp_m(\{\tau_1,\dots,\tau_q\})}\frac{\delta^{m+1}\mathcal{F}_1[\cdots]}{\delta u(\tau_{q+1})\prod_{\tau\in A}\delta u(\tau)}\frac{\delta^{q-m}\mathcal{F}_2[\cdots]}{\prod_{\tau\notin A}\delta u(\tau)}+$$

$$+\sum_{m=0}^{q}\sum_{A\in\wp_m(\{\tau_1,\dots,\tau_q\})}\frac{\delta^m\mathcal{F}_1[\cdots]}{\prod_{\tau\in A}\delta u(\tau)}\frac{\delta^{q+1-m}\mathcal{F}_2[\cdots]}{\delta u(\tau_{q+1})\prod_{\tau\notin A}\delta u(\tau)}. \qquad \text{(B.39)}$$

The first double sum in the rightmost side of Eq. (B.39) is further evaluated as

$$\sum_{m=0}^{q} \sum_{A\in \wp_m(\{\tau_1,\dots,\tau_q\})} \frac{\delta^{m+1} \mathcal{F}_1[\cdots]}{\delta u(\tau_{q+1}) \prod_{\tau\in A} \delta u(\tau)} \frac{\delta^{q-m} \mathcal{F}_2[\cdots]}{\prod_{\tau\notin A} \delta u(\tau)} =$$

$$= \sum_{m=0}^{q} \sum_{\substack{A\in \wp_{m+1}(\{\tau_1,\dots,\tau_{q+1}\}) \\ \text{with } \tau_{q+1}\in A}} \frac{\delta^{m+1} \mathcal{F}_1[\cdots]}{\prod_{\tau\in A} \delta u(\tau)} \frac{\delta^{q+1-(m+1)} \mathcal{F}_2[\cdots]}{\prod_{\tau\notin A} \delta u(\tau)} = \quad \text{[change in outer sum index]}$$

$$= \sum_{m=1}^{q+1} \sum_{\substack{A\in \wp_{m}(\{\tau_1,\dots,\tau_{q+1}\}) \\ \text{with } \tau_{q+1}\in A}} \frac{\delta^{m} \mathcal{F}_1[\cdots]}{\prod_{\tau\in A} \delta u(\tau)} \frac{\delta^{q+1-m} \mathcal{F}_2[\cdots]}{\prod_{\tau\notin A} \delta u(\tau)} =$$

[and since all $A\in \wp_{q+1}(\{\tau_1,\dots,\tau_{q+1}\})$ necessarily contain $\tau_{q+1}$]

$$= \sum_{m=1}^{q} \sum_{\substack{A\in \wp_{m}(\{\tau_1,\dots,\tau_{q+1}\}) \\ \text{with } \tau_{q+1}\in A}} \frac{\delta^{m} \mathcal{F}_1[\cdots]}{\prod_{\tau\in A} \delta u(\tau)} \frac{\delta^{q+1-m} \mathcal{F}_2[\cdots]}{\prod_{\tau\notin A} \delta u(\tau)} + \\ + \sum_{A\in \wp_{q+1}(\{\tau_1,\dots,\tau_{q+1}\})} \frac{\delta^{m} \mathcal{F}_1[\cdots]}{\prod_{\tau\in A} \delta u(\tau)} \frac{\delta^{q+1-m} \mathcal{F}_2[\cdots]}{\prod_{\tau\notin A} \delta u(\tau)}. \tag{B.40}$$

Similarly, the second double sum in the rightmost side of Eq. (B.39) is further evaluated as

$$\sum_{m=0}^{q} \sum_{A\in \wp_m(\{\tau_1,\dots,\tau_q\})} \frac{\delta^{m} \mathcal{F}_1[\cdots]}{\prod_{\tau\in A} \delta u(\tau)} \frac{\delta^{q+1-m} \mathcal{F}_2[\cdots]}{\delta u(\tau_{q+1}) \prod_{\tau\notin A} \delta u(\tau)} =$$

$$= \sum_{m=0}^{q} \sum_{\substack{A\in \wp_{m}(\{\tau_1,\dots,\tau_{q+1}\}) \\ \text{with } \tau_{q+1}\notin A}} \frac{\delta^{m} \mathcal{F}_1[\cdots]}{\prod_{\tau\in A} \delta u(\tau)} \frac{\delta^{q+1-m} \mathcal{F}_2[\cdots]}{\prod_{\tau\notin A} \delta u(\tau)} =$$

[and since all $A\in \wp_{0}(\{\tau_1,\dots,\tau_{q+1}\})$ necessarily do not contain $\tau_{q+1}$]

$$= \sum_{A\in \wp_{0}(\{\tau_1,\dots,\tau_{q+1}\})} \frac{\delta^{m} \mathcal{F}_1[\cdots]}{\prod_{\tau\in A} \delta u(\tau)} \frac{\delta^{q+1-m} \mathcal{F}_2[\cdots]}{\prod_{\tau\notin A} \delta u(\tau)} + \\ + \sum_{m=1}^{q} \sum_{\substack{A\in \wp_{m}(\{\tau_1,\dots,\tau_{q+1}\}) \\ \text{with } \tau_{q+1}\notin A}} \frac{\delta^{m} \mathcal{F}_1[\cdots]}{\prod_{\tau\in A} \delta u(\tau)} \frac{\delta^{q+1-m} \mathcal{F}_2[\cdots]}{\prod_{\tau\notin A} \delta u(\tau)}. \tag{B.41}$$

By substituting Eqs. (B.40), (B.41) into Eq. (B.39), we obtain

$$\frac{\delta^{q+1}\left[\mathcal{F}_1[\cdots]\mathcal{F}_2[\cdots]\right]}{\delta u(\tau_1)\cdots\delta u(\tau_{q+1})} = \sum_{A\in \wp_{0}(\{\tau_1,\dots,\tau_{q+1}\})} \frac{\delta^{m} \mathcal{F}_1[\cdots]}{\prod_{\tau\in A} \delta u(\tau)} \frac{\delta^{q+1-m} \mathcal{F}_2[\cdots]}{\prod_{\tau\notin A} \delta u(\tau)} +$$

$$= \sum_{m=1}^{q} \left[ \sum_{\substack{A\in \wp_{m}(\{\tau_1,\dots,\tau_{q+1}\}) \\ \text{with } \tau_{q+1}\in A}} \frac{\delta^{m} \mathcal{F}_1[\cdots]}{\prod_{\tau\in A} \delta u(\tau)} \frac{\delta^{q+1-m} \mathcal{F}_2[\cdots]}{\prod_{\tau\notin A} \delta u(\tau)} + \sum_{\substack{A\in \wp_{m}(\{\tau_1,\dots,\tau_{q+1}\}) \\ \text{with } \tau_{q+1}\notin A}} \frac{\delta^{m} \mathcal{F}_1[\cdots]}{\prod_{\tau\in A} \delta u(\tau)} \frac{\delta^{q+1-m} \mathcal{F}_2[\cdots]}{\prod_{\tau\notin A} \delta u(\tau)} \right] +$$

$$+ \sum_{A\in \wp_{q+1}(\{\tau_1,\dots,\tau_{q+1}\})} \frac{\delta^{m} \mathcal{F}_1[\cdots]}{\prod_{\tau\in A} \delta u(\tau)} \frac{\delta^{q+1-m} \mathcal{F}_2[\cdots]}{\prod_{\tau\notin A} \delta u(\tau)} \Rightarrow$$

$$\frac{\delta^{q+1}\left[\mathcal{F}_1[\cdots]\,\mathcal{F}_2[\cdots]\right]}{\delta u(\tau_1)\cdots\delta u(\tau_{q+1})} = \sum_{m=0}^{q+1}\;\sum_{A\in\wp_m(\{\tau_1,\ldots,\tau_{q+1}\})} \frac{\delta^m \mathcal{F}_1[\cdots]}{\prod\limits_{\tau\in A}\delta u(\tau)}\,\frac{\delta^{q+1-m}\mathcal{F}_2[\cdots]}{\prod\limits_{\tau\notin A}\delta u(\tau)}, \tag{B.42}$$

Eq. (B.42) is Eq. (B.5) for $q+1$. ■

## List of double summation rearrangement formulae

**1.**
$$\sum_{p=0}^{\infty}\sum_{m=0}^{\min\{p,\,n\}} K_{p,m} = \sum_{m=0}^{n}\sum_{p=m}^{\infty} K_{p,m}. \tag{B.43}$$

**Proof.** In order to be able to treat the upper limit $\min\{p,\,n\}$ of $k-$sum, we split $p-$sum:

$$\sum_{p=0}^{\infty}\sum_{m=0}^{\min\{p,\,n\}} K_{p,m} = \sum_{p=0}^{n}\sum_{m=0}^{p} K_{p,m} + \sum_{p=n+1}^{\infty}\sum_{m=0}^{n} K_{p,m}. \tag{B.44}$$

The summations in the second double sum in the right-hand side of Eq. (B.44) can be interchanged. The first double sum is over $0\le p\le n$, $0\le m\le p$. By substituting $p=n$ into the double inequality for $m$, we define the maximum range of $m$; $0\le m\le n$. Also, for a given $m$, $m\le p$, and so $m\le p\le n$. Thus, Eq. (B.44) is expressed as

$$\sum_{p=0}^{\infty}\sum_{m=0}^{\min\{p,\,n\}} K_{p,m} = \sum_{m=0}^{n}\sum_{p=m}^{n} K_{p,m} + \sum_{m=0}^{n}\sum_{p=n+1}^{\infty} K_{p,m} = \sum_{m=0}^{n}\sum_{p=m}^{\infty} K_{p,m}. \quad ■$$

**2.**
$$\sum_{p=0}^{\infty}\sum_{m=0}^{\min\{2p,\,n\}} K_{p,m} = \sum_{m=0}^{n}\sum_{p=\lceil m/2\rceil}^{\infty} K_{p,m}\,,. \tag{B.45}$$

where $\lceil\bullet\rceil$ denotes the ceiling function.

**Proof.** We distinguish the cases with $n$ even and $n$ odd.

<u>For $n$ even</u>; $n=2\ell$: First, we split the $p-$sum in order to be able to treat the upper limit $\min\{2p,\,2\ell\}$ of $k-$sum:

$$\sum_{p=0}^{\infty}\sum_{m=0}^{\min\{2p,\,2\ell\}} K_{p,m} = \sum_{p=0}^{\ell}\sum_{m=0}^{2p} K_{p,m} + \sum_{p=\ell+1}^{\infty}\sum_{m=0}^{2\ell} K_{p,m}. \tag{B.46}$$

The summations in the second double sum in the right-hand side of Eq. (B.46) can be interchanged. The first double sum is over $0\le p\le \ell$, $0\le m\le 2p$. By substituting $p=\ell$ into the double inequality for $m$, we define the maximum range of $m$; $0\le m\le 2\ell$. Also, for a given $m$, $m\le 2p \Rightarrow \lceil m/2\rceil\le p$, and so $\lceil m/2\rceil\le p\le \ell$. Thus, Eq. (B.46) is expressed as

$$\sum_{p=0}^{\infty}\sum_{m=0}^{\min\{2p,\,2\ell\}} K_{p,m} = \sum_{m=0}^{2\ell}\sum_{p=\lceil m/2\rceil}^{\ell} K_{p,m} + \sum_{m=0}^{2\ell}\sum_{p=\ell+1}^{\infty} K_{p,m} = \sum_{m=0}^{2\ell}\sum_{p=\lceil m/2\rceil}^{\infty} K_{p,m}. \tag{B.47}$$

For $n$ odd; $n = 2\ell + 1$, and with similar arguments

$$\sum_{p=0}^{\infty} \sum_{m=0}^{\min\{2p,\, 2\ell+1\}} K_{p,m} = \sum_{p=0}^{\ell} \sum_{m=0}^{2p} K_{p,m} + \sum_{p=\ell+1}^{\infty} \sum_{m=0}^{2\ell+1} K_{p,m} = \text{[change the order of sums as previously]}$$

$$= \sum_{m=0}^{2\ell} \sum_{p=\lceil m/2 \rceil}^{\ell} K_{p,m} + \sum_{m=0}^{2\ell+1} \sum_{p=\ell+1}^{\infty} K_{p,m} = \quad \text{[separate the } m = 2\ell+1 \text{ term of the last double sum]}$$

$$= \sum_{m=0}^{2\ell} \sum_{p=\lceil m/2 \rceil}^{\ell} K_{p,m} + \sum_{m=0}^{2\ell} \sum_{p=\ell+1}^{\infty} K_{p,m} + \sum_{p=\ell+1}^{\infty} K_{p,2\ell+1} = \quad \text{[merge the two double sums]}$$

$$= \sum_{m=0}^{2\ell} \sum_{p=\lceil m/2 \rceil}^{\infty} K_{p,m} + \sum_{p=\ell+1}^{\infty} K_{p,2\ell+1} .$$

By observing that $\lceil (2\ell+1)/2 \rceil = \ell+1$, the single sum is identified as the $m = 2\ell+1$ term of the double sum, resulting into

$$\sum_{p=0}^{\infty} \sum_{m=0}^{\min\{2p,\, 2\ell+1\}} K_{p,m} = \sum_{m=0}^{2\ell+1} \sum_{p=\lceil m/2 \rceil}^{\infty} K_{p,m} . \tag{B.48}$$

Eqs. (B.47), (B.48) are expressed in a unifying way via Eq. (A.45). ■

**3.** $$\sum_{p=\lceil m/2 \rceil}^{\infty} \sum_{k=m-\min\{p,\, m\}}^{\lfloor m/2 \rfloor} K_{p,k} = \sum_{k=0}^{\lfloor m/2 \rfloor} \sum_{p=m-k}^{\infty} K_{p,k} , \tag{B.49}$$

where $\lfloor \bullet \rfloor$, $\lceil \bullet \rceil$ denote the floor and ceiling functions respectively.

**Proof.** First, we split the $p-$sum in order to be able to treat the lower limit $m - \min\{p,\, m\}$ of $k-$sum:

$$\sum_{p=\lceil m/2 \rceil}^{\infty} \sum_{k=m-\min\{p,\, m\}}^{\lfloor m/2 \rfloor} K_{p,k} = \sum_{p=\lceil m/2 \rceil}^{m} \sum_{k=m-p}^{\lfloor m/2 \rfloor} K_{p,k} + \sum_{p=m+1}^{\infty} \sum_{k=0}^{\lfloor m/2 \rfloor} K_{p,k} . \tag{B.50}$$

In the right-hand side of Eq. (B.50), the order of summations in the second double sum can be interchanged. The first double sum is over $\lceil m/2 \rceil \le p \le m$, $m - p \le k \le \lfloor m/2 \rfloor$. By substituting $p = m$ into the double inequality for $k$, we define the maximum range of $k$; $0 \le k \le \lfloor m/2 \rfloor$. Also, for a given $k$, $m - p \le k \Rightarrow m - k \le p$, and so $m - k \le p \le m$. Thus, Eq. (B.50) is expressed as

$$\sum_{p=\lceil m/2 \rceil}^{\infty} \sum_{k=m-\min\{p,\, m\}}^{\lfloor m/2 \rfloor} K_{p,k} = \sum_{k=0}^{\lfloor m/2 \rfloor} \sum_{p=m-k}^{m} K_{p,k} + \sum_{k=0}^{\lfloor m/2 \rfloor} \sum_{p=m+1}^{\infty} K_{p,k} . \tag{B.51}$$

By merging the two double sums in the right-hand side of Eq. (B.51), Eq. (B.49) is obtained. ■

# Appendix C: Proof of Lemma 3.7

For proving Eq. (3.89) of Chapter 3, that gives the sum $E_{13}+E_2$, we begin with term $E_2$, given by Eq. (3.83):

$$
\begin{aligned}
E_2 = & \sum_{I_1 \cup I_2 \cup I_3 = I^{(n)}} \prod_{i_1 \in I_1} m_{\Xi}(s_{i_1}) \prod_{i_2 \in I_2} C_{X_0 \Xi}(s_{i_2}) \times \\
& \times \sum_{k=0}^{\lfloor |I_3|/2 \rfloor} \sum_{\{P,S\} \in \wp_k(I_3)} \sum_{\ell \in S} \prod_{\{j_1, j_2\} \in P \cup \{\ell, n+1\}} C_{\Xi\Xi}(s_{j_1}, s_{j_2}) \times \\
& \times \int_{t_0}^{t} \overset{(|I_3|-2k-1)}{\cdots} \int_{t_0}^{t} \prod_{i_3 \in S \setminus \{\ell\}} C_{\Xi\Xi}(s_{i_3}, \tau_{i_3}) \, \Xi^{\theta} \left[ \frac{\partial^{|I_2|} \delta^{|I_3|-2k-1} \mathcal{F}[\cdots]}{\partial X_0^{|I_2|}(\theta) \prod_{i_3 \in S \setminus \{\ell\}} \delta \Xi(\tau_{i_3};\theta)} \right] \prod_{i_3 \in S \setminus \{\ell\}} d\tau_{i_3} .
\end{aligned}
\tag{C.1}
$$

**Remark C.1.** Since $\sum_{\ell \in \varnothing} \bullet = 0$ (assumption made before Eq. (3.81)), terms with $S=\varnothing$ should be excluded from the right-hand side of Eq. (C.1). The first, trivial case of such term is for $I_3=\varnothing$. Furthermore, for $I_3 \neq \varnothing$, a case with $S=\varnothing$ exists for even $|I_3|$ and for $k=\lfloor |I_3|/2 \rfloor$, since, in this case, all elements of $I_3$ are paired. Thus, $k-$sum in the right-hand side of Eq. (C.1) should have as upper limit $\lfloor |I_3|/2 \rfloor - 1$ for even $|I_3|$, and $\lfloor |I_3|/2 \rfloor$ for odd $|I_3|$. It can be easily proved that the two said values can be written in a unified way as $\lfloor (|I_3|-1)/2 \rfloor$, resulting thus into

$$
\begin{aligned}
E_2 = & \sum_{\substack{I_1 \cup I_2 \cup I_3 = I^{(n)} \\ |I_3| \geq 1}} \prod_{i_1 \in I_1} m_{\Xi}(s_{i_1}) \prod_{i_2 \in I_2} C_{X_0 \Xi}(s_{i_2}) \times \\
& \times \sum_{k=0}^{\lfloor (|I_3|-1)/2 \rfloor} \sum_{\{P,S\} \in \wp_k(I_3)} \sum_{\ell \in S} \prod_{\{j_1, j_2\} \in P \cup \{\ell, n+1\}} C_{\Xi\Xi}(s_{j_1}, s_{j_2}) \times \\
& \times \int_{t_0}^{t} \overset{(|I_3|-2k-1)}{\cdots} \int_{t_0}^{t} \prod_{i_3 \in S \setminus \{\ell\}} C_{\Xi\Xi}(s_{i_3}, \tau_{i_3}) \, \Xi^{\theta} \left[ \frac{\partial^{|I_2|} \delta^{|I_3|-2k-1} \mathcal{F}[\cdots]}{\partial X_0^{|I_2|}(\theta) \prod_{i_3 \in S \setminus \{\ell\}} \delta \Xi(\tau_{i_3};\theta)} \right] \prod_{i_3 \in S \setminus \{\ell\}} d\tau_{i_3} .
\end{aligned}
$$

By also performing the change of index $\kappa = k+1$, we finally obtain

$$
E_2 = \sum_{\substack{I_1 \cup I_2 \cup I_3 = I^{(n)} \\ |I_3| \geq 1}} \prod_{i_1 \in I_1} m_{\Xi}(s_{i_1}) \prod_{i_2 \in I_2} C_{X_0 \Xi}(s_{i_2}) \times
$$

$$\times \sum_{\kappa=1}^{\lfloor (|I_3|+1)/2 \rfloor} \sum_{\{P,S\} \in \wp_{\kappa-1}(I_3)} \sum_{\ell \in S} \prod_{\{j_1, j_2\} \in P \cup \{\ell, n+1\}} C_{\Xi\Xi}(s_{j_1}, s_{j_2}) \times \tag{C.2}$$
$$\times \overbrace{\int_{t_0}^{t} \cdots \int_{t_0}^{t}}^{(|I_3|+1-2\kappa)} \prod_{i_3 \in S \setminus \{\ell\}} C_{\Xi\Xi}(s_{i_3}, \tau_{i_3}) \, \Xi^{\theta} \left[ \frac{\partial^{|I_2|} \delta^{|I_3|+1-2\kappa} \mathcal{F}[\cdots]}{\partial X_0^{|I_2|}(\theta) \prod_{i_3 \in S \setminus \{\ell\}} \delta \Xi(\tau_{i_3}; \theta)} \right] \prod_{i_3 \in S \setminus \{\ell\}} d\tau_{i_3} .$$

Regarding the upper limit of $\kappa$ – sum, it can be easily verified, by considering the cases of $|I_3|$ being even and odd separately, that $\lfloor (|I_3|-1)/2 \rfloor + 1 = \lfloor (|I_3|+1)/2 \rfloor$. By adding $E_2$, given by Eq. (C.2), and $E_{13}$, given by Eq. (3.87), we obtain the relation:

$$\begin{aligned} E_{13} + E_2 = & \sum_{I_1 \cup I_2 \cup I_3 = I^{(n)}} \prod_{i_1 \in I_1} m_{\Xi}(s_{i_1}) \prod_{i_2 \in I_2} C_{X_0 \Xi}(s_{i_2}) \sum_{k=0}^{\lfloor |I_3|/2 \rfloor} A_k^{(13)}(I_3) + \\ & + \sum_{\substack{I_1 \cup I_2 \cup I_3 = I^{(n)} \\ |I_3| \geq 1}} \prod_{i_1 \in I_1} m_{\Xi}(s_{i_1}) \prod_{i_2 \in I_2} C_{X_0 \Xi}(s_{i_2}) \sum_{k=1}^{\lfloor (|I_3|+1)/2 \rfloor} A_{k-1}^{(2)}(I_3), \end{aligned} \tag{C.3}$$

where

$$A_k^{(13)}(I_3) = \sum_{\{P,S\} \in \wp_k(I_3)} \prod_{\{j_1, j_2\} \in P} C_{\Xi\Xi}(s_{j_1}, s_{j_2}) \times$$
$$\times \overbrace{\int_{t_0}^{t} \cdots \int_{t_0}^{t}}^{(|I_3|+1-2k)} \prod_{i_3 \in S \cup \{n+1\}} C_{\Xi\Xi}(s_{i_3}, \tau_{i_3}) \, \Xi^{\theta} \left[ \frac{\partial^{|I_2|} \delta^{|I_3|+1-2k} \mathcal{F}[\cdots]}{\partial X_0^{|I_2|}(\theta) \prod_{i_3 \in S \cup \{n+1\}} \delta \Xi(\tau_{i_3}; \theta)} \right] \prod_{i_3 \in S \cup \{n+1\}} d\tau_{i_3} , \tag{C.4}$$

and

$$A_{k-1}^{(2)}(I_3) = \sum_{\{P,S\} \in \wp_{k-1}(I_3)} \sum_{\ell \in S} \prod_{\{j_1, j_2\} \in P \cup \{\ell, n+1\}} C_{\Xi\Xi}(s_{j_1}, s_{j_2}) \times$$
$$\times \overbrace{\int_{t_0}^{t} \cdots \int_{t_0}^{t}}^{(|I_3|+1-2k)} \prod_{i_3 \in S \setminus \{\ell\}} C_{\Xi\Xi}(s_{i_3}, \tau_{i_3}) \, \Xi^{\theta} \left[ \frac{\partial^{|I_2|} \delta^{|I_3|+1-2k} \mathcal{F}[\cdots]}{\partial X_0^{|I_2|}(\theta) \prod_{i_3 \in S \setminus \{\ell\}} \delta \Xi(\tau_{i_3}; \theta)} \right] \prod_{i_3 \in S \setminus \{\ell\}} d\tau_{i_3} . \tag{C.5}$$

The upper indices, (13) and (2), in $A_k$'s are reminiscent of the terms $E_{13}$, $E_2$, from which $A_k^{(13)}(I_3)$, $A_{k-1}^{(2)}(I_3)$ came. Also, each $A_k$ is defined as a sum over all partitions of $I_3$ with a certain number of pairs. This number is denoted by the lower index in $A_k^{(13)}(I_3)$, $A_{k-1}^{(2)}(I_3)$.

As we see, the right-hand side of Eq. (C.3) is expressed in terms of partitions of set $I_3$, with the additional element $\{n+1\}$ being joined with the singletons, in $A_k^{(13)}(I_3)$'s, or considered paired, in $A_{k-1}^{(2)}(I_3)$'s. The essential part of the proof of the present lemma is to express the right-hand side of Eq. (C.3) in terms of partitions of the set $I_3 \cup \{n+1\}$.

First, in Eq. (C.3), the term with $I_3 = \varnothing$ is only one, coming from $E_{13}$:

$$\Sigma_0 = \sum_{\substack{I_1 \cup I_2 \cup I_3 = I^{(n)} \\ I_3 = \varnothing}} \prod_{i_1 \in I_1} m_{\Xi}(s_{i_1}) \prod_{i_2 \in I_2} C_{X_0\Xi}(s_{i_2})\, A_0^{(13)}(\varnothing) =$$

$$= \sum_{I_1 \cup I_2 \cup \varnothing = I^{(n)}} \prod_{i_1 \in I_1} m_{\Xi}(s_{i_1}) \prod_{i_2 \in I_2} C_{X_0\Xi}(s_{i_2}) \int_{t_0}^{t} C_{\Xi\Xi}(s_{n+1}, \tau_{n+1})\, \Xi^{\theta}\left[\frac{\partial^{|I_2|}\delta \mathcal{F}[\cdots]}{\partial X_0^{|I_2|}(\theta)\, \delta\Xi(\tau_{n+1};\theta)}\right] d\tau_{n+1}. \tag{C.6}$$

Since $\lfloor (|I_3|+1)/2 \rfloor = 0$ for $|I_3| = 0$, Eq. (C.6) is trivially expressed as

$$\Sigma_0 = \sum_{\substack{I_1 \cup I_2 \cup I_3 = I^{(n)} \\ |I_3| = 0}} \prod_{i_1 \in I_1} m_{\Xi}(s_{i_1}) \prod_{i_2 \in I_2} C_{X_0\Xi}(s_{i_2}) \sum_{k=0}^{\lfloor (|I_3|+1)/2 \rfloor} \sum_{\{P,S\} \in \wp_k(I_3 \cup \{n+1\})} \prod_{\{j_1, j_2\} \in P} C_{\Xi\Xi}(s_{j_1}, s_{j_2}) \times$$

$$\times \int_{t_0}^{t} \overset{(|I_3|+1-2k)}{\cdots} \int_{t_0}^{t} \prod_{i_3 \in S} C_{\Xi\Xi}(s_{i_3}, \tau_{i_3})\, \Xi^{\theta}\left[\frac{\partial^{|I_2|}\delta^{|I_3|+1-2k} \mathcal{F}[\cdots]}{\partial X_0^{|I_2|}(\theta)\, \delta\Xi(\tau_{n+1};\theta) \prod_{i_3 \in S} \delta\Xi(\tau_{i_3};\theta)}\right] \prod_{i_3 \in S} d\tau_{i_3}. \tag{C.7}$$

Similarly, all terms with $|I_3| = 1$ in the right-hand side of Eq. (C.3) are grouped in

$$\Sigma_1 = \sum_{\substack{I_1 \cup I_2 \cup I_3 = I^{(n)} \\ |I_3| = 1}} \prod_{i_1 \in I_1} m_{\Xi}(s_{i_1}) \prod_{i_2 \in I_2} C_{X_0\Xi}(s_{i_2}) \left(A_0^{(13)}(I_3) + A_0^{(2)}(I_3)\right) =$$

$$= \sum_{\ell \in I^{(n)}} \sum_{I_1 \cup I_2 \cup \{\ell\} = I^{(n)}} \prod_{i_1 \in I_1} m_{\Xi}(s_{i_1}) \prod_{i_2 \in I_2} C_{X_0\Xi}(s_{i_2}) \left( C_{\Xi\Xi}(s_{\ell}, s_{n+1})\, \Xi^{\theta}\left[\frac{\partial^{|I_2|} \mathcal{F}[\cdots]}{\partial X_0^{|I_2|}(\theta)}\right] + \right. \tag{C.8}$$

$$\left. + \int_{t_0}^{t}\int_{t_0}^{t} \prod_{i_3 \in \{\ell, n+1\}} C_{\Xi\Xi}(s_{i_3}, \tau_{i_3})\, \Xi^{\theta}\left[\frac{\partial^{|I_2|}\delta^{2} \mathcal{F}[\cdots]}{\partial X_0^{|I_2|}(\theta) \prod_{i_3 \in \{\ell, n+1\}} \delta\Xi(\tau_{i_3};\theta)}\right] \prod_{i_3 \in \{\ell, n+1\}} d\tau_{i_3} \right).$$

Since $\lfloor (|I_3|+1)/2 \rfloor = 1$ for $|I_3| = 1$, Eq. (C.8) is easily written equivalently as

$$\Sigma_1 = \sum_{\substack{I_1 \cup I_2 \cup I_3 = I^{(n)} \\ |I_3| = 1}} \prod_{i_1 \in I_1} m_{\Xi}(s_{i_1}) \prod_{i_2 \in I_2} C_{X_0\Xi}(s_{i_2}) \sum_{k=0}^{\lfloor (|I_3|+1)/2 \rfloor} \sum_{\{P,S\} \in \wp_k(I_3 \cup \{n+1\})} \prod_{\{j_1, j_2\} \in P} C_{\Xi\Xi}(s_{j_1}, s_{j_2}) \times$$

$$\times \int_{t_0}^{t} \overset{(|I_3|+1-2k)}{\cdots} \int_{t_0}^{t} \prod_{i_3 \in S} C_{\Xi\Xi}(s_{i_3}, \tau_{i_3})\, \Xi^{\theta}\left[\frac{\partial^{|I_2|}\delta^{|I_3|+1-2k} \mathcal{F}[\cdots]}{\partial X_0^{|I_2|}(\theta)\, \delta\Xi(\tau_{n+1};\theta) \prod_{i_3 \in S} \delta\Xi(\tau_{i_3};\theta)}\right] \prod_{i_3 \in S} d\tau_{i_3}. \tag{C.9}$$

By taking into account Eqs. (C.7), (C.9), and since $\lfloor |I_3|/2 \rfloor = \lfloor (|I_3|+1)/2 \rfloor$ for even $|I_3|$, and $\lfloor |I_3|/2 \rfloor + 1 = \lfloor (|I_3|+1)/2 \rfloor$ for odd $|I_3|$, Eq. (C.3) is expressed equivalently as

$$E_{13} + E_2 = \Sigma_0 + \Sigma_1 + \Sigma_2 + \Sigma_3, \tag{C.10}$$

with

$$\Sigma_2 = \sum_{\substack{I_1 \cup I_2 \cup I_3 = I^{(n)} \\ |I_3| \geq 2,\, |I_3| \text{ even}}} \prod_{i_1 \in I_1} m_\Xi(s_{i_1}) \prod_{i_2 \in I_2} C_{X_0 \Xi}(s_{i_2}) \left[ A_0^{(13)}(I_3) + \sum_{k=1}^{\lfloor |I_3|/2 \rfloor} \left( A_k^{(13)}(I_3) + A_{k-1}^{(2)}(I_3) \right) \right], \tag{C.11}$$

$$\Sigma_3 = \sum_{\substack{I_1 \cup I_2 \cup I_3 = I^{(n)} \\ |I_3| \geq 3,\, |I_3| \text{ odd}}} \prod_{i_1 \in I_1} m_\Xi(s_{i_1}) \prod_{i_2 \in I_2} C_{X_0 \Xi}(s_{i_2}) \times \left[ A_0^{(13)}(I_3) + \sum_{k=1}^{\lfloor |I_3|/2 \rfloor} \left( A_k^{(13)}(I_3) + A_{k-1}^{(2)}(I_3) \right) + A_{\lfloor |I_3|/2 \rfloor}^{(2)}(I_3) \right]. \tag{C.12}$$

Now, by using Eq. (C.4), $A_0^{(13)}(I_3)$ is specified into

$$A_0^{(13)}(I_3) = \overbrace{\int_{t_0}^{t} \cdots \int_{t_0}^{t}}^{(|I_3|+1)} \prod_{i_3 \in I_3 \cup \{n+1\}} C_{\Xi\Xi}(s_{i_3}, \tau_{i_3})\, \Xi^\theta \left[ \frac{\partial^{|I_2|} \delta^{|I_3|+1} \mathcal{F}[\cdots]}{\partial X_0^{|I_2|}(\theta) \prod_{i_3 \in I_3 \cup \{n+1\}} \delta \Xi(\tau_{i_3}; \theta)} \right] \prod_{i_3 \in I_3 \cup \{n+1\}} d\tau_{i_3},$$

which is expressed in terms of $\wp_0(I_3 \cup \{n+1\})$ (all elements are singletons), as

$$A_0^{(13)}(I_3) = \sum_{\{P,S\} \in \wp_0(I_3 \cup \{n+1\})} \overbrace{\int_{t_0}^{t} \cdots \int_{t_0}^{t}}^{(|I_3|+1)} \prod_{i_3 \in S} C_{\Xi\Xi}(s_{i_3}, \tau_{i_3})\, \Xi^\theta \left[ \frac{\partial^{|I_2|} \delta^{|I_3|+1} \mathcal{F}[\cdots]}{\partial X_0^{|I_2|}(\theta) \prod_{i_3 \in S} \delta \Xi(\tau_{i_3}; \theta)} \right] \prod_{i_3 \in S} d\tau_{i_3}. \tag{C.13}$$

Second, by using Eqs. (C.4) and (C.5), we obtain, for $k = 1, \ldots, \lfloor |I_3|/2 \rfloor$:

$$\begin{aligned} A_k^{(13)}(I_3) + A_{k-1}^{(2)}(I_3) &= \sum_{\{P,S\} \in \wp_k(I_3)} \prod_{\{j_1, j_2\} \in P} C_{\Xi\Xi}(s_{j_1}, s_{j_2}) \times \\ &\times \overbrace{\int_{t_0}^{t} \cdots \int_{t_0}^{t}}^{(|I_3|+1-2k)} \prod_{i_3 \in S \cup \{n+1\}} C_{\Xi\Xi}(s_{i_3}, \tau_{i_3})\, \Xi^\theta \left[ \frac{\partial^{|I_2|} \delta^{|I_3|+1-2k} \mathcal{F}[\cdots]}{\partial X_0^{|I_2|}(\theta) \prod_{i_3 \in S \cup \{n+1\}} \delta \Xi(\tau_{i_3}; \theta)} \right] \prod_{i_3 \in S \cup \{n+1\}} d\tau_{i_3} \\ &+ \sum_{\{P,S\} \in \wp_{k-1}(I_3)} \sum_{\ell \in S} \prod_{\{j_1, j_2\} \in P \cup \{\ell, n+1\}} C_{\Xi\Xi}(s_{j_1}, s_{j_2}) \times \\ &\times \overbrace{\int_{t_0}^{t} \cdots \int_{t_0}^{t}}^{(|I_3|+1-2k)} \prod_{i_3 \in S \setminus \{\ell\}} C_{\Xi\Xi}(s_{i_3}, \tau_{i_3})\, \Xi^\theta \left[ \frac{\partial^{|I_2|} \delta^{|I_3|+1-2k} \mathcal{F}[\cdots]}{\partial X_0^{|I_2|}(\theta) \prod_{i_3 \in S \setminus \{\ell\}} \delta \Xi(\tau_{i_3}; \theta)} \right] \prod_{i_3 \in S \setminus \{\ell\}} d\tau_{i_3}. \end{aligned} \tag{C.14}$$

**Remark C.2.** The set $\wp_k(I_3 \cup \{n+1\})$ of all partitions of $I_3 \cup \{n+1\}$, $|I_3| \geq 2$, with $1 \leq k \leq \lfloor |I_3|/2 \rfloor$ pairs, consists of: **i)** all partitions of $I_3$ with $k$ pairs, in which the additional element $n+1$ is added as a singleton, and, **ii)** all partitions of $I_3$ with $k-1$ pairs, with the additional element $n+1$ paired with one of the singletons of the partition. Since we observe that the sum of $A_k^{(13)}(I_3)$ extends over all partitions of i) category, while $A_{k-1}^{(2)}(I_3)$ extends over all partitions of ii) category, Eq. (C.14) can be expressed as

$$A_k^{(13)}(I_3) + A_{k-1}^{(2)}(I_3) = \sum_{\{P,S\}\in\wp_k(I_3\cup\{n+1\})} \prod_{\{j_1,j_2\}\in P} C_{\Xi\Xi}(s_{j_1}, s_{j_2}) \int_{t_0}^{t} \overset{(|I_3|+1-2k)}{\cdots} \int_{t_0}^{t} \prod_{i_3\in S} C_{\Xi\Xi}(s_{i_3}, \tau_{i_3}) \times$$
$$\times\ \mathbb{E}^{\theta}\left[\frac{\partial^{|I_2|}\delta^{|I_3|+1-2k}\mathcal{F}[\cdots]}{\partial X_0^{|I_2|}(\theta)\prod_{i_3\in S}\delta\Xi(\tau_{i_3};\theta)}\right]\prod_{i_3\in S} d\tau_{i_3}, \qquad k = 1,\ldots,\lfloor |I_3|/2 \rfloor. \tag{C.15}$$

Last, for sets $I_3$ with odd cardinality, we have the additional term in $\Sigma_3$:

$$A_{\lfloor |I_3|/2\rfloor}^{(2)}(I_3) = \sum_{\{P,S\}\in\wp_{\lfloor |I_3|/2\rfloor}(I_3)} \sum_{\ell\in S} \prod_{\{j_1,j_2\}\in P\cup\{\ell,n+1\}} C_{\Xi\Xi}(s_{j_1}, s_{j_2})\, \mathbb{E}^{\theta}\left[\frac{\partial^{|I_2|}\mathcal{F}[\cdots]}{\partial X_0^{|I_2|}(\theta)}\right]. \tag{C.16}$$

Since $|I_3|$ is odd, in all partitions $\{P, S\}\in\wp_{\lfloor |I_3|/2\rfloor}(I_3)$, the set of singletons $S$ contains only one element. In the right-hand side of Eq. (C16), for each partition, its only singleton is paired with the element $\{n+1\}$. Thus, the right-hand side of Eq. (C.16) is expressed equivalently in terms of $\wp_{\lfloor (|I_3|+1)/2\rfloor}(I_3\cup\{n+1\})$ (all elements of the set $I_3\cup\{n+1\}$, whose cardinality is even, are paired), as

$$A_{\lfloor |I_3|/2\rfloor}^{(2)}(I_3) = \sum_{\{P,S\}\in\wp_{\lfloor (|I_3|+1)/2\rfloor}(I_3\cup\{n+1\})} \prod_{\{j_1,j_2\}\in P} C_{\Xi\Xi}(s_{j_1}, s_{j_2})\, \mathbb{E}^{\theta}\left[\frac{\partial^{|I_2|}\mathcal{F}[\cdots]}{\partial X_0^{|I_2|}(\theta)}\right]. \tag{C.17}$$

By substituting Eq. (C.13), (C.15) into Eq. (C.11), and using that $\lfloor |I_3|/2\rfloor = \lfloor (|I_3|+1)/2\rfloor$ for even $|I_3|$, we obtain

$$\Sigma_2 = \sum_{\substack{I_1\cup I_2\cup I_3 = I^{(n)}\\ |I_3|\geq 2,\, |I_3|\text{ even}}} \prod_{i_1\in I_1} m_{\Xi}(s_{i_1}) \prod_{i_2\in I_2} C_{X_0\Xi}(s_{i_2}) \sum_{k=0}^{\lfloor (|I_3|+1)/2\rfloor} \sum_{\{P,S\}\in\wp_k(I_3\cup\{n+1\})} \prod_{\{j_1,j_2\}\in P} C_{\Xi\Xi}(s_{j_1}, s_{j_2}) \times$$
$$\times \int_{t_0}^{t} \overset{(|I_3|+1-2k)}{\cdots} \int_{t_0}^{t} \prod_{i_3\in S} C_{\Xi\Xi}(s_{i_3}, \tau_{i_3})\, \mathbb{E}^{\theta}\left[\frac{\partial^{|I_2|}\delta^{|I_3|+1-2k}\mathcal{F}[\cdots]}{\partial X_0^{|I_2|}(\theta)\prod_{i_3\in S}\delta\Xi(\tau_{i_3};\theta)}\right]\prod_{i_3\in S} d\tau_{i_3}. \tag{C.18}$$

By substituting Eq. (C.13), (C.15), (C.17) into Eq. (C.12), and using that $\lfloor |I_3|/2\rfloor + 1 = \lfloor (|I_3|+1)/2\rfloor$ for odd $|I_3|$, we also obtain

$$\Sigma_3 = \sum_{\substack{I_1\cup I_2\cup I_3 = I^{(n)}\\ |I_3|\geq 3,\, |I_3|\text{ odd}}} \prod_{i_1\in I_1} m_{\Xi}(s_{i_1}) \prod_{i_2\in I_2} C_{X_0\Xi}(s_{i_2}) \sum_{k=0}^{\lfloor (|I_3|+1)/2\rfloor} \sum_{\{P,S\}\in\wp_k(I_3\cup\{n+1\})} \prod_{\{j_1,j_2\}\in P} C_{\Xi\Xi}(s_{j_1}, s_{j_2}) \times$$
$$\times \int_{t_0}^{t} \overset{(|I_3|+1-2k)}{\cdots} \int_{t_0}^{t} \prod_{i_3\in S} C_{\Xi\Xi}(s_{i_3}, \tau_{i_3})\, \mathbb{E}^{\theta}\left[\frac{\partial^{|I_2|}\delta^{|I_3|+1-2k}\mathcal{F}[\cdots]}{\partial X_0^{|I_2|}(\theta)\prod_{i_3\in S}\delta\Xi(\tau_{i_3};\theta)}\right]\prod_{i_3\in S} d\tau_{i_3}. \tag{C.19}$$

Last, by substituting Eqs. (C.7) for $\Sigma_0$, Eq. (C.9) for $\Sigma_1$, and Eqs. (C.18), (C.19) for $\Sigma_2$, $\Sigma_3$ respectively, into Eq. (C.10), we obtain

$$E_{13} + E_2 = \sum_{I_1 \cup I_2 \cup I_3 = I^{(n)}} \prod_{i_1 \in I_1} m_\Xi(s_{i_1}) \prod_{i_2 \in I_2} C_{X_0 \Xi}(s_{i_2}) \times$$
$$\times \sum_{k=0}^{\lfloor (|I_3|+1)/2 \rfloor} \sum_{\{P,S\} \in \wp_k(I_3 \cup \{n+1\})} \prod_{\{j_1, j_2\} \in P} C_{\Xi\Xi}(s_{j_1}, s_{j_2}) \times \tag{C.20}$$
$$\times \int_{t_0}^{t} \overset{(|I_3|+1-2k)}{\cdots} \int_{t_0}^{t} \prod_{i_3 \in S} C_{\Xi\Xi}(s_{i_3}, \tau_{i_3}) \, \mathbb{E}^{\theta} \left[ \frac{\partial^{|I_2|} \delta^{|I_3|+1-2k} \mathcal{F}[\cdots]}{\partial X_0^{|I_2|}(\theta) \prod_{i_3 \in S} \delta \Xi(\tau_{i_3};\theta)} \right] \prod_{i_3 \in S} d\tau_{i_3} ,$$

which is Eq. (3.89) of Chapter 3. Thus, the proof of Lemma 3.7 is completed.

# Appendix D: Direct formulation and solution of the moment problem for a linear, additively-excited RDE

The random differential equation for the linear RDE with additive colored noise reads as

$$\dot{X}(t;\theta) = \eta X(t;\theta) + \kappa\, \Xi(t;\theta), \qquad X(t_0;\theta) = X_0(\theta). \tag{D.1a,b}$$

In this Appendix, we will derive, from Eqs. (D.1a,b), the corresponding deterministic initial value problems (IVPs) for the first and second moments of the response $X(t;\theta)$, as well as the cross-correlation functions of the response with excitation $\Xi(t;\theta)$, and initial value $X_0(\theta)$. The derivation of these moment equations will be performed by multiplying both sides of Eqs. (D.1a,b) with the appropriate, for each problem, random function, and then by taking the average $\mathbb{E}^{\theta}[\bullet]$ of both sides of the equation. In these moment IVPs, moments of the initial value and excitations, as well as their joint moments, are considered known as data of the problem. Since the original RIVP, Eqs. (D.1a,b) is a linear RDE, the corresponding moment equations will be linear ODEs of the general form

$$\frac{dx(t)}{dt} = a\,x(t) + b\,y(t), \quad x(t_0) = x_0, \tag{i}$$

whose solution is

$$x(t) = x_0 e^{a(t-t_0)} + b\int_{t_0}^{t} y(\tau)\, e^{a(t-\tau)}\, d\tau\,. \tag{ii}$$

**D1. IVP for mean value of the response**

We construct an equation for mean value $m_X(t)$ by averaging both sides of Eqs. (D.1a,b):

$$\left.\begin{aligned} \mathbb{E}^{\theta}\left[\dot{X}(t;\theta)\right] &= \eta\mathbb{E}^{\theta}\left[X(t;\theta)\right] + \kappa\,\mathbb{E}^{\theta}\left[\Xi(t;\theta)\right] \\ \mathbb{E}^{\theta}\left[X(t_0;\theta)\right] &= \mathbb{E}^{\theta}\left[X_0(\theta)\right] \end{aligned}\right\} \Rightarrow$$

$$\left.\begin{aligned} \frac{d}{dt}\mathbb{E}^{\theta}\left[X(t;\theta)\right] &= \eta\mathbb{E}^{\theta}\left[X(t;\theta)\right] + \kappa\,\mathbb{E}^{\theta}\left[\Xi(t;\theta)\right] \\ \mathbb{E}^{\theta}\left[X(t_0;\theta)\right] &= \mathbb{E}^{\theta}\left[X_0(\theta)\right], \end{aligned}\right\}$$

that is,

$$\frac{d\, m_X(t)}{dt} = \eta\, m_X(t) + \kappa\, m_\Xi(t), \quad m_X(t_0) = m_{X_0}. \tag{D.2a,b}$$

Solution to the initial value problem (D.2a,b), by using Eqs. (i), (ii), is determined into

$$m_X(t) = m_{X_0}\, e^{\eta(t-t_0)} + \kappa \int_{t_0}^{t} m_\Xi(\tau)\, e^{\eta(t-\tau)}\, d\tau \tag{D.3}$$

**D2. IVP for the two times response-excitation cross-correlation**

We will construct an equation for the cross-correlation function $R_{X\Xi}(t, s)$, by multiplying both sides of Eq. (D.1a) with $\Xi(s;\theta)$ and taking the average:

$$\dot X(t;\theta)\, \Xi(s;\theta) = \eta\, X(t;\theta)\, \Xi(s;\theta) + \kappa\, \Xi(t;\theta)\, \Xi(s;\theta) \;\Rightarrow$$

$$\mathbb{E}^\theta\left[\dot X(t;\theta)\, \Xi(s;\theta)\right] = \eta\, \mathbb{E}^\theta\left[X(t;\theta)\, \Xi(s;\theta)\right] + \kappa\, \mathbb{E}^\theta\left[\Xi(t;\theta)\, \Xi(s;\theta)\right] \;\Rightarrow$$

$$\frac{\partial}{\partial t}\, \mathbb{E}^\theta\left[X(t;\theta)\, \Xi(s;\theta)\right] = \eta\, \mathbb{E}^\theta\left[X(t;\theta)\, \Xi(s;\theta)\right] + \kappa\, \mathbb{E}^\theta\left[\Xi(t;\theta)\, \Xi(s;\theta)\right] \;\Rightarrow$$

$$\frac{\partial R_{X\Xi}(t, s)}{\partial t} = \eta\, R_{X\Xi}(t, s) + \kappa\, R_{\Xi\Xi}(t, s). \tag{D.4a}$$

The same procedure is applied also to the initial condition

$$X(t_0;\theta)\, \Xi(s;\theta) = X_0(\theta)\, \Xi(s;\theta) \;\Rightarrow$$

$$\mathbb{E}^\theta\left[X(t_0;\theta)\, \Xi(s;\theta)\right] = \mathbb{E}^\theta\left[X_0(\theta)\, \Xi(s;\theta)\right] \;\Rightarrow$$

$$\mathbb{E}^\theta\left[X(t_0;\theta)\, \Xi(s;\theta)\right] = \mathbb{E}^\theta\left[X_0(\theta)\, \Xi(s;\theta)\right] \;\Rightarrow$$

$$R_{X\Xi}(t_0, s) = R_{X_0\Xi}(s). \tag{D.4b}$$

Solution to initial value problem (D.4a,b) with respect to $t$ using Eqs. (i), (ii) is written

$$R_{X\Xi}(t, s) = R_{X_0\Xi}(s)\, e^{\eta(t-t_0)} + \kappa \int_{t_0}^{t} R_{\Xi\Xi}(\tau, s)\, e^{\eta(t-\tau)}\, d\tau. \tag{D.5}$$

**D3. IVP for the initial value-response cross-correlation**

We also construct an equation for the initial value-response cross-correlation $R_{XX_0}(t)$ as:

$$X_0(\theta)\, \dot X(t;\theta) = \eta\, X_0(\theta)\, X(t;\theta) + \kappa\, X_0(\theta)\, \Xi(t;\theta) \;\Rightarrow$$

$$\mathbb{E}^\theta\left[X_0(\theta)\, \dot X(t;\theta)\right] = \eta\, \mathbb{E}^\theta\left[X_0(\theta)\, X(t;\theta)\right] + \kappa\, \mathbb{E}^\theta\left[X_0(\theta)\, \Xi(t;\theta)\right] \;\Rightarrow$$

$$\frac{d}{dt}\, \mathbb{E}^\theta\left[X_0(\theta)\, X(t;\theta)\right] = \eta\, \mathbb{E}^\theta\left[X_0(\theta)\, X(t;\theta)\right] + \kappa\, \mathbb{E}^\theta\left[X_0(\theta)\, \Xi(t;\theta)\right] \;\Rightarrow$$

$$\frac{d\, R_{XX_0}(t)}{dt} = \eta\, R_{XX_0}(t) + \kappa\, R_{X_0\Xi}(t), \tag{D.6a}$$

with the initial condition

$$
\begin{aligned}
& X(t_0;\theta)\, X_0(\theta) \;=\; X_0(\theta)\, X_0(\theta) \;\Rightarrow \\
& \Xi^{\theta}\Big[\, X(t_0;\theta)\, X_0(\theta)\Big] \;=\; \mathbb{E}^{\theta}\Big[\, X_0(\theta)\, X_0(\theta)\Big] \;\Rightarrow \\
& R_{X X_0}(t_0) \;=\; R_{X_0 X_0} \,. \qquad\qquad \text{(D.6b)}
\end{aligned}
$$

Thus, the solution to initial value problem (D.6a,b), using Eqs. (i), (ii), is

$$
R_{X X_0}(t) \;=\; R_{X_0 X_0}\, e^{\eta(t-t_0)} + \kappa \int_{t_0}^{t} R_{X_0 \Xi}(\tau)\, e^{\eta(t-\tau)}\, d\tau \,. \qquad \text{(D.7)}
$$

**D4. IVP for the two-time autocorrelation of the response**

We construct an equation for the autocorrelation of the response $R_{XX}\left(s,t\right)$ as:

$$
\begin{aligned}
& X(s;\theta)\, \dot{X}(t;\theta) \;=\; \eta\, X(s;\theta)\, X(t;\theta) + \kappa\, X(s;\theta)\, \Xi(t;\theta) \;\Rightarrow \\
& \mathbb{E}^{\theta}\Big[\, X(s;\theta)\, \dot{X}(t;\theta)\Big] \;=\; \eta\, \mathbb{E}^{\theta}\big[X(s;\theta)\, X(t;\theta)\big] + \kappa\, \mathbb{E}^{\theta}\big[X(s;\theta)\, \Xi(t;\theta)\big] \;\Rightarrow \\
& \frac{\partial}{\partial t}\, \mathbb{E}^{\theta}\big[X(s;\theta)\, X(t;\theta)\big] \;=\; \eta\, \mathbb{E}^{\theta}\big[X(s;\theta)\, X(t;\theta)\big] + \kappa\, \mathbb{E}^{\theta}\big[X(s;\theta)\, \Xi(t;\theta)\big] \;\Rightarrow \\
& \frac{\partial R_{XX}\left(s,t\right)}{\partial t} \;=\; \eta\, R_{XX}\left(s,t\right) + \kappa\, R_{X\Xi}\left(s,t\right), \qquad\qquad \text{(D.8a)}
\end{aligned}
$$

and the initial condition:

$$
\begin{aligned}
& X(s;\theta)\, X(t_0;\theta) \;=\; X(s;\theta)\, X_0(\theta) \;\Rightarrow \\
& \mathbb{E}^{\theta}\Big[\, X(s;\theta)\, X(t_0;\theta)\Big] \;=\; \mathbb{E}^{\theta}\Big[\, X(s;\theta)\, X_0(\theta)\Big] \;\Rightarrow \\
& R_{XX}\Big(s, t_0\Big) \;=\; R_{X X_0}(s) \,. \qquad\qquad \text{(D.8b)}
\end{aligned}
$$

Solution to initial value problem (D.8a,b), by using Eqs. (i), (ii), is

$$
R_{XX}\left(s,t\right) \;=\; R_{X X_0}(s)\, e^{\eta(t-t_0)} + \kappa \int_{t_0}^{t} R_{X\Xi}\left(s,\tau\right) e^{\eta(t-\tau)}\, d\tau \,. \qquad \text{(D.9)}
$$

In Eq. (D.9), moments $R_{X\Xi}\left(s,\tau\right)$, $R_{X X_0}(s)$ are determined by Eqs. (D.5) and (D.7) respectively. In Eq. (D.7) we substitute $t$ with $s$:

$$
R_{X X_0}(s) \;=\; R_{X_0 X_0}\, e^{\eta(s-t_0)} + \kappa \int_{t_0}^{s} R_{X_0 \Xi}(\tau)\, e^{\eta(s-\tau)}\, d\tau \,. \qquad \text{(D.10a)}
$$

In Eq. (D.5) we interchange the two arguments

$$
R_{X\Xi}\left(s,t\right) \;=\; R_{X_0 \Xi}(t)\, e^{\eta(s-t_0)} + \kappa \int_{t_0}^{s} R_{\Xi\Xi}\left(u,t\right) e^{\eta(s-u)}\, du
$$

and set $t = \tau$:

$$
R_{X\Xi}\left(s,\tau\right) \;=\; R_{X_0 \Xi}(\tau)\, e^{\eta(s-t_0)} + \kappa \int_{t_0}^{s} R_{\Xi\Xi}\left(u,\tau\right) e^{\eta(s-u)}\, du \,. \qquad \text{(D.10b)}
$$

By substituting Eqs. (D.10a,b) into Eq. (D.9), we obtain

$$R_{XX}(s,t) = \left[ R_{X_0X_0}\, e^{\eta(s-t_0)} + \kappa \int_{t_0}^{s} R_{X_0\Xi}(\tau)\, e^{\eta(s-\tau)}\, d\tau \right] e^{\eta(t-t_0)} +$$

$$+ \kappa \int_{t_0}^{t} \left[ R_{X_0\Xi}(\tau)\, e^{\eta(s-t_0)} + \kappa \int_{t_0}^{s} R_{\Xi\Xi}(u,\tau)\, e^{\eta(s-u)}\, du \right] e^{\eta(t-\tau)}\, d\tau \Rightarrow$$

$$R_{XX}(s,t) = R_{X_0X_0}\, e^{\eta(t+s-2t_0)} + \kappa \int_{t_0}^{s} R_{X_0\Xi}(\tau)\, e^{\eta(t+s-\tau-t_0)}\, d\tau +$$

$$+ \kappa \int_{t_0}^{t} R_{X_0\Xi}(\tau)\, e^{\eta(t+s-\tau-t_0)}\, d\tau + \kappa^2 \int_{t_0}^{t}\int_{t_0}^{s} R_{\Xi\Xi}(u,\tau)\, e^{\eta(t+s-\tau-u)}\, du\, d\tau. \qquad \text{(D.11)}$$

## D5. IVP for the one-time autocorrelation of the response

In Eq. (D.11) we substitute $s = t$, taking thus the value of autocorrelation $R_{XX}(t,t)$ of the response in one time. This substitution is legitimate since the right-hand side of Eq. (D.11) is continuous with respect to the two arguments $s$, $t$.

$$R_{XX}(t,t) = R_{X_0X_0}\, e^{2\eta(t-t_0)} + 2\kappa\, e^{2\eta t} \int_{t_0}^{t} R_{X_0\Xi}(\tau)\, e^{-\eta(\tau+t_0)}\, d\tau +$$

$$+ \kappa^2\, e^{2\eta t} \int_{t_0}^{t}\int_{t_0}^{t} R_{\Xi\Xi}(u,\tau)\, e^{-\eta(\tau+u)}\, du\, d\tau. \qquad \text{(D.12)}$$

By differentiating Eq. (D.12) with respect to $t$, we obtain

$$\frac{d\, R_{XX}(t,t)}{dt} = 2\eta\, R_{X_0X_0}\, e^{2\eta(t-t_0)} + 2\kappa\, R_{X_0\Xi}(t)\, e^{\eta(t-t_0)} +$$

$$+ 4\eta\kappa\, e^{2\eta t} \int_{t_0}^{t} R_{X_0\Xi}(\tau)\, e^{-\eta(\tau+t_0)}\, d\tau + 2\kappa^2 \int_{t_0}^{t} R_{\Xi\Xi}(u,t)\, e^{\eta(t-u)}\, du +$$

$$+ 2\eta\kappa^2\, e^{2\eta t} \int_{t_0}^{t}\int_{t_0}^{t} R_{\Xi\Xi}(u,\tau)\, e^{-\eta(\tau+u)}\, du\, d\tau \;\Rightarrow$$

[identifying $R_{XX}(t,t)$, as given by Eq. (12), in the right-hand side]

$$\frac{1}{2}\,\frac{d\, R_{XX}(t,t)}{dt} = \eta\, R_{XX}(t,t) + \kappa\, R_{X_0\Xi}(t)\, e^{\eta(t-t_0)} + \kappa^2 \int_{t_0}^{t} R_{\Xi\Xi}(u,t)\, e^{\eta(t-u)}\, du\,. \qquad \text{(D.13)}$$

Eq. (D.13), along with the initial condition, which is determined by the data of the problem,

$$R_{XX}(t_0,t_0) = R_{X_0X_0}, \qquad \text{(D.14)}$$

constitutes the IVP for the *one-time* autocorrelation function $R_{XX}(t,t)$. Note that this IVP for $R_{XX}(t,t)$ cannot be obtained by simply substituting $s=t$ in the IVP for $R_{XX}(s,t)$, since Eqs. (D.8a), under the substitution of Eq. (D.5) for $R_{X\Xi}(s,t)$, is written as

$$\frac{\partial R_{XX}(s,t)}{\partial t} = \eta R_{XX}(s,t) + \kappa R_{X_0\Xi}(t)\, e^{\eta(s-t_0)} + \kappa^2 \int_{t_0}^{s} R_{\Xi\Xi}(u,t)\, e^{\eta(s-u)}\, du. \quad \text{(D.15)}$$

The multiplying factor $1/2$ in the left-hand side of Eq. (D.13) cannot be obtained from Eq. (D.15) for $s=t$.

## D6. Formulae for central moments

First, we define the second-order central moments of the data of the problem, i.e. the initial value and excitation

initial value variance: $\sigma^2_{X_0} = C_{X_0X_0} = R_{X_0X_0} - m^2_{X_0}$, (D.16)

two-time excitation autocovariance: $C_{\Xi\Xi}(t,s) = R_{\Xi\Xi}(t,s) - m_\Xi(t)m_\Xi(s)$, (D.17)

initial value-excitation cross-covariance: $C_{X_0\Xi}(t) = R_{X_0\Xi}(t) - m_{X_0}\, m_\Xi(t)$. (D.18)

Also, we define the second-order central moments that include the response, i.e.

initial value-response cross-covariance: $C_{X_0X}(t) = R_{X_0X}(t) - m_{X_0}\, m_X(t)$, (D.19)

two-time response excitation cross-covariance: $C_{X\Xi}(t,s) = R_{X\Xi}(t,s) - m_X(t)m_\Xi(s)$, (D.20)

two-time response autocovariance: $C_{XX}(t,s) = R_{XX}(t,s) - m_X(t)m_X(s)$, (D.21)

By substituting now, the relations: (D.3) for $m_X(t)$, (D.5) for $R_{X\Xi}(t,s)$, (D.7) for $R_{X_0X}(t)$ and (D.11) for $R_{XX}(t,s)$, into definition relations (D.19)-(D.21), and employing the relations (D.16)-(D.18), we obtain

$$C_{X\Xi}(t,s) = C_{X_0\Xi}(s)\, e^{\eta(t-t_0)} + \kappa \int_{t_0}^{t} C_{\Xi\Xi}(\tau,s)\, e^{\eta(t-\tau)}\, d\tau, \quad \text{(D.22)}$$

$$C_{X_0X}(t) = C_{X_0X_0}\, e^{\eta(t-t_0)} + \kappa \int_{t_0}^{t} C_{X_0\Xi}(\tau)\, e^{\eta(t-\tau)}\, d\tau, \quad \text{(D.23)}$$

$$\begin{aligned} C_{XX}(t,s) = {} & C_{X_0X_0}\, e^{\eta(t+s-2t_0)} + \kappa \int_{t_0}^{t} C_{X_0\Xi}(\tau)\, e^{\eta(t+s-\tau-t_0)}\, d\tau + \\ & + \kappa \int_{t_0}^{s} C_{X_0\Xi}(\tau)\, e^{\eta(t+s-\tau-t_0)}\, d\tau + \kappa^2 \int_{t_0}^{s}\int_{t_0}^{t} C_{\Xi\Xi}(u,\tau)\, e^{\eta(t+s-\tau-u)}\, du\, d\tau. \end{aligned} \quad \text{(D.24)}$$

Last, for the variance of the response, we set $s=t$, in Eq. (D.24):

$$\sigma_X^2(t) = C_{XX}\left(t,t\right) = \sigma_{X_0}^2\, e^{2\eta(t-t_0)} + \\ + 2\kappa\, e^{\eta(t-t_0)} \int_{t_0}^{t} C_{X_0\Xi}(\tau)\, e^{\eta(t-\tau)}\, d\tau + \kappa^2 \int_{t_0}^{t}\int_{t_0}^{t} C_{\Xi\Xi}\left(\tau,u\right) e^{\eta(2t-\tau-u)}\, du\, d\tau. \tag{D.25}$$

Since the integrand of the double integral is symmetric with respect to the two integration variables, $u$, $\tau$,

$$\int_{t_0}^{t}\int_{t_0}^{t} C_{\Xi\Xi}\left(\tau,u\right) e^{\eta(2t-\tau-u)}\, du\, d\tau = 2\int_{t_0}^{t}\int_{t_0}^{\tau} C_{\Xi\Xi}\left(\tau,u\right) e^{\eta(2t-\tau-u)}\, du\, d\tau.$$

Thus, Eq. (D.25) is expressed equivalently as

$$\sigma_X^2(t) = \sigma_{X_0}^2\, e^{2\eta(t-t_0)} + 2\kappa\, e^{\eta(t-t_0)} \int_{t_0}^{t} \left( C_{X_0\Xi}(\tau)\, e^{\eta(t-\tau)} + \kappa \int_{t_0}^{\tau} C_{\Xi\Xi}\left(\tau,u\right) e^{\eta(2t-\tau-u)}\, du \right) d\tau,$$

and after some algebraic manipulations

$$\sigma_X^2(t) = \sigma_{X_0}^2\, e^{2\eta(t-t_0)} + 2\kappa \int_{t_0}^{t} \left( C_{X_0\Xi}(\tau)\, e^{\eta(\tau-t_0)} + \kappa \int_{t_0}^{\tau} C_{\Xi\Xi}\left(\tau,u\right) e^{\eta(\tau-u)}\, du \right) e^{2\eta\,(t-\tau)}\, d\tau. \tag{D.26}$$

By employing now Eq. (D.22), Eq. (D.26) reads as

$$\sigma_X^2(t) = \sigma_{X_0}^2\, e^{2\eta(t-t_0)} + 2\kappa \int_{t_0}^{t} C_{X\Xi}\left(\tau,\tau\right) e^{2\eta\,(t-\tau)}\, d\tau. \tag{D.27}$$

# Appendix E: On the solvability of the response pdf evolution equation of a linear RDE[41]

In this appendix we study the solvability (uniqueness and existence of solution) of the exact pdf evolution Eq. (4.20), governing the one-time response pdf $f_{X(t)}(x)$ of a linear RDE under additive, Gaussian, coloured excitation. This issue seems to be trivial from the practical point of view, since in this case the response pdfs of all orders are known (Gaussian). Nevertheless, it is significant from the foundational point of view, since it ensures the well-posedness of a new equation, as well as its consistency with known results. In addition, after its theoretical justification, this equation provides us with a useful benchmark case for the assessment of the accuracy of the numerical methods for solving the various pdf evolution equations, see Chapter 6, which are more complicated variants of the same type, that is, advection-diffusion equations.

## E1. Statement of the problem

To facilitate the reader, we start by giving a complete yet concise statement of what has been already proved in paragraph 4.1.2:

Under the conditions that: **i)** the excitation $\Xi(t;\theta)$ is a smoothly correlated (coloured) Gaussian process, with mean value $m_{\Xi}(t)$ and autocovariance $C_{\Xi\Xi}(t,\tau)$,
**ii)** the initial value $X_0(\theta)$ is a Gaussian random variable, with mean value $m_{X_0}$ and variance $\sigma^2_{X_0} \equiv C_{X_0 X_0}$, and
**iii)** $X_0(\theta)$ and $\Xi(t;\theta)$ are jointly Gaussian with cross-covariance $C_{X_0\Xi}(t)$,

the one-time response pdf $f(x,t) = f_{X(t)}(x)$ of the linear random initial-value problem (RIVP):

$$\dot{X}(t;\theta) = \eta X(t;\theta) + \kappa \Xi(t;\theta), \tag{E.1a}$$

$$X(t_0;\theta) = X_0(\theta) \ , \tag{E.1b}$$

satisfies the following initial-value problem (IVP) (of a partial differential equation):

$$\frac{\partial f(x,t)}{\partial t} + \frac{\partial}{\partial x}\Big[\big(\eta x + \kappa m_{\Xi}(t)\big) f(x,t)\Big] - D(t)\frac{\partial^2 f(x,t)}{\partial x^2} = 0, \qquad t \ge t_0, \quad -\infty < x < +\infty, \tag{E.2a}$$

$$f(x,t_0) = f_0(x) = given, \tag{E.2b}$$

[41] The present Appendix was initially published in the electronic supplementary material of (Mamis et al., 2019).

where the coefficient $D(t) = D^{\text{eff}}(t)$ (the effective noise intensity) is given by

$$D(t) = \kappa\, e^{\eta(t-t_0)}\, C_{X_0\Xi}(t) + \kappa^2 \int_{t_0}^{t} e^{\eta_1(t-s)}\, C_{\Xi\Xi}(t,s)\, ds \,. \qquad \text{(E.2c)}$$

The IVP (E.2) will be studied under the following plausible assumptions:

**A1)** The function $x \to f(x,t)$ belongs to $C^2(\mathbb{R})$ for each $t > t_0$,
The function $t \to f(x,t)$ belongs to $C^1\big((t_0,\infty)\big)$ for each $x \in \mathbb{R}$,
The function $(x,t) \to f(x,t)$ is jointly continuous on $\mathbb{R} \times [t_0,\infty)$,

**A2)** The integrals $J(t) = \int_{-\infty}^{+\infty} f(x,t)\, dx$, $m_{X(\cdot)}(t) = \int_{-\infty}^{+\infty} x\, f(x,t)\, dx$,
$I(t) = \int_{-\infty}^{+\infty} f^2(x,t)\, dx$ and $P(t) = \int_{-\infty}^{+\infty} \left( \dfrac{\partial f(x,t)}{\partial x} \right)^2 dx$ <u>are finite</u>,

**A3)** $|x|\, f^2(x,t) \to 0$ as $|x| \to \infty$,

**A4)** $D(t) > 0$ for $t > t_0$.

A solution of the IVP (E.2) satisfying the above assumptions is called a *classical solution*.

A first, straightforward consequence from Eq. (E.2a) and the assumptions **A1)** and **A2)** is that the integral $J(t)$ is an invariant of the motion, which implies that $J(t) = J(t_0) = 1$, as it should be, since $f(x,t)$ is a probability density function. To prove this assertion, we integrate Eq. (E.2a) over the real axis and observe that the quantities appearing as end terms in the integration by pats tend to zero as $|x| \to \infty$.

## E2. Uniqueness of solution

**Theorem E1: Uniqueness of the classical solution.** The IVP (B2) has at most one solution satisfying the assumptions **A1)** – **A4)**.

The main tool for proving Theorem 1 is the following

**Lemma E1.** Under the assumptions **A1)** – **A3)**, the integrals $I(t)$ and $P(t)$ of every classical solution of Eq. (B2a) satisfy the following identity, for any $t > t_0$:

$$\frac{1}{2}\frac{d}{dt} I(t) + \frac{1}{2}\eta I(t) + D(t)\, P(t) = 0. \qquad \text{(E.3)}$$

**Proof.** Multiplying both members of Eq. (E.2a) by $f(x,t)$ and integrating over the whole $x-$axis, we obtain

$$\int_{-\infty}^{+\infty} \left( \frac{\partial f(x,t)}{\partial t} + \frac{\partial}{\partial x}\Big[\big(\eta x + \kappa\, m(t)\big)\, f(x,t)\Big] - D(t)\, \frac{\partial^2 f(x,t)}{\partial x^2} \right) f(x,t)\, dx = 0 \,. \qquad \text{(E.4a)}$$

All integrals appearing in the above equation are well defined on the basis of the stated assumptions. Eq. (E.4a) is also written in the form

$$I_1(t) + I_2(t) + I_3(t) = 0, \tag{E.4b}$$

where

$$I_1(t) = \int_{-\infty}^{+\infty} \frac{\partial f(x,t)}{\partial t} f(x,t)\,dx = \frac{1}{2}\frac{d}{dt}\int_{-\infty}^{+\infty} f^2(x,t)\,dx, \tag{E.5a}$$

$$\begin{aligned} I_2(t) &= \int_{-\infty}^{+\infty} \frac{\partial}{\partial x}\left[\left(\eta x + \kappa m(t)\right) f(x,t)\right] f(x,t)\,dx = \\ &= \eta_1 \int_{-\infty}^{+\infty} f^2(x,t)dx + \int_{-\infty}^{+\infty}\left(\eta_1 x + \kappa m(t)\right) f(x,t)\frac{\partial}{\partial x} f(x,t)\,dx = \\ &= \eta_1 \int_{-\infty}^{+\infty} f^2(x,t)dx + \frac{1}{2}\int_{-\infty}^{+\infty}\left(\eta_1 x + \kappa m(t)\right)\frac{\partial f^2(x,t)}{\partial x}dx, \end{aligned} \tag{E.5b}$$

$$I_3(t) = -D(t)\int_{-\infty}^{+\infty} \frac{\partial^2 f(x,t)}{\partial x^2} f(x,t)\,dx. \tag{E.5c}$$

Making an integration by parts in the last integral of the rightmost member of Eq. (E.5b) and the integral in the right-hand side of Eq. (E.5c), and observing that the end terms appearing after this integration by parts are zero because of the assumptions **A1) – A3)**, we find

$$\int_{-\infty}^{+\infty}\left(\eta x + \kappa m(t)\right)\frac{\partial f^2(x,t)}{\partial x}dx = -\eta\int_{-\infty}^{+\infty} f^2(x,t)\,dx, \tag{E.6a}$$

$$\int_{-\infty}^{+\infty} \frac{\partial^2 f(x,t)}{\partial x^2} f(x,t)\,dx = -\int_{-\infty}^{+\infty}\left(\frac{\partial f(x,t)}{\partial x}\right)^2 dx. \tag{E.6b}$$

Substituting the results (E.6a,b) into Eqs. (E.5b,c) we get

$$I_2(t) = \frac{1}{2}\eta\int_{-\infty}^{+\infty} f^2(x,t)\,dx \tag{E.7a}$$

and

$$I_3(t) = D(t)\int_{-\infty}^{+\infty}\left(\frac{\partial f(x,t)}{\partial x}\right)^2 dx. \tag{E.7b}$$

Using Eqs. (E.5a) and (E.7a,b), Eq. (E.4b) takes the form

$$\frac{1}{2}\frac{d}{dt}\int_{-\infty}^{+\infty} f^2(x,t)\,dx + \frac{1}{2}\eta_1\int_{-\infty}^{+\infty} f^2(x,t)\,dx + D(t)\int_{-\infty}^{+\infty}\left(\frac{\partial f(x,t)}{\partial x}\right)^2 dx = 0,$$

which is identical with Eq. (E.3). This completes the proof of the lemma. ■

We now proceed to the:

**Proof of Theorem E1.** Assume that the IVP (E.2) has two solutions, $f^{(1)}(x,t)$ and $f^{(2)}(x,t)$. Then, their difference $f^*(x,t) = f^{(1)}(x,t) - f^{(2)}(x,t)$ satisfies Eq. (E.2a) and the homogeneous initial condition $f^*(x,t_0) = 0$. The integrals $I^*(t)$ and $P^*(t)$ (that is, the integrals $I(t)$ and $P(t)$ with $f$ substituted by $f^*$) will satisfy identity (E.3), which is now written in the form

$$\frac{d}{dt} I^*(t) + \eta I^*(t) = -2D(t)\, P^*(t), \qquad t > t_0. \tag{E.8}$$

Clearly,

$$P^*(t) \ge 0, \quad I^*(t) \ge 0 \quad \text{and} \quad I^*(t_0) = 0. \tag{E.9a,b,c}$$

Considering Eq. (E.8) as a differential equation with respect to $I^*(t)$, with initial condition (E.9c), we obtain

$$I^*(t) = -2\int_{t_0}^{t} D(\tau)\, P^*(\tau)\, \exp\big(-\eta\,(t-\tau)\big)\, d\tau. \tag{E.10}$$

Since $D(t) > 0$ and $P^*(t) \ge 0$, Eq. (E.10) tells us that $I^*(t) \le 0$. The latter inequality, in conjunction with inequality (E.9b), implies that

$$I^*(t) \equiv \int_{-\infty}^{+\infty} f^{*2}(x,t)\, dx = 0.$$

Thus, $f^*(x,t) = 0$, which means that the two solutions, $f^{(1)}(x,t)$ and $f^{(2)}(x,t)$ are identical. This completes the proof of uniqueness theorem. ■

**E3. Existence of solution and consistency with known results**

We start by invoking the well-known result, that the response process of any linear system to an additive Gaussian excitation (either coloured or white) is also a Gaussian process. Thus, the response pdf $f_{X(t)}(x)$ of the linear RIVP (E.1) is given by the formula

$$f_{X(t)}(x) = \frac{1}{\sqrt{2\pi\sigma_X^2(t)}} \exp\left[-\frac{1}{2}\frac{\big(x - m_X(t)\big)^2}{\sigma_X^2(t)}\right], \tag{E.11}$$

where $m_X(t)$ is the mean value and $\sigma_X^2(t)$ is the variance of the response $X(t;\theta)$. The deterministic functions $m_X(t)$ and $\sigma_X^2(t)$ are defined as solutions of the corresponding moment equations which, in the present case, take the form (see e.g. (Sun, 2006) ch. 8, or (Athanassoulis et al., 2015)):

$$\dot{m}_X(t) = \eta\, m_X(t) + \kappa\, m_\Xi(t), \qquad m_X(t_0) = m_{X_0}, \tag{E.12a,b}$$

$$\frac{1}{2}\dot{\sigma}_X^2(t) = \eta\sigma_X^2(t) + \kappa C_{X_0\Xi}(t) e^{\eta(t-t_0)} + \kappa^2 \int_{t_0}^{t} C_{\Xi\Xi}(u,t)\, e^{\eta(t-u)}\, du, \tag{E.13a}$$

$$\sigma_X^2(t_0) = \sigma_{X_0}^2. \tag{E.13b}$$

Observe that, because of Eq. (E.2c), Eq. (E.13a) can be equivalently written as

$$\frac{1}{2}\dot{\sigma}_X^2(t) = \eta\sigma_X^2(t) + D(t). \tag{E.13a′}$$

Taking advantage of the above results, we shall now verify that the (unique) solution of the IVP (E.2) is given by Eq. (E.11). This establishes existence of solution, and consistency with existing results.

In order to verify that the function $f_{X(t)}(x)$, as defined by Eq. (E.11), satisfies the IVP (E.2), the temporal and spatial derivatives of it are needed. The calculation of these derivatives is straightforward (details are omitted). The useful results are collected below:

$$\frac{\partial}{\partial t} f_{X(t)}(x) = \frac{f_{X(t)}(x)}{\left(\sigma_X^2(t)\right)^2} \mathrm{A}\left(x\,;\, m_X(t),\, \sigma_X^2(t)\right), \tag{E.14}$$

$$\frac{\partial}{\partial x} f_{X(t)}(x) = \frac{f_{X(t)}(x)}{\left(\sigma_X^2(t)\right)^2} \mathrm{B}\left(x\,;\, m_X(t),\, \sigma_X^2(t)\right), \tag{E.15}$$

$$\frac{\partial^2}{\partial x^2} f_{X(t)}(x) = \frac{f_{X(t)}(x)}{\left(\sigma_X^2(t)\right)^2} \Gamma\left(x\,;\, m_X(t),\, \sigma_X^2(t)\right), \tag{E.16}$$

where

$$\begin{aligned} \mathrm{A} = \mathrm{A}\left(x\,;\, m_X(t),\, \sigma_X^2(t)\right) &= \frac{1}{2}\dot{\sigma}_X^2(t)\, x^2 + \left(\sigma_X^2(t)\,\dot{m}_X(t) - m_X(t)\,\dot{\sigma}_X^2(t)\right) x + \\ &+ \frac{1}{2} m_X^2(t)\,\dot{\sigma}_X^2(t) - m_X(t)\,\sigma_X^2(t)\,\dot{m}_X(t) - \frac{1}{2}\sigma_X^2(t)\,\dot{\sigma}_X^2(t). \end{aligned} \tag{E.17}$$

$$\mathrm{B} = \mathrm{B}\left(x\,;\, m_X(t),\, \sigma_X^2(t)\right) = -\sigma_X^2(t)\, x + m_X(t)\,\sigma_X^2(t). \tag{E.18}$$

$$\Gamma = \Gamma\left(x\,;\, m_X(t),\, \sigma_X^2(t)\right) = x^2 - 2\, m_X(t)\, x + m_X^2(t) - \sigma_X^2(t). \tag{E.19}$$

Using Eqs. (E.14)-(E.16), the pdf evolution (advection-diffusion) Eq. (E.2a) takes the form

$$\mathrm{A} + \eta\left(\sigma_X^2(t)\right)^2 + \left(\eta\, x + \kappa\, m_\Xi(t)\right)\mathrm{B} - D(t)\,\Gamma = 0 \quad . \tag{E.20}$$

**Lemma E2:** Eq. (E.20) is satisfied for all $x \in \Re$ and $t > t_0$ if and only if the moment equations (E.12a) and (E.13a′) are satisfied for all $t > t_0$.

**Proof.** Substituting the expressions (E.17)–(E.19), for the quantities $\mathrm{A}, \mathrm{B}, \Gamma$, into Eq. (E.20), and grouping together the coefficients of the same powers of $x$, results in

$$\begin{aligned} &\left(\frac{1}{2}\dot{\sigma}_X^2(t) - \eta\sigma_X^2(t) - D(t)\right) x^2 + \\ &+ \left[\sigma_X^2(t)\left(\dot{m}_X(t) - \kappa m_\Xi(t)\right) - m_X(t)\left(\dot{\sigma}_X^2(t) - \eta\sigma_X^2(t) - 2D(t)\right)\right] x + \\ &\quad + \left(\frac{1}{2}\dot{\sigma}_X^2(t) - D(t)\right) m_X^2(t) + \left(\kappa m_\Xi(t) - \dot{m}_X(t)\right) m_X(t)\,\sigma_X^2(t) + \\ &\qquad + \left(-\frac{1}{2}\dot{\sigma}_X^2(t) + \eta\sigma_X^2(t) + D(t)\right)\sigma_X^2(t) = 0. \end{aligned} \tag{E.21}$$

This means that Eq. (E.20) will be satisfied for all $x \in \Re$ and $t > t_0$ if and only if the (three) coefficients of $x^2$, $x$ and $x^0$ (the $x-$independent term), in Eq. (E.21), are zero for all $t > t_0$. We shall now show that the latter condition is equivalent with the satisfaction of the two moment equations (E.12a) and (E.13a′).

Assume that Eqs. (E.12a) and (E.13a′) are satisfied. Then, the coefficient of $x^2$ in Eq. (E.21) becomes zero, and the same happens with the last term of the coefficient of $x^0$. By substituting $\dot{\sigma}_X^2(t) = 2\eta\sigma_X^2(t) + 2D(t)$, from Eq. (E.13a′), in the coefficient of $x$, we obtain

$$\text{Coefficient of } x: \ \sigma_X^2(t)\left(\dot{m}_{X(\bullet)}(t) - \eta m_X(t) - \kappa m_\Xi(t)\right), \tag{E.22}$$

which vanishes because of Eq. (E.12a). Finally, making the same substitution of $\dot{\sigma}_X^2(t)$ into the remaining part of the $x-$independent term, in Eq. (E.21), we find

$$\text{Coefficient of } x^0: \left(\eta m_X(t) + \kappa m_\Xi(t) - \dot{m}_X(t)\right) m_X(t)\, \sigma_X^2(t), \tag{E.23}$$

which also vanishes because of Eq. (E.12a). This completes the proof that, if Eqs. (E.12a) and (E.13a′) are satisfied, then the coefficients of $x^2$, $x$ and $x^0$, in Eq. (E.21), are zero and thus Eq. (E.21) becomes an identity.

Assume now that the three coefficients of $x^2$, $x$ and $x^0$, in Eq. (E.21), are zero. From the coefficient of $x^2$ we obtain Eq. (E.13a′). Using the latter, we transform the coefficient of $x$ in the form of Eq. (E.22), from which we conclude that Eq. (E.12a) is also satisfied. Finally, the coefficients of $x^0$ takes again the form (E.23), being zero without providing any new information. The proof of Lemma E2 is completed. ■

**Theorem E2: Existence and consistency of solution.** The unique solution of the IVP (E.2) is given by the Gaussian pdf, Eq. (E.11).

**Proof:** According to Lemma E2, and the discussion preceding it, the function $f_{X(t)}(x)$, as defined by Eq. (E.11), with mean value $m_X(t)$ and variance $\sigma_X^2(t)$ defined through the moment equations (E.12a) and (E.13a), satisfies Eq. (E.2a). Further, the initial Gaussian pdf $f_0(x)$ is identical with $f_{X(t_0)}(x)$, since both have the same mean value $m_{X_0}$ and the same variance $\sigma_{X_0}^2$; see Eqs. (E.12b) and (E.13b). ■

# Appendix F: Solution of exact response(-excitation) pdf equations via reduction to heat equation

In the first Section of this Appendix, we consider a one-dimensional drift-diffusion equation, whose drift is linear with respect to state variable $x$ (possibly with time-varying coefficients), and the diffusion coefficient is $x-$independent and time-varying. This equation is solved by using the ansatz and transformation of variables presented in (Shtelen & Stogny, 1989), which reduces the said drift-diffusion equation to the one-dimensional heat equation. As a first example of this solution method, we consider the exact response pdf evolution Eq. (4.20) corresponding to a scalar, linear, and additively-excited RDE. In the second Section, the Shtelen & Stogny ansatz is generalized, in order to be able to solve a two-dimensional degenerate drift-diffusion equation of the form of the exact response excitation pdf Eq. (4.85).

### F1. Shtelen & Stogny solution method

Let us consider the drift-diffusion equation

$$\frac{\partial f(x,t)}{\partial t} + \frac{\partial}{\partial x}\Big[\big(a(t)x + b(t)\big)\, f(x,t)\Big] = c(t)\,\frac{\partial^2 f(x,t)}{\partial x^2}. \tag{F.1}$$

Following (Shtelen & Stogny, 1989) we introduce the ansatz

$$f(x,t) = g(t)\, w\big(y(x,t), \tau(t)\big), \qquad \text{with} \tag{F.2}$$

$$g(t) = \exp\left(-\int_{t_0}^{t} a(s)\, ds\right), \tag{F.3}$$

$$y(x,t) = \exp\left(-\int_{t_0}^{t} a(s)\, ds\right) x - \int_{t_0}^{t} b(s)\exp\left(-\int_{t_0}^{s} a(\sigma)\, d\sigma\right) ds, \tag{F.4}$$

$$\tau(t) = \int_{t_0}^{t} c(s)\exp\left(-2\int_{t_0}^{s} a(\sigma)\, d\sigma\right) ds. \tag{F.5}$$

By substituting representation (F.2) into Eq. (F.1), we obtain

$$g'(t)\, w(y,\tau) + g(t)\, \frac{\partial\big(w(y,\tau)\big)}{\partial y}\, \frac{\partial y}{\partial t} + g(t)\, \frac{\partial\big(w(y,\tau)\big)}{\partial \tau}\, \frac{\partial \tau}{\partial t} +$$

$$+\, a(t) g(t) w(y,\tau) + \big(a(t)x + b(t)\big) g(t) \frac{\partial w(y,\tau)}{\partial y}\, \frac{\partial y}{\partial x} = c(t) g(t)\, \frac{\partial}{\partial x}\left[\frac{\partial w(y,\tau)}{\partial y}\, \frac{\partial y}{\partial x}\right] \Rightarrow$$

$$\big[g'(t) + a(t)\, g(t)\big]\, w(y,\tau) + g(t)\, \frac{\partial \tau}{\partial t}\, \frac{\partial\big(w(y,\tau)\big)}{\partial \tau} +$$

$$+\, g(t)\left[\frac{\partial y}{\partial t} + \big(a(t)x + b(t)\big)\, \frac{\partial y}{\partial x} - c(t)\, \frac{\partial^2 y}{\partial x^2}\right] \frac{\partial w(y,\tau)}{\partial y} = \qquad \text{(F.6)}$$

$$= g(t)\, c(t) \left(\frac{\partial y}{\partial x}\right)^2 \frac{\partial^2 w(y,\tau)}{\partial y^2}.$$

Using now Eqs. (F.3)-(F.5) we calculate

$$g'(t) + a(t)\, g(t) = -a(t) \exp\left(-\int_{t_0}^{t} a(s)\, ds\right) + a(t) \exp\left(-\int_{t_0}^{t} a(s)\, ds\right) = 0\,, \qquad \text{(F.7)}$$

$$\frac{\partial \tau}{\partial t} = \frac{\partial}{\partial t}\int_{t_0}^{t} c(s) \exp\left(-2\int_{t_0}^{s} a(\sigma)\, d\sigma\right) ds = c(t) \exp\left(-2\int_{t_0}^{t} a(s)\, ds\right), \qquad \text{(F.8)}$$

$$\frac{\partial y}{\partial x} = \frac{\partial}{\partial x}\left[\exp\left(-\int_{t_0}^{t} a(s)\, ds\right) x - \int_{t_0}^{t} b(s) \exp\left(-\int_{t_0}^{s} a(\sigma)\, d\sigma\right) ds\right] = \exp\left(-\int_{t_0}^{t} a(s)\, ds\right), \qquad \text{(F.9)}$$

$$\frac{\partial^2 y}{\partial x^2} = 0\,. \qquad \text{(F.10)}$$

$$\frac{\partial y}{\partial t} = -a(t) \exp\left(-\int_{t_0}^{t} a(s)\, ds\right) x - b(t) \exp\left(-\int_{t_0}^{t} a(s)\, ds\right) =$$

$$= -\big(a(t)x + b(t)\big) \exp\left(-\int_{t_0}^{t} a(s)\, ds\right). \qquad \text{(F.11)}$$

Substituting Eqs. (F.7)-(F.10) into Eq. (F.6) we obtain

$$c(t) \exp\left(-2\int_{t_0}^{t} a(s)\, ds\right) \frac{\partial\big(w(y,\tau)\big)}{\partial \tau} +$$

$$+\left[\big(a(t)x + b(t) - a(t)x - b(t)\big) \exp\left(-\int_{t_0}^{t} a(s)\, ds\right)\right] \frac{\partial w(y,\tau)}{\partial y} =$$

$$= c(t)\left(\exp\left(-\int_{t_0}^{t} a(s)\, ds\right)\right)^2 \frac{\partial^2 w(y,\tau)}{\partial y^2} \Rightarrow$$

$$\frac{\partial\big(w(y,\tau)\big)}{\partial\tau} = \frac{\partial^2 w(y,\tau)}{\partial y^2}. \tag{F.12}$$

This is the one-dimensional heat equation. We supplement Eq. (F.12) with the Gaussian initial condition

$$w(x,0) = f(x,t_0) = \frac{1}{\sqrt{2\pi\sigma_0^2}} \exp\left(-\frac{(y-\mu_0)^2}{2\sigma_0^2}\right), \tag{F.13}$$

since $f(x,t_0) = g(t_0)\, w\big(y(x,t_0), \tau(t_0)\big)$, $g(t_0)=1$, $\tau(t_0)=0$, and $y(x,t_0)=x$. Solution to initial value problem (F.12), (F.13) is expressed, by employing the fundamental solution to heat equation (Evans, 2010, sec. 2.3), as

$$w(y,\tau) = \frac{1}{\sqrt{4\pi\tau}} \int_{-\infty}^{+\infty} w(z,0) \exp\left(-\frac{(z-y)^2}{4\tau}\right) dz. \tag{F.14}$$

Since now both the fundamental solution, $e^{-(z-y)^2/(4\tau)}/\sqrt{4\pi\tau}$, and $w(z,0)$ are Gaussian distributions, we may use the following lemma from (Petersen & Pedersen, 2008, sec. 8.1.8):

**Lemma F1:** Let $\mathcal{N}_x(\boldsymbol{m},\boldsymbol{\Sigma})$ be a (multidimensional in general) normal distribution of variable $\boldsymbol{x}$, mean value $\boldsymbol{m}$ and autocovariance $\boldsymbol{\Sigma}$. Then:

$$\mathcal{N}_x(\boldsymbol{m}_1,\boldsymbol{\Sigma}_1)\,\mathcal{N}_x(\boldsymbol{m}_2,\boldsymbol{\Sigma}_2) = \mathcal{N}_{\boldsymbol{m}_1}\big(\boldsymbol{m}_2, (\boldsymbol{\Sigma}_1+\boldsymbol{\Sigma}_2)\big)\,\mathcal{N}_x(\boldsymbol{m}_3,\boldsymbol{\Sigma}_3),$$

with $\boldsymbol{m}_3 = (\boldsymbol{\Sigma}_1^{-1}+\boldsymbol{\Sigma}_2^{-1})^{-1}\big(\boldsymbol{\Sigma}_1^{-1}\boldsymbol{m}_1+\boldsymbol{\Sigma}_2^{-1}\boldsymbol{m}_2\big)$ and $\boldsymbol{\Sigma}_3 = (\boldsymbol{\Sigma}_1^{-1}+\boldsymbol{\Sigma}_2^{-1})^{-1}$. ■

Since also in the right-hand side of Eq. (F.14) the said product is integrated with regard to $z$, we obtain the following result

$$w(y,\tau) = \mathcal{N}_{\mu_0}\big(y, (\sigma_0^2+2\tau)\big) = \frac{1}{\sqrt{2\pi(\sigma_0^2+2\tau)}} \exp\left(-\frac{(y-\mu_0)^2}{2(\sigma_0^2+2\tau)}\right).$$

And thus

$$f(x,t) = g(t)\, w\big(y(x,t), \tau(t)\big) = \\ = \frac{g(t)}{\sqrt{2\pi\big(\sigma_0^2+2\tau(t)\big)}} \exp\left(-\frac{\big(y(x,t)-\mu_0\big)^2}{2\big(\sigma_0^2+2\tau(t)\big)}\right). \tag{F.15}$$

Let us now consider the ***exact response pdf evolution Eq. (4.20)***:

$$\frac{\partial}{\partial t} f_{X(t)}(x) + \frac{\partial}{\partial x}\Big[\big(\eta x + \kappa m_\Xi(t)\big) f_{X(t)}(x)\Big] = D^{\mathrm{eff}}(t) \frac{\partial^2}{\partial x^2} f_{X(t)}(x). \tag{F.16}$$

In this case, we identify; $\alpha(t)=\eta$, $b(t)=\kappa m_\Xi(t)$, and $c(t)=D^{\mathrm{eff}}(t)$, and thus

$$g(t) = e^{-\eta(t-t_0)}, \tag{F.17}$$

$$y(x,t) = e^{-\eta(t-t_0)}\, x - \kappa \int_{t_0}^{t} m_\Xi(s)\, e^{-\eta(s-t_0)}\, ds \;\Rightarrow$$

$$y(x,t) = e^{-\eta(t-t_0)} \left[ x - \kappa e^{-\eta(t-t_0)} \int_{t_0}^{t} m_\Xi(s)\, e^{-\eta(s-t_0)}\, ds \right] \Rightarrow$$

$$y(x,t) = e^{-\eta(t-t_0)} \left[ x - \kappa \int_{t_0}^{t} m_\Xi(s)\, e^{\eta(t-s)}\, ds \right], \tag{F.18}$$

$$\tau(t) = \int_{t_0}^{t} D^{\text{eff}}(s)\, e^{-2\eta(s-t_0)}\, ds = e^{-2\eta(t-t_0)} \int_{t_0}^{t} D^{\text{eff}}(s)\, e^{2\eta(t-s)}\, ds. \tag{F.19}$$

From the above relations, we calculate:

$$(y-\mu_0) = e^{-\eta(t-t_0)} \left[ x - \mu_0\, e^{-\eta(t-t_0)} - \kappa \int_{t_0}^{t} m_\Xi(s)\, e^{\eta(t-s)}\, ds \right], \tag{F.20}$$

and

$$2\tau + \sigma_0^2 = 2e^{-2\eta(t-t_0)} \int_{t_0}^{t} D^{\text{eff}}(s)\, e^{2\eta(t-s)}\, ds + \sigma_0^2 =$$

$$= e^{-2\eta(t-t_0)} \left[ \sigma_0^2\, e^{2\eta(t-t_0)} + 2\int_{t_0}^{t} D^{\text{eff}}(s)\, e^{2\eta(t-s)}\, ds \right]. \tag{F.21}$$

For $\mu_0 = m_{X_0}$, $\sigma_0^2 = \sigma_{X_0}^2$, and by employing the verified relations (4.23), (4.27) for the response moments $m_X(t)$, $\sigma_X^2(t)$, Eqs. (F.20), (F.21) are expressed equivalently as

$$(y-\mu_0) = e^{-\eta(t-t_0)} \big(x - m_X(t)\big), \quad 2\tau + \sigma_0^2 = e^{-2\eta(t-t_0)}\, \sigma_X^2(t). \tag{F.22a,b}$$

Substitution of Eqs. (F.17) and (F.22a,b) into ansatz (F.2) results in

$$f_{X(t)}(x) = g(t)\, w\big(y(x,t), \tau(t)\big) = \frac{e^{-\eta(t-t_0)}}{\sqrt{2\pi\big(2\tau + \sigma_0^2\big)}} \exp\left[ -\frac{1}{2} \frac{(y-\mu_0)^2}{2\tau + \sigma_0^2} \right] =$$

$$= \frac{e^{-\eta(t-t_0)}}{\sqrt{2\pi e^{-2\eta(t-t_0)}\, \sigma_X^2(t)}} \exp\left[ -\frac{1}{2} \frac{e^{-2\eta(t-t_0)} \big(x - m_X(t)\big)^2}{e^{-2\eta(t-t_0)}\, \sigma_X^2(t)} \right] \Rightarrow$$

$$f_{X(t)}(x) = \frac{1}{\sqrt{2\pi \sigma_X^2(t)}} \exp\left[ -\frac{1}{2} \frac{\big(x - m_X(t)\big)^2}{\sigma_X^2(t)} \right]. \tag{F.23}$$

Eq. (F.23) is the expected Gaussian solution.

**F2. Generalization of solution method for two-dimensional, degenerate drift-diffusion equations**

Let us now consider the exact response-excitation pdf evolution Eq. (4.85) corresponding to a scalar, linear and additively excited RDE:

$$\frac{\partial f_{X(t)\Xi(t)}(x,u)}{\partial t}+\frac{\partial}{\partial x}\Big[\big(\eta x+\kappa u\big)\, f_{X(t)\Xi(t)}(x,u)\Big]+\dot{m}_{\Xi}(t)\,\frac{\partial f_{X(t)\Xi(t)}(x,u)}{\partial u}= \\ = G(t)\,\frac{\partial^2 f_{X(t)\Xi(t)}(x,u)}{\partial x\,\partial u}+\frac{1}{2}\dot{\sigma}^2_{\Xi}(t)\,\frac{\partial^2 f_{X(t)\Xi(t)}(x,u)}{\partial u^2}. \tag{F.24}$$

The drift-diffusion operator of Eq. (F.24) is degenerate, since the term with the second-order $x$ – derivative is missing. For the above equation, we seek a solution of the form

$$f_{X(t)\Xi(t)}(x,u) = q(t)\, w\big(y(x,u,t),\tau(t)\big)\, p(u,t), \qquad \text{with} \tag{F.25}$$

$$q(t) = e^{-\eta(t-t_0)}, \tag{F.26}$$

$$y(x,u,t) = e^{-\eta(t-t_0)}x-\frac{D(t)}{\sigma^2_{\Xi}(t)}\,e^{-\eta(t-t_0)}\,u+ \\ +D(t)\,\frac{m_{\Xi}(t)}{\sigma^2_{\Xi}(t)}e^{-\eta(t-t_0)}-D(t_0)\,\frac{m_{\Xi}(t_0)}{\sigma^2_{\Xi}(t_0)}-\kappa\int_{t_0}^{t} m_{\Xi}(s)\,e^{-\eta(s-t_0)}\,ds, \tag{F.27}$$

$$\tau(t) = -\frac{1}{2}\,\frac{D^2(t)}{\sigma^2_{\Xi}(t)}\,e^{-2\eta(t-t_0)}+\frac{1}{2}\,\frac{D^2(t_0)}{\sigma^2_{\Xi}(t_0)}+\kappa\int_{t_0}^{t} D(s)\,e^{-2\eta(s-t_0)}\,ds, \qquad \text{where} \tag{F.28}$$

$$D(t) = C_{X_0\Xi}(t_0)+\int_{t_0}^{t}\big(G(s)+\kappa\sigma^2_{\Xi}(s)\big)\,e^{\eta(t-s)}\,ds. \tag{F.29}$$

Substituting ansatz (F.25) into Eq. (F.24) we obtain:

$$\big[q'(t)+\eta q(t)\big]\,w(y,\tau)\,p(u,t)+ \\ +q(t)\,w(y,\tau)\left[\frac{\partial p(u,t)}{\partial t}+\dot{m}_{\Xi}(t)\,\frac{\partial p(u,t)}{\partial u}-\frac{1}{2}\dot{\sigma}^2_{\Xi}(t)\,\frac{\partial^2 p(u,t)}{\partial u^2}\right]+ \\ +q(t)\,p(u,t)\,\frac{\partial\tau}{\partial t}\,\frac{\partial w(y,\tau)}{\partial\tau}+ \\ +q(t)\,p(u,t)\left[\begin{array}{l}\dfrac{\partial y}{\partial t}+\left(\eta x+\kappa u-G(t)\,\dfrac{1}{p(u,t)}\dfrac{\partial p(u,t)}{\partial u}\right)\dfrac{\partial y}{\partial x}+ \\ +\left(\dot{m}_{\Xi}(t)-\dot{\sigma}^2_{\Xi}(t)\,\dfrac{1}{p(u,t)}\dfrac{\partial p(u,t)}{\partial u}\right)\dfrac{\partial y}{\partial u}- \\ -G(t)\,\dfrac{\partial^2 y}{\partial x\partial u}-\dfrac{1}{2}\dot{\sigma}^2_{\Xi}(t)\,\dfrac{\partial^2 y}{\partial u^2}\end{array}\right]\frac{\partial w(y,\tau)}{\partial y}= \\ = q(t)\,p(u,t)\left[G(t)\,\frac{\partial y}{\partial u}\,\frac{\partial y}{\partial x}\,\frac{\partial^2 w(y,\tau)}{\partial y^2}+\frac{1}{2}\dot{\sigma}^2_{\Xi}(t)\left(\frac{\partial y}{\partial u}\right)^2\right]\frac{\partial^2 w(y,\tau)}{\partial y^2}. \tag{F.30}$$

By employing Eq. (F.26), the first term of the left-hand side of Eq. (F.30) vanishes, and Eq. (F.30) is simplified into

$$w(y,\tau)\left[\frac{\partial p(u,t)}{\partial t}+\dot{m}_\Xi(t)\frac{\partial p(u,t)}{\partial u}-\frac{1}{2}\dot{\sigma}_\Xi^2(t)\frac{\partial^2 p(u,t)}{\partial u^2}\right]+p(u,t)\frac{\partial \tau}{\partial t}\frac{\partial w(y,\tau)}{\partial \tau}+$$

$$+\,p(u,t)\begin{bmatrix}\dfrac{\partial y}{\partial t}+\left(\eta x+\kappa u-G(t)\dfrac{1}{p(u,t)}\dfrac{\partial p(u,t)}{\partial u}\right)\dfrac{\partial y}{\partial x}+\\ +\left(\dot{m}_\Xi(t)-\dot{\sigma}_\Xi^2(t)\dfrac{1}{p(u,t)}\dfrac{\partial p(u,t)}{\partial u}\right)\dfrac{\partial y}{\partial u}-\\ -G(t)\dfrac{\partial^2 y}{\partial x\partial u}-\dfrac{1}{2}\dot{\sigma}_\Xi^2(t)\dfrac{\partial^2 y}{\partial u^2}\end{bmatrix}\frac{\partial w(y,\tau)}{\partial y}=$$

$$=p(u,t)\left[G(t)\frac{\partial y}{\partial u}\frac{\partial y}{\partial x}\frac{\partial^2 w(y,\tau)}{\partial y^2}+\frac{1}{2}\dot{\sigma}_\Xi^2(t)\left(\frac{\partial y}{\partial u}\right)^2\right]\frac{\partial^2 w(y,\tau)}{\partial y^2}. \tag{F.31}$$

Now, in order that the first term in the right-hand side of Eq. (F.31) to be equal to zero, the following one-dimensional drift-diffusion equation for $p(u,t)$ should be satisfied:

$$\frac{\partial p(u,t)}{\partial t}+\dot{m}_\Xi(t)\frac{\partial p(u,t)}{\partial u}=\frac{1}{2}\dot{\sigma}_\Xi^2(t)\frac{\partial^2 p(u,t)}{\partial u^2}. \tag{F.32}$$

In order to solve Eq. (F.32), its initial condition should be also determined. By considering that $f_{X(t_0)\Xi(t_0)}(x,u)=f_{X_0\Xi(t_0)}(x,u)$ is a bivariate Gaussian, it can be decomposed as

$$f_{X(t)\Xi(t)}(x,u)=\frac{1}{\sqrt{2\pi\,\sigma_\Xi^2(t_0)}}\exp\left[-\frac{1}{2}\frac{\left(u-m_\Xi(t_0)\right)^2}{\sigma_\Xi^2(t_0)}\right]\times$$

$$\times\frac{1}{\sqrt{2\pi\left(\sigma_{X_0}^2-C_{X_0\Xi}^2(t_0)/\sigma_\Xi^2(t_0)\right)}}\exp\left[-\frac{1}{2}\frac{\left(x-m_{X_0}-C_{X_0\Xi}(t_0)\left(u-m_\Xi(t_0)\right)/\sigma_\Xi^2(t_0)\right)^2}{\sigma_{X_0}^2-C_{X_0\Xi}^2(t_0)/\sigma_\Xi^2(t_0)}\right]. \tag{F.33}$$

From Eq. (F.33) we deduce that

$$p(u,t_0)=\frac{1}{\sqrt{2\pi\,\sigma_\Xi^2(t_0)}}\exp\left[-\frac{1}{2}\frac{\left(u-m_\Xi(t_0)\right)^2}{\sigma_\Xi^2(t_0)}\right], \tag{F.34}$$

and since, from Eq. (F.27),

$$y(x,u,t_0)=x-\frac{D(t_0)}{\sigma_\Xi^2(t_0)}u=x-\frac{\sigma_{X_0\Xi}(t_0)}{\sigma_\Xi^2(t_0)}u, \tag{F.35}$$

$$w(y(x,u,t_0),\tau(t_0)) = w(y_0,0) =$$

$$\frac{1}{\sqrt{2\pi\left(\sigma_{X_0}^2 - C_{X_0\Xi}^2(t_0)/\sigma_\Xi^2(t_0)\right)}} \exp\left[-\frac{1}{2}\frac{\left(y_0 - m_{X_0} + C_{X_0\Xi}(t_0)\, m_\Xi(t_0)/\sigma_\Xi^2(t_0)\right)^2}{\sigma_{X_0}^2 - C_{X_0\Xi}^2(t_0)/\sigma_\Xi^2(t_0)}\right]. \quad \text{(F.36)}$$

Eq. (F.32), supplemented with initial condition (F.34) is solved by employing the method presented in the Section F1, resulting in the Gaussian solution

$$p(u,t) = \frac{1}{\sqrt{2\pi}\;\sigma_\Xi(t)} \exp\left[-\frac{1}{2}\frac{\left(u - m_\Xi(t)\right)^2}{\sigma_\Xi^2(t)}\right]. \quad \text{(F.37)}$$

From solution (F.37) we also calculate

$$\frac{\partial p(u,t)}{\partial u} = -\frac{\left(u - m_\Xi(t)\right)}{\sigma_\Xi^2(t)}\, p(u,t) = \frac{m_\Xi(t) - u}{\sigma_\Xi^2(t)}\, p(u,t). \quad \text{(F.38)}$$

Substitution of Eqs. (F.37), (F.38) into Eq. (F.31) results in

$$\frac{\partial \tau}{\partial t}\frac{\partial w(y,\tau)}{\partial \tau} + \left[\begin{array}{l} \dfrac{\partial y}{\partial t} + \left(\eta x + \left(\kappa + \dfrac{G(t)}{\sigma_\Xi^2(t)}\right) u - G(t)\,\dfrac{m_\Xi(t)}{\sigma_\Xi^2(t)}\right)\dfrac{\partial y}{\partial x} + \\ + \left(\dfrac{\dot\sigma_\Xi^2(t)}{\sigma_\Xi^2(t)}\, u + \dot m_\Xi(t) - \dot\sigma_\Xi^2(t)\,\dfrac{m_\Xi(t)}{\sigma_\Xi^2(t)}\right)\dfrac{\partial y}{\partial u} - \\ - G(t)\,\dfrac{\partial^2 y}{\partial x \partial u} - \dfrac{1}{2}\dot\sigma_\Xi^2(t)\,\dfrac{\partial^2 y}{\partial u^2} \end{array}\right]\frac{\partial w(y,\tau)}{\partial y} =$$

$$= \left[G(t)\,\frac{\partial y}{\partial u}\frac{\partial y}{\partial x} + \frac{1}{2}\dot\sigma_\Xi^2(t)\left(\frac{\partial y}{\partial u}\right)^2\right]\frac{\partial^2 w(y,\tau)}{\partial y^2}. \quad \text{(F.39)}$$

Finally, by using the definition relations (F.27), (F.28) for $y$, $\tau$, Eq. (F.39) is simplified into the one-dimensional heat equation

$$\frac{\partial w(y,\tau)}{\partial \tau} = \frac{\partial^2 w(y,\tau)}{\partial y^2}. \quad \text{(F.40)}$$

We use now solution (F.14) for heat equation and Lemma F1:

$$w(y,\tau) = \frac{1}{\sqrt{4\pi\tau}} \int_{-\infty}^{+\infty} w(z,0) \exp\left(-\frac{(y-z)^2}{4\tau}\right) dz =$$

$$= \frac{1}{\sqrt{2\pi\left(\sigma_{X_0}^2 - C_{X_0\Xi}^2(t_0)/\sigma_\Xi^2(t_0) + 2\tau\right)}} \exp\left(-\frac{1}{2}\frac{\left(y - m_{X_0} + C_{X_0\Xi}(t_0)\, m_\Xi(t_0)/\sigma_\Xi^2(t_0)\right)^2}{\left(\sigma_{X_0}^2 - C_{X_0\Xi}^2(t_0)/\sigma_\Xi^2(t_0) + 2\tau\right)}\right), \quad \text{(F.41)}$$

where $w(z,0)$ was specified by initial condition (F.36). Finally, we return to the original variables $x$, $u$, $t$, by performing the following calculations. In them, we employ the fact that $D(t)$ is identified as the one-time, response-excitation cross-covariance $C_{X\Xi}(t,t)$ (compare definition relation (F.29) of $D(t)$ with Eq. (D.22) of Appendix D for cross-covariance) as well as formulae (D.3), (D.27) for response mean value $m_X(t)$ and variance $\sigma_X^2(t)$ respectively:

$$y - m_{X_0} + \frac{C_{X_0\Xi}(t_0)\, m_\Xi(t_0)}{\sigma_\Xi^2(t_0)} =$$

$$= e^{-\eta(t-t_0)}x - \frac{D(t)}{\sigma_\Xi^2(t)}\, e^{-\eta(t-t_0)}u + D(t)\frac{m_\Xi(t)}{\sigma_\Xi^2(t)}\, e^{-\eta(t-t_0)} - D(t_0)\frac{m_\Xi(t_0)}{\sigma_\Xi^2(t_0)} -$$

$$-\kappa\int_{t_0}^{t} m_\Xi(s)\, e^{-\eta(s-t_0)}\, ds - m_{X_0} + \frac{C_{X_0\Xi}(t_0)\, m_\Xi(t_0)}{\sigma_\Xi^2(t_0)} =$$

$$= e^{-\eta(t-t_0)}\left( x - m_X(t) - \frac{C_{X\Xi}(t,t)}{\sigma_\Xi^2(t)}\big(u - m_\Xi(t)\big)\right), \tag{F.42}$$

and

$$\sigma_{X_0}^2 - \frac{\sigma_{X_0\Xi}^2(t_0)}{\sigma_\Xi^2(t_0)} + 2\tau =$$

$$= \sigma_{X_0}^2 - \frac{\sigma_{X_0\Xi}^2(t_0)}{\sigma_\Xi^2(t_0)} - \frac{D^2(t)}{\sigma_\Xi^2(t)}\, e^{-2\eta(t-t_0)} + \frac{D^2(t_0)}{\sigma_\Xi^2(t_0)} + 2\kappa\int_{t_0}^{t} D(s)\, e^{-2\eta(s-t_0)}\, ds =$$

$$= e^{-2\eta(t-t_0)}\left[\sigma_{X_0}^2\, e^{2\eta(t-t_0)} - \frac{D^2(t)}{\sigma_\Xi^2(t)} + 2\kappa\int_{t_0}^{t} D(s)\, e^{2\eta(t-s)}\, ds\right] =$$

$$= e^{-2\eta(t-t_0)}\left(\sigma_X^2(t) - \frac{C_{X\Xi}^2(t,t)}{\sigma_\Xi^2(t)}\right). \tag{F.43}$$

Under Eqs. (F.42), (F.43), Eq. (F.41) is expressed as

$$w(y(x,u,t),\tau(t)) =$$

$$= \frac{e^{\eta(t-t_0)}}{\sqrt{2\pi\left(\sigma_X^2(t) - C_{X\Xi}^2(t,t)/\sigma_\Xi^2(t)\right)}} \exp\left[-\frac{1}{2}\,\frac{\left(x - m_X(t) - C_{X\Xi}(t,t)\left(u - m_\Xi(t)\right)/\sigma_\Xi^2(t)\right)^2}{\sigma_X^2(t) - C_{X\Xi}^2(t,t)/\sigma_\Xi^2(t)}\right]. \tag{F.44}$$

By substituting now Eq. (F.26) for $q(t)$, Eq. (F.37) for $p(u,t)$, as well as Eq. (F.44) into ansatz (F.25), we obtain, after some algebraic manipulations, the expected Gaussian solution $f_{X(t)\Xi(t)}(x,u)$ for Eq. (F.24).

# Appendix G: Calculations for paragraph 4.6.1

In paragraph 4.6.1 in Chapter 4, we have defined the following functionals:

$$\mathcal{E}[X(\bullet|_s^t;\theta)] = \exp\left(\int_s^t h'\big(X(u;\theta)\big)\,du\right), \qquad s \le t, \tag{G.1}$$

$$\mathcal{I}[X(\bullet|_s^t;\theta)] = \int_s^t h''\big(X(u;\theta)\big)\,\mathcal{E}[X(\bullet|_s^u;\theta)]\,du, \qquad s \le t, \tag{G.2}$$

$$\begin{aligned}\mathcal{G}_1[X(\bullet|_{s_*}^t;\theta)] = {} & \frac{\partial\delta(x-X(t;\theta))}{\partial X(t;\theta)}\,\mathcal{E}[X(\bullet|_{s_1}^t;\theta)]\,\mathcal{I}[X(\bullet|_{s_2}^t;\theta)] + \\ & + \frac{\partial^2\delta(x-X(t;\theta))}{\partial X^2(t;\theta)}\,\mathcal{E}[X(\bullet|_{s_1}^t;\theta)]\,\mathcal{E}[X(\bullet|_{s_2}^t;\theta)].\end{aligned} \tag{G.3}$$

Note that, for $t < s$, $\mathcal{E}[X(\bullet|_s^t;\theta)]$, $\mathcal{I}[X(\bullet|_s^t;\theta)]$ equal to zero. Also, it has been derived (Eqs. (4.128), (4.129) of paragraph 4.6.1) that

$$\begin{aligned}\frac{\delta^2\mathcal{G}_1[X[\Xi(\bullet|_{t_0}^t;\theta)]]}{\delta\Xi(\tau_1;\theta)\,\delta\Xi(\tau_2;\theta)} = {} & 2\kappa\,\delta(\tau_2-\tau_1)\int_{t_0}^t \frac{\delta\mathcal{G}_1[X(\bullet|_{s_*}^t;\theta)]}{\delta X(u;\theta)}\,\mathcal{E}[X(\bullet|_{\tau_1}^u;\theta)]\,du + \\ & + 4\kappa^2\,\Xi(\tau_1;\theta)\,\Xi(\tau_2;\theta)\,\mathcal{G}_2[X(\bullet|_{\tau_1}^t;\theta)],\end{aligned} \tag{G.4}$$

with

$$\begin{aligned}\mathcal{G}_2[X(\bullet|_{\tau_1}^t;\theta)] = {} & \int_{t_0}^t \frac{\delta\mathcal{G}_1[X(\bullet|_{s_*}^t;\theta)]}{\delta X(u;\theta)}\,\mathcal{E}[X(\bullet|_{\tau_1}^u;\theta)]\,\mathcal{I}[X(\bullet|_{\tau_2}^u;\theta)]\,du + \\ & + \int_{t_0}^t\int_{t_0}^t \frac{\delta^2\mathcal{G}_1[X(\bullet|_{s_*}^t;\theta)]}{\delta X(u_2;\theta)\,\delta X(u_1;\theta)}\,\mathcal{E}[X(\bullet|_{\tau_2}^{u_2};\theta)]\,\mathcal{E}[X(\bullet|_{\tau_1}^{u_1};\theta)]\,du_1\,du_2.\end{aligned} \tag{G.5}$$

The goal of this Appendix is to express the integral in Eq. (G4), as well as $\mathcal{G}_2[X(\bullet|_{\tau_*}^t;\theta)]$ in terms of $\mathcal{E}[X(\bullet|_s^t;\theta)]$, $\mathcal{I}[X(\bullet|_s^t;\theta)]$. First, we can easily calculate the following Volterra derivatives with respect to $X(u;\theta)$ for $u \in [s,t]$:

$$\frac{\delta\mathcal{E}[X(\bullet|_s^t;\theta)]}{\delta X(u;\theta)} = h''\big(X(u;\theta)\big)\,\mathcal{E}[X(\bullet|_s^t;\theta)], \tag{G.6}$$

$$\frac{\delta \mathcal{I}[X(\bullet|_{s}^{t};\theta)]}{\delta X(u;\theta)} = h'''\big(X(u;\theta)\big)\,\mathcal{E}[X(\bullet|_{s}^{u};\theta)] + h''\big(X(u;\theta)\big)\,\mathcal{I}[X(\bullet|_{u}^{t};\theta)]. \qquad \text{(G.7)}$$

By using its definition, Eq. (G.3), the first Volterra derivative of $\mathcal{G}_1[X(\bullet|_{s_*}^{t};\theta)]$ with respect to $X(u;\theta)$ can be easily determined using the product rule. By also grouping together the terms according to the order of random delta function derivative appearing, we have

$$\begin{aligned}
\frac{\delta \mathcal{G}_1[X(\bullet|_{s_*}^{t};\theta)]}{\delta X(u;\theta)} &= \frac{\partial^3 \delta(x-X(t;\theta))}{\partial X^3(t;\theta)}\,\delta(t-u)\,\mathcal{E}[X(\bullet|_{s_1}^{t};\theta)]\,\mathcal{E}[X(\bullet|_{s_2}^{t};\theta)] + \\
&+ \frac{\partial^2 \delta(x-X(t;\theta))}{\partial X^2(t;\theta)}\left[\frac{\delta \mathcal{E}[X(\bullet|_{s_1}^{t};\theta)]}{\delta X(u;\theta)}\,\mathcal{E}[X(\bullet|_{s_2}^{t};\theta)] + \mathcal{E}[X(\bullet|_{s_1}^{t};\theta)]\,\frac{\delta \mathcal{E}[X(\bullet|_{s_2}^{t};\theta)]}{\delta X(u;\theta)} + \right. \\
&\qquad \left. + \,\delta(t-u)\,\mathcal{E}[X(\bullet|_{s_1}^{t};\theta)]\,\mathcal{I}[X(\bullet|_{s_2}^{t};\theta)]\right] + \\
&+ \frac{\partial \delta(x-X(t;\theta))}{\partial X(t;\theta)}\left[\frac{\delta \mathcal{E}[X(\bullet|_{s_1}^{t};\theta)]}{\delta X(u;\theta)}\,\mathcal{I}[X(\bullet|_{s_2}^{t};\theta)] + \mathcal{E}[X(\bullet|_{s_1}^{t};\theta)]\,\frac{\delta \mathcal{I}[X(\bullet|_{s_2}^{t};\theta)]}{\delta X(u;\theta)}\right],
\end{aligned} \qquad \text{(G.8)}$$

By substituting Eq. (G.8) into the integral of Eq. (G.4) we obtain

$$\begin{aligned}
&\int_{t_0}^{t} \frac{\delta \mathcal{G}_1[X(\bullet|_{s_*}^{t};\theta)]}{\delta X(u;\theta)}\,\mathcal{E}[X(\bullet|_{\tau_1}^{u};\theta)]\,du = \\
&= \frac{\partial^3 \delta(x-X(t;\theta))}{\partial X^3(t;\theta)}\,\mathcal{E}[X(\bullet|_{s_1}^{t};\theta)]\,\mathcal{E}[X(\bullet|_{s_2}^{t};\theta)]\,\mathcal{E}[X(\bullet|_{\tau_1}^{t};\theta)] + \\
&+ \frac{\partial^2 \delta(x-X(t;\theta))}{\partial X^2(t;\theta)}\left[\mathcal{E}[X(\bullet|_{s_2}^{t};\theta)]\int_{t_0}^{t}\frac{\delta \mathcal{E}[X(\bullet|_{s_1}^{t};\theta)]}{\delta X(u;\theta)}\,\mathcal{E}[X(\bullet|_{\tau_1}^{u};\theta)]\,du + \right. \\
&\qquad + \mathcal{E}[X(\bullet|_{s_1}^{t};\theta)]\int_{t_0}^{t}\frac{\delta \mathcal{E}[X(\bullet|_{s_2}^{t};\theta)]}{\delta X(u;\theta)}\,\mathcal{E}[X(\bullet|_{\tau_1}^{u};\theta)]\,du + \\
&\qquad \left. + \,\mathcal{E}[X(\bullet|_{s_1}^{t};\theta)]\,\mathcal{I}[X(\bullet|_{s_2}^{t};\theta)]\,\mathcal{E}[X(\bullet|_{\tau_1}^{t};\theta)]\right] + \\
&+ \frac{\partial \delta(x-X(t;\theta))}{\partial X(t;\theta)}\left[\mathcal{I}[X(\bullet|_{s_2}^{t};\theta)]\int_{t_0}^{t}\frac{\delta \mathcal{E}[X(\bullet|_{s_1}^{t};\theta)]}{\delta X(u;\theta)}\,\mathcal{E}[X(\bullet|_{\tau_1}^{u};\theta)]\,du + \right. \\
&\qquad \left. + \,\mathcal{E}[X(\bullet|_{s_1}^{t};\theta)]\int_{t_0}^{t}\frac{\delta \mathcal{I}[X(\bullet|_{s_2}^{t};\theta)]}{\delta X(u;\theta)}\,\mathcal{E}[X(\bullet|_{\tau_1}^{u};\theta)]\,du\right].
\end{aligned} \qquad \text{(G.9)}$$

Adjusting the lower limits of the integrals in the right-hand side of Eq. (G.9), due to the causality principle that $\delta \mathcal{F}[X(\bullet|_s^t;\theta)]/\delta X(u;\theta) = 0$ for $u < s$, and substituting the derivatives by Eqs. (G.6), (G.7) results in

$$\int_{t_0}^{t} \frac{\delta \mathcal{G}_1[X(\bullet|_{s*}^{t};\theta)]}{\delta X(u;\theta)} \mathcal{E}[X(\bullet|_{\tau_1}^{u};\theta)]\, du =$$

$$= \frac{\partial^3 \delta(x-X(t;\theta))}{\partial X^3(t;\theta)} \mathcal{E}[X(\bullet|_{s_1}^{t};\theta)]\,\mathcal{E}[X(\bullet|_{s_2}^{t};\theta)]\,\mathcal{E}[X(\bullet|_{\tau_1}^{t};\theta)] +$$

$$+ \frac{\partial^2 \delta(x-X(t;\theta))}{\partial X^2(t;\theta)} \mathcal{E}[X(\bullet|_{s_1}^{t};\theta)] \Bigg[ \mathcal{I}[X(\bullet|_{s_2}^{t};\theta)]\,\mathcal{E}[X(\bullet|_{\tau_1}^{t};\theta)] +$$

$$+ \mathcal{E}[X(\bullet|_{s_2}^{t};\theta)] \left( \int_{s_1}^{t} h''\big(X(u;\theta)\big)\,\mathcal{E}[X(\bullet|_{\tau_1}^{u};\theta)]\,du + \int_{s_2}^{t} h''\big(X(u;\theta)\big)\,\mathcal{E}[X(\bullet|_{\tau_1}^{u};\theta)]\,du \right) \Bigg] +$$

$$+ \frac{\partial \delta(x-X(t;\theta))}{\partial X(t;\theta)} \mathcal{E}[X(\bullet|_{s_1}^{t};\theta)] \Bigg[ \mathcal{I}[X(\bullet|_{s_2}^{t};\theta)] \int_{s_1}^{t} h''\big(X(u;\theta)\big)\,\mathcal{E}[X(\bullet|_{\tau_1}^{u};\theta)]\,du +$$

$$+ \int_{s_2}^{t} h''\big(X(u;\theta)\big)\,\mathcal{I}[X(\bullet|_{u}^{t};\theta)]\,\mathcal{E}[X(\bullet|_{\tau_1}^{u};\theta)]\,du + \int_{s_2}^{t} h'''\big(X(u;\theta)\big)\,\mathcal{E}[X(\bullet|_{s_2}^{u};\theta)]\,\mathcal{E}[X(\bullet|_{\tau_1}^{u};\theta)]\,du \Bigg].$$

(G.10)

Last, by noting that $\mathcal{E}[X(\bullet|_s^t;\theta)] = 0$ for $t < s$, we have another adjustment of the lower limit of integrals in the right-hand side of Eq. (G.10):

$$\int_{t_0}^{t} \frac{\delta \mathcal{G}_1[X(\bullet|_{s*}^{t};\theta)]}{\delta X(u;\theta)} \mathcal{E}[X(\bullet|_{\tau_1}^{u};\theta)]\, du =$$

$$= \frac{\partial^3 \delta(x-X(t;\theta))}{\partial X^3(t;\theta)} \mathcal{E}[X(\bullet|_{s_1}^{t};\theta)]\,\mathcal{E}[X(\bullet|_{s_2}^{t};\theta)]\,\mathcal{E}[X(\bullet|_{\tau_1}^{t};\theta)] +$$

$$+ \frac{\partial^2 \delta(x-X(t;\theta))}{\partial X^2(t;\theta)} \mathcal{E}[X(\bullet|_{s_1}^{t};\theta)] \Bigg[ \mathcal{I}[X(\bullet|_{s_2}^{t};\theta)]\,\mathcal{E}[X(\bullet|_{\tau_1}^{t};\theta)] + \mathcal{E}[X(\bullet|_{s_2}^{t};\theta)] \times$$

$$\times \left( \int_{\max\{s_1,\tau_1\}}^{t} h''\big(X(u;\theta)\big)\,\mathcal{E}[X(\bullet|_{\tau_1}^{u};\theta)]\,du + \int_{\max\{s_2,\tau_1\}}^{t} h''\big(X(u;\theta)\big)\,\mathcal{E}[X(\bullet|_{\tau_1}^{u};\theta)]\,du \right) \Bigg] +$$

$$+ \frac{\partial \delta(x-X(t;\theta))}{\partial X(t;\theta)} \mathcal{E}[X(\bullet|_{s_1}^{t};\theta)] \Bigg[ \mathcal{I}[X(\bullet|_{s_2}^{t};\theta)] \int_{\max\{s_1,\tau_1\}}^{t} h''\big(X(u;\theta)\big)\,\mathcal{E}[X(\bullet|_{\tau_1}^{u};\theta)]\,du +$$

$$+ \int_{\max\{s_2,\tau_1\}}^{t} \mathcal{E}[X(\bullet|_{\tau_1}^{u};\theta)] \Big( h''\big(X(u;\theta)\big)\,\mathcal{I}[X(\bullet|_{u}^{t};\theta)] + h'''\big(X(u;\theta)\big)\,\mathcal{E}[X(\bullet|_{s_2}^{u};\theta)] \Big)\,du \Bigg].$$

(G.11)

Substitution of Eq. (G.11) into SLE (4.131) of paragraph 4.6.1:

$$\frac{\partial f_{X(t)}(x)}{\partial t} + \frac{\partial}{\partial x}\Big[\big(h(x) + \kappa\, C_{\Xi\Xi}(t,t)\big)\, f_{X(t)}(x)\Big] =$$

$$+\,2\kappa^2 \frac{\partial^2}{\partial x^2}\int_{t_0}^{t} C_{\Xi\Xi}^2(t,s)\, \mathbb{E}^\theta\Big[\delta(x - X(t;\theta))\, \mathcal{E}[X(\bullet|_s^t;\theta)]\Big]\, ds\, -$$

$$-\,4\kappa^3 \frac{\partial}{\partial x}\int_{t_0}^{t}\int_{t_0}^{t} C_{\Xi\Xi}(t,s_1)\, C_{\Xi\Xi}(s_1,s_2)\, C_{\Xi\Xi}(s_2,t)\, \mathbb{E}^\theta\Big[\mathcal{G}_1[X(\bullet|_{s_*}^t;\theta)]\Big]\, ds_1 ds_2\, -$$

$$-\,8\kappa^4 \frac{\partial}{\partial x}\int_{t_0}^{t}\int_{t_0}^{t}\int_{t_0}^{t} C_{\Xi\Xi}(t,s_1)\, C_{\Xi\Xi}(s_1,\tau)\, C_{\Xi\Xi}(\tau,s_2)\, C_{\Xi\Xi}(s_2,t)\, \times$$

$$\times\, \mathbb{E}^\theta\left[\int_{t_0}^{t} \frac{\delta \mathcal{G}_1[X(\bullet|_{s_*}^t;\theta)]}{\delta X(u;\theta)}\, \mathcal{E}[X(\bullet|_\tau^u;\theta)]\, du\right] d\tau\, ds_1 ds_2\, -$$

$$-\,16\kappa^5 \frac{\partial}{\partial x}\int_{t_0}^{t}\int_{t_0}^{t}\int_{t_0}^{t}\int_{t_0}^{t} C_{\Xi\Xi}(t,s_1)\, C_{\Xi\Xi}(s_1,\tau_1)\, C_{\Xi\Xi}(\tau_2,s_2)\, C_{\Xi\Xi}(s_2,t)\, \times$$

$$\times\, \mathbb{E}^\theta\Big[\Xi(\tau_1;\theta)\, \Xi(\tau_2;\theta)\, \mathcal{G}_2[X(\bullet|_{\tau_*}^t;\theta)]\Big]\, d\tau_1 d\tau_2 ds_1 ds_2\,, \qquad \text{(G.12)}$$

and omission of the term containing $\mathcal{G}_2[X(\bullet|_{\tau_*}^t;\theta)]$ results in the approximate SLE (4.132)

$$\frac{\partial f_{X(t)}(x)}{\partial t} + \frac{\partial}{\partial x}\Big[\big(h(x) + \kappa\, C_{\Xi\Xi}(t,t)\big)\, f_{X(t)}(x)\Big] =$$
$$= \frac{\partial^2}{\partial x^2}\, \mathbb{E}^\theta\Big[\delta(x - X(t;\theta))\big(2\kappa^2 \mathcal{N}_{2,1} + 4\kappa^3 \mathcal{N}_{2,2} + 8\kappa^4 \mathcal{N}_{2,3}\big)\Big] -$$
$$-\,\frac{\partial^3}{\partial x^3}\, \mathbb{E}^\theta\Big[\delta(x - X(t;\theta))\big(4\kappa^3 \mathcal{N}_{3,1} + 8\kappa^4 \mathcal{N}_{3,2}\big)\Big] + \qquad \text{(G.13)}$$
$$+\,\frac{\partial^4}{\partial x^4}\, \mathbb{E}^\theta\Big[\delta(x - X(t;\theta))\big(8\kappa^4 \mathcal{N}_{4,1}\big)\Big],$$

with

$$\mathcal{N}_{2,1}[X(\bullet|_{t_0}^t;\theta)] = \int_{t_0}^{t} C_{\Xi\Xi}^2(t,s)\, \mathcal{E}[X(\bullet|_s^t;\theta)]\, ds\,, \qquad \text{(G.14)}$$

$$\mathcal{N}_{2,2}[X(\bullet|_{t_0}^t;\theta)] = \int_{t_0}^{t}\int_{t_0}^{t} C_{\Xi\Xi}(t,s_1)\, C_{\Xi\Xi}(s_1,s_2)\, C_{\Xi\Xi}(s_2,t)\, \mathcal{E}[X(\bullet|_{s_1}^t;\theta)]\, \mathcal{I}[X(\bullet|_{s_2}^t;\theta)]\, ds_1 ds_2$$
$$\text{(G.15)}$$

$$\mathcal{N}_{3,1}[X(\bullet|_{t_0}^t;\theta)] = \int_{t_0}^{t}\int_{t_0}^{t} C_{\Xi\Xi}(t,s_1)\, C_{\Xi\Xi}(s_1,s_2)\, C_{\Xi\Xi}(s_2,t)\, \mathcal{E}[X(\bullet|_{s_1}^t;\theta)]\, \mathcal{E}[X(\bullet|_{s_2}^t;\theta)]\, ds_1 ds_2$$
$$\text{(G.16)}$$

$$\mathcal{N}_{2,3}[X(\bullet|_{t_0}^{t};\theta)] = \int_{t_0}^{t}\int_{t_0}^{t}\int_{t_0}^{t} C_{\Xi\Xi}(t,s_1)\, C_{\Xi\Xi}(s_1,\tau)\, C_{\Xi\Xi}(\tau,s_2)\, C_{\Xi\Xi}(s_2,t)\, \mathcal{E}[X(\bullet|_{s_1}^{t};\theta)]\times$$
$$\times\Bigg[\mathcal{I}[X(\bullet|_{s_2}^{t};\theta)]\int_{\max\{s_1,\tau\}}^{t} h''\big(X(u;\theta)\big)\,\mathcal{E}[X(\bullet|_{\tau}^{u};\theta)]\,du +$$
$$+\int_{\max\{s_2,\tau\}}^{t} \mathcal{E}[X(\bullet|_{\tau}^{u};\theta)]\Big(h''\big(X(u;\theta)\big)\,\mathcal{I}[X(\bullet|_{u}^{t};\theta)] + h'''\big(X(u;\theta)\big)\,\mathcal{E}[X(\bullet|_{s_2}^{u};\theta)]\Big)\,du\Bigg]\,d\tau\,ds_1 ds_2 \tag{G.17}$$

$$\mathcal{N}_{3,2}[X(\bullet|_{t_0}^{t};\theta)] = \int_{t_0}^{t}\int_{t_0}^{t}\int_{t_0}^{t} C_{\Xi\Xi}(t,s_1)\, C_{\Xi\Xi}(s_1,\tau)\, C_{\Xi\Xi}(\tau,s_2)\, C_{\Xi\Xi}(s_2,t)\, \mathcal{E}[X(\bullet|_{s_1}^{t};\theta)]\times$$
$$\times\Bigg[\mathcal{I}[X(\bullet|_{s_2}^{t};\theta)]\,\mathcal{E}[X(\bullet|_{\tau}^{t};\theta)] + \mathcal{E}[X(\bullet|_{s_2}^{t};\theta)]\times$$
$$\times\left(\int_{\max\{s_1,\tau\}}^{t} h''\big(X(u;\theta)\big)\,\mathcal{E}[X(\bullet|_{\tau}^{u};\theta)]\,du + \int_{\max\{s_2,\tau\}}^{t} h''\big(X(u;\theta)\big)\,\mathcal{E}[X(\bullet|_{\tau}^{u};\theta)]\,du\right)\Bigg]\,d\tau\,ds_1 ds_2 \tag{G.18}$$

$$\mathcal{N}_{4,1}[X(\bullet|_{t_0}^{t};\theta)] = \int_{t_0}^{t}\int_{t_0}^{t}\int_{t_0}^{t} C_{\Xi\Xi}(t,s_1)\, C_{\Xi\Xi}(s_1,\tau)\, C_{\Xi\Xi}(\tau,s_2)\, C_{\Xi\Xi}(s_2,t)\,\times$$
$$\times\,\mathcal{E}[X(\bullet|_{s_1}^{t};\theta)]\,\mathcal{E}[X(\bullet|_{s_2}^{t};\theta)]\,\mathcal{E}[X(\bullet|_{\tau}^{t};\theta)]\,d\tau\,ds_1 ds_2\,. \tag{G.19}$$